\documentclass[]{aa}
\usepackage{graphicx}
\usepackage{natbib}
\usepackage{amssymb}

\begin{document}

\title{The evolution of the luminosity functions in the FORS Deep
  Field from low to high redshift: I. The blue bands \thanks{Based on
    observations collected with the VLT on Cerro Paranal (Chile) and
    the NTT on La Silla (Chile) operated by the European Southern
    Observatory in the course of the observing proposals 63.O-0005,
    64.O-0149, 64.O-0158, 64.O-0229, 64.P-0150, 65.O-0048, 65.O-0049,
    66.A-0547, 68.A-0013, and 69.A-0014.}
}

\author{
Armin Gabasch\inst{1,2}
\and
Ralf Bender\inst{1,2}
\and
Stella Seitz\inst{1}
\and
Ulrich Hopp\inst{1,2}
\and
Roberto~P. Saglia\inst{1,2}
\and
Georg Feulner\inst{1}
\and
Jan Snigula\inst{1}
\and
Niv Drory\inst{3}
\and
Immo Appenzeller\inst{4}
\and
Jochen Heidt\inst{4} 
\and
D\"orte Mehlert\inst{4}
\and
Stefan Noll\inst{4}
\and
Asmus B\"ohm\inst{5}
\and
Klaus J\"ager\inst{5}
\and
Bodo Ziegler\inst{5}
\and
Klaus J. Fricke\inst{5}
}

\institute{
Universit\"ats-Sternwarte M\"unchen, Scheinerstr. 1, D$-$81679 M\"unchen,
Germany 
\and
Max-Planck-Institut f\"ur Extraterrestrische Physik,
Giessenbachstra\ss e, D-85748 Garching b. M\"unchen, Germany
\and
McDonald Observatory, University of Texas at Austin, Austin, Texas 78712
\and 
Landessternwarte Heidelberg, K\"onigstuhl,
D$-$69117 Heidelberg, Germany
\and 
Universit\"ats-Sternwarte G\"ottingen, Geismarlandstr. 11, D$-$37083
G\"ottingen, Germany
}

\authorrunning{Gabasch et al.}
\titlerunning{The evolution of the luminosity functions in the FDF: I. The blue bands}

\offprints{A.~Gabasch}
\mail{gabasch@usm.uni-muenchen.de}
\date{Received --; accepted --}

\abstract{ We use the very deep and homogeneous I-band selected
  dataset of the FORS Deep Field (FDF) to trace the evolution of the
  luminosity function over the redshift range \mbox{$0.5 < z < 5.0$}.
   We show that the FDF I-band selection down to $I_{AB}=26.8$
    misses of the order of 10~\% of the galaxies that would be
    detected in a K-band selected survey with magnitude limit
    $K_{AB}=26.3$ (like FIRES). 
  Photometric redshifts for 5558 galaxies are estimated based on the
  photometry in 9 filters (U, B, Gunn g, R, I, SDSS z, J, K and a
  special filter centered at 834~nm).  A comparison with 362
  spectroscopic redshifts shows that the achieved accuracy of the
  photometric redshifts is \mbox{$\Delta z / (z_{spec}+1) \le 0.03$}
  with only $\sim 1$\% outliers.
   This allows us to derive luminosity functions with a
    reliability similar to spectroscopic surveys.  
  In addition, the luminosity functions can be traced to objects of
  lower luminosity which generally are not accessible to spectroscopy.
  We investigate the evolution of the luminosity functions evaluated
  in the restframe UV (1500~\AA\ and 2800~\AA), u', B, and g' bands.
  Comparison with results from the literature shows the reliability of
  the derived luminosity functions.
   Out to redshifts of $z\sim 2.5$ the data are consistent with a
    slope of the luminosity function approximately constant with
    redshift, at a value of $-1.07 \pm 0.04$ in the UV (1500~\AA\ ,
    2800~\AA) as well as u', and $-1.25 \pm 0.03$ in the blue (g', B).
    We do not see evidence for a very steep slope (\mbox{$\alpha \le
      -1.6$}) in the UV at \mbox{$\langle z \rangle\sim 3.0$} and
    \mbox{$\langle z \rangle\sim 4.0$} favoured by other authors.
    There  may be a tendency for the faint-end slope to become
    shallower with increasing redshift but the effect is marginal.  
  We find a brightening of M$^\ast$ and a decrease of $\phi^\ast$ with
  redshift for all analyzed wavelengths. The effect is systematic and
  much stronger than what can be expected to be caused by cosmic
  variance seen in the FDF. The evolution of M$^\ast$ and $\phi^\ast$
  from $z=0$ to $z=5$ is well described by the simple approximations
  \mbox{$M^\ast(z)= M^\ast_0 + {a} \ln(1+z)$} and \mbox{$\phi^\ast(z)=
    \phi^\ast_0 (1+z)^{b}$} for $M^\ast$ and $\phi^\ast$.  The
  evolution is very pronounced at shorter wavelengths
  (\mbox{$a=-2.19$, and $b=-1.76$ for 1500~\AA\ rest wavelength}) and
  decreases systematically with increasing wavelength, but is also
  clearly visible at the longest wavelength investigated here
  (\mbox{$a=-1.08$, and $b=-1.29$ for g'}).  Finally we show a
  comparison with semi-analytical galaxy formation models.
 \keywords{Galaxies:
    luminosity function -- Galaxies: fundamental parameters --
    Galaxies: high-redshift -- Galaxies: distances and redshifts --
    Galaxies: evolution} }

\maketitle
\section{Introduction}
\label{sec:intro}


Observational constraints on galaxy formation have improved
significantly over the last years and it has become possible to study
the evolution of global galaxy properties up to very high redshifts.
A crucial step to probe the properties of galaxies up to the highest
redshifts was the work of \citet{steidel_lbg:1} and
\citet{steidel_lbg:2} who used color selection to discriminate
between low redshift and high redshift galaxies.  Although the
Lyman-break technique is very efficient in selecting high redshift
galaxies (see \citealt{blaizot} for a detailed discussion) with a
minimum of photometric data, it has the disadvantage that it does not
sample galaxies homogeneously in redshift space and may select against
certain types of objects.  With the advent of deep multi-band
photometric surveys (Hubble Deep Field North (HDFN; \citealt{HDF96}),
NTT SUSI deep Field (NDF; \citealt{arnouts_ntt}), Hubble Deep Field
South (HDFS; \citealt{HDFS00,HDFS00a}), Chandra Deep Field South
(CDFS; \citealt{arnouts_cdfs}), William Herschel Deep Field (WHDF;
\citealt{mccracken:1,metcalf:1}), Subaru Deep Field/Survey (SDF;
\citealt{maihara:1,ouchi:2}), 
 
The Great Observatories Origins Deep Survey (GOODS;
\citealt{giavalisco:1}))
the photometric redshift technique
(essentially a generalization of the drop-out technique) has
increasingly been used to identify high-redshift galaxies. Several
methods have been described in the literature to derive photometric
redshifts \citep{baum:1, koo:1, brunner:1, Soto:1, benitez:1,
  borgne:1, firth:1}.

Based on either spectroscopic redshifts, drop-out techniques, or
photometric redshifts, it has been possible to derive luminosity
functions at different redshifts in the ultraviolet (UV)
\citep{treyer:98,steidel:1,cowie:1, adelberger:1, Cohen:1, sullivan:1,
  ouchi:1, poli:1, wilson:1, combo17:1, rowan:1, kashikawa:1,
  ouchi:2, iwata:1} and in the blue bands \citep{lilly:2,heyl:1,
  lin:1, sawicki:1, small:1, zucca:1, loveday:1, marinoni:1, fried:1,
  cross:1, im:1, marinoni:2, norberg:1, bell:1, lapparent:1, liske,
  poli:3, prezgonz:1}.  Within the uncertainties given by IMF and dust
content, the flux in the UV allows to trace the star formation rate
(SFR; \citealt{mad_poz_dick1}) in the galaxies, while the optical
luminosities provide constraints on more evolved stellar populations
\citep{franx:1}.

Locally, the 2dF Galaxy Redshift Survey (2dFGRS; \citealt{colless:1}),
the Sloan Digital Sky Survey (SDSS; \citealt{stoughton:1}) and the
2MASS survey \citep{2MASS} have provided superb reference points for
galaxy luminosity functions over a large wavelength range (see
\citealt{norberg:1} for 2dFGRS, \citealt{blanton:1, blanton:2} for the
SDSS, and \citealt{Kochaneketal01, Coleetal2001} for  2MASS).

In parallel to the observational effort, theoretical models have been
developed within the framework of the cold dark matter cosmology.
Most notably, semi-analytic models (SAMs) \citep{kauffmann:1, cole:1,
  sommerville:1, kauffmann:2, poli:2, wu:1, cole:2, menci:1, menci:2}
and simulations based on smoothed-particle hydrodynamics (SPH)
\citep{dave:1, weinberg:1, nagamine:1, nagamine:2} have made testable
predictions.  Starting with the mass function of dark matter halos and
their merging history, SAMs use simplified recipes to describe the
baryonic physics (gas cooling, photoionization, star formation,
feedback processes, etc., see \citealt{benson:1}) to derive stellar mass
and luminosity functions.

Ideally, a comparison between observations and models should be done
with deep multiwavelengths datasets that also cover a large area.  The
dataset has to be sufficiently deep in order to be able to derive the
faint-end slope of the luminosity function. On the other hand, one
also needs as large an area as possible to overcome cosmic variance
and to quantify the density of rare bright galaxies, which define the 
cut-off of the luminosity function.

The FORS Deep Field \citep{fdf_data} has a depth close to the HDFs but
an area of \mbox{8 -- 10} times the area of the HDFN.  This depth
allows us to detect galaxies at $z > 2$ which would be missed by
Lyman-break studies which usually reach only $R_{AB} < 25.5$ 
(see also \citealt{franx:1} and \citealt{dokkum:1}).

Very reliable photometric redshifts are crucial for the analysis of
the evolution of the luminosity functions in the FDF. Photometric
redshifts have been determined with a template matching algorithm
described in \citet{bender:1} that applies Bayesian statistics and
uses semi-empirical template spectra matched to broad band photometry.
We achieved an accuracy of \mbox{$\Delta z / (z_{spec}+1) \le 0.03$}
with only $\sim 1$\% extreme outliers (numbers based on a comparison
with 362 spectroscopic redshifts).  Redshifts of galaxies that are
several magnitudes fainter than typical spectroscopic limits could be
determined reliably and thus allowed better constraints on the
faint-end slope of the luminosity functions.

In this paper we present the redshift evolution of the luminosity
function evaluated in the restframe UV-range (1500~\AA, 2800~\AA), u'
(SDSS), B, and g' (SDSS) bands in the redshift range \mbox{$0.5 < z <
  5.0$}.  Luminosity functions at longer wavelengths as well as the
evolution of the luminosity density and the star formation rate will
be presented in future papers (Gabasch et al., in preparation).
We provide a short description of the FDF in Sect.~\ref{sec:fdf} where
we also present the selection criteria of our galaxies. 
In Sect.~\ref{sec:lumfkt:selection} we investigate possible selection
effects due to our purely I-band selected catalogue.
In Sect.~\ref{sec:photred} we discuss the accuracy of the photometric
redshifts as well as the redshift distribution of the selected
galaxies.  In Sect.~\ref{sec:lumfkt} and in the Appendix we show
luminosity functions at different wavelengths and redshifts. In
Sect.~\ref{sec:evol_parameter}, a parameterization of the redshift
evolution of the Schechter \citep{schechter:1} parameters M$^\ast$ and
$\phi^\ast$ is given.  We compare our results with previous
observational results in Sect.~\ref{sec:lit}, and with model
predictions in Sect.~\ref{sec:model}, before we summarize this work in
Sect.~\ref{sec:summary_conclusion}.

We use AB magnitudes and adopt a $\Lambda$ cosmology throughout the paper
with \mbox{$\Omega_M=0.3$}, \mbox{$\Omega_\Lambda=0.7$}, and 
\mbox{$H_0=70 \, \mathrm{km} \, \mathrm{s}^{-1} \, \mathrm{Mpc}^{-1}$}.

\section{The FORS deep field}
\label{sec:fdf}
The FORS Deep Field \citep{appenzeller:1} is a multi-color 
photometric and spectroscopic survey of a $7' \times 7'$  region
near the south galactic pole including the QSO Q 0103-260 at redshift
$z = 3.36$. The data have been taken with FORS1 and FORS2 
at the ESO VLT and SofI at the NTT.

The data in the U, B, g, R, I, J and Ks filters were reduced and
calibrated (including the correction for galactic extinction) as
described in \citet{fdf_data}.  The reduction of the images in the 
z-band and the special filter centered at 834~nm follows the same
recipe, except for additional de-fringing in the z-band.

The images were stacked with weights to get optimal signal to noise
for point-like faint objects.  The formal 50\% completeness limits for
point sources are 26.5, 27.6, 26.9, 26.9, 26.8, \mbox{$\sim 25.5$},
\mbox{$\sim 25.8$}, 23.8, 22.6 in U, B, g, R, I, 834~nm, z, J and Ks,
respectively.  The seeing varied from 0.5 arcsec in the I and z band
to 1.0 arcsec in the U-band. Because the depth of the images decreases
towards the borders, we limited our analysis to the inner 39.81
arcmin$^2$ of our field. The signal-to-noise ratio (S/N) in this
`deep' region is at least 90~\% of the best S/N in every filter. This
prevents a possible bias of the photometric redshifts (see
Sect.~\ref{sec:photred}) due to a not completely homogeneous dataset.

Object detection was done in the I-band image using SExtractor
\citep{bertin}, and the catalogue for this `deep' part of the FDF
includes 5636 objects.  To avoid contamination from stars, we rely on
three sources of information: The star-galaxy classifier of the
detection software SExtractor, the goodness of fit for galaxy objects
of the photometric redshift code and, if available, on the
spectroscopic information.  We first exclude all bright ($I < 22^m$)
starlike objects (SExtractor star galaxy classifier $>0.95$).  Then we
exclude all objects whose best fitting stellar spectral energy
distribution (SED) -- according to the photometric redshift code --
gives a better match to the flux in the different wavebands than any
galaxy template ($2\ \chi_{star}^2 < \chi_{galaxy}^2$). These objects
are subsequently flagged as star and removed from our catalogue.
Further inspection of the images confirms, that none of these flagged
objects are extended. 
Finally, we reject all objects spectroscopically classified as stars.
We checked the influence of misidentified or missed stars on the
luminosity functions.  If stars are fitted by galaxy templates their
redshifts are mostly very small ($z<0.15$, especially if they are G
and K stars) and, therefore, did not enter the analysis.  M stars
interpreted as galaxies tend to be distributed more evenly in redshift
space but they do not contribute significantly to the number density in
any redshift interval. Even if all stars were included as galaxies in
the sample, they would not affect the derived luminosity functions at a
noticeable level.

In total 78 objects were classified as stars and removed from our
sample. Our final I-band selected catalogue comprises therefore 5558
objects.

\begin{figure}[b!]
\includegraphics[width=0.5\textwidth]{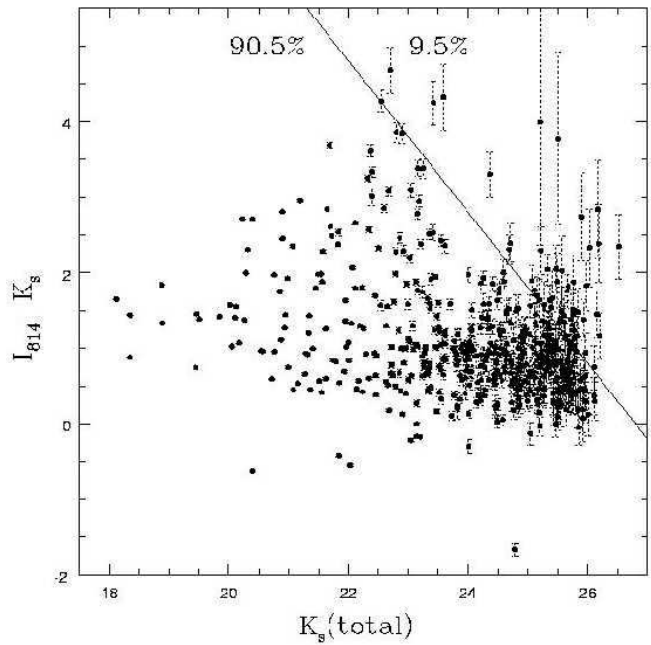}
\caption{\label{fig:selection_fires}
  \mbox{$I_{814}-K_s$} versus $K_s$ color-magnitude relation for
  $K_s$-selected objects of the HDF-S as given by \citet{labbe:1}.
  Following \citet{labbe:1} only sources with a minimum of 20\% of
  the total exposure time in all bands are included and shown as
  filled symbols. Colors are plotted with $1 \sigma$ error bars. The
  solid line corresponds to the limiting magnitude of the FDF
  (I=26.8).  Only the objects on the right of the solid line are
  beyond our I-band limit.
}
\end{figure}

\section{I selection versus K selection}
\label{sec:lumfkt:selection}

We use the ultradeep near-infrared ISAAC observations of the Hubble
Deep Field South \citep{labbe:1} for a more quantitative analysis of
possible selection effects between K and I band selected samples.

In Fig.~\ref{fig:selection_fires} we show the \mbox{$I_{814}-K_s$}
versus $K_s$ color-magnitude relation for $K_s$-selected objects of
the HDF-S as given by \citet{labbe:1} (data were taken from:
http://www.strw.leidenuniv.nl/$\sim$fires/).  Following
\citet{labbe:1}, only sources with a minimum of 20\% of the total
exposure time in all bands are included and shown as filled symbols.
Colors are plotted with $1 \sigma$ error bars. The solid line
corresponds to the 50~\% completeness limiting magnitude of the FDF in the I-band
($I\sim 26.8$).  The figure clearly shows, that although we selected in I,
we miss  only about 10~\% of the objects that would have been
detected in deep K-band images (with a  50~\% completeness limiting
magnitude of $K_{AB} \sim 26.3$).
All objects on the left of the solid line would have been detected in
the I-selected FDF catalogue as well.  Therefore we conclude that only
a small fraction ($ \sim 10$~\%) of galaxies is missed in deep
I-band selected samples relative to deep K-band selected samples,
provided the I-band images are about 0.5 AB-magnitudes deeper than the
K-band images. 
Of course, this holds only for galaxies at redshift
below 6. At higher redshifts no signal is detectable in the 
I-band, due to the Lyman break and intervening intergalactic
absorption. 

Another indication that we are unlikely to miss a
large population of high redshift red galaxies comes from
Fig.~\ref{mag_z_phot_dis} (left panel). 
Out to
redshifts of about 1.5, red galaxies define the bright end of the
luminosity function. Beyond $z\sim 1.5$ bluer star-forming galaxies
take over. Red galaxies could still be detected at $z > 1.5$ but seem to be
largely absent. In any case, even if we
missed a few objects, the evolution of
luminosity functions that we discuss below will not be affected.

As a side remark we note that also a B-band selected FDF catalogue
delivers similar conclusions on the evolution of the luminosity
functions out to redshift $\sim 3$. Again, above this redshift no
signal is detectable in the B-band due to the Lyman break and
intervening intergalactic absorption.

\section{Photometric redshifts}
\label{sec:photred}

\begin{figure*}[ht]\centering
\includegraphics[width=0.9\textwidth]{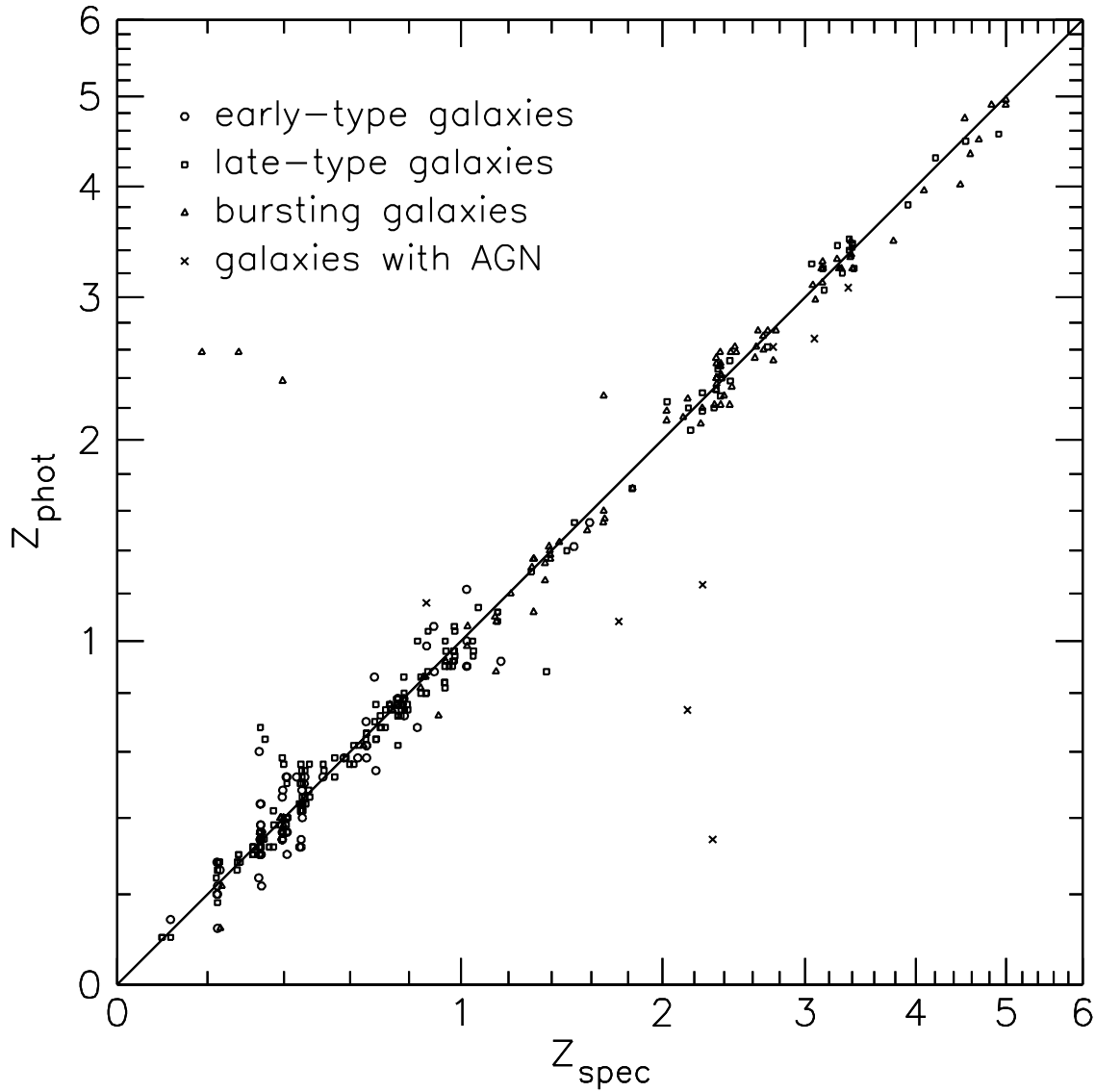}\hfill
\caption{\label{z_spec_z_phot_fdf} 
  Comparison of spectroscopic \citep{noll:1, boehm:1} and
  photometric redshifts for different galaxy types and quasars in the
  FDF (362 objects).}
\end{figure*}
\begin{figure*}[ht]
\includegraphics[width=0.5\textwidth]{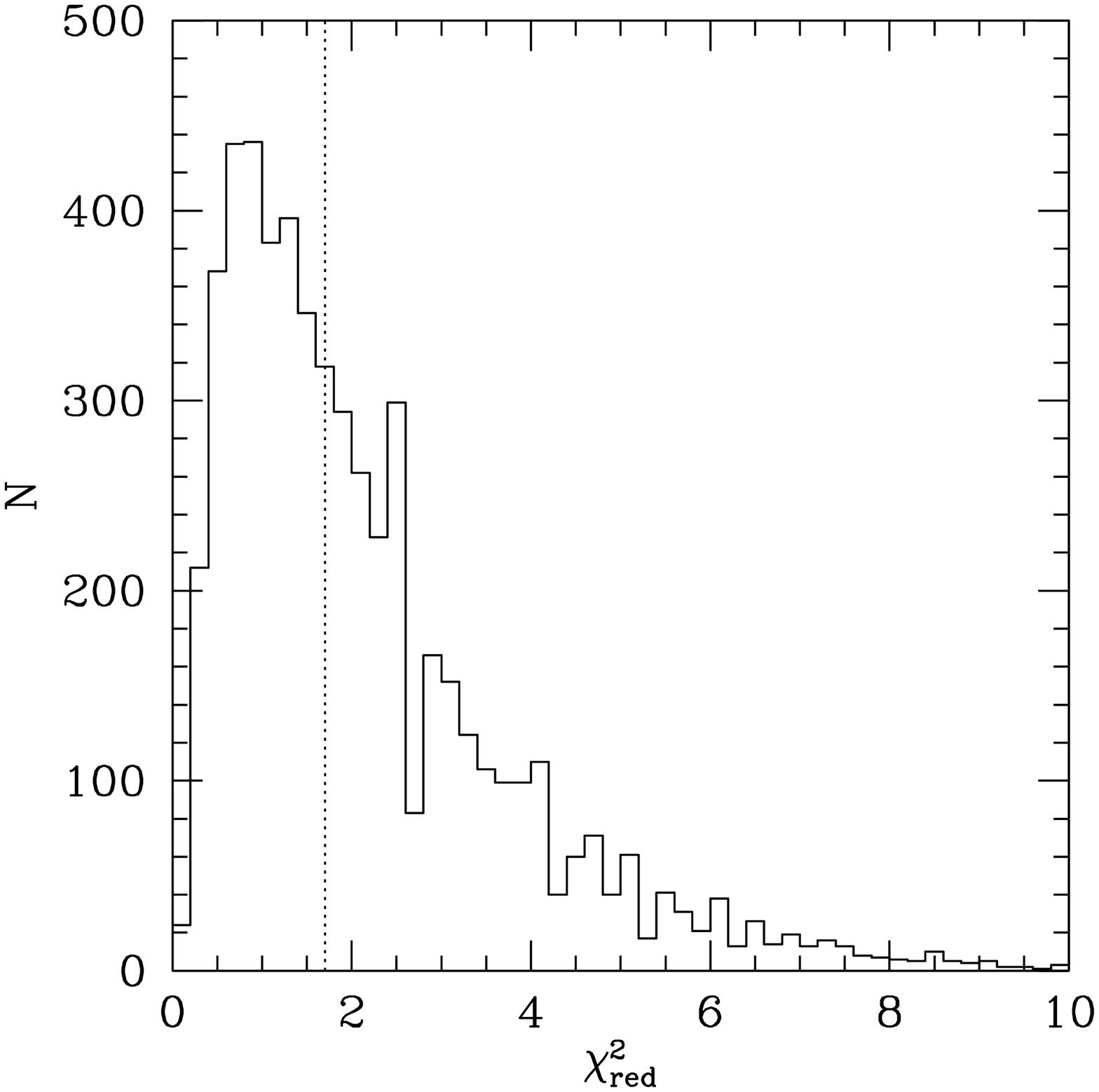}\hfill
\includegraphics[width=0.5\textwidth]{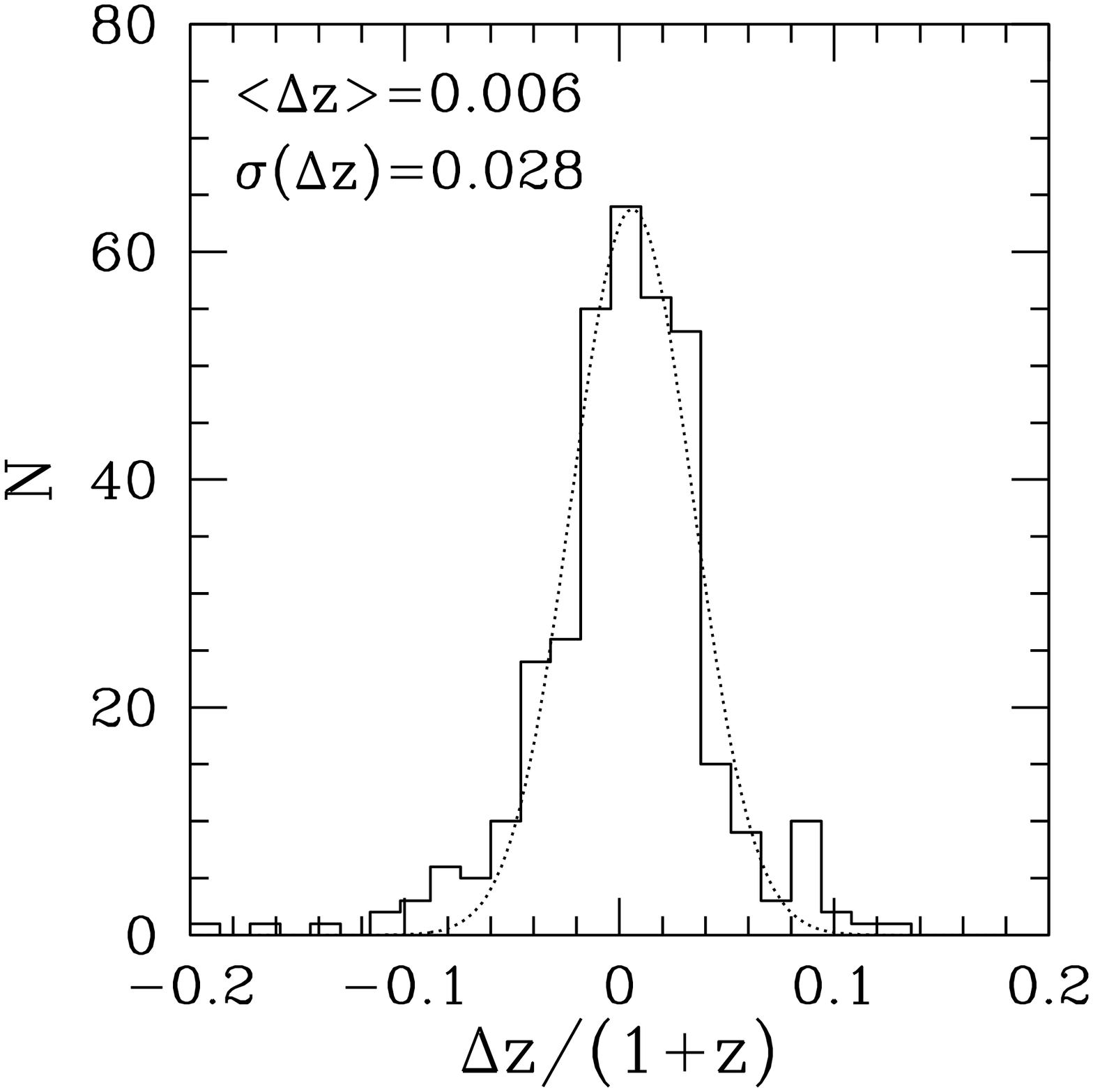}\hfill
\caption{\label{z_spec_z_phot_fdf_error_dis} 
Left panel: Histogram of the reduced $\chi^2$ for all galaxies in the
FDF as obtained for the best fitting
template and redshift. The dotted vertical line indicates the median
reduced  $\chi^2$. Right panel: Histogram of the photometric 
redshift errors. The error distribution can be approximated by a Gaussian centered at
0.006 with an rms of 0.028 (dotted line).  
 }
\end{figure*}
\begin{figure*}[ht]
\includegraphics[width=0.5\textwidth]{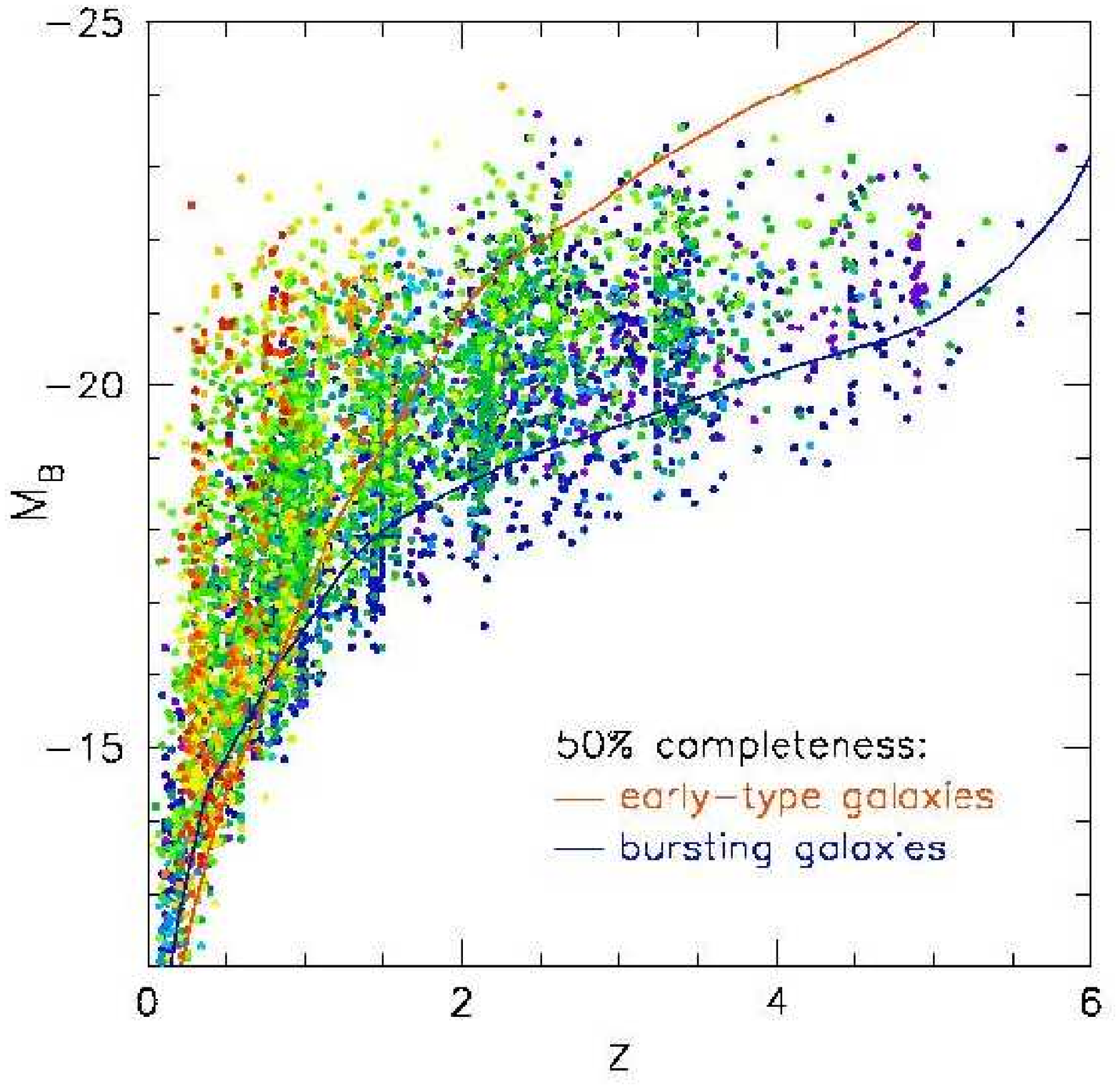}\hfill
\includegraphics[width=0.5\textwidth]{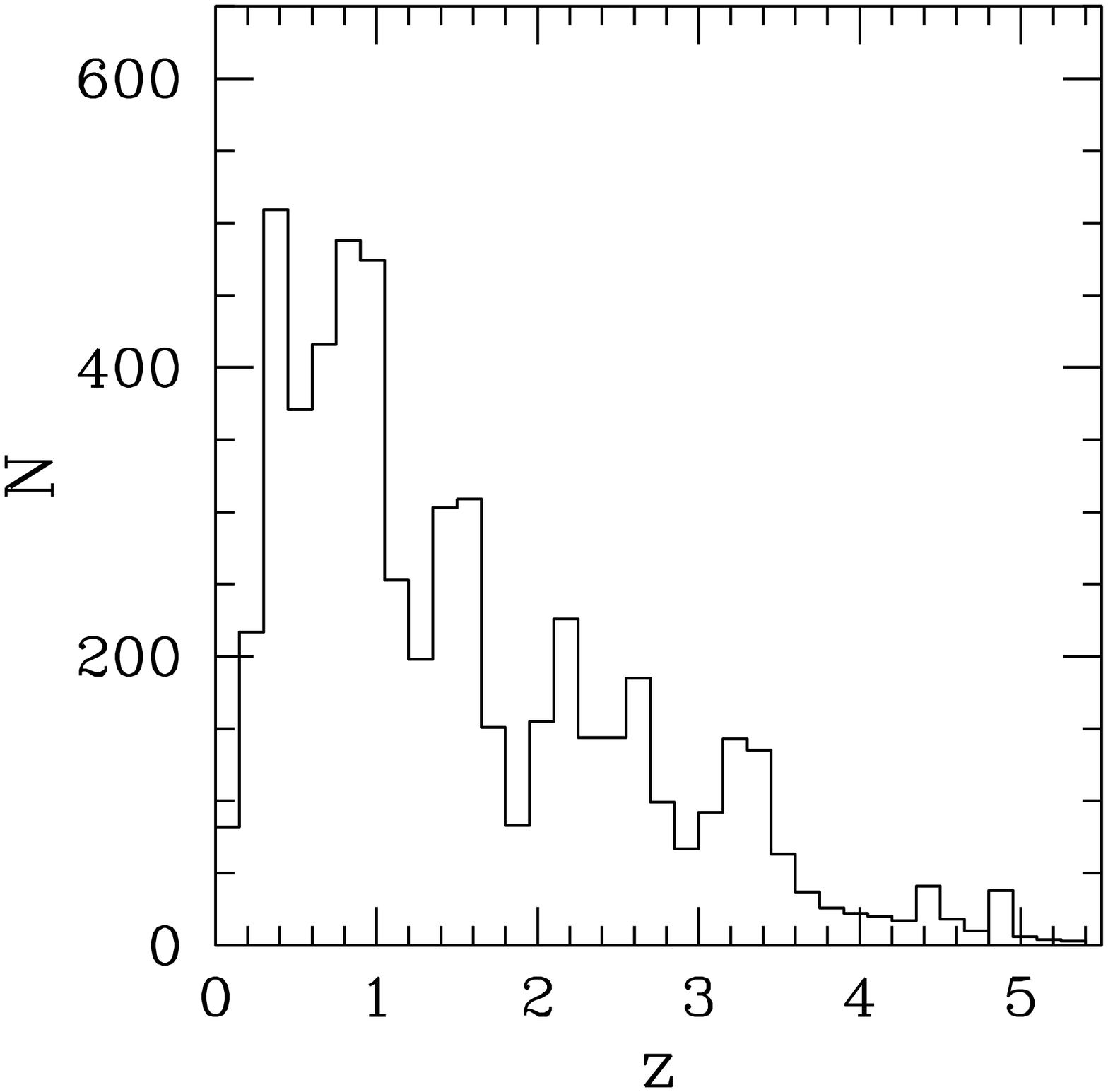}
\caption{\label{mag_z_phot_dis} 
  Left panel: Absolute B magnitudes of galaxies in the FDF against
  redshift. Colors indicate spectral types (red to blue: old to
  young).  The two lines indicate the 50\% completeness limit for a
  red and a blue spectral energy distribution corresponding to an
  I-band limiting magnitude of 26.8.  Right panel: Redshift
  number distribution of all galaxies in the FDF sample. The
  clustering observed in photometric redshift space (both panels) is
  probably mostly real, as we see clustered {\em spectroscopic}
  redshifts at $z=0.22$, $z=0.33$, $z=0.39$, $z=0.45$, $z=0.77$,
  $z=2.35$ and possibly at $z=0.95$, $z=3.15$, and $z=3.4$.  }

\end{figure*}

A brief summary of the photometric redshift technique used to derive
the distances to the galaxies in the FDF can be found in
\citet{bender:1}, a more detailed description will be published in a
future paper (Bender et al. 2004).  Well determined colors of the
objects which implies very precise zeropoints in all filters are
crucial to derive accurate photometric redshifts. Therefore we checked
and fine-tuned the calibration of our zeropoints by means of
color-color plots of stars.  We compared the colors of FDF stars with
the colors of stellar templates from the library of \citet{pickles:1}
converted to the FORS filter system.  In general, corrections to the
photometric zeropoints of only a few hundredth of a magnitude were
needed to obtain an optimal match to the stars and best results for
the photometric redshifts.  In order to avoid contamination from
close-by objects, we derived object fluxes for a fixed aperture of
1.5'' ($1.5 \times$ seeing) from images which had been convolved to
the same point spread function. A redshift probability function P(z)
was then determined for each object by matching the object's fluxes to
a set of 30 template spectra redshifted between $z=0$ and $z=10$ and
covering a wide range of ages and star formation histories. As
templates we used (a) local galaxy templates from \citet{mannucci:1},
and \citet{kinney:1} and (b) semi-empirical templates more appropriate
for modest to high redshift galaxies.  The semi-empirical templates
were constructed by fitting combinations of theoretical spectral
energy distributions of different ages from \citet{maraston:1} and
\citet{bruzual:1} with variable reddening \citep{kinney:2} to the
observed broad band colors of about 100 galaxies in the Hubble Deep
Field and about 180 galaxies from the FDF with spectroscopic
redshifts. The remaining 180 galaxies in the FDF with spectroscopic
redshift were used as an independent control sample.  Lyman forest
absorption was parameterized following \citet{madau:1} and references
therein.

In Fig.~\ref{z_spec_z_phot_fdf} we compare the photometric and
spectroscopic redshifts of 362 galaxies and QSOs in the FDF (see
\citealt{noll:1, boehm:1} for the spectroscopic redshifts). The
agreement is very good and we have only 6 outliers with a redshift
error larger than $\Delta z > 1$ among 362 objects.  Three of the
outliers are quasars or galaxies with a strong power-law AGN component
(crosses). The others are very blue objects with an almost
feature-less continuum (triangles).
Fig.~\ref{z_spec_z_phot_fdf_error_dis} (left panel) presents the
$\chi^2$ distribution for the best fitting template and photometric
redshifts. Note that to calculate the $\chi^2$, we have used the
observational photometric errors and, in addition, have assumed that
the templates have an intrinsic uncertainty of typically 5\% in the
optical bands and 20\% in the infrared bands. 
The larger errors for the near-IR take into account the slightly lower
quality of the infrared data if compared to the optical.
Allowing for this
intrinsic uncertainty turns a discrete set of templates into a template-continuum.
Observational errors and intrinsic `errors' were added in quadrature.
The median value of the reduced $\chi^2$ is below 1.7 and demonstrates
that the galaxy templates describe the vast majority of galaxies in
the FDF very well.  The right panel of
Fig.~\ref{z_spec_z_phot_fdf_error_dis} shows the distribution of the
redshift errors. It is nearly Gaussian and scatters around zero with
an rms error of \mbox{$\Delta z / (z_{spec}+1) \approx 0.03 $.}  In
Fig.~\ref{mag_z_phot_dis} (left panel), we plot the absolute B-band
magnitudes against the photometric redshifts of the objects. Colors
from red to blue indicate increasingly bluer spectral energy
distributions. The two lines indicate the 50\% completeness limit for
a red and a blue spectral energy distribution corresponding to an
I-band limiting magnitude of 26.8.  The redshift histogram of all
objects in the FDF is shown in the right panel of
Fig.~\ref{mag_z_phot_dis} (see also Table~\ref{tab:binning}). Most
if not all peaks in the distribution are due to real clustering in
redshift space. From the 362 spectroscopic redshifts, we have
identified clusters, groups or filaments of galaxies with more than 10
identical or almost identical redshifts at $z=0.22$, $z=0.33$,
$z=0.39$, $z=0.45$, $z=0.77$, $z=2.35$.  Other structures (with only a
few identical spectroscopic redshifts) are possibly present at
$z=0.95$, $z=3.15$, and $z=3.4$.

\begin{table}[]
\caption[]{\label{tab:binning}Galaxy distribution in the FDF for the
  redshift intervals used for computing the luminosity function. Note
  that we derive the luminosity function in all redshift bins, but
  exclude the lowest ($z < 0.45$) and highest redshift bin ($z >
  5.01$) from our analysis of the luminosity function evolution, since
  the lowest redshift bin corresponds to a too small volume while the
  $z>5.01$ bin suffers from incompleteness.}
\begin{center}
\begin{tabular}{c|cr}
\hline
redshift &  number     & fraction\\
interval & of galaxies & of galaxies\\
\hline
 0.00 -  0.45  &    808 & 14.54 \%    \\
 0.45 -  0.81  &    998 & 17.96 \%    \\
 0.81 -  1.11  &    885 & 15.92 \%    \\
 1.11 -  1.61  &    898 & 16.16 \%    \\
 1.61 -  2.15  &    504 &  9.07 \%    \\
 2.15 -  2.91  &    746 & 13.42 \%    \\
 2.91 -  4.01  &    549 &  9.88 \%    \\
 4.01 -  5.01  &    150 &  2.70 \%    \\
 $>5.01$       &    18  &  0.32 \%    \\
 unknown       &    2   &  0.04 \%    \\
\hline
\end{tabular}
\end{center}
\end{table}

\section{Luminosity functions}
\label{sec:lumfkt}

\subsection{The method}
\label{sec:lumfkt:method}

We compute the absolute magnitudes of our galaxies using the I-band
selected catalogue as described in Sect.~\ref{sec:fdf} and the
photometric redshifts described in Sect.~\ref{sec:photred}.  To derive
the absolute magnitude for a given band we use the best fitting SED as
determined by the photometric redshift code and convolve it with the
appropriate filter function. As the SED fits all 9 observed-frame
wavebands simultaneously, possible systematic errors which could be
introduced by using K-corrections applied to a single observed
magnitude are reduced. Since the photometric redshift code works with
1.5'' aperture fluxes, we only need to correct to total luminosities
by applying an object dependent scale factor. For this scale factor we
used the ratio of the I-band aperture flux to the total flux as
provided by SExtractor (MAG\_APER and MAG\_AUTO). We have chosen the
I-band because (a) our I-band data are very deep, (b) all objects were
detected and selected in the I-band, and (c) high redshift galaxies
have only poorly determined or no flux at shorter wavelengths.
This procedure may introduce a slight bias, as galaxies are more
  compact or knotty in the rest-frame UV bands (tracing HII regions)
  than at longer wavelengths.  
However, scaling factors derived in the deep B-band turned out to be
similar (for low enough redshifts).

In a given redshift interval, the luminosity function is computed by
dividing the number of galaxies in each magnitude bin by the volume
$V_\mathrm{bin}$ of the redshift interval. To account for the fact that
some fainter galaxies are not visible in the whole survey volume we
perform a $V/V_{max}$ \citep{Schmidt1} correction.  Using the best
fitting SED we calculate the maximum redshift $z_{max}$ at which the
object could have been observed given the magnitude limit of our
field.  We weight each object by $V_{bin}/V_{max}$ where $V_{bin}$ is
the volume of our redshift bin enclosed by $z_{low}$ and $z_{high}$
and $V_{max}$ is the volume enclosed between $[z_{low},
\textrm{min}(z_{high},z_{max})]$.

To derive reliable Schechter parameters we limit our analysis of the
luminosity function to the bin where the $V/V_{max}$ begins to
contribute at most by a factor of 3 (we also show the uncorrected
luminosity function in the various plots as open circles).  The
redshift binning was chosen such that we have good statistics in every
redshift bin and that the influence of redshift clustering was
minimized.  The redshift binning and the number of galaxies in
every bin is shown in Table~\ref{tab:binning}.

The errors of the luminosity functions are calculated by means of
Monte-Carlo simulations as follows. The photometric redshift code
provides redshift probability distributions P(z) for each single
galaxy. In each Monte-Carlo realization, we randomly pick a new
redshift for each object from a sample of redshifts distributed like
P(z) and calculate the corresponding luminosity. This we repeat 250
times which allows us to derive the dispersion of the galaxy number
density $\phi(M,z)$ for each magnitude and redshift bin due to the
finite width of P(z) for each galaxy.  The total error in $\phi$ is
finally obtained by adding in quadrature the error from the
Monte-Carlo simulations and the Poissonian error derived from the
number of objects in the bin.

Photometric redshift errors may, in principle, affect the shape of the
luminosity function at the bright end: By scattering objects to higher
redshifts they let the steep fall-off at high luminosities appear
shallower \citep{drory:1}. However, in the case of the FDF the
redshift errors are so small that the influence on the shape of the
luminosity function is negligible.

\subsection{The slope of the luminosity function}
\label{sec:lumfkt:slope}

\begin{table*}[]
\setlength{\tabcolsep}{1mm}
\caption[]{\label{tab:slope_single}Slope of the luminosity function
  for all wavelengths and all redshifts as derived from 3-parameter
  Schechter fits.}
\begin{center}
\begin{tabular}{c||c|c|c|c|c}
$z$ & $\alpha$ (1500~\AA) & $\alpha$ (2800~\AA) & $\alpha$ (u') & $\alpha$ (g') & $\alpha$ (B) \\ 
\hline
 $[0.45, 0.81]$  & $-$1.14 (+0.08 $-$0.07) &     $-$1.23 (+0.08 $-$0.07) &     $-$1.27 (+0.06 $-$0.05) &     $-$1.34 (+0.05 $-$0.03) &     $-$1.30 (+0.05 $-$0.03)\\
 $[0.81, 1.11]$  & $-$0.96 (+0.13 $-$0.10) &     $-$0.99 (+0.10 $-$0.08) &     $-$0.93 (+0.09 $-$0.07) &     $-$1.16 (+0.07 $-$0.04) &     $-$1.21 (+0.07 $-$0.04)\\
 $[1.11, 1.61]$  & $-$1.05 (+0.18 $-$0.16) &     $-$1.03 (+0.13 $-$0.11) &     $-$0.95 (+0.10 $-$0.09) &     $-$1.13 (+0.11 $-$0.09) &     $-$1.12 (+0.09 $-$0.07)\\
 $[1.61, 2.15]$  & $-$0.81 (+0.48 $-$0.45) &     $-$0.97 (+0.32 $-$0.28) &     $-$0.80 (+0.31 $-$0.27) &     $-$1.29 (+0.24 $-$0.21) &     $-$1.33 (+0.27 $-$0.20)\\
 $[2.15, 2.91]$  & $-$0.38 (+0.21 $-$0.15) &     $-$0.67 (+0.18 $-$0.15) &     $-$0.70 (+0.16 $-$0.16) &     $-$0.89 (+0.22 $-$0.15) &     $-$0.70 (+0.24 $-$0.21)\\
 $[2.91, 4.01]$  & $-$0.98 (+0.28 $-$0.24) &     $-$0.95 (+0.19 $-$0.17) &     $-$1.25 (+0.19 $-$0.14) &     $-$1.24 (+0.23 $-$0.20) &     $-$1.30 (+0.27 $-$0.20)\\
 $[4.01, 5.01]$  & $-$0.77 (+0.38 $-$0.26) &     $-$1.03 (+0.46 $-$0.35) &     $-$1.09 (+0.54 $-$0.27) &     $-$1.18 (+0.37 $-$0.21) &     $-$0.77 (+0.49 $-$0.39)\\
\end{tabular}
\end{center}
\end{table*}

We first investigate the redshift evolution of the faint-end slope of
the luminosity function by fitting all three parameters
of the Schechter function (M$^\ast$, $\phi^\ast$, and $\alpha$). 
The best fitting $\alpha$ and the corresponding $1\sigma$ errors for all wavebands and
redshifts are listed in Table~\ref{tab:slope_single}.

Despite of the depth of the FDF, Table~\ref{tab:slope_single} shows
that it is only possible to obtain reasonably tight constraints on the
slope $\alpha$ for $z< 1.5$.  In addition, strong parameter coupling
is observed between M$^\ast$ and $\alpha$ (see
Fig.~\ref{fig:contour_3param_alpha} in the Appendix
\ref{sec:3_para_fit_slope}). We find only marginal evidence for a
change of $\alpha$ with redshift for all wavebands. The lowest
redshift bin ($0.15<z<0.45$), which we excluded from the fit because
of poor number statistics in bright objects, generally shows the
steepest faint-end slope.  Beyond redshift 0.5, all data are
consistent with a constant and shallow faint-end slope.  

We obtain as best error-weighted values for all redshifts between 0.45
and 5.0 the numbers given in Table~\ref{tab:slope_fixed} (upper part),
assuming that $\alpha$ does not depend on redshift. The slopes in
the 1500~\AA , 2800~\AA , and u' band are very similar. The same
applies for the slope in the g' and B band.  Therefore, we combined
the data for the 1500~\AA , 2800~\AA , and u' band as well as for the
g' and B band and derived combined slopes with an error weighted fit
to the data of Table~\ref{tab:slope_single}. The results are also
listed in Table~\ref{tab:slope_fixed} (lower part).

\begin{table}[]
\caption[]{\label{tab:slope_fixed}Slope $\alpha$ of the luminosity
  functions for the different wavebands as determined from an
  error-weighted fit to the data in Table~\ref{tab:slope_single} under
  the assumption that $\alpha(z)=\mathrm{const.}$ (upper part). In the
  lower part of the Table we show the best values of $\alpha$ after
  combining the UV bands and the blue optical bands.}
\begin{center}
\begin{tabular}{l|c}
filter  & $\alpha(z)=const$\\
\hline
\rule[+3mm]{-1.4mm}{2mm}
\rule[-3mm]{0mm}{2mm}{1500~\AA\ } & $-1.01 \pm 0.08$\\
\rule[-3mm]{0mm}{2mm}{2800~\AA\ } & $-1.06 \pm 0.07$\\
\rule[-3mm]{0mm}{2mm}{u' }        & $-1.10 \pm 0.08$\\
\rule[-3mm]{0mm}{2mm}{g' }        & $-1.26 \pm 0.04$\\
\rule[-3mm]{0mm}{2mm}{B  }        & $-1.24 \pm 0.04$\\
\hline
\rule[+3mm]{-1.4mm}{2mm}
\rule[-3mm]{0mm}{2mm}{1500~\AA\ \& 2800~\AA\ \& u'} & $-1.07 \pm 0.04$\\
\rule[-3mm]{0mm}{2mm}{g' \& B }                     & $-1.25 \pm 0.03$\\
\end{tabular}
\end{center}
\end{table}

Almost all of the slopes listed in Table~\ref{tab:slope_single} are
compatible within $2\sigma$ with the slopes in
Table~\ref{tab:slope_fixed}. Therefore, we fixed the slope to these
values for further analysis. This simplification is also justified by 
the fact that for  all subsequent fits with fixed 
slope the reduced $\chi^2$ was generally close to 1.

As a last test, we investigated the influence of the redshift binning on
the slope $\alpha$. We enlarged our first two redshift bins to $0.41 <z
\le 0.91$ (1433 galaxies) and $0.91 <z\le 1.61$ (1438 galaxies) which
allowed us to determine luminosity functions with lower errors in all
wavebands. The slopes derived in these two larger bins were compatible
with our previously derived fixed slope in every waveband.

\subsection{The restframe luminosity functions}
\label{sec:lumfkt:restframe}

\begin{figure*}[]
\includegraphics[width=0.33\textwidth]{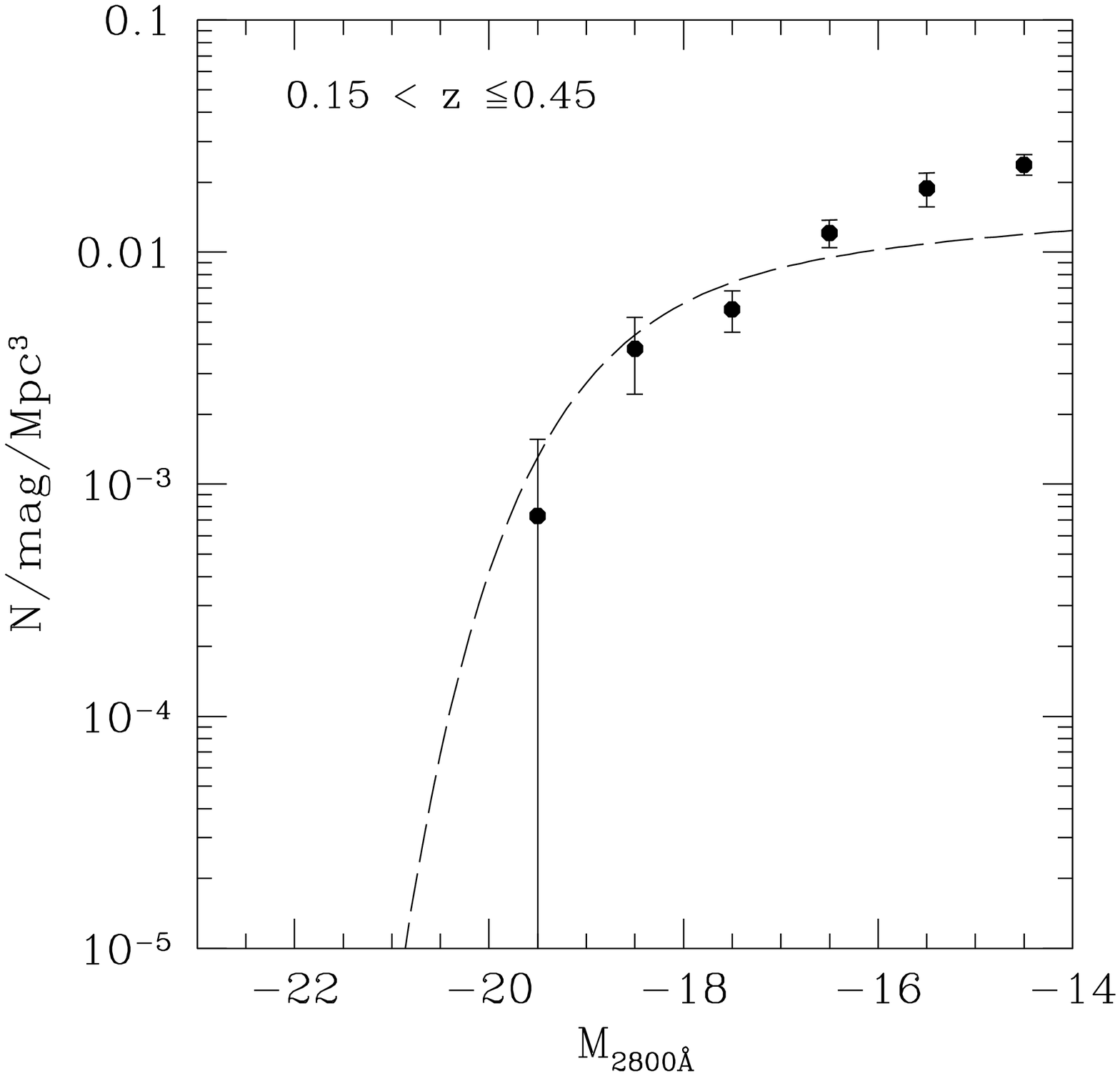}
\includegraphics[width=0.33\textwidth]{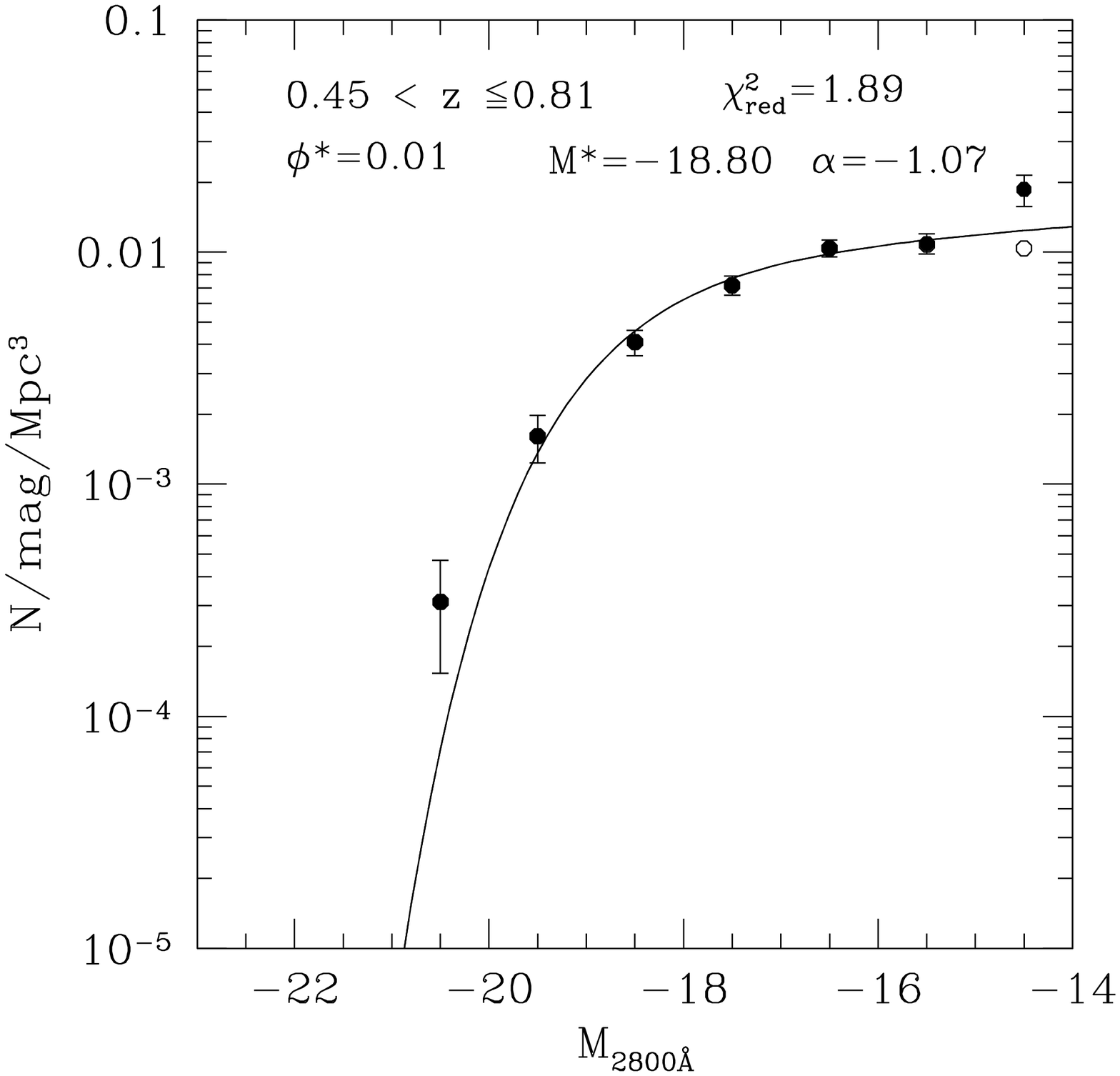}
\includegraphics[width=0.33\textwidth]{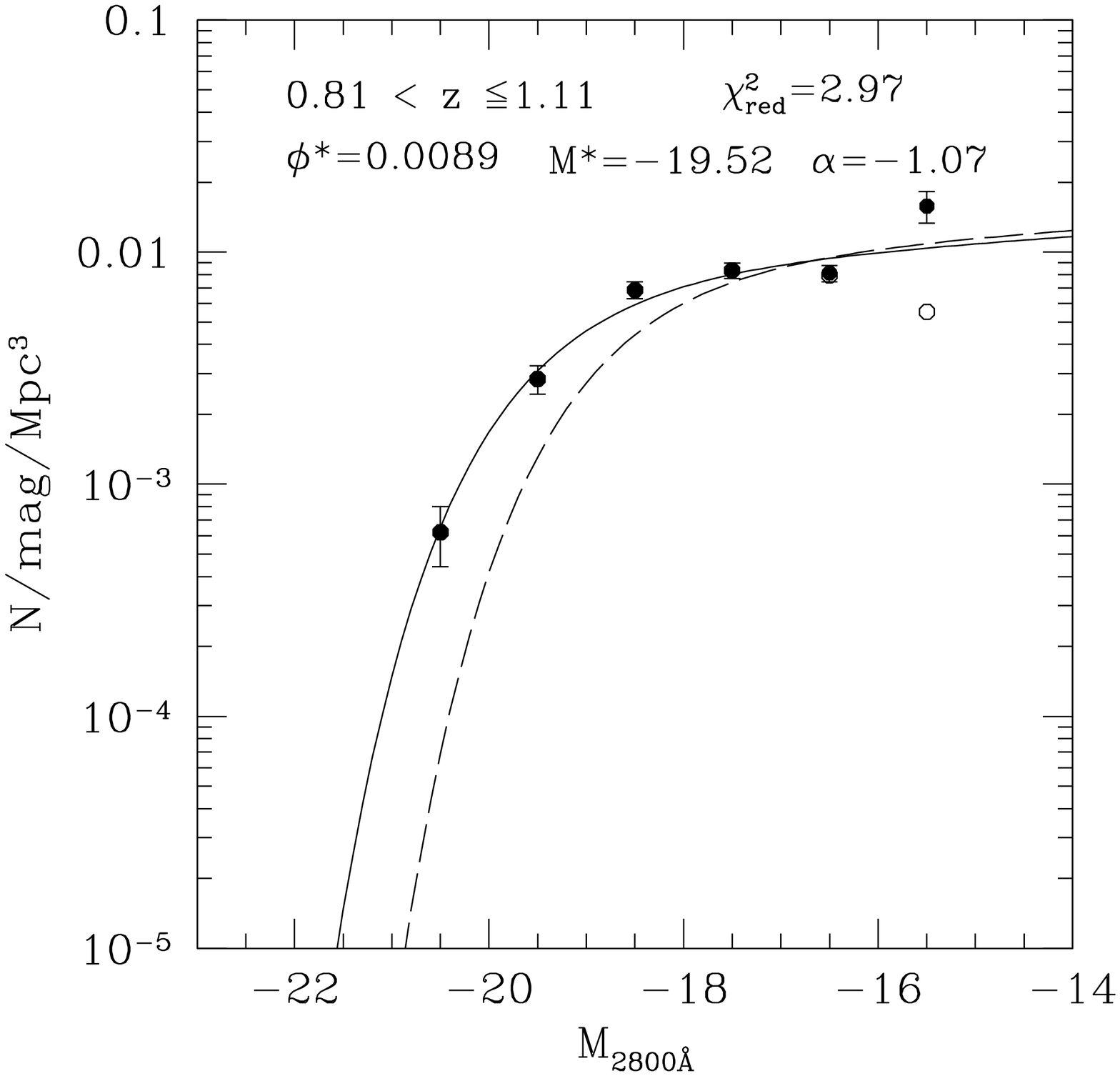}
\includegraphics[width=0.33\textwidth]{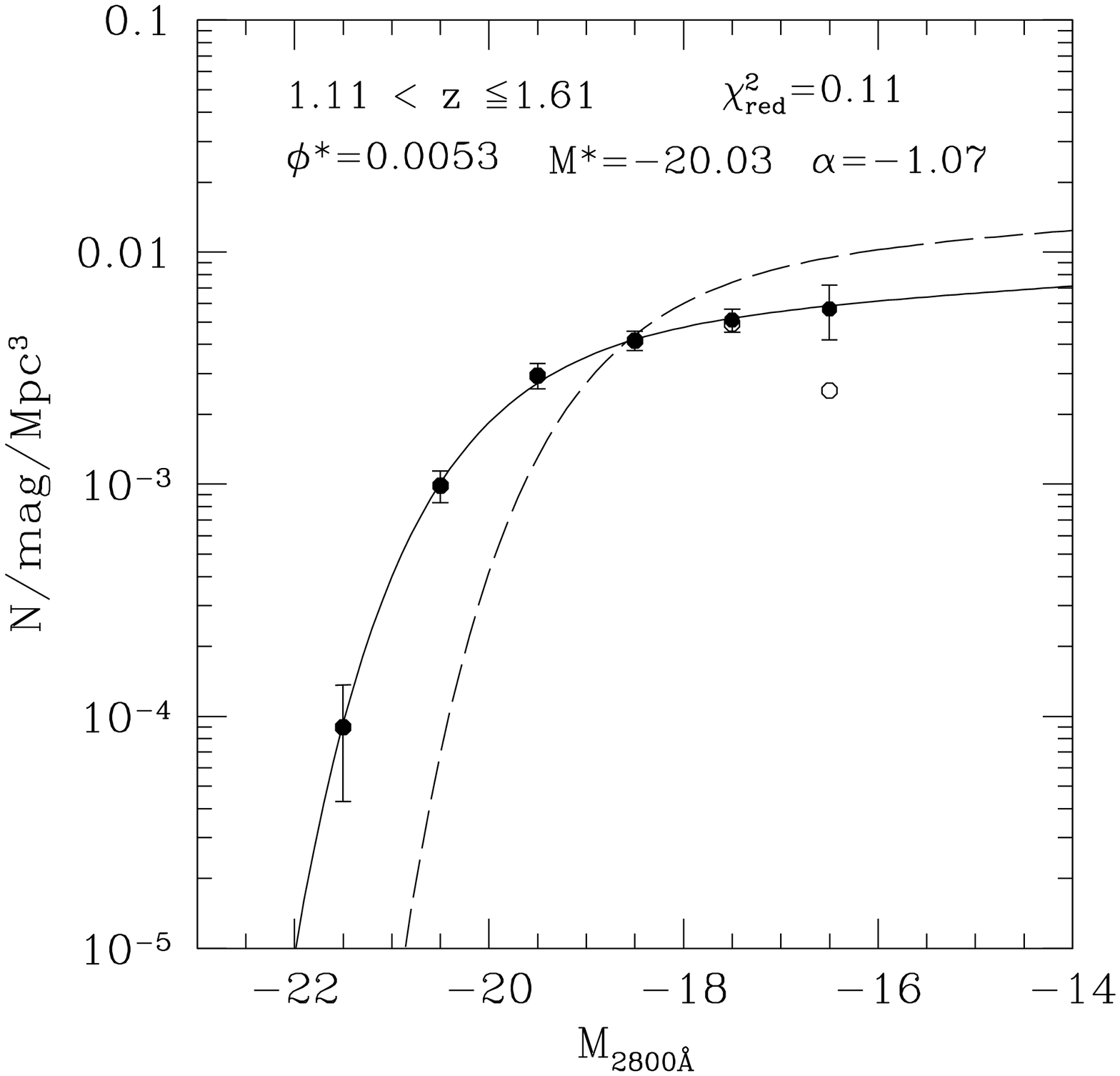}
\includegraphics[width=0.33\textwidth]{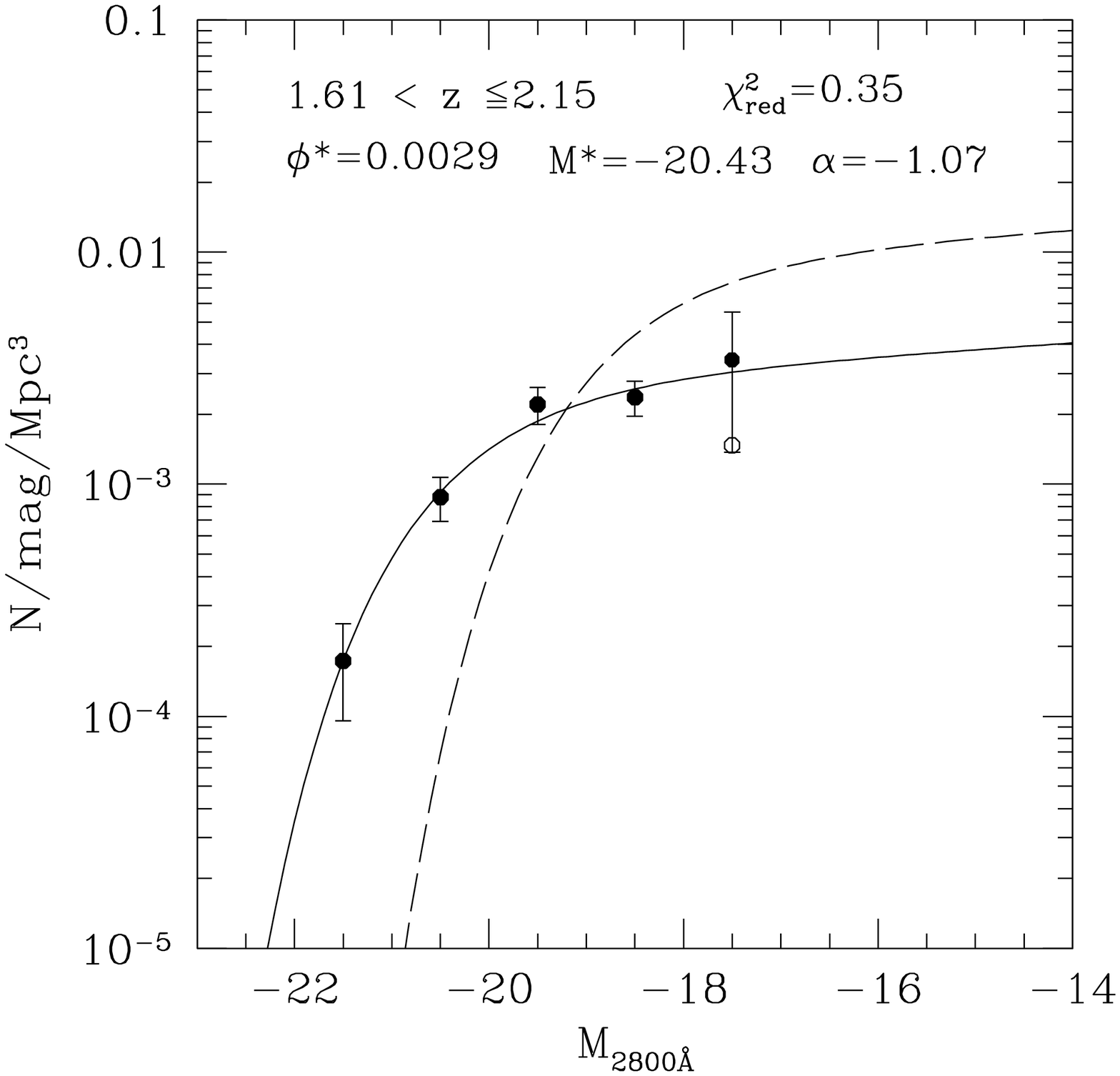}
\includegraphics[width=0.33\textwidth]{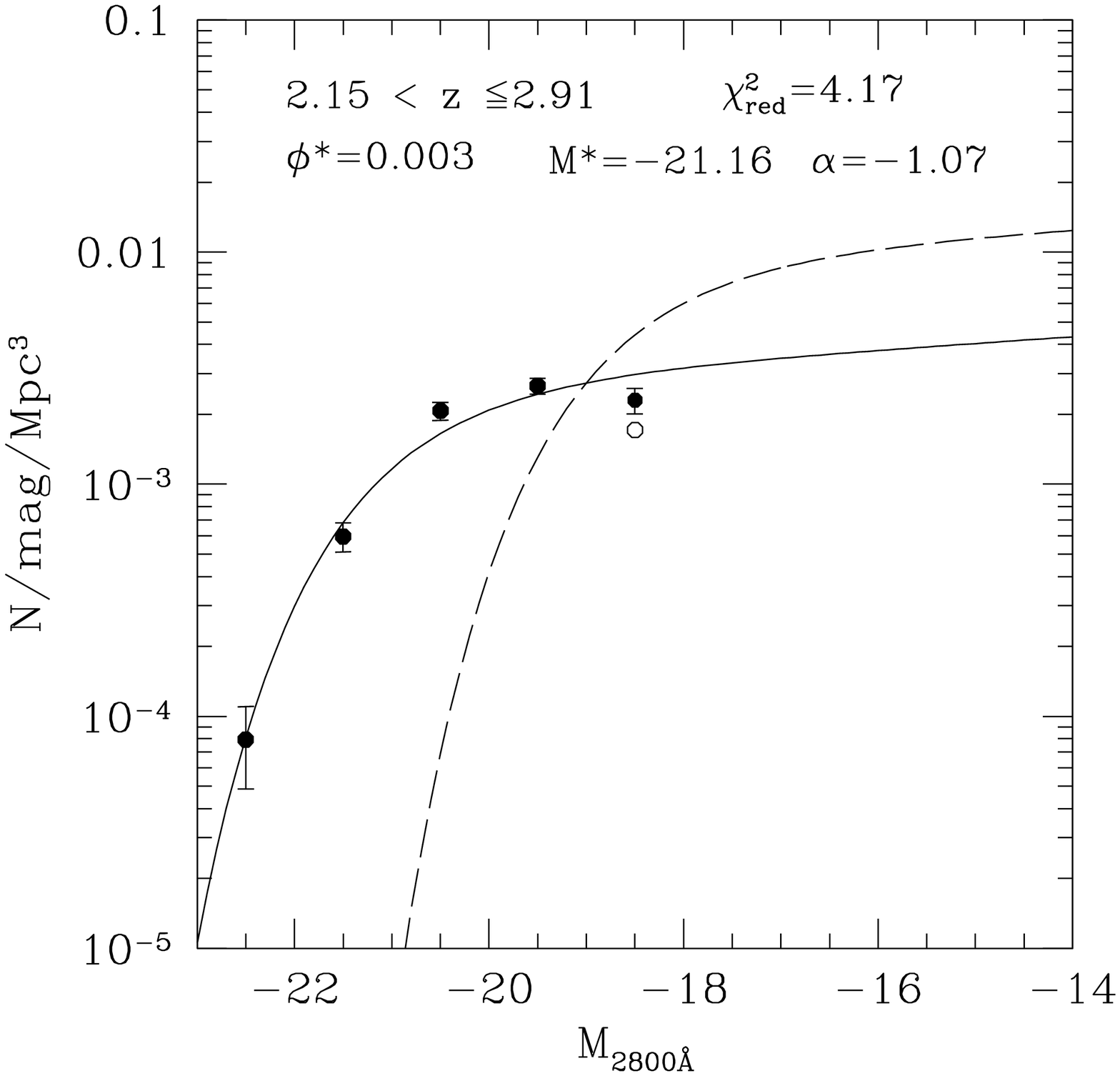}
\includegraphics[width=0.33\textwidth]{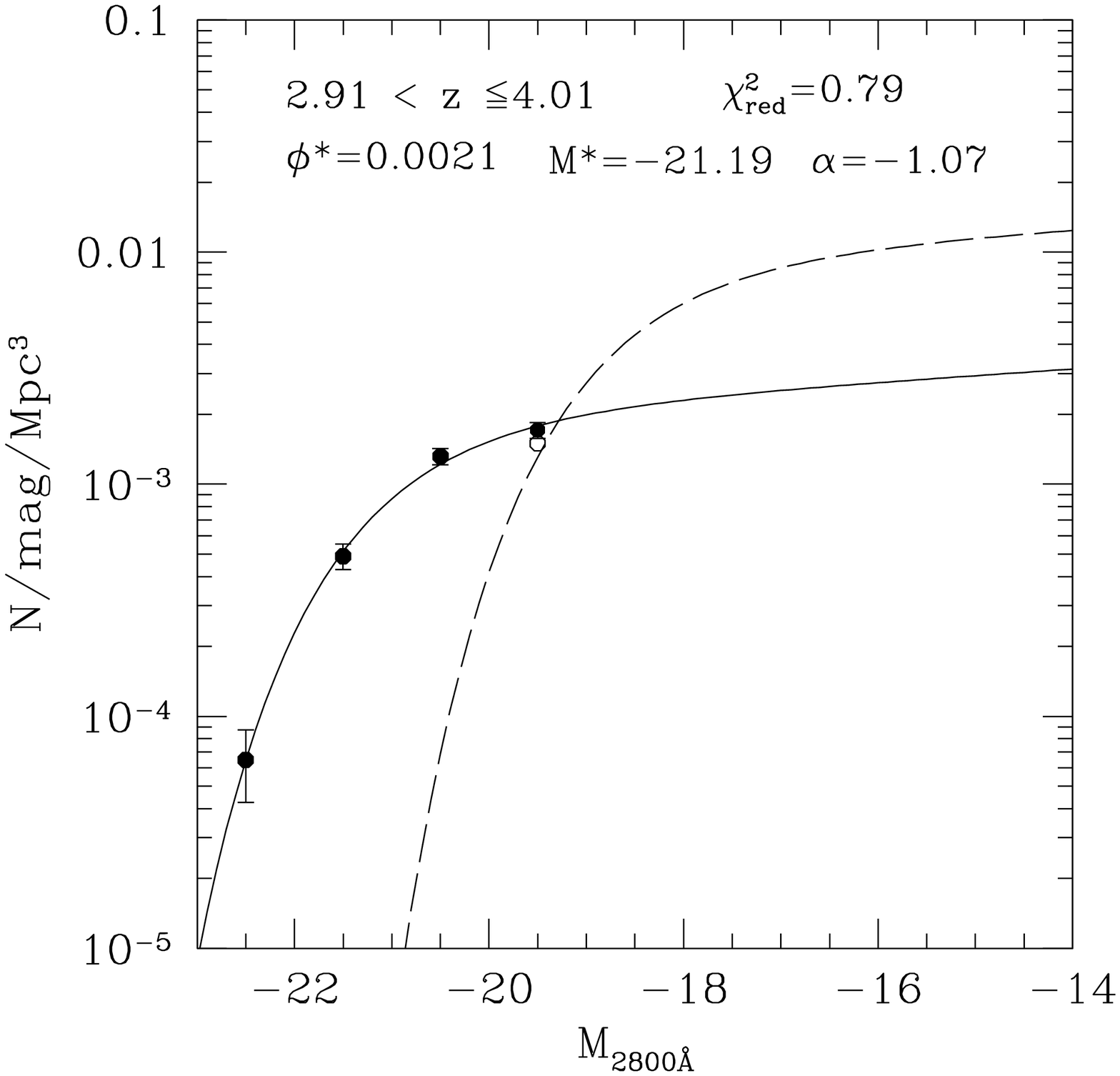}
\includegraphics[width=0.33\textwidth]{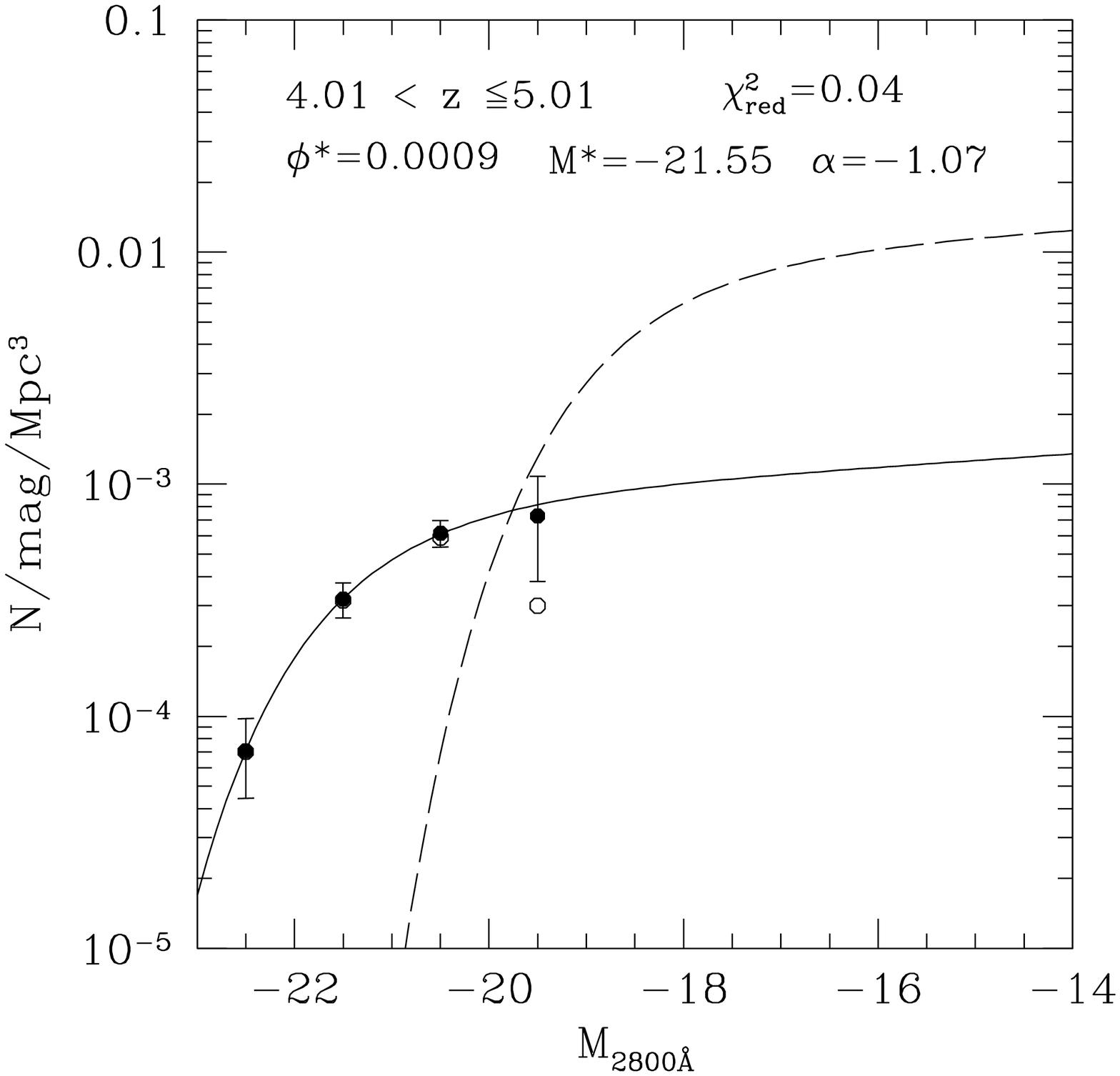}
\includegraphics[width=0.33\textwidth]{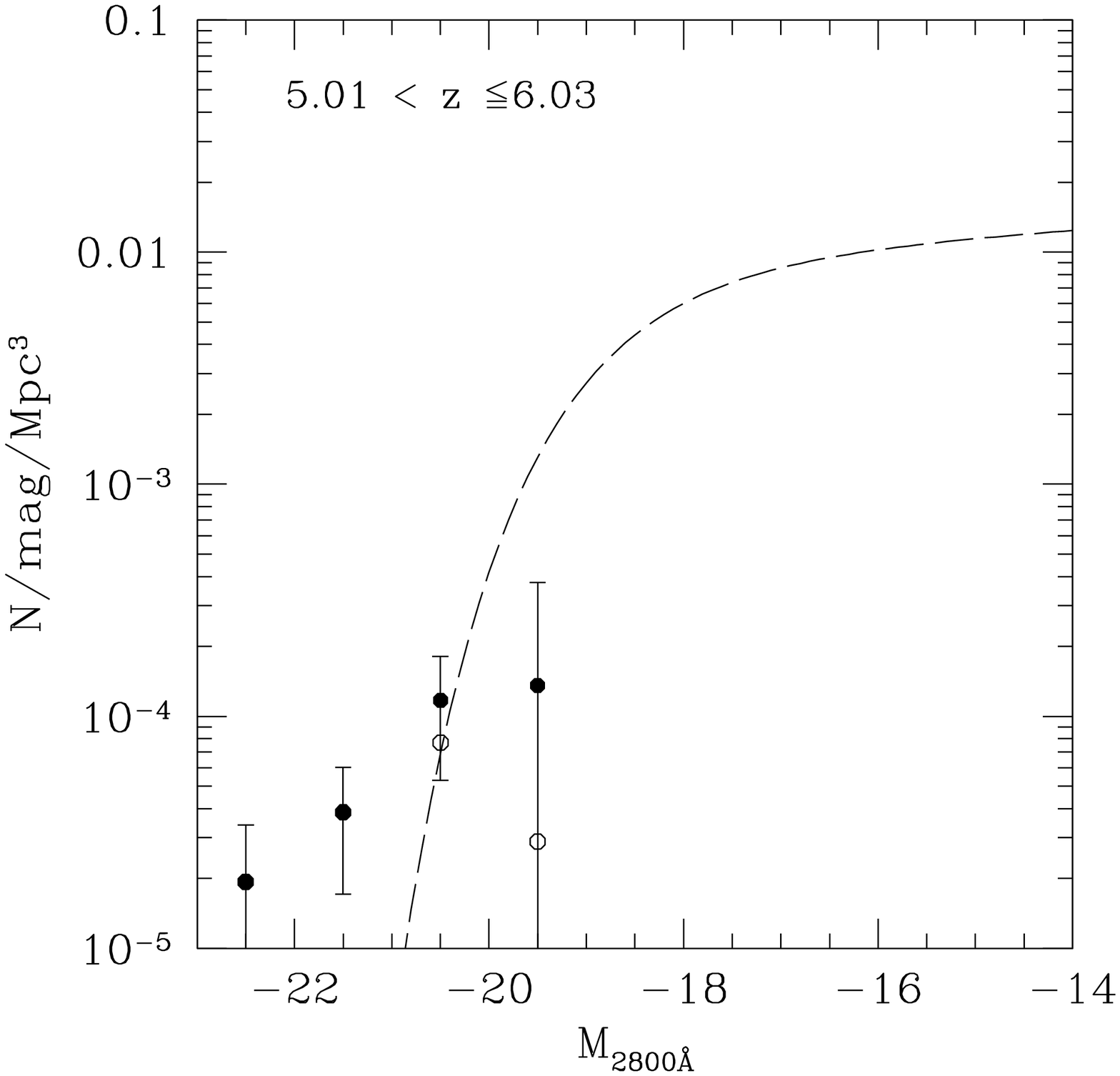}
\caption{\label{fig:lumfkt_fdf_2800}  
  Luminosity functions at \textit{2800~\AA } from low redshift
  (\mbox{$\langle z\rangle=0.3$}, upper left panel) to high redshift
  (\mbox{$\langle z\rangle=5.5$}, lower right panel). The filled
  (open) symbols show the luminosity function corrected (uncorrected)
  for $V/V_{max}$. The fitted Schechter functions for a fixed slope
  $\alpha$ are shown as solid lines. Note that we only fit the
  luminosity functions from $\langle z\rangle=0.6$ to $\langle
  z\rangle=4.5$. The parameters of the Schechter functions are given
  in Table~\ref{tab:schechter_fit_2800}.  The Schechter fit for
  redshift $\langle z\rangle=0.6$ is indicated as dashed line in all
  panels.}
\end{figure*}

\begin{figure*}
\includegraphics[width=0.33\textwidth]{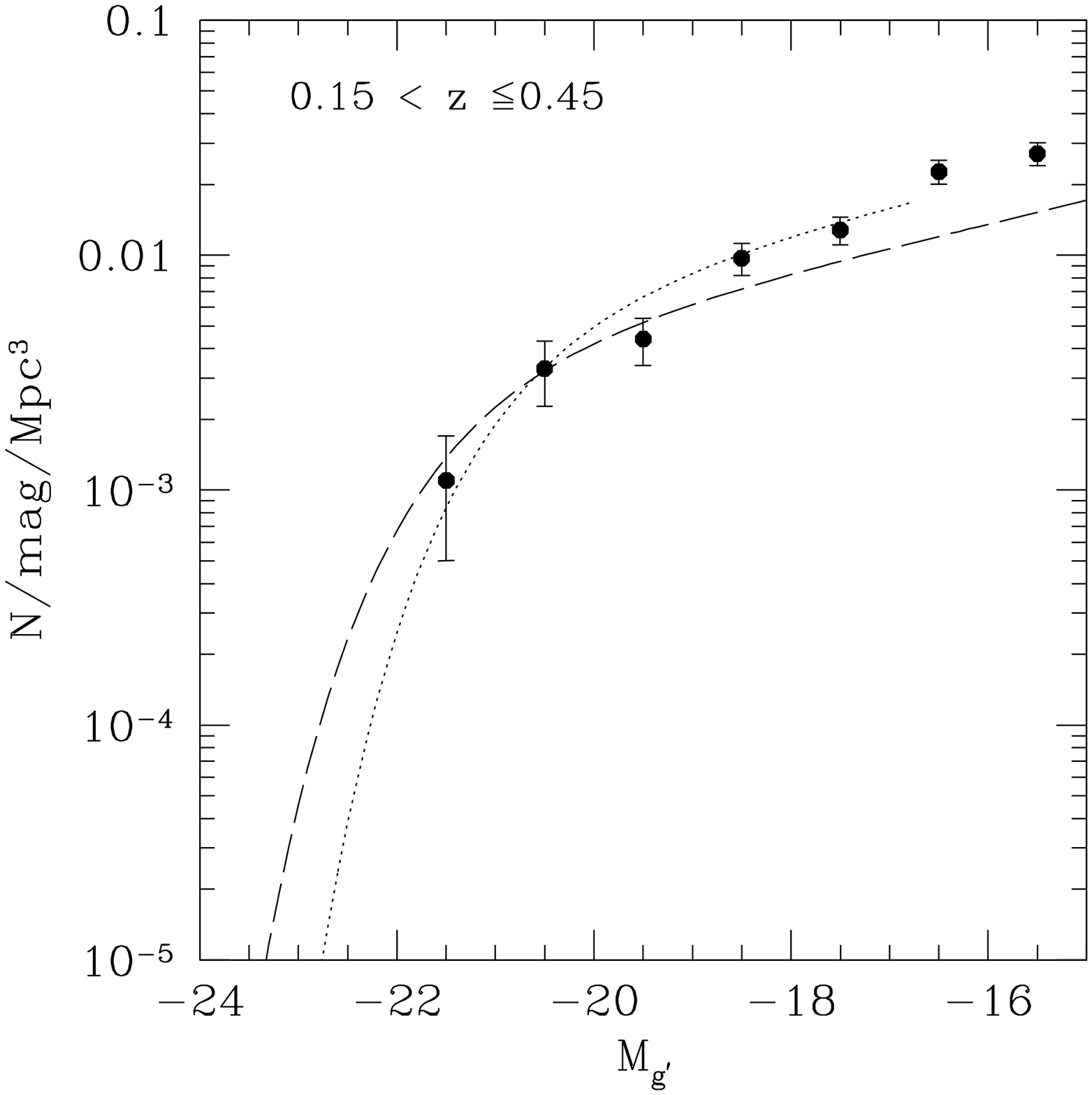}
\includegraphics[width=0.33\textwidth]{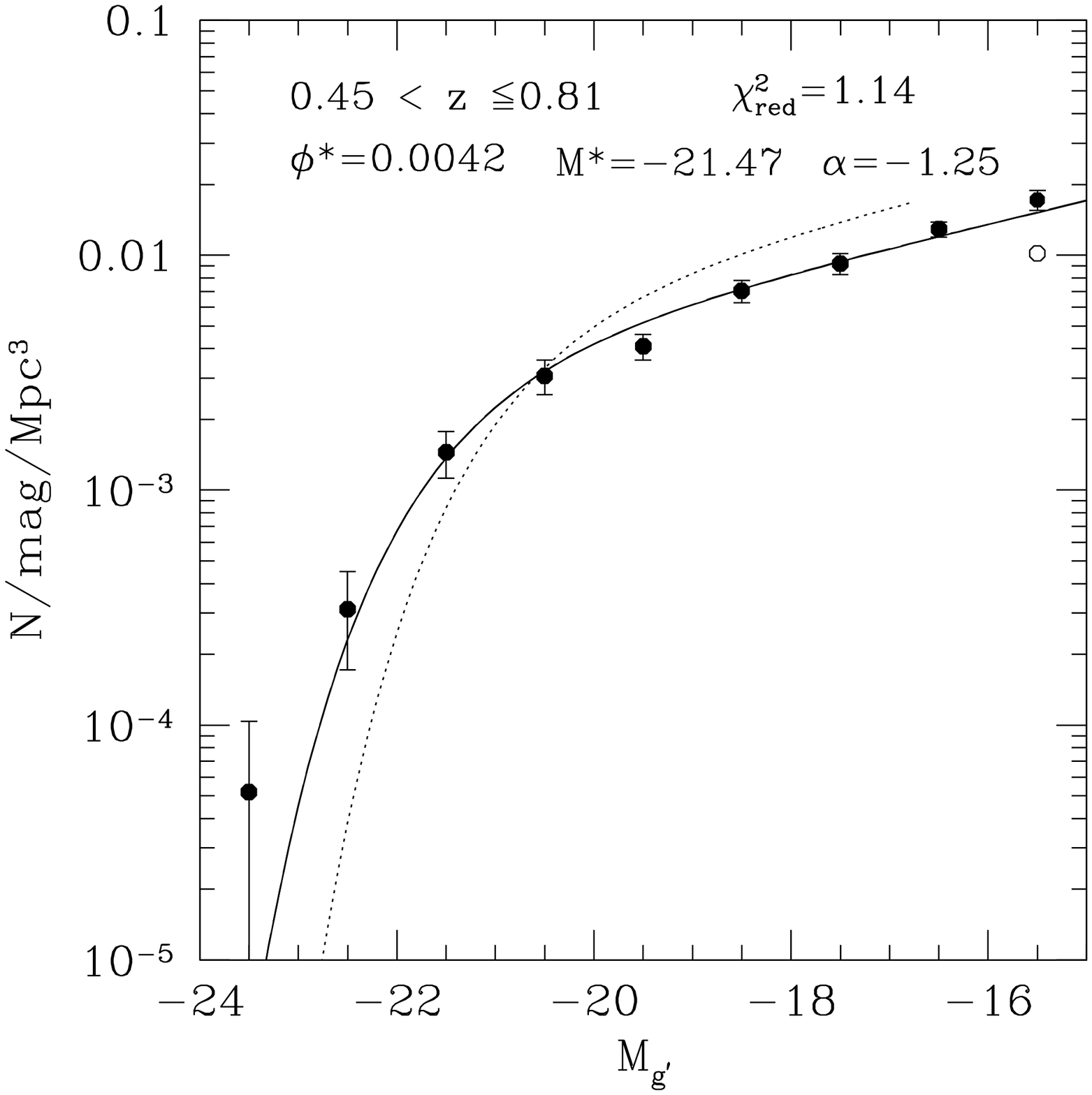}
\includegraphics[width=0.33\textwidth]{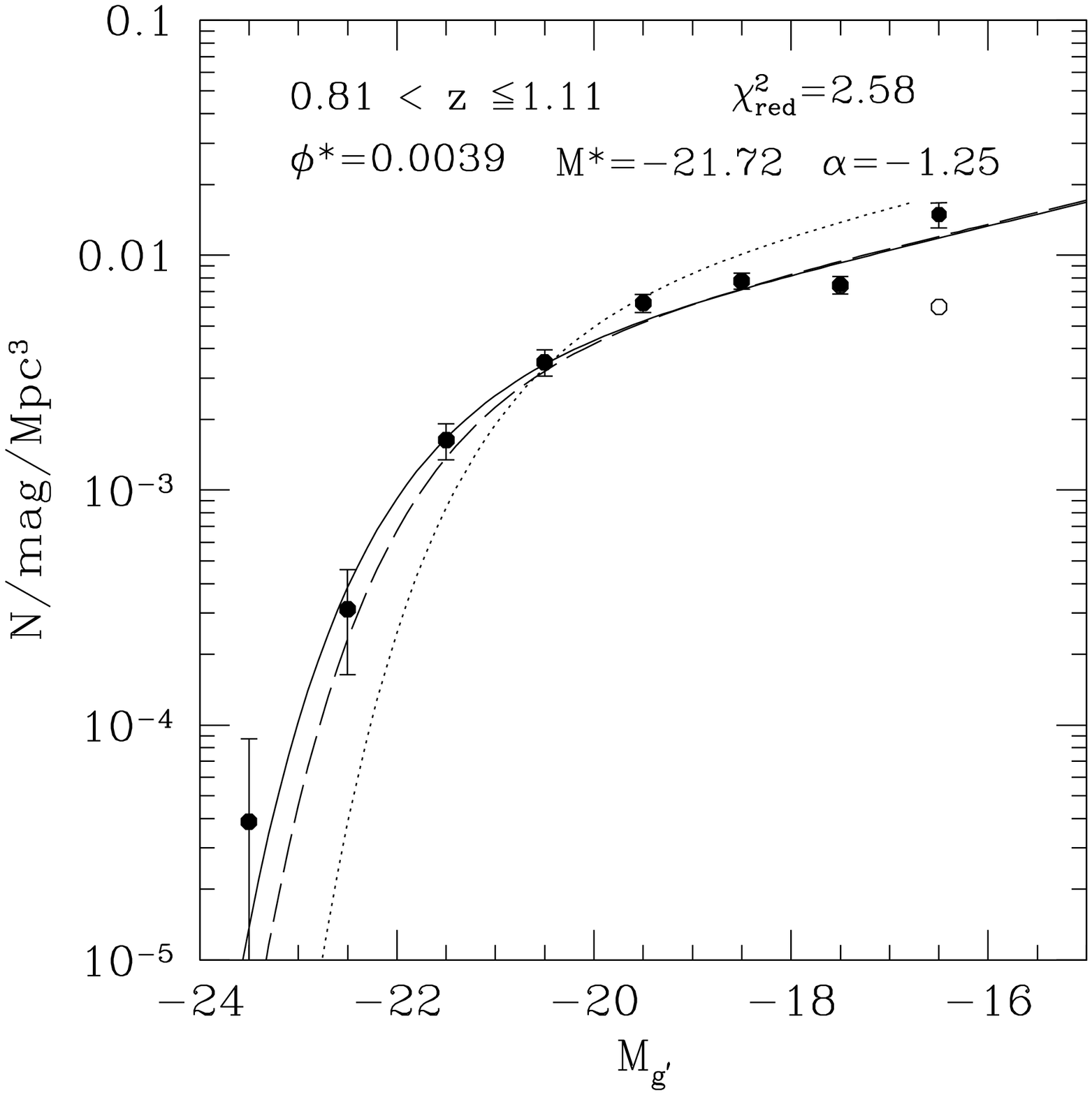}
\includegraphics[width=0.33\textwidth]{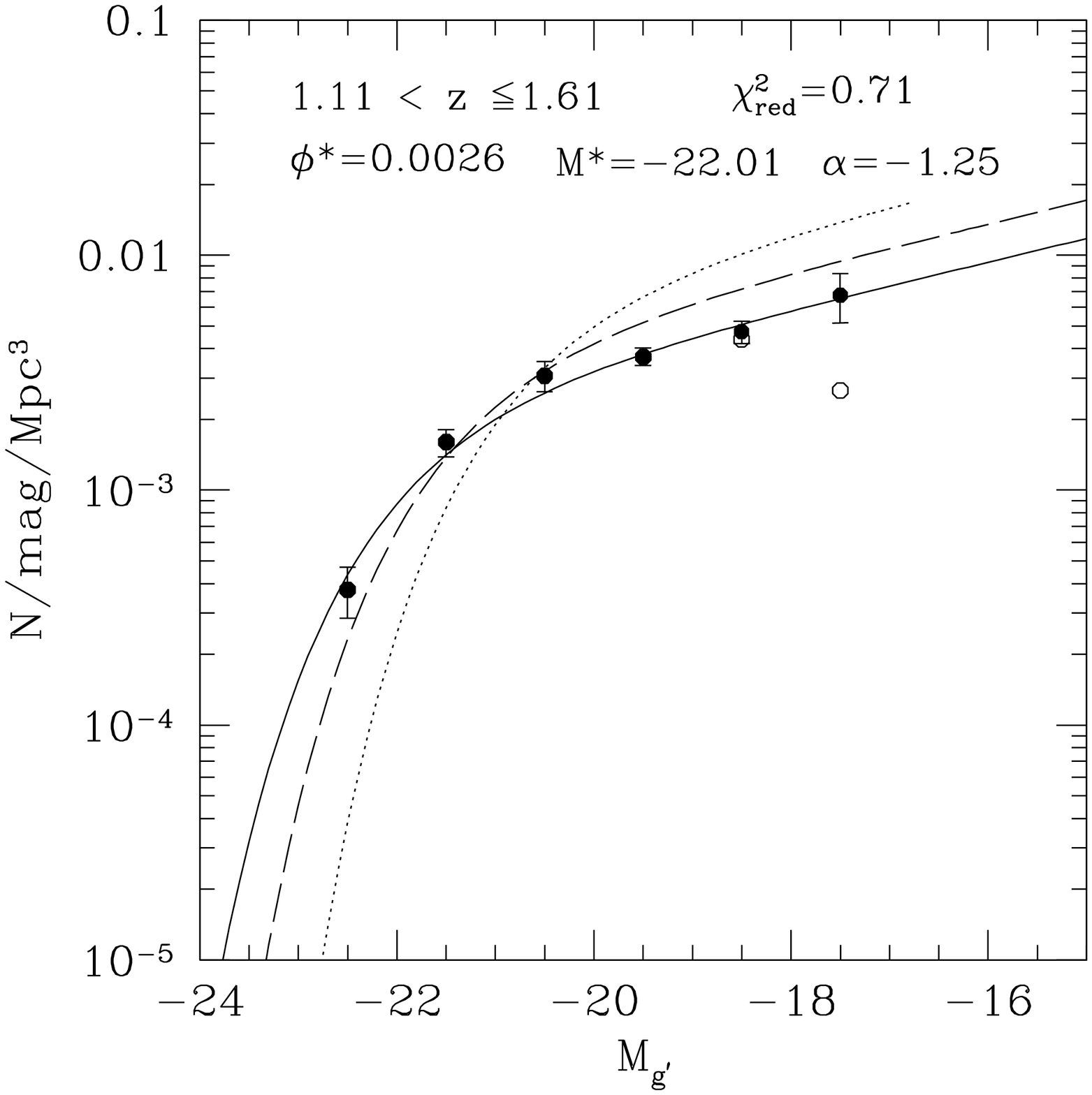}
\includegraphics[width=0.33\textwidth]{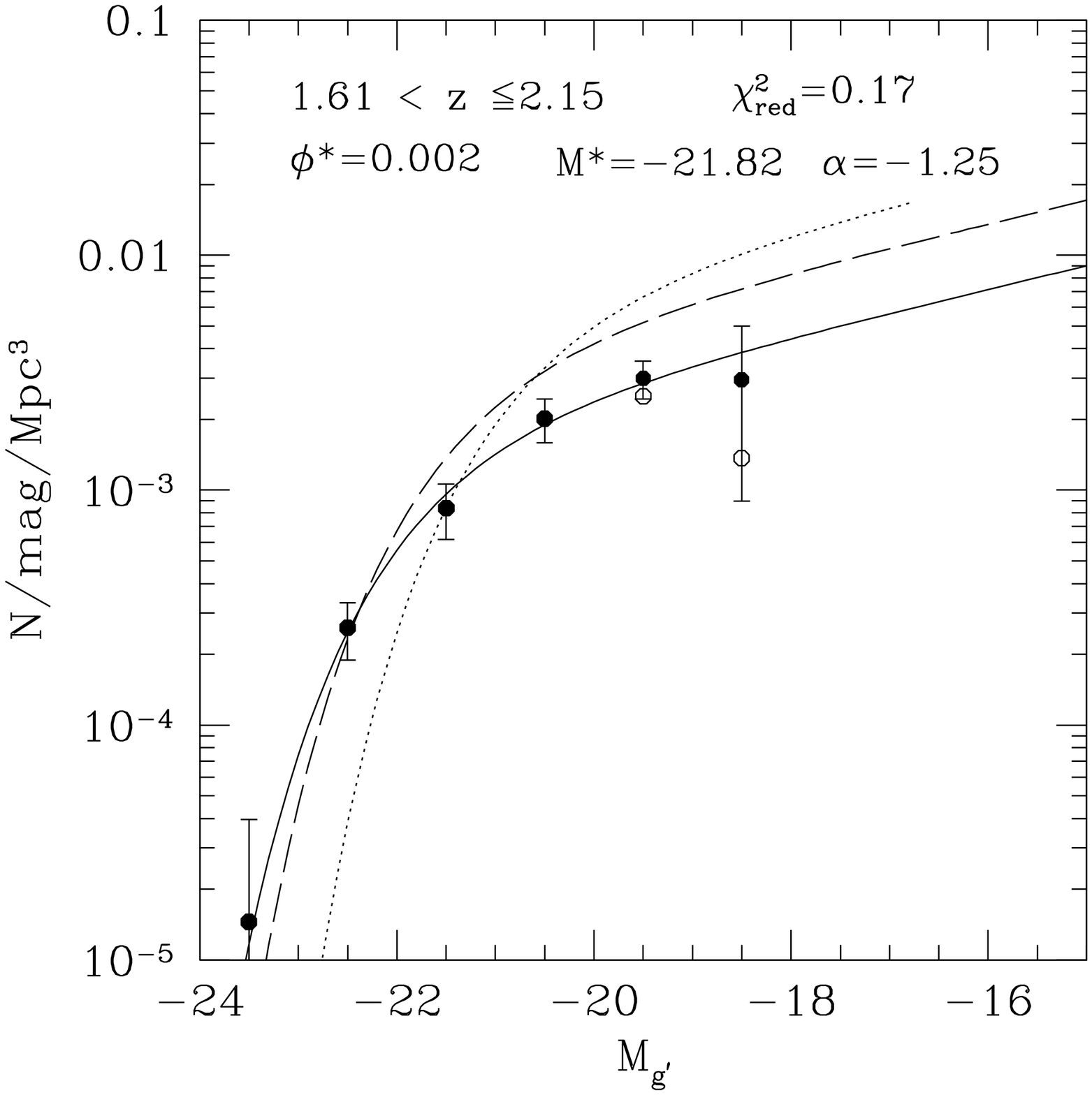}
\includegraphics[width=0.33\textwidth]{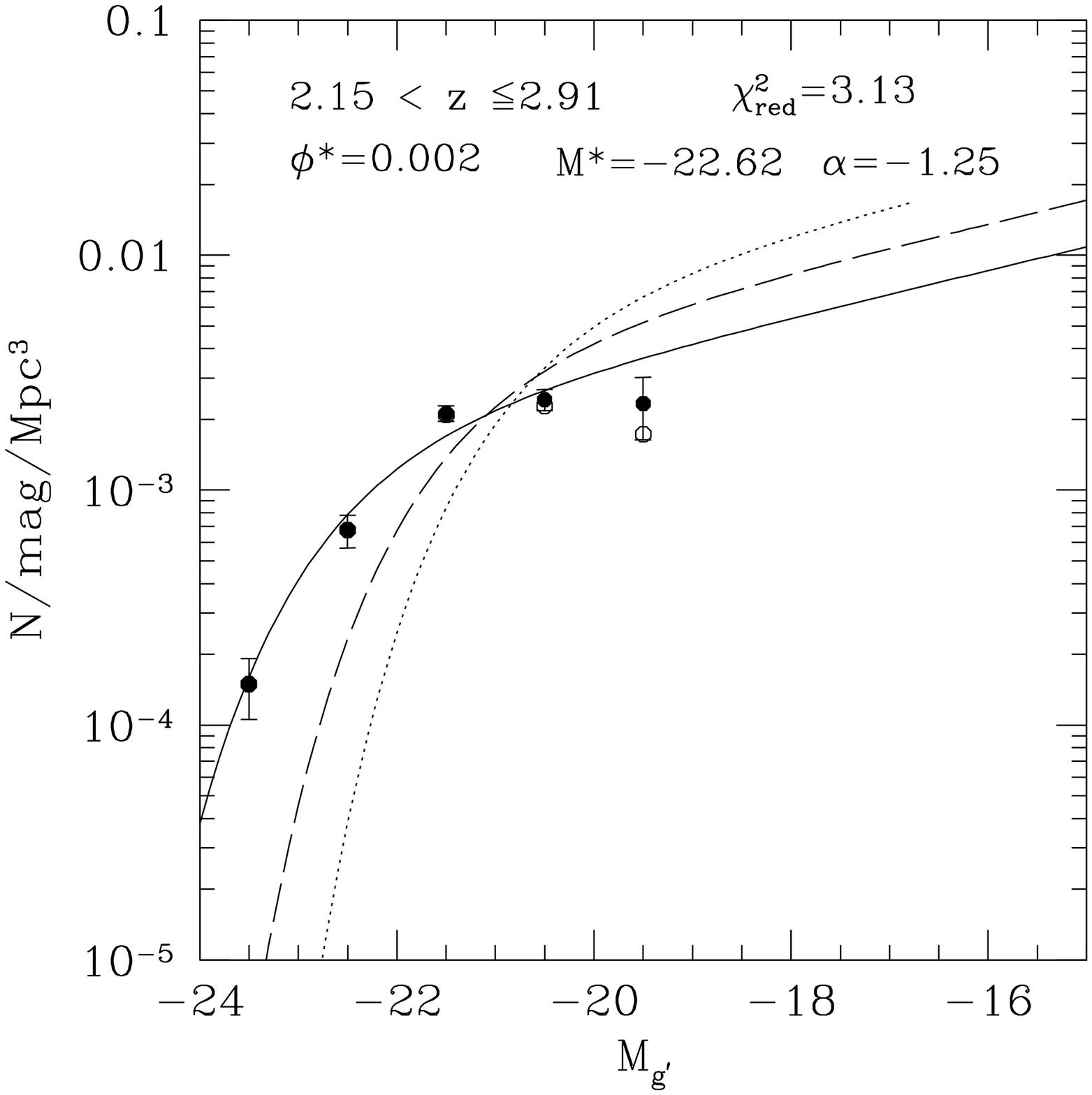}
\includegraphics[width=0.33\textwidth]{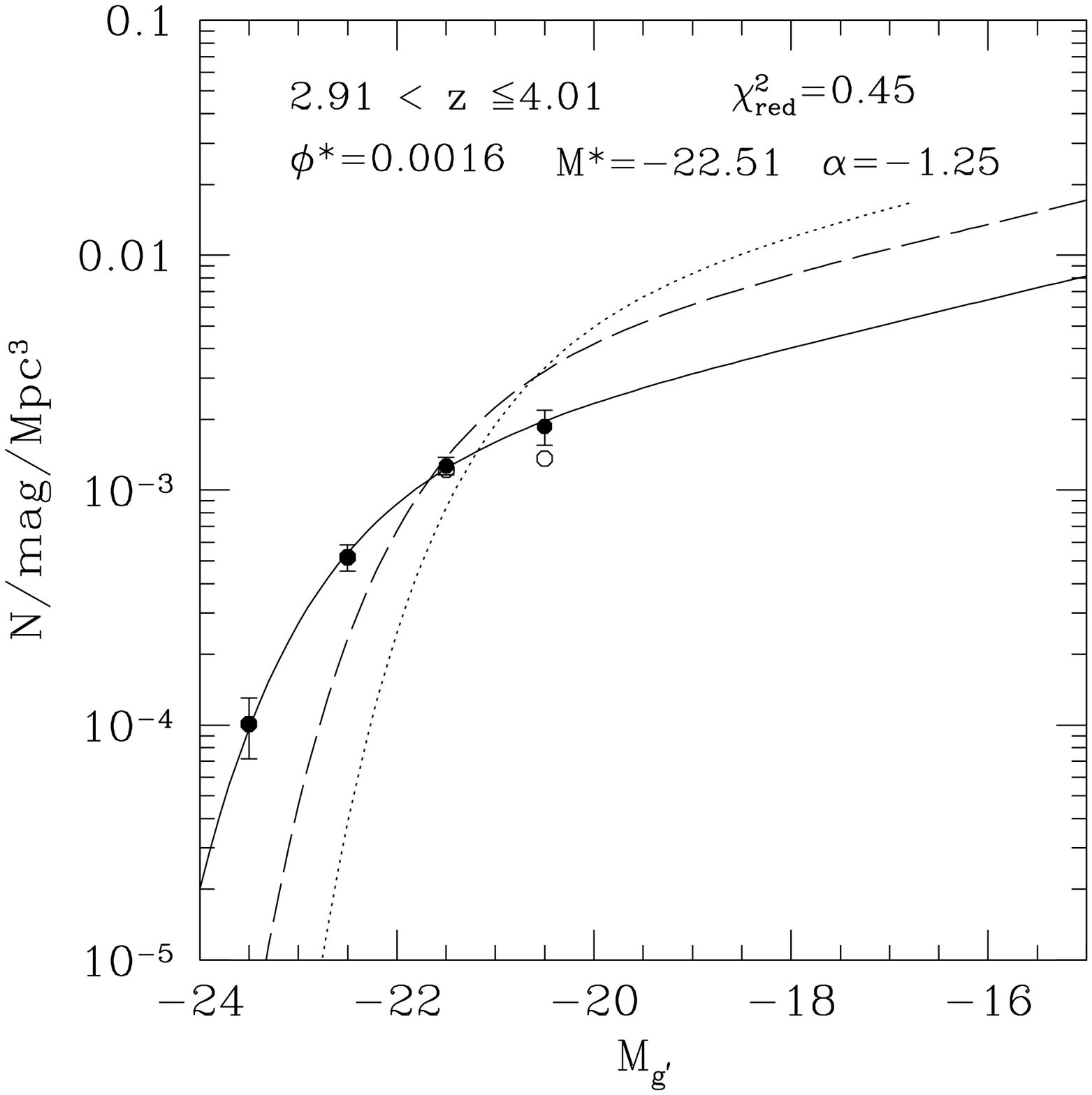}
\includegraphics[width=0.33\textwidth]{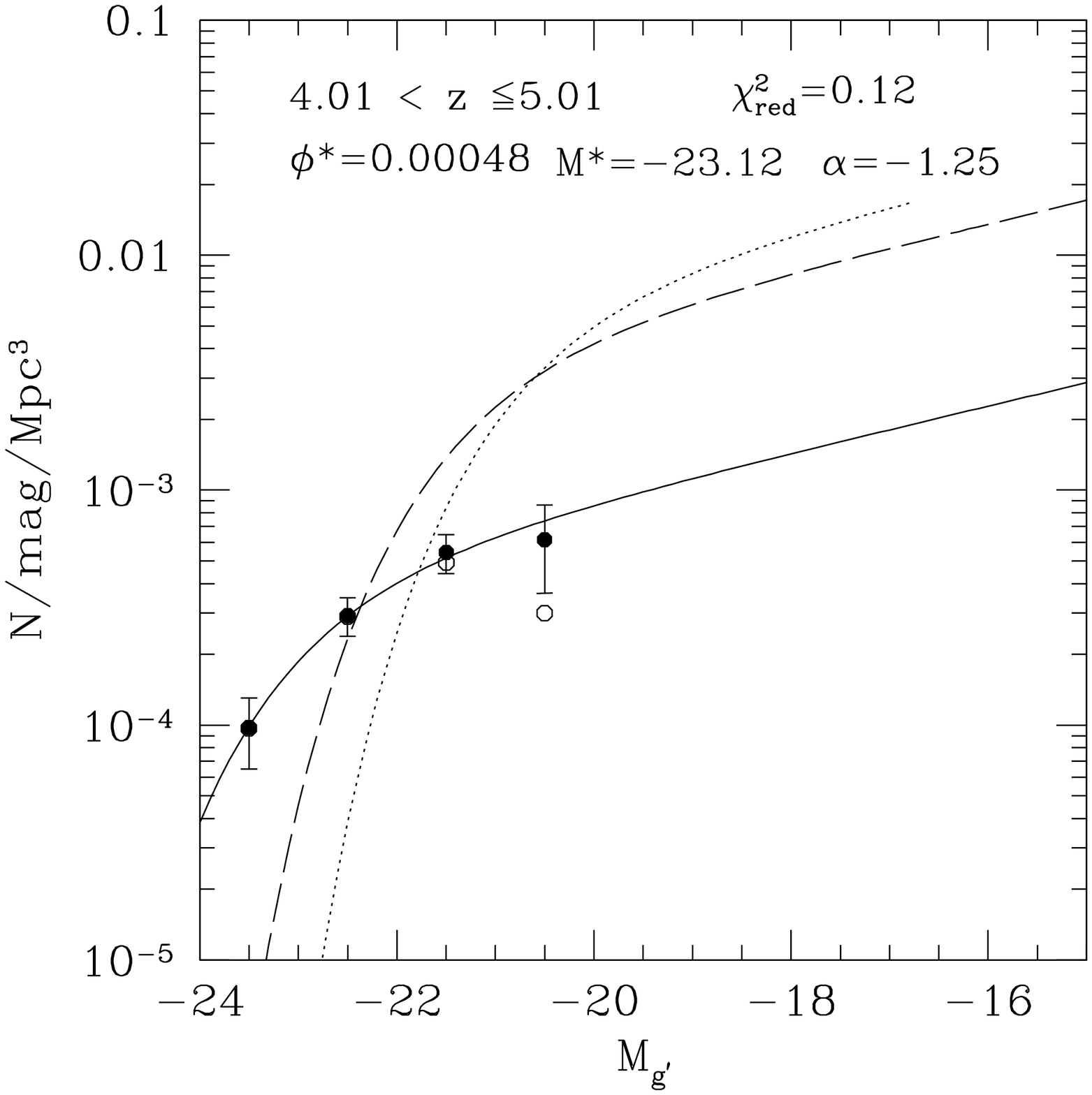}
\includegraphics[width=0.33\textwidth]{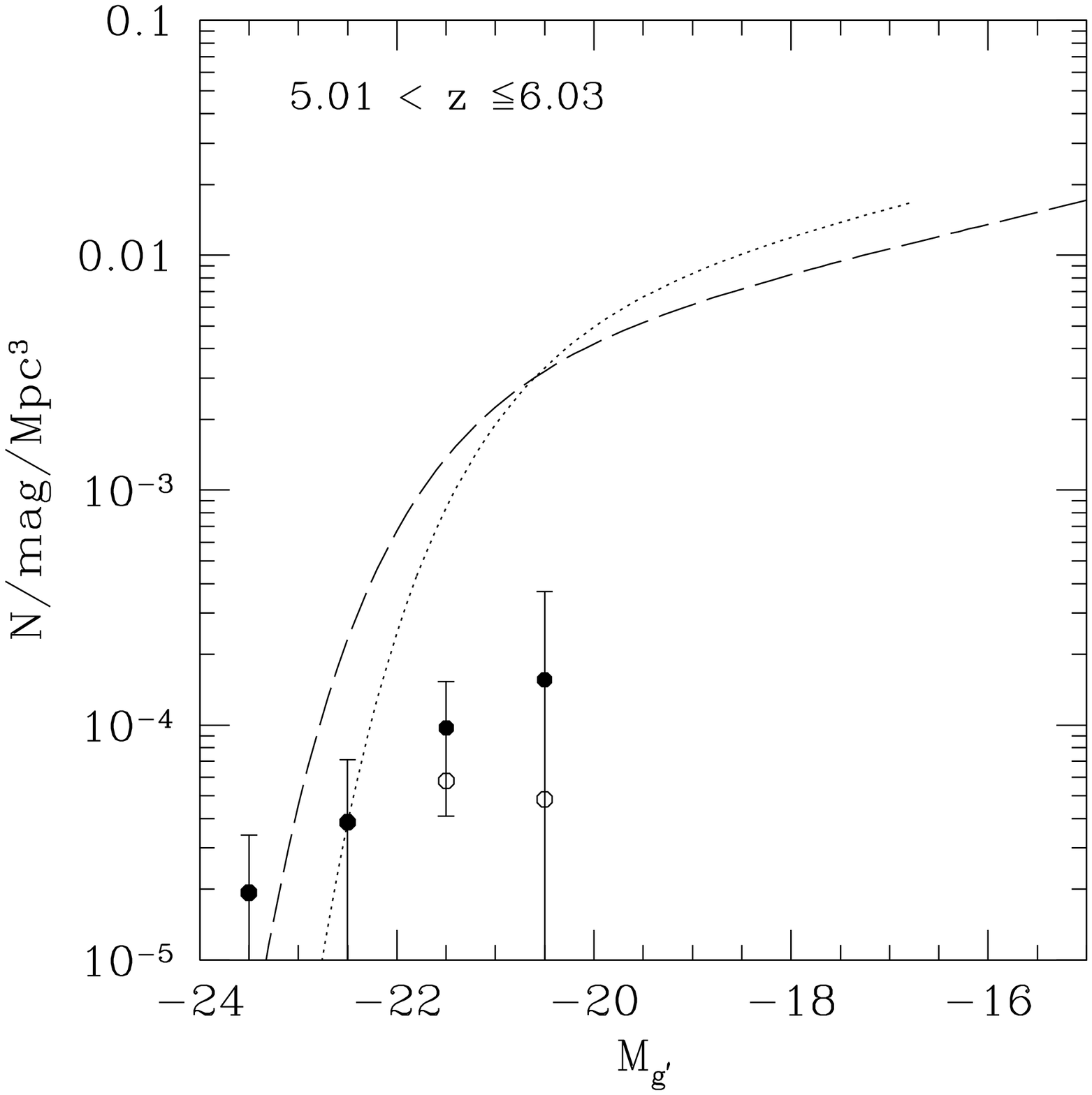}
\caption{\label{fig:lumfkt_fdf_g}
  Luminosity functions in the \textit{g'-band} from low redshift
  (\mbox{$\langle z\rangle=0.3$}, upper left panel) to high redshift
  (\mbox{$\langle z\rangle=5.5$}, lower right panel). The filled
  (open) symbols show the luminosity function corrected (uncorrected)
  for $V/V_{max}$. The fitted Schechter functions for a fixed slope
  $\alpha$ are shown as solid lines. Note that we only fit the
  luminosity functions from $\langle z\rangle=0.6$ to $\langle
  z\rangle=4.5$. The parameters of the Schechter functions can be
  found in Table~\ref{tab:schechter_fit_g}. The dotted line represents
  the local g'-band luminosity function derived from the SDSS
  \citep{blanton:1}. The Schechter fit for redshift $\langle
  z\rangle=0.6$ is indicated as dashed line in all panels.}
\end{figure*}

In this section we analyze the luminosity function in the UV (1500~\AA
, 2800~\AA ), u', g', and B band by means of a Schechter
function fit with fixed slope (see Sect.~\ref{sec:lumfkt:slope}).

In the UV, we evaluate the luminosity function in two rectangular
filters centered at $1500 \pm 100$~\AA \ and $2800 \pm 100$~\AA.
There are three reasons to analyze both wavelengths.  First, for our
lowest redshift bin (\mbox{$\langle z \rangle\sim 0.6$}) the restframe
magnitude derived at 2800~\AA \ is more robust than the one at 1500
\AA\ because the restframe wavelength of 2800~\AA\ corresponds to the
observed U and does not need extrapolation to shorter
wavelength. Second, we also include the 1500~\AA\ luminosity function
as it corresponds to a frequently used reference wavelength and is
very well determined beyond redshifts of 2.5. Third, we want to show
that the galaxy luminosity functions at both wavelengths are very
similar and show the same redshift evolution.

\begin{figure*}[t!]
\centering
\includegraphics[width=0.45\textwidth]{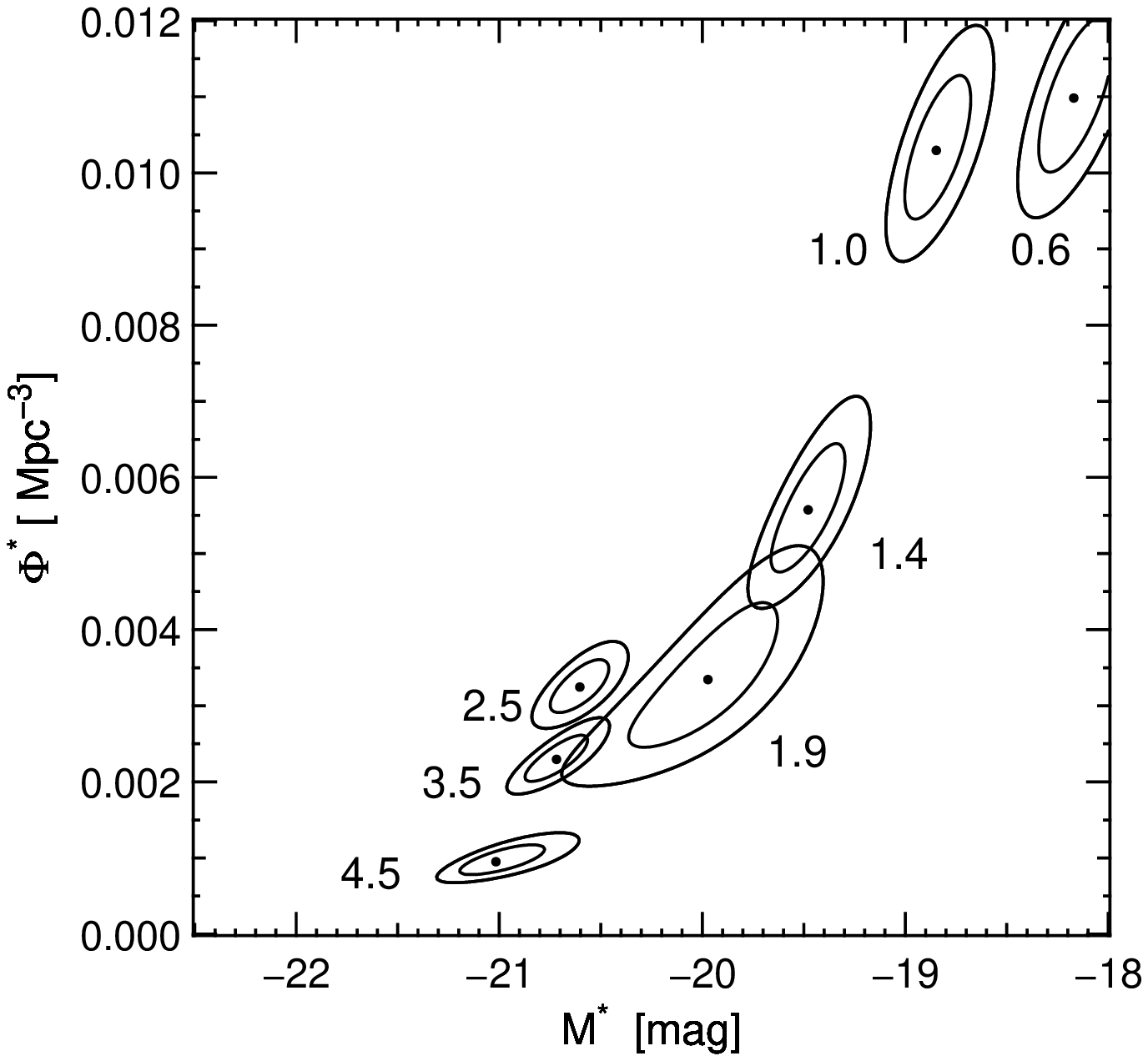}
\hspace*{5mm}
\includegraphics[width=0.45\textwidth]{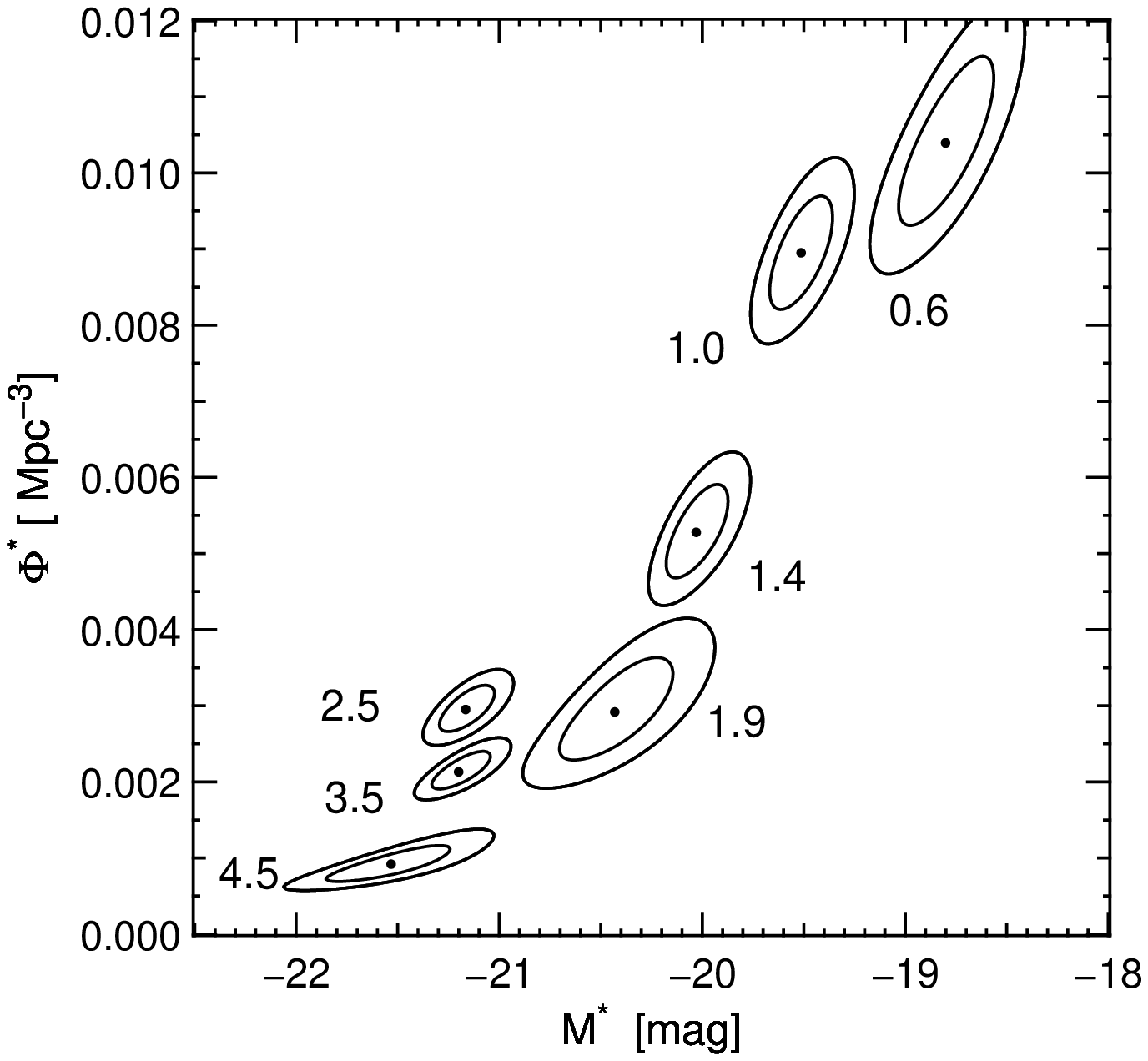}
\includegraphics[width=0.45\textwidth]{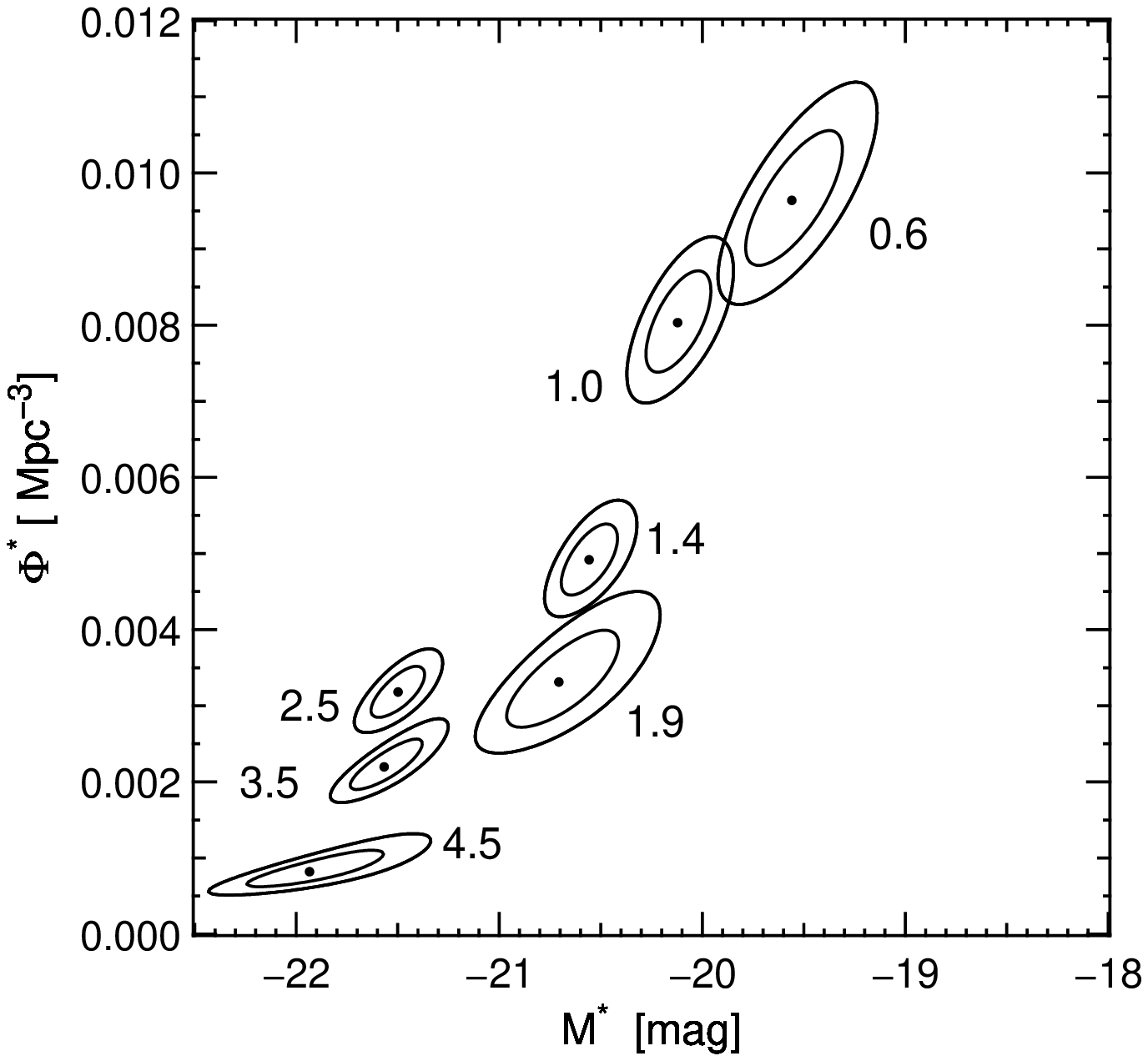}
\hspace*{5mm}
\includegraphics[width=0.44\textwidth]{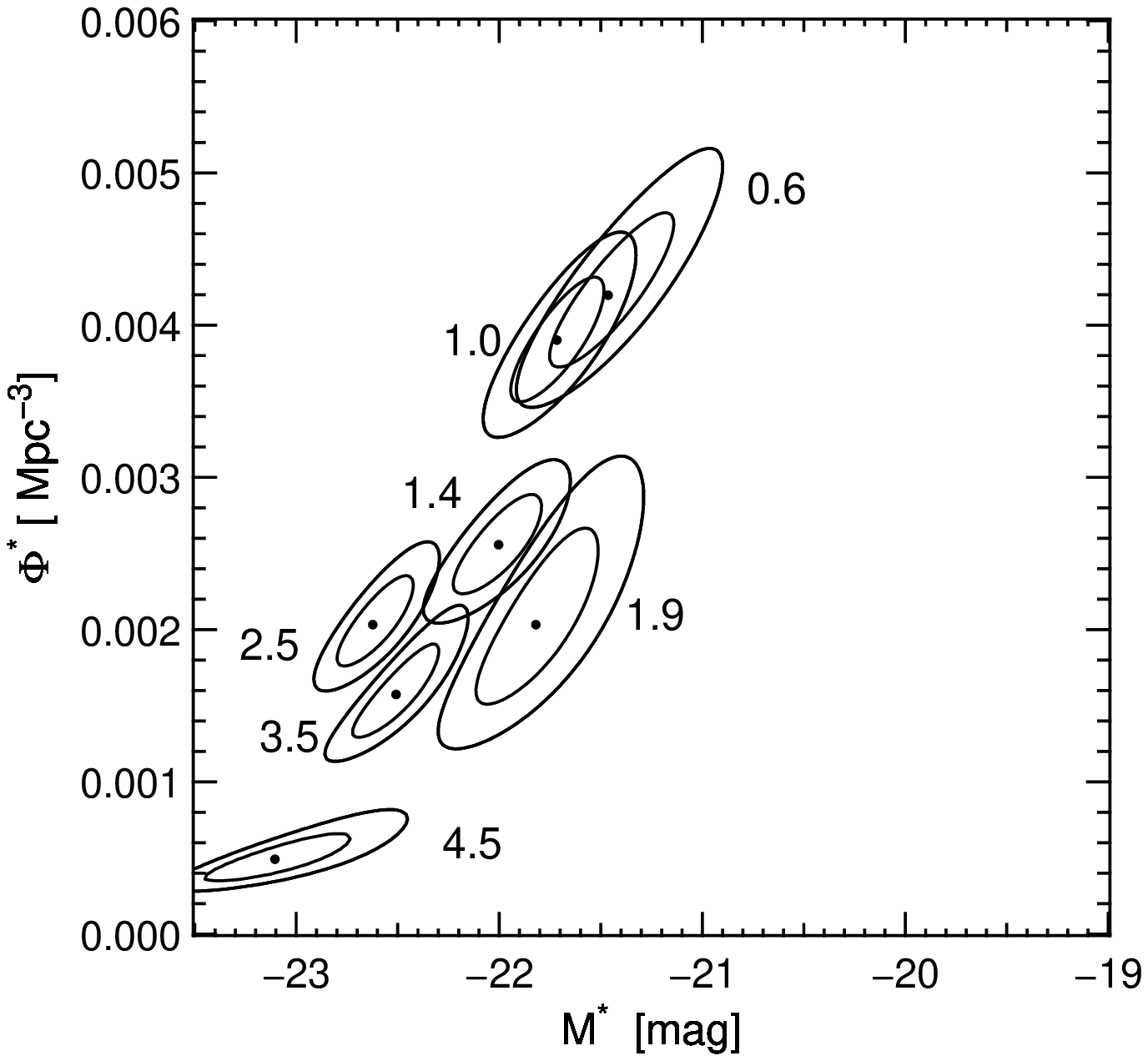}
\caption{\label{fig:likelihood_fdf_all} 
  \mbox{$1\sigma$} and \mbox{$2\sigma$} confidence levels in Schechter
  parameter space for the different redshift bins, labeled by their
  mean redshift.  A Schechter function with a fixed slope (see
  Table~\ref{tab:slope_fixed}) has been fitted to the luminosity
  function at 1500~\AA \ (\textit{upper left panel}), 2800~\AA \ 
  (\textit{upper right panel}), in the u'-band (\textit{lower left
    panel}) and in the g'-band (\textit{lower right panel}). The
  parameters of the Schechter function can be found in Table
  \ref{tab:schechter_fit_1500}, Table~\ref{tab:schechter_fit_2800}
  Table~\ref{tab:schechter_fit_u} and Table~\ref{tab:schechter_fit_g}.
  }
\end{figure*}

In the optical bands, we calculated the evolution of the luminosity
functions in the u' and g' bands (g' of SDSS, see \citealt{fukugita:1},
not to be confused with Gunn g which was part of the filter set with
which we observed the FDF). 
Because many authors have already published
luminosity functions in the Johnson B-band, we include also this
filter in our analysis.

In Figs.~\ref{fig:lumfkt_fdf_2800} and \ref{fig:lumfkt_fdf_g} we
present the luminosity functions at 2800~\AA\ and in the g' band,
while the results at 1500~\AA\ as well as for the u' and B bands can
be found in Figs.~\ref{fig:lumfkt_fdf_1500}, \ref{fig:lumfkt_fdf_u}
and \ref{fig:lumfkt_fdf_b} in appendix~\ref{sec:schechter_parameter}.
The filled (open) symbols denote the luminosity function with
(without) completeness correction.

Even without fitting Schechter functions to the data, it is obvious
that there is strong evolution in characteristic luminosity and number
density between redshifts 0.6 and 4.5.

The solid lines show the Schechter function fitted to the luminosity
function. The best fitting Schechter parameter, the redshift binning
as well as the reduced $\chi^2$ are also listed.  The reduced $\chi^2$
are quite good for all but one redshift bin ($2.15 <z\le 2.91$). The
slope we adopted is not suitable for that bin and increases the
$\chi^2$.  The depth of the FDF allows us to trace the luminosity
function over a relatively large magnitude range. Even in our highest
redshift bin ($4.01 <z\le 5.01$) the luminosity function spans an
interval of 4 magnitudes.

In Fig.~\ref{fig:likelihood_fdf_all} we show the \mbox{$1\sigma$} and
\mbox{$2\sigma$} confidence contours of M$^\ast$ and $\phi^\ast$ for
the different redshift bins, illustrating the correlation of the two
Schechter parameters. The contours correspond to $\Delta \chi^2 =
2.30$ and $\Delta \chi^2 = 6.17$ above the minimum $\chi^2$.  The best
fitting Schechter parameters and their $1\sigma$ errors are summarized
in Tables~\ref{tab:schechter_fit_1500}, \ref{tab:schechter_fit_2800},
\ref{tab:schechter_fit_u}, \ref{tab:schechter_fit_g} and
\ref{tab:schechter_fit_b} for the 1500~\AA , 2800~\AA , u', g' and B
bands, respectively.  The \mbox{$1\sigma$} errorbars of the single
parameters are derived from the projections of the two-dimensional
contours using $\Delta \chi^2 = 1$.

\textit{We find a systematic brightening of M$^\ast$ and a systematic
  decrease of $\phi^\ast$ from low to high redshift}. The evolution is
very strong at 1500~\AA \ (upper left panel), 2800~\AA \ (upper right
panel) and in the u'-band (lower left panel) and moderately strong in
the g'-band (lower right panel).  We do not show the B-band results as
they behave almost identical as the g'-band.  Although the variation
of M$^\ast$ and $\phi^\ast$ between adjacent redshift bins is in part
influenced by large scale structure, the overall trend in the
evolution of M$^\ast$ and $\phi^\ast$ is very robust.

  Since the integral of the luminosity function in the UV is strongly
  related to the star-formation rate (SFR) \citep{mad_poz_dick1}, we
  can derive the star-formation history from the evolution of the
  luminosity function. The brightening of M$^\ast$ and decrease of
  $\phi^\ast$ in the UV leads to an increase of the SFR between $0.5 <
  z < 1.5 $, whereas it remains approximately constant between $1.5 <
  z < 4.0 $. A detailed analysis of the star-formation history will be
  presented in a future paper (Gabasch et al., in preparation),
  preliminary results are published in \citet{gabasch_mykonos:1}.

\section{Parameterizing the evolution of the luminosity function}
\label{sec:evol_parameter}

In order to quantify the redshift evolution of M$^\ast$  and  $\phi^\ast$
we assume the simple relations of the form: 

\begin{eqnarray}
  \nonumber
   M^\ast  \, (z) & = & M^\ast_0 \, + \, a \ln (1+z) , \; \\
  \phi^\ast \, (z)& = & \phi^\ast_0 \: \left( 1 \, + \, z
   \right)^b, \; \mathrm{and}   
\label{e:evol_mstar_phistar}\\
  \nonumber
  \alpha \, (z) & = & \alpha_0 \;\; \equiv \;\; \mathrm{const} .
\end{eqnarray}
Parameterizing \mbox{$M^\ast(z)= M^\ast_0 + a \ln(1+z)$} is equivalent to
a dependence  of \mbox{$L^\ast(z)= L^\ast_0 (1+z)^{ \xi}$} with
\mbox{$\xi=-0.4 \ln(10) a \approx -0.921a$}.

\begin{figure}[t!]
\includegraphics[width=0.5\textwidth]{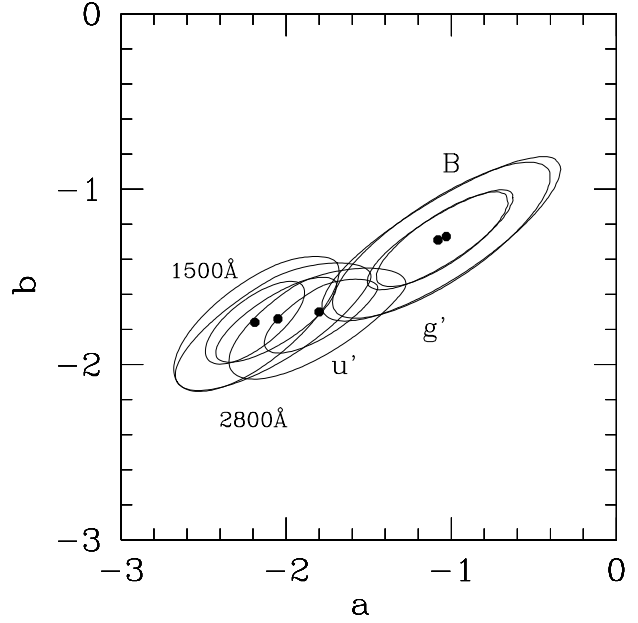}
\caption{\label{fig:evol_param}
  \mbox{$1\sigma$} and \mbox{$2\sigma$} confidence levels of the
  parameters $a$ and $b$ in different bands (1500~\AA, 2800~\AA, u',
  g' and B) for the evolutionary model described in the text. The
  values for $a$ and $b$ can be found in Table~\ref{tab:evo_param}. }
\end{figure}

The best fitting values for $a$, $b$, $M^\ast_0$, and $\phi^\ast_0$ are
derived by minimizing

\begin{eqnarray}
  \nonumber
  \chi^2 & = & \chi^2 \, (a, b, M^\ast_0, \phi^\ast_0) \\
  & = & \sum_{j=1}^{N_j} \sum_{i=1}^{N_i} \, \frac{\left[
  \phi (M_{ij}) - \Psi (M_{ij}, z_j, a, b, M^\ast_0,\phi^\ast_0)
  \right]^2}{\sigma_{ij}^2} ,
  \label{e:chi}
\end{eqnarray}

for the galaxy number densities in all magnitude and redshift bins
\textit{simultaneously}.  $\phi (M_{ij})$ is the number density of
galaxies in magnitude bin $i$ at redshift $z_j$; $\Psi (M_{ij}, z_j,
a, b, M^\ast_0, \phi^\ast_0)$ is the Schechter function evolved to the
redshift $z_j$ according to the evolution model defined in
equation~(\ref{e:evol_mstar_phistar}), and $\sigma_{ij}$ is the rms
error of the luminosity function value. The resulting values for $a$,
$b$, $M^\ast_0$, and $\phi^\ast_0$ can be found in
Table~\ref{tab:evo_param}.

The \mbox{$1\sigma$} and \mbox{$2\sigma$} confidence levels of the
evolution parameters $a$ and $b$ are shown for the different filters
in Fig.~\ref{fig:evol_param}. These contours were derived by
projecting the four-dimensional $\chi^2$ distribution to the $a$-$b$
plane, i.e.\ for given $a$ and $b$ we use the value of $M^\ast_0$ and
$\phi^\ast_0$ which minimizes the $\chi^2(a,b)$.

\begin{figure*}
\includegraphics[width=0.50\textwidth]{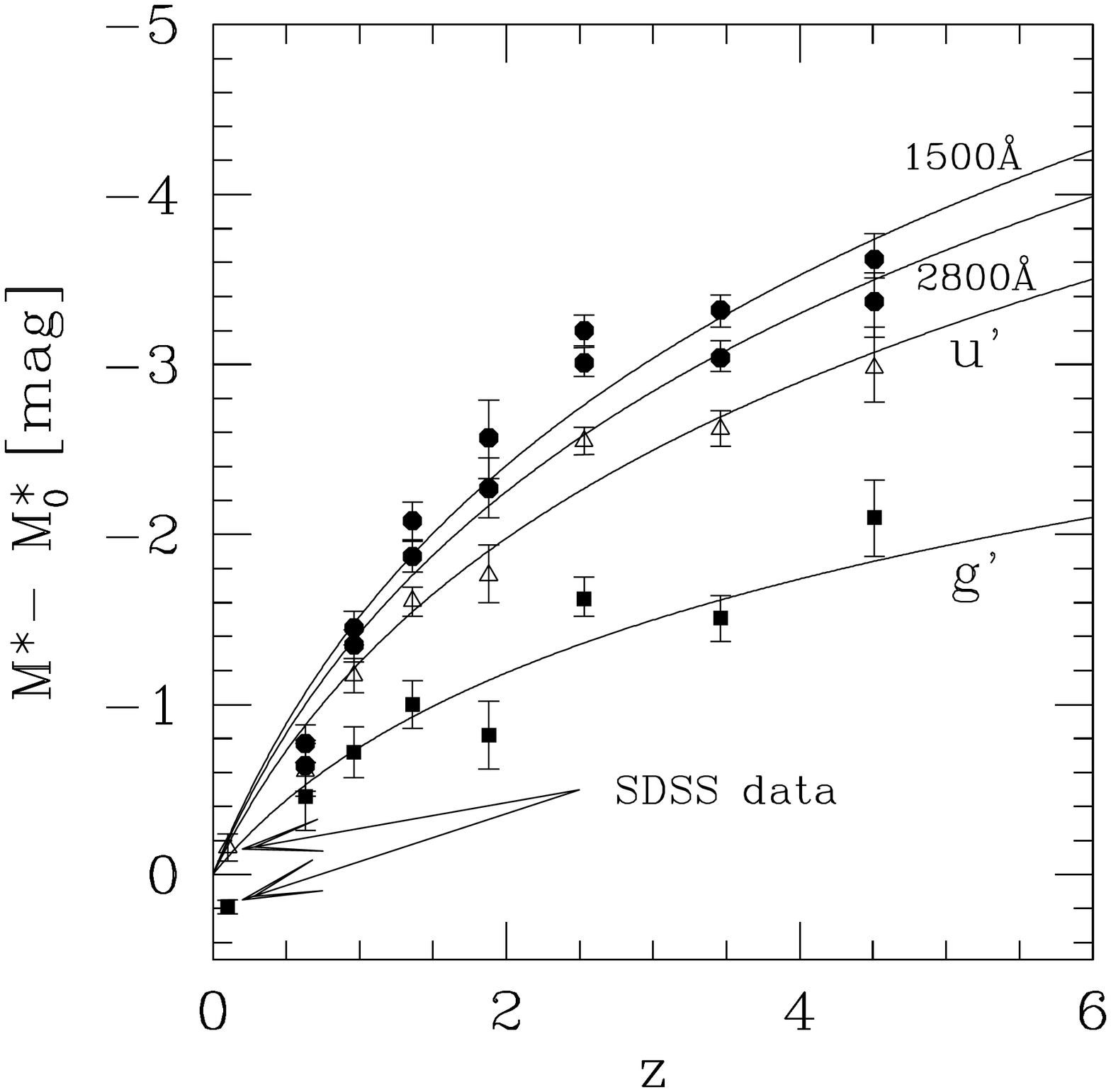}
\includegraphics[width=0.50\textwidth]{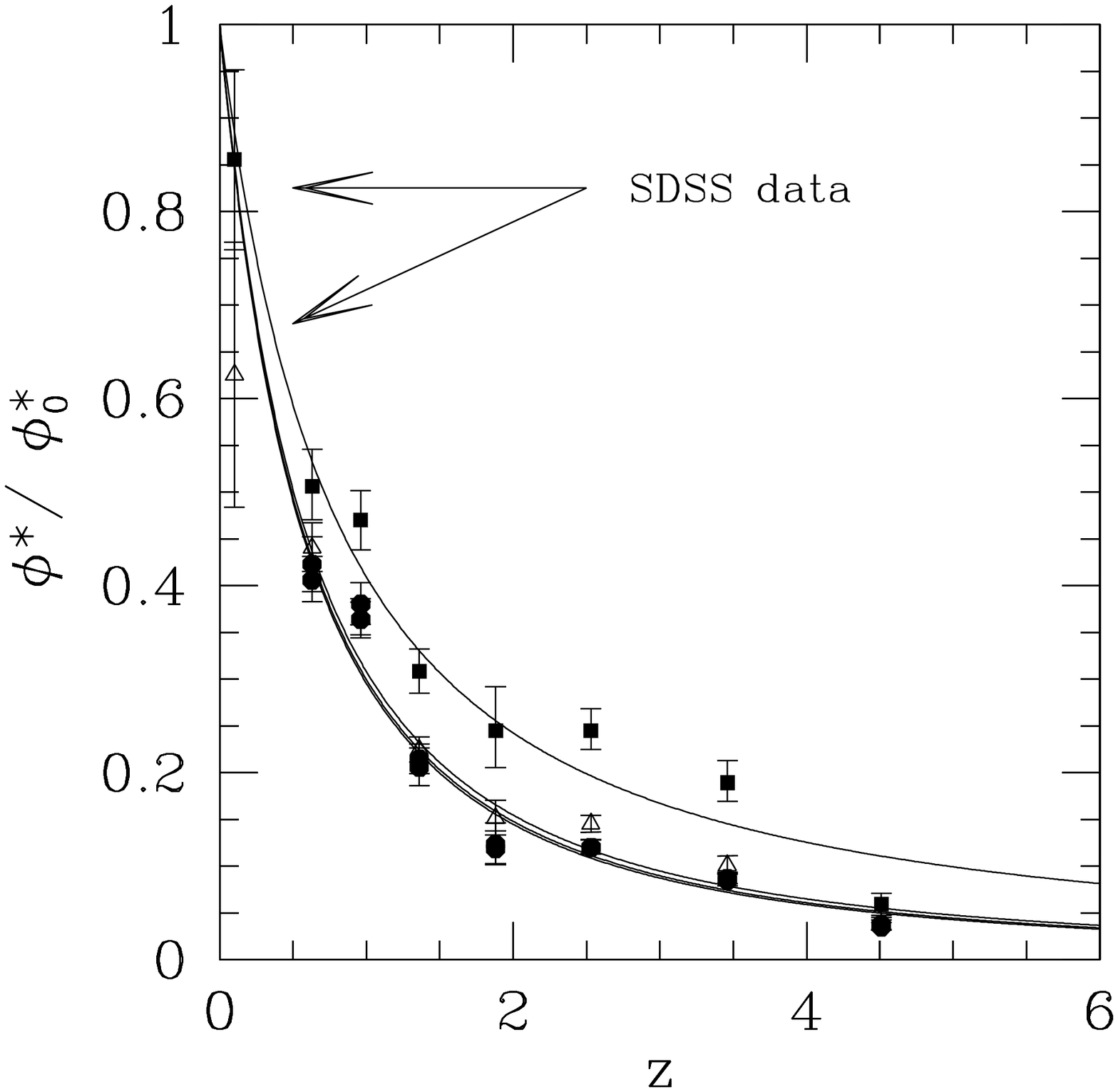}
\caption{\label{fig:evol_lumfkt_fdf}
  Redshift evolution of M$^\ast$ (left panel) and $\phi^\ast$ (right
  panel) for the filters g' (filled squares), u' (open triangles) and
  the two UV bands at 2800~\AA\ and 1500~\AA\ (filled circles).  The
  arrows mark the values for M$^\ast$ and $\phi^\ast$ as derived in
  the SDSS \citep{blanton:1}.}
\end{figure*}

In Fig.~\ref{fig:evol_lumfkt_fdf} we show the relative redshift
evolution of $M^\ast$ (left panel) and $\phi^\ast$ (right panel) in
the chosen filters. The Schechter parameters are the ones given in the
tables in Appendix~\ref{sec:schechter_parameter}. The solid lines show
the relative change according to our evolutionary model with the
parameters from Table~\ref{tab:evo_param}.

Note that $a$, $b$, $M^\ast_0$, and $\phi^\ast_0$ were derived by
minimizing equation~(\ref{e:chi}) and not the differences between the
(best fitting) lines and the data points in
Fig.~\ref{fig:evol_lumfkt_fdf}.

Fig.~\ref{fig:evol_lumfkt_fdf} shows that the simple parameterization
we have chosen with equation~(\ref{e:evol_mstar_phistar}) describes
the evolution of the galaxy luminosity functions very well. Still, the
reduced $\chi^2_\nu$ values are somewhat larger than unity ($\sim 4$),
because our approximations for evolution and faint-end slope may not
be adequate for every redshift bin and because of the influence of
large scale structure. Nevertheless, as there are no stringent
theoretical predictions for the evolution of $M^\ast$ and $\phi^\ast$
we do not want to increase the number of free parameters, but increase
the errors of $a$, $b$, $M^\ast_0$, and $\phi^\ast_0$ instead. We do
this by an appropriate scaling of the errors $\sigma_{ij}$ of
equation~(\ref{e:chi}) to obtain a $\chi^2_\nu$ of unity.

For comparison, we also show in Fig.~\ref{fig:evol_lumfkt_fdf} the
local values from the SDSS \citep{blanton:1}. There is good agreement
in the u'-band for both $M^\ast$ and $\phi^\ast$ between our
extrapolated values and the SDSS values.  In the g'-band the value of
$M^\ast$ is lower than the predicted one, but still within the
$1\sigma$ error of the $M^\ast_0$.

\begin{table}[]
\setlength{\tabcolsep}{1mm}
  \caption[]{\label{tab:evo_param}Evolution parameters according to
    equation~(\ref{e:evol_mstar_phistar})}
  \begin{center}
    \begin{tabular}{c|cccc}
      filter& a  & b & M$^\ast_0$ & $\phi^\ast_0 $   \\
      &  &  & (mag) & ($10^{-2} \, \mathrm{Mpc}^{-3}$)   \\

      \hline
      \rule[+3mm]{-1.4mm}{3mm}
      \rule[-3mm]{0mm}{3mm}1500~\AA & $-2.19^{+0.19}_{-0.19}$ & $-1.76^{+0.15}_{-0.15}$ & $-17.40^{+0.25}_{-0.22} $ & $2.71^{+0.47}_{-0.38}$\\ 
      \rule[-3mm]{0mm}{3mm}2800~\AA & $-2.05^{+0.23}_{-0.24}$ & $-1.74^{+0.15}_{-0.16}$ & $-18.16^{+0.27}_{-0.26} $ & $2.46^{+0.39}_{-0.37}$\\ 
      \rule[-3mm]{0mm}{3mm}u'       & $-1.80^{+0.24}_{-0.21}$ & $-1.70^{+0.14}_{-0.15}$ & $-18.95^{+0.24}_{-0.26} $ & $2.19^{+0.37}_{-0.28}$\\ 
      \rule[-3mm]{0mm}{3mm}g'       & $-1.08^{+0.30}_{-0.28}$ & $-1.29^{+0.18}_{-0.18}$ & $-21.00^{+0.32}_{-0.31} $ & $0.83^{+0.15}_{-0.12}$\\ 
      \rule[-3mm]{0mm}{3mm}B        & $-1.03^{+0.23}_{-0.28}$ & $-1.27^{+0.16}_{-0.19}$ & $-20.92^{+0.32}_{-0.25} $ & $0.82^{+0.14}_{-0.12}$\\ 
    \end{tabular}
  \end{center}
\end{table}

\begin{figure*}
\includegraphics[width=0.33\textwidth]{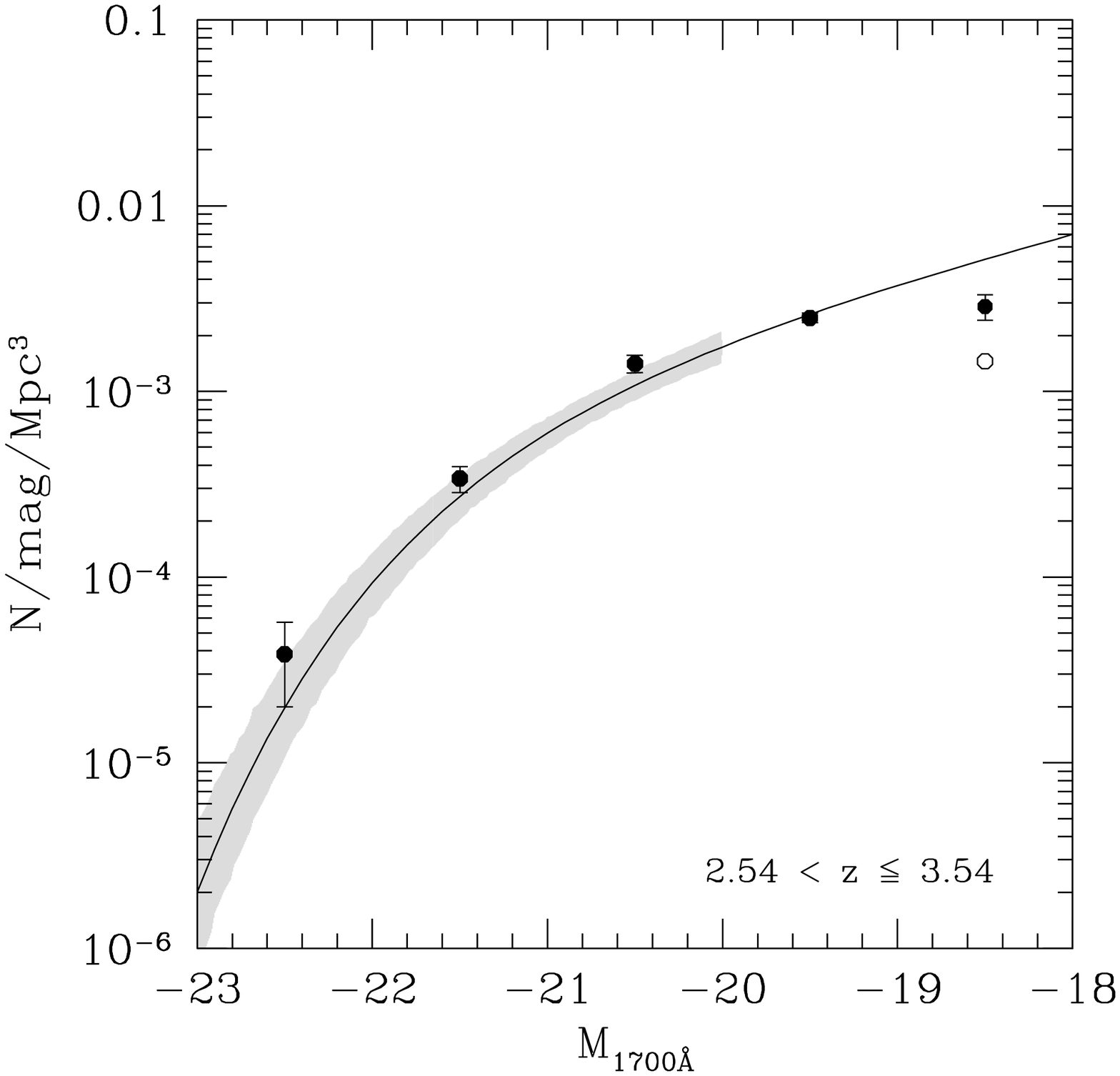}
\includegraphics[width=0.33\textwidth]{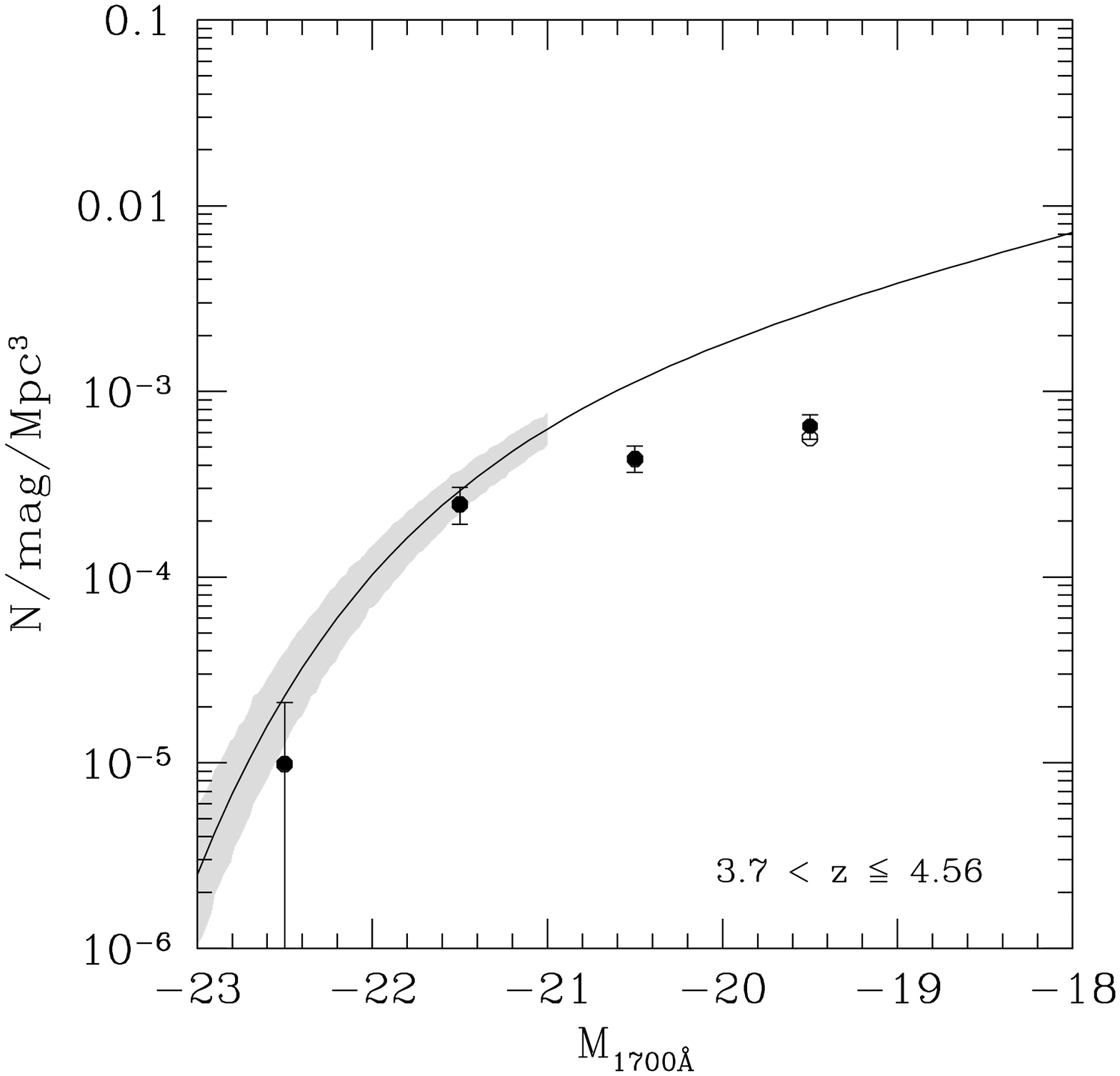}
\includegraphics[width=0.33\textwidth]{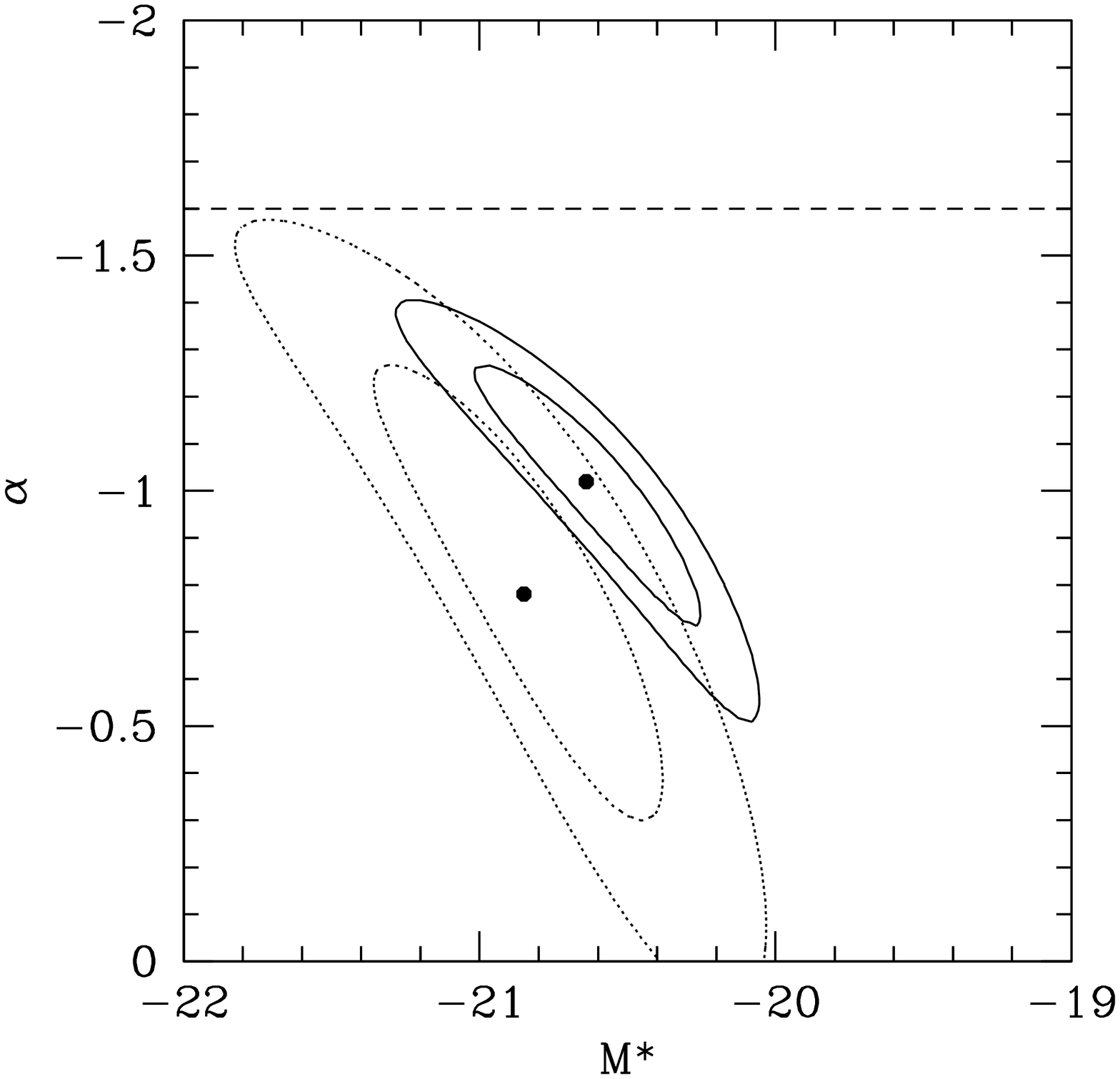}
\caption{\label{fig:lit_uv_steidel}
  Comparison of the luminosity function at 1700~\AA\ of the FDF with
  the Schechter function derived in \textit{\cite{steidel:1}}:
  \mbox{$\langle z \rangle\sim 3.04$} (left panel) and \mbox{$\langle
    z \rangle\sim 4.13$} (middle panel). The filled (open) symbols
  show the luminosity function corrected (uncorrected) for
  $V/V_{max}$.  The shaded region in all plots is based only on
  $\Delta$M$^\ast$, and $\Delta\alpha$ (a detailed discussion
  concerning the errors $\Delta$M$^\ast$, and $\Delta\alpha$ can be
  found in Sect.~\ref{sec:lit}), where the cut-off at low luminosity
  indicates the limiting magnitude of the sample.  In the right panel
  the \mbox{$1\sigma$} and \mbox{$2\sigma$} confidence levels for
  M$^\ast$ and $\alpha$ for a 3 parameter Schechter fit as derived in
  the FDF in the redshift interval \mbox{$2.54 <z\le 3.54$} (solid
  contours) and \mbox{$3.70 <z\le 4.56$} (dotted contours) are shown.
  The horizontal dashed line marks the slope $\alpha=-1.6$ as derived
  in \cite{steidel:1}.  }
\end{figure*}

\section{Comparison with the literature}
\label{sec:lit}

In this section we compare the luminosity functions derived in the FDF
with the luminosity functions of other surveys. As the cosmology and
the wavebands in which the luminosity functions were determined are
different from ours for most of the surveys we chose the following
approach.  \textit{First} we convert results from the literature to
our cosmology (\mbox{$\Omega_M=0.3$}, \mbox{$\Omega_\Lambda=0.7$} and
\mbox{$H_0=70 \, \mathrm{km} \, \mathrm{s}^{-1} \,
\mathrm{Mpc}^{-1}$}). Note that this conversion may not be perfect,
because we can only transform number densities and magnitudes but lack
the knowledge of the individual magnitudes and redshifts of the
galaxies.  Nevertheless, the errors introduced in this way are not
large and the method is suitable for our purpose.  \textit{Second}, in
order to avoid uncertainties due to conversion between different
filter bands, we always use the same band as the survey we want to
compare with.  \textit{Third}, we also try to use the same redshift
binning if possible. In addition, if the number of galaxies in the
FDF is too small to derive a well sampled luminosity function we
increase the binning. 

To visualize the errors of the literature luminosity functions we
perform Monte-Carlo simulations using the $\Delta$M$^\ast$,
$\Delta\phi^\ast$, and $\Delta\alpha$ given in the papers. In cases
where not all of these values could be found in the paper, this is
mentioned in the figure caption.  We do not take into account any
correlation between the Schechter parameters and assume a Gaussian
distribution of the errors $\Delta$M$^\ast$, $\Delta\phi^\ast$, and
$\Delta\alpha$.  From 1000 simulated Schechter functions we derive the
region where \mbox{68.8 \%} of the realizations lie.  The resulting
region, roughly corresponding to 1$\sigma$ errors, is shaded in the
figures.  The luminosity functions derived in the FDF are also shown
as filled and open circles. The filled circles are completeness
corrected whereas the open circles are not corrected. The redshift
binning used to derive the luminosity function in the FDF is given in
the lower right part of every figure.  Moreover, the limiting
magnitude of the respective survey is indicated by the low-luminosity
cut-off of the shaded region in all figures.  If the limiting
magnitude was not explicitly given it was estimated from the figures
in the literature.

We first compare our luminosity functions in the UV to the
results of \citet{steidel:1} and the spectroscopic studies of
\citet{wilson:1}.

Fig.~\ref{fig:lit_uv_steidel} shows a comparison of the 1700~\AA\ 
luminosity function derived by \citet{steidel:1} at redshift
\mbox{$\langle z \rangle\sim 3.04$} (left panel) and \mbox{$\langle z
  \rangle\sim 4.13$} (middle panel) with the luminosity function in
the FDF.  The galaxy sample of \citet{steidel:1} is based on a R-band
(\mbox{$\langle z \rangle\sim 3.04$}) and an I-band (\mbox{$\langle z
  \rangle\sim 4.13$}) selected catalogue and therefore similar to our
I-band selected sample. 
Candidate galaxies were identified with the
Lyman-break technique and most of them spectroscopically confirmed
(564  galaxies of the
\mbox{$\langle z \rangle\sim 3.04$} and 46 of the  
\mbox{$\langle z \rangle\sim 4.13$} sample, respectively).

To derive the associated errors (shaded region) of the Schechter
functions derived by \citet{steidel:1} we use the errors of M$^\ast$
and $\alpha$ of the \mbox{$\langle z \rangle\sim 3.04$} sample as
given in Fig.~8 of their paper.  As there are no errors reported for
the \mbox{$\langle z \rangle\sim 4.13$} sample we assume the same
errors as for the \mbox{$\langle z \rangle\sim 3.04$} sample.
Therefore, the shaded region in Fig.~\ref{fig:lit_uv_steidel} (middle
panel) is probably a lower limit.  For the luminosity function in the
FDF we use a redshift binning of $2.54 <z\le 3.54$ (789 galaxies),
and $3.70 <z\le 4.56$ (144 galaxies) with the mean redshift of
\mbox{$\langle z \rangle\sim 3.04$} and \mbox{$\langle z \rangle\sim
  4.13$} to be as close as possible to \citet{steidel:1}'s mean
redshifts.

Fig.~\ref{fig:lit_uv_steidel} (left and middle panel) shows that there
is very good agreement between the results derived in the FDF and
the luminosity function of \citet{steidel:1} if we focus only on the
luminosity function brighter than the limiting magnitudes 
(shaded regions).  On the other hand, because
of the depth of the FDF we can trace the luminosity function 2
magnitudes deeper and therefore give better constraints on the slope
of the Schechter function.  We show in Fig.~\ref{fig:lit_uv_steidel}
(right panel) the \mbox{$1\sigma$} and \mbox{$2\sigma$} confidence
levels for M$^\ast$ and $\alpha$ for a 3 parameter Schechter fit as
derived from the FDF in the redshift interval \mbox{$2.54 <z\le 3.54$}
(solid line) and \mbox{$3.70 <z\le 4.56$} (dotted line).  The steep
slope $\alpha=-1.6$ derived by \citet{steidel:1} as marked by the
horizontal dashed line can be excluded on a \mbox{$2\sigma$} level.\\


\begin{figure*}
\includegraphics[width=0.33\textwidth]{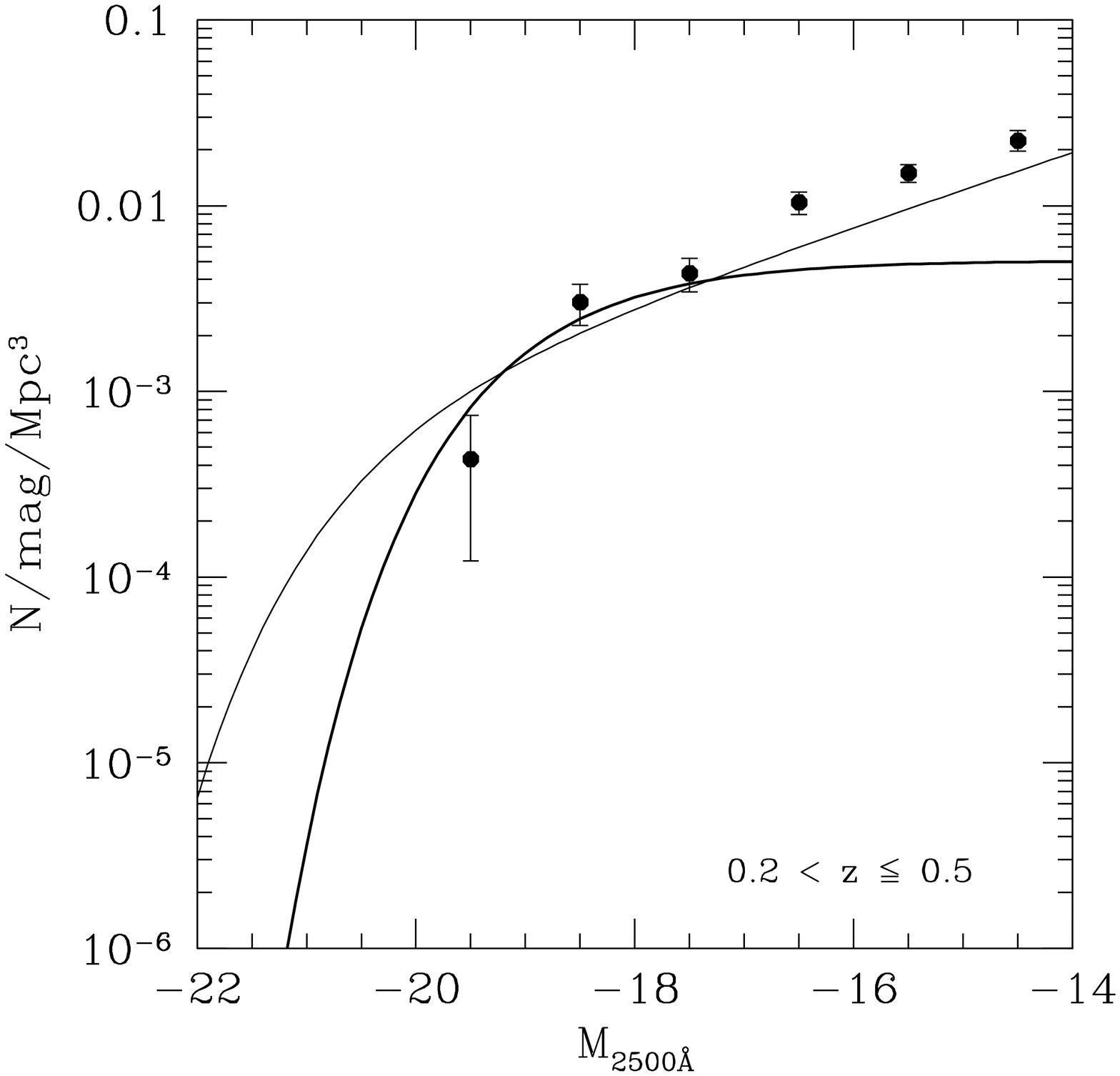}
\includegraphics[width=0.33\textwidth]{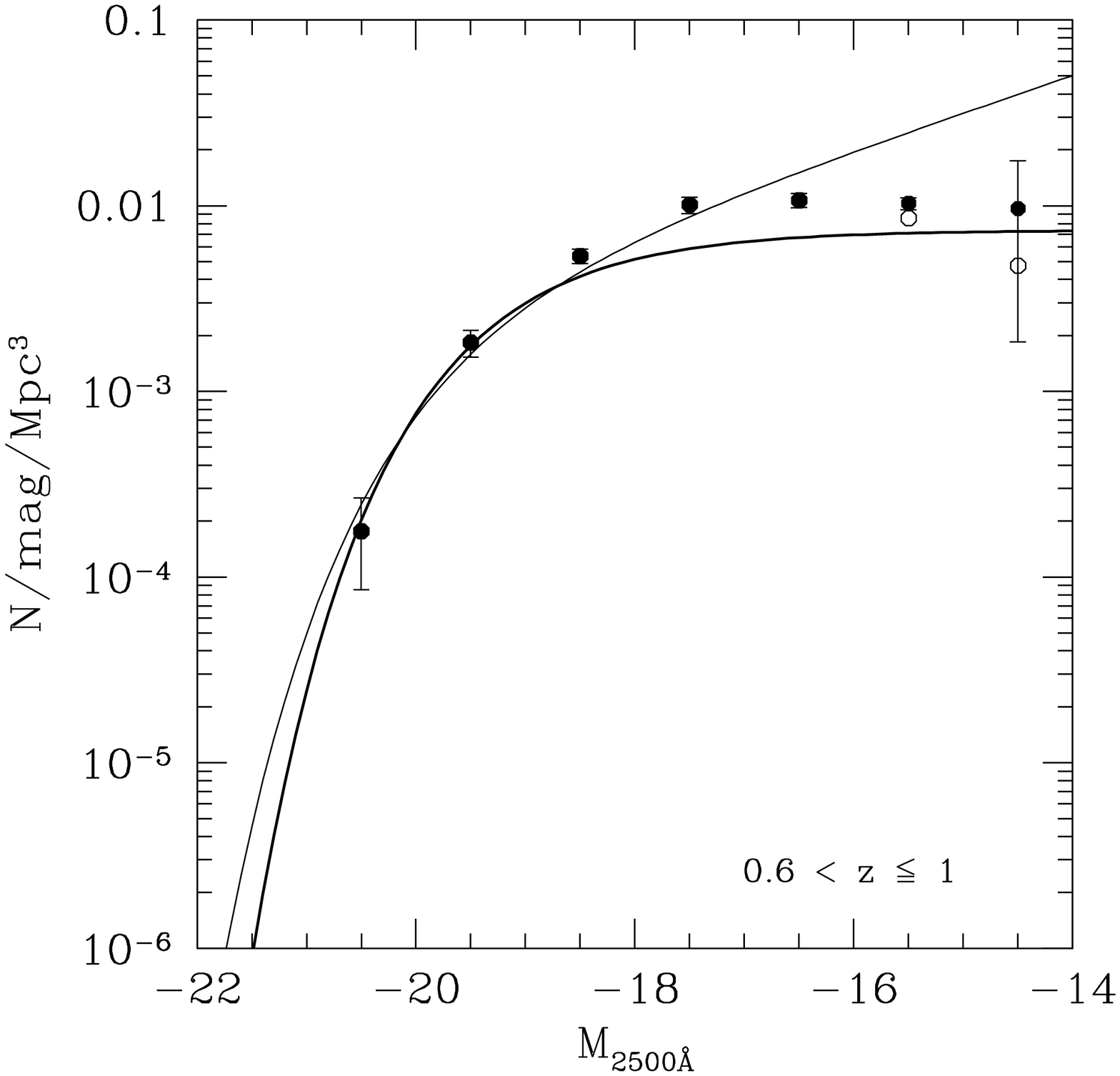}
\includegraphics[width=0.33\textwidth]{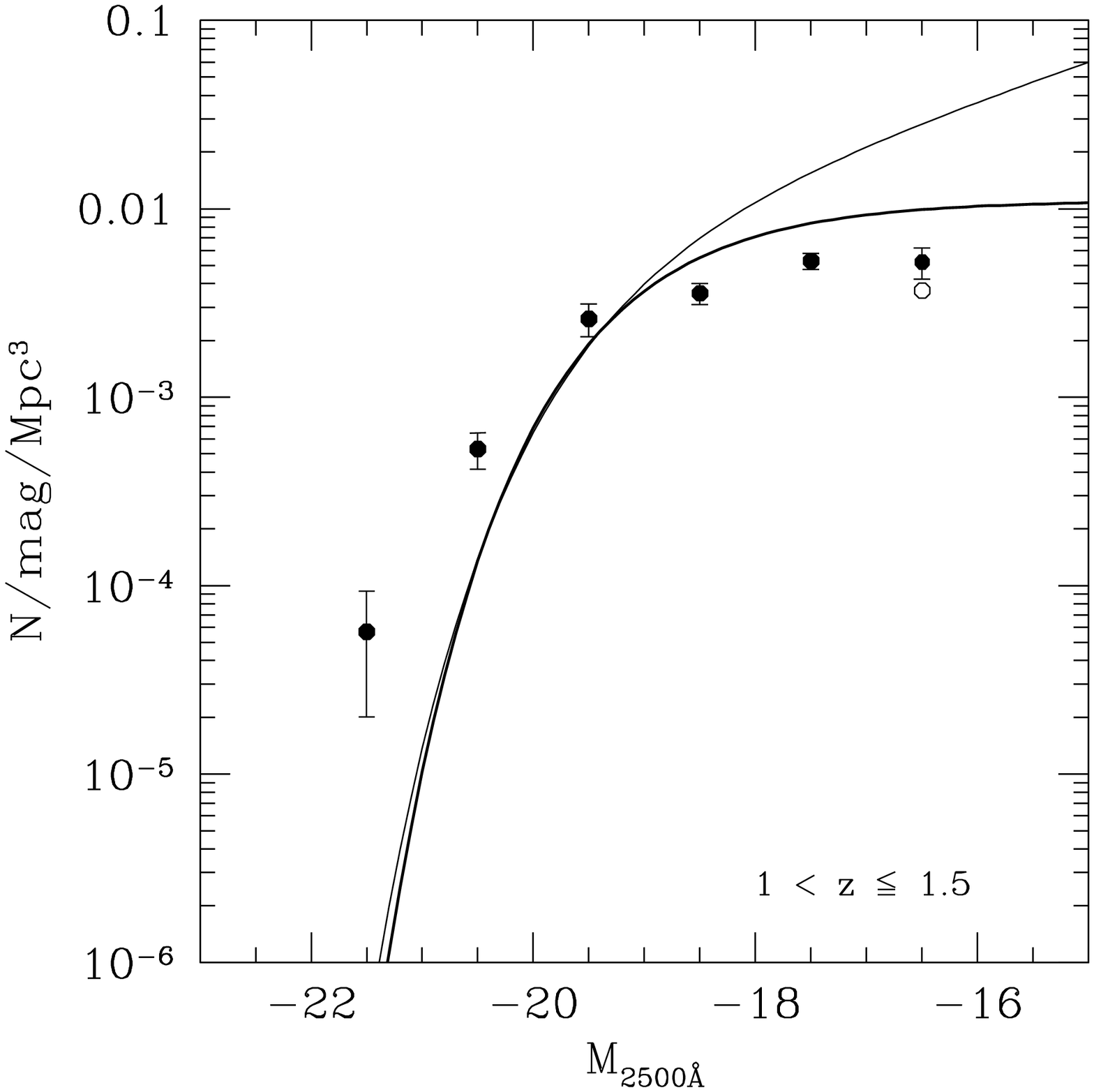}
\caption{\label{fig:lit_uv_wilson}
Comparison of the luminosity function at 2500~\AA\ of the FDF with the
Schechter function derived in 
\textit{\citet{wilson:1}}: 
\mbox{$0.2 <z\le 0.5$} (left panel), 
\mbox{$0.6 <z\le 1.0$} (middle panel), and   
\mbox{$1.0 <z\le 1.5$} (right panel).
\citet{wilson:1} fixed the slope to $\alpha=-1.0$ (thick line) and
$\alpha=-1.5$ (thin line) and used only M$^\ast$ and $\phi^\ast$ as
free parameters to determine the Schechter functions. 
}
\end{figure*}

\citet{wilson:1} used galaxies selected in the restframe UV with
spectroscopic redshifts to derive the luminosity function at 2500~\AA\ 
in 3 redshift bins: $0.2 <z\le 0.5$ (U'-selected; 403 galaxies), $0.6
<z\le 1.0$ (B-selected; 414 galaxies) and $1.0 <z\le 1.5$ (V-selected;
518 galaxies).  As the sample is not deep enough to constrain the
slope of the Schechter function \citet{wilson:1} used two fixed slopes
of $\alpha=-1.0$ and $\alpha=-1.5$ to derive the best-fitting
Schechter parameters. Since the errors of those parameters are not
reported in the paper we can only make qualitative statements about
the consistency of their and our luminosity functions:
Fig.~\ref{fig:lit_uv_wilson} shows that in the low and intermediate
redshift bin there is reasonable 
agreement with our data, while in contrast to our result, the
Schechter functions of \citet{wilson:1} do not show a significant
brightening of $M^\ast$ in their highest redshift bin.\\


Comparison of the FDF luminosity function with the Schechter functions
derived in
\citet{sullivan:1},
\citet{combo17:1}, 
\citet{kashikawa:1}, 
\citet{poli:1},
\citet{iwata:1}, 
\citet{ouchi:3},
\citet{blanton:1},
\citet{blanton:2}, and
\citet{poli:3} 
are presented in Appendix~\ref{sec:comp_lit}. In general, we find good
agreement at the bright end, where literature datasets are complete.
Differences in the faint-end slope in some cases can be attributed to
the shallower limiting magnitudes of most of the other surveys.

\begin{figure*}
\includegraphics[width=0.33\textwidth]{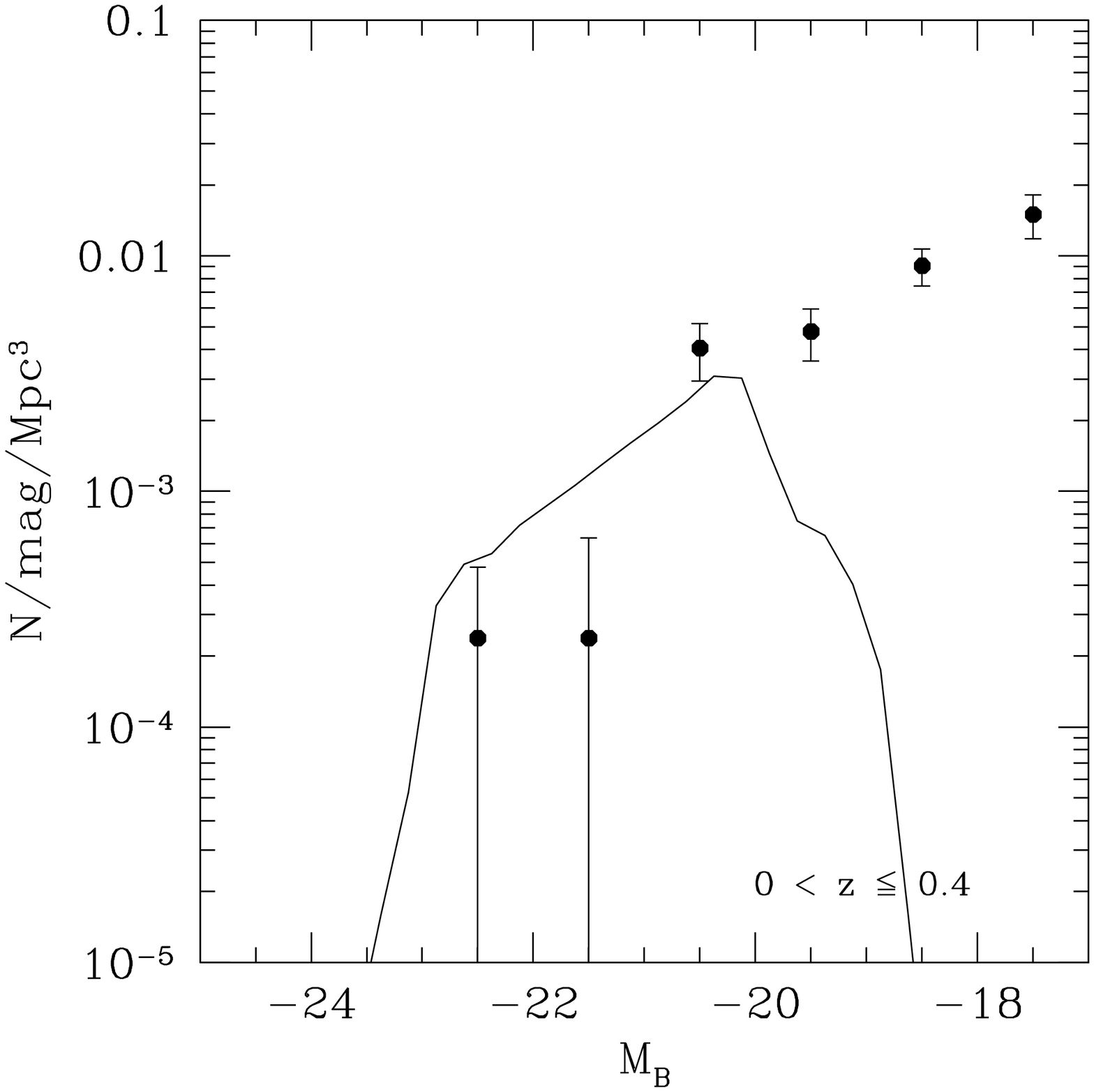}
\includegraphics[width=0.33\textwidth]{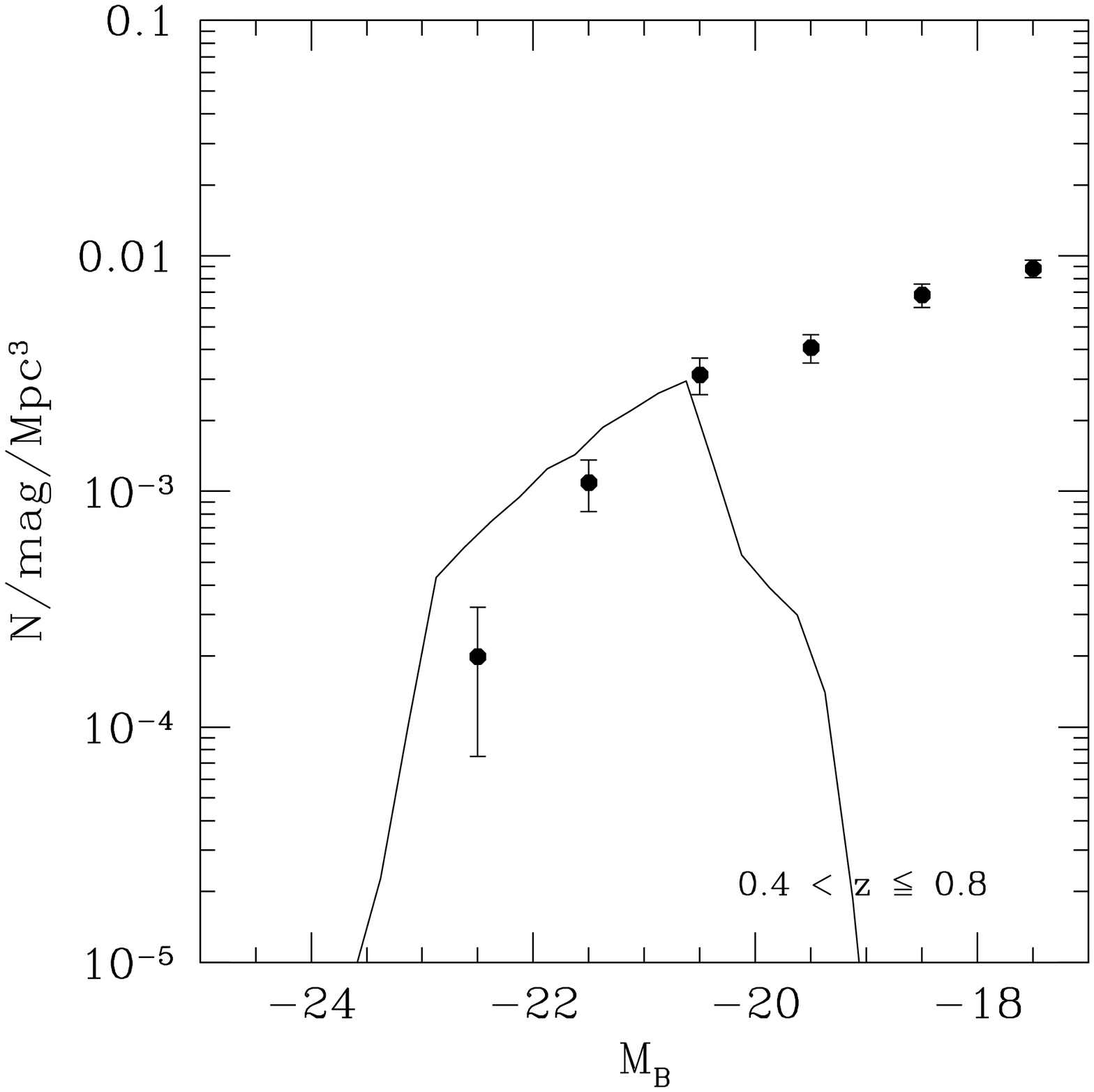}
\includegraphics[width=0.33\textwidth]{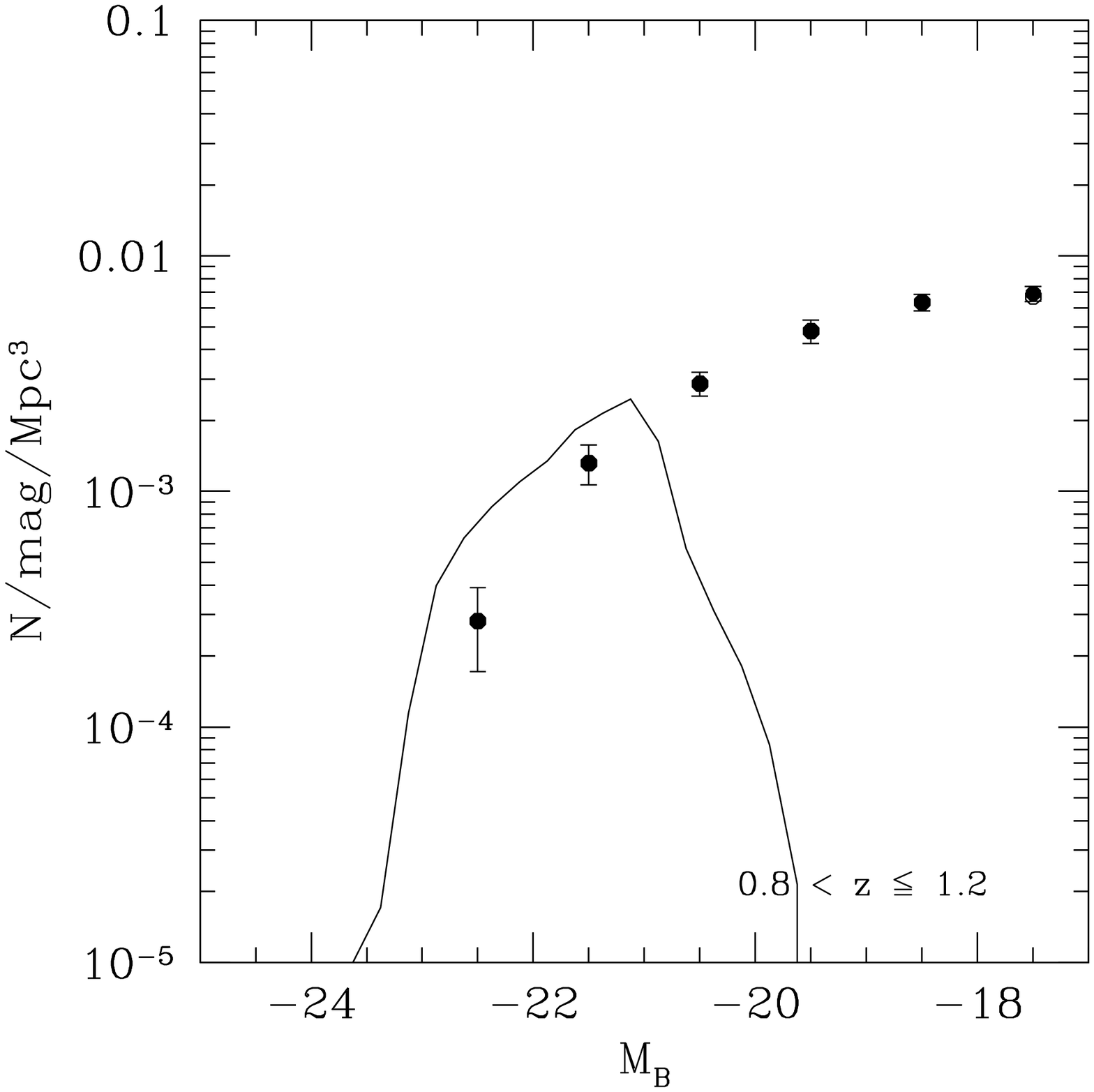}
\includegraphics[width=0.33\textwidth]{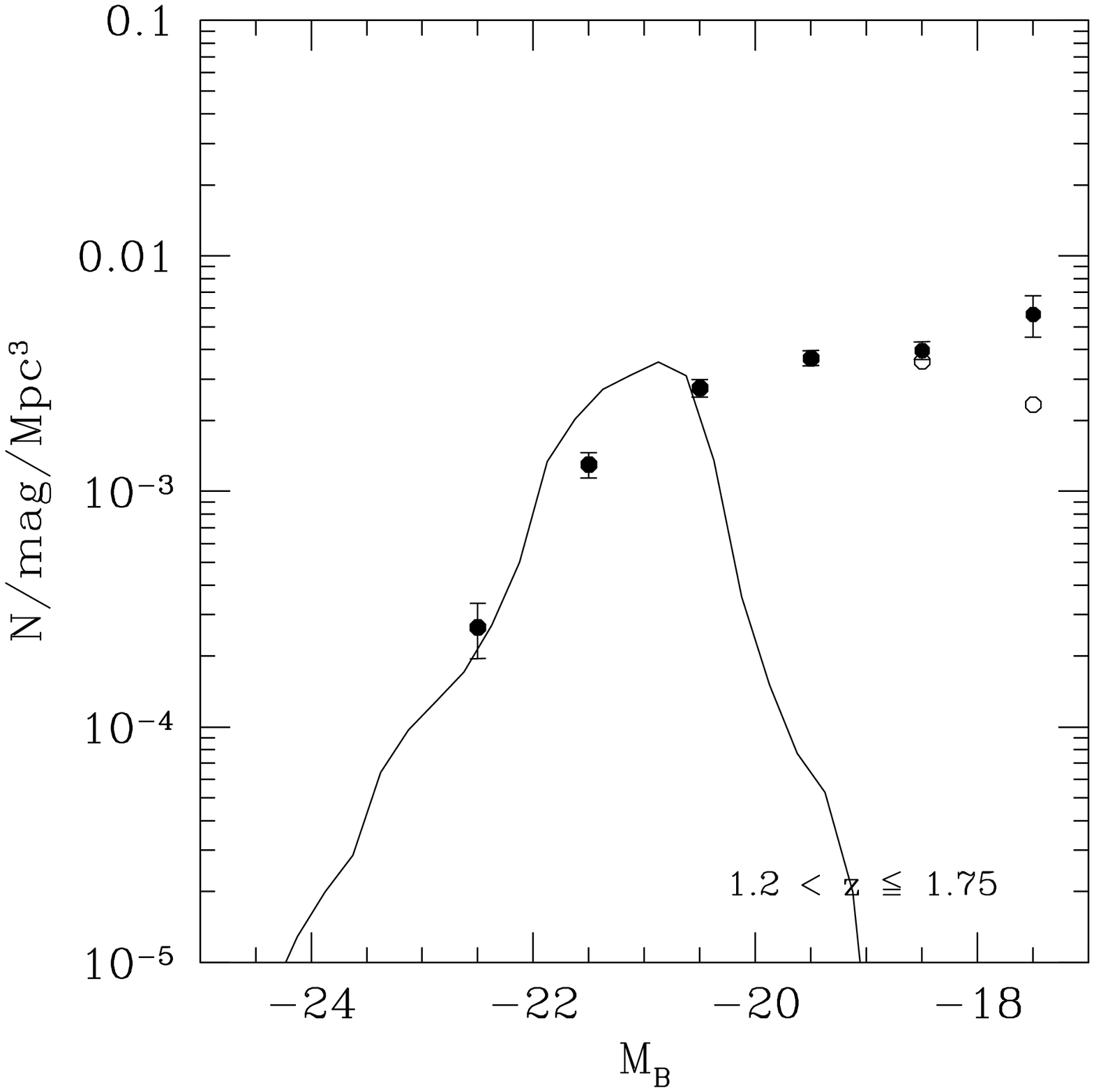}
\includegraphics[width=0.33\textwidth]{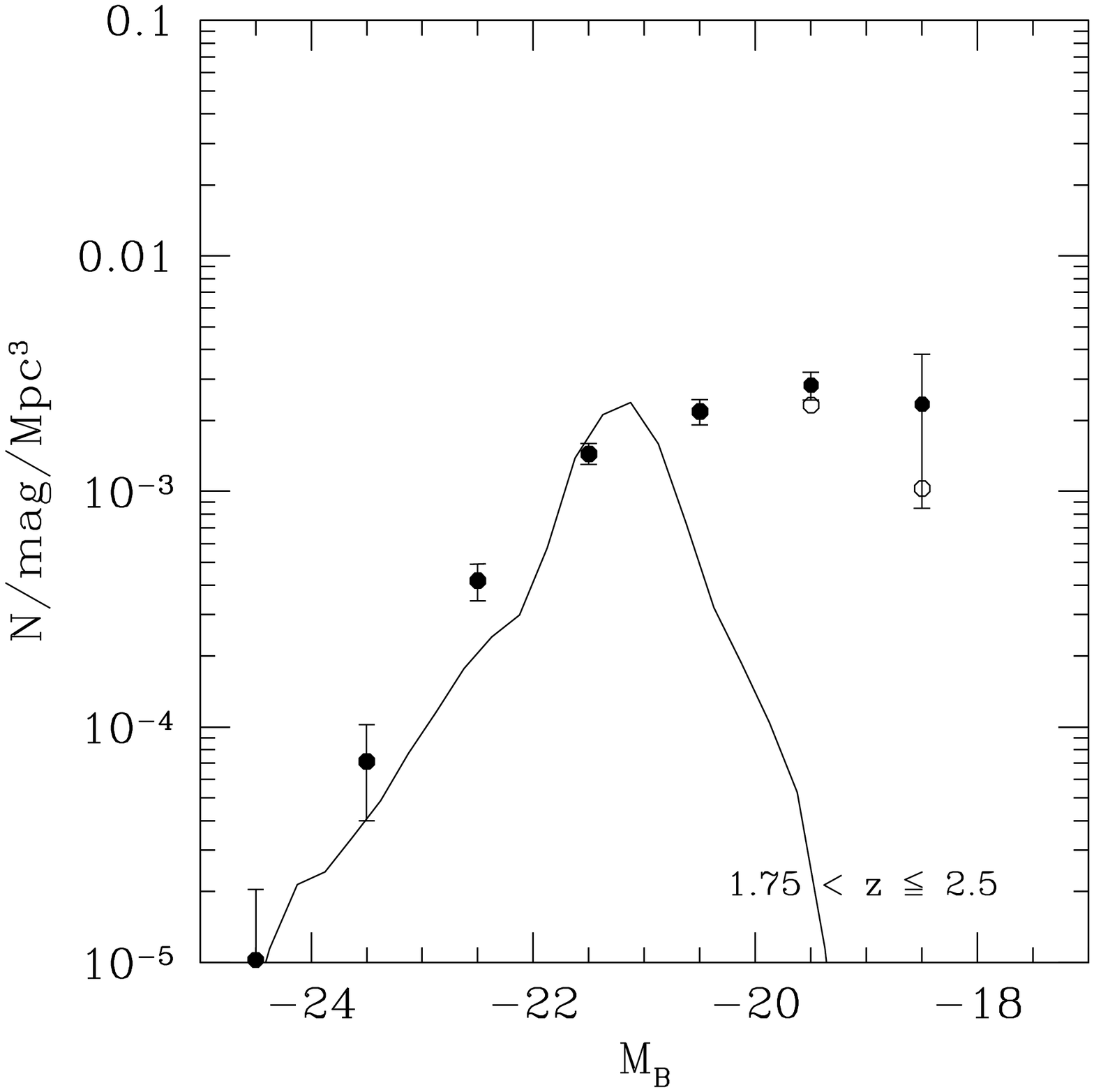}
\includegraphics[width=0.33\textwidth]{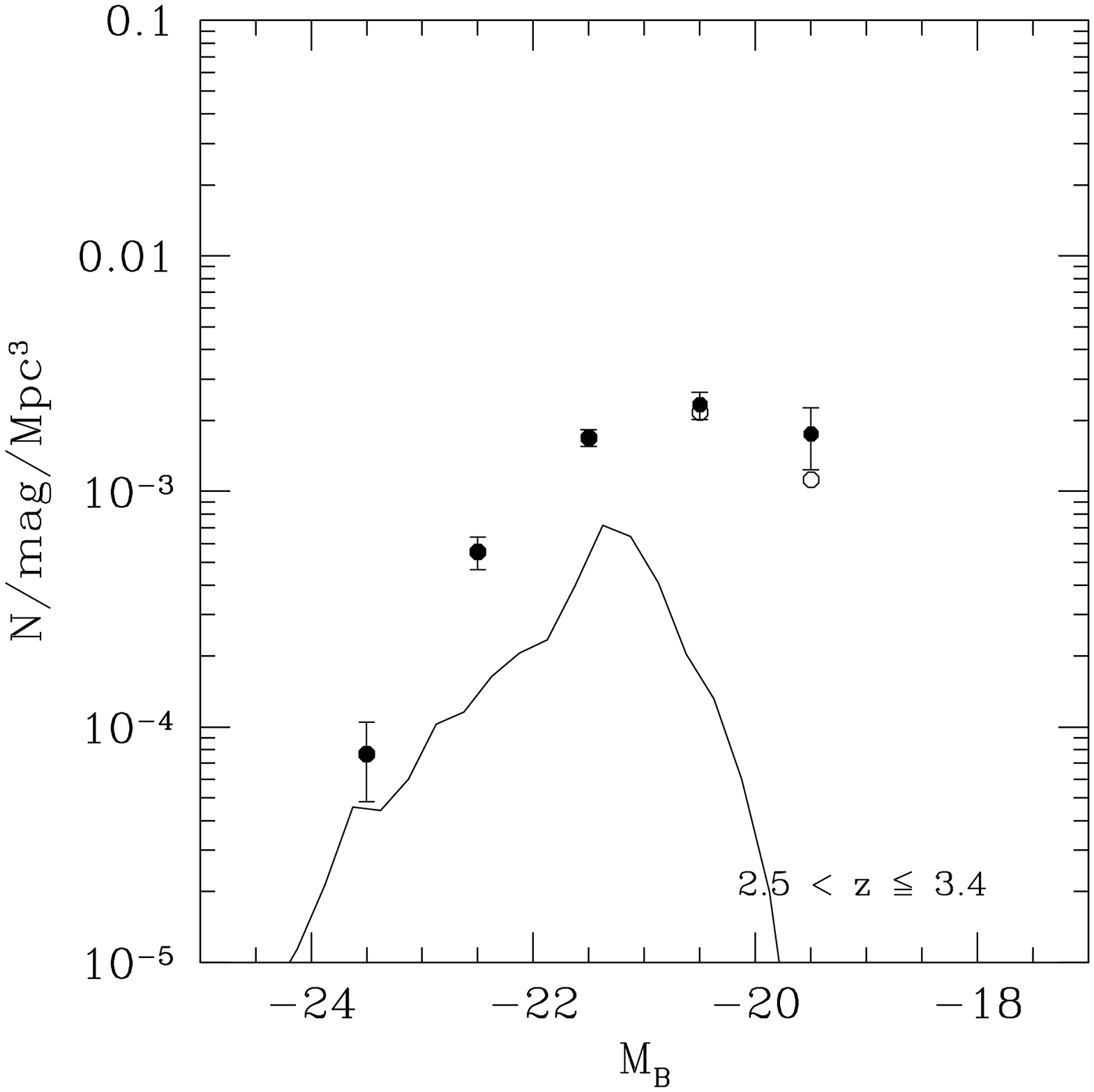}
\caption{\label{fig:lit_B_kauffmann}
Comparison of the B-band luminosity function of the FDF with
predictions based on 
\textit{\citet{kauffmann:2}}(solid line):
\mbox{$\langle z \rangle\sim 0.20 $},
\mbox{$\langle z \rangle\sim 0.62 $},
\mbox{$\langle z \rangle\sim 1.05 $},
\mbox{$\langle z \rangle\sim 1.46 $},
\mbox{$\langle z \rangle\sim 2.12 $}, and 
\mbox{$\langle z \rangle\sim 2.97 $}
(from upper left to lower right panel).  The filled (open) symbols
show the luminosity function corrected (uncorrected) for $V/V_{max}$.
The drops of the theoretical curves towards the faint-end is caused by
the limited mass resolution of the models, see \citet{kauffmann:2} for
details.}
\end{figure*}

\section{Comparison with model predictions}
\label{sec:model}

As discussed in Sect.~\ref{sec:intro}, key physical processes are
involved in shaping the bright and the faint-end of the galaxy luminosity
function. Therefore, it is interesting to compare luminosity
functions predicted by models with observational results to
better constrain those processes. In this section we compare the
B-band luminosity function in different redshift bins with model
predictions of \citet{kauffmann:2} and \citet{menci:1}.\\

\noindent\textit{\citet{kauffmann:2}:}\\
In Fig.~\ref{fig:lit_B_kauffmann} we show the B-band luminosity
function of the FDF together with the semi-analytic model predictions
by \citet{kauffmann:2} \footnote{The models
  were taken from:\\
  http://www.mpa-garching.mpg.de/Virgo/data\_download.html} for the
following redshifts:
\mbox{$\langle z \rangle\sim 0.20 $},
\mbox{$\langle z \rangle\sim 0.62 $},
\mbox{$\langle z \rangle\sim 1.05 $},
\mbox{$\langle z \rangle\sim 1.46 $},
\mbox{$\langle z \rangle\sim 2.12 $}, and 
\mbox{$\langle z \rangle\sim 2.97 $}.

There seems to be reasonably good agreement between the models (solid
lines) and the luminosity functions derived in the FDF up to redshift
\mbox{$\langle z \rangle\sim 2.12 $}.
Of course at $z\approx 0$ the
model is tuned to reproduce the data.  At $z\sim 3$, the discrepancy
increases as the model does not contain enough bright galaxies. 
Unfortunately, the models only
predict luminosities for massive galaxies and, therefore, they do not
predict galaxy number densities below M$^\ast$.\\

\begin{figure*}
\includegraphics[width=0.33\textwidth]{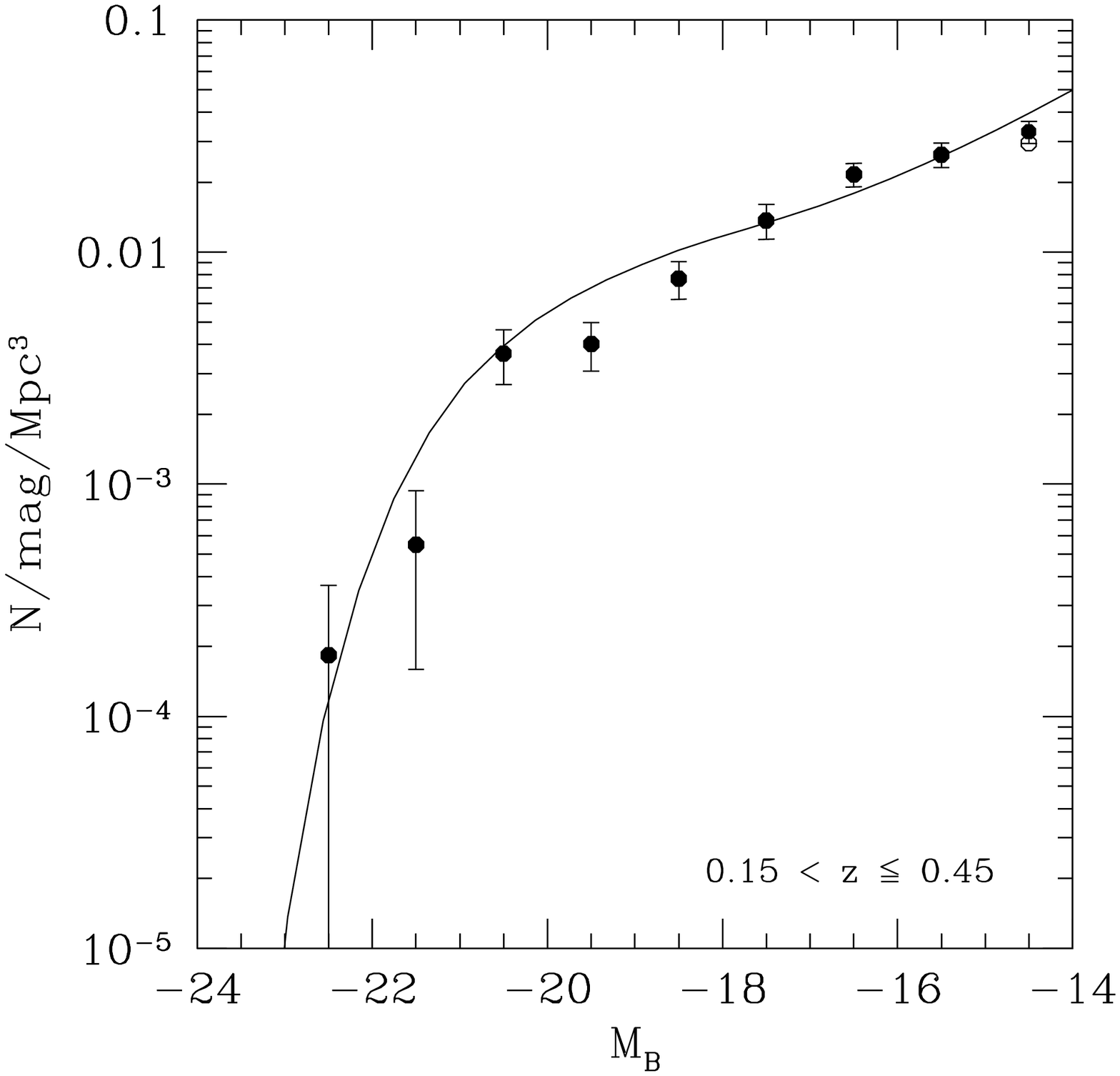}
\includegraphics[width=0.33\textwidth]{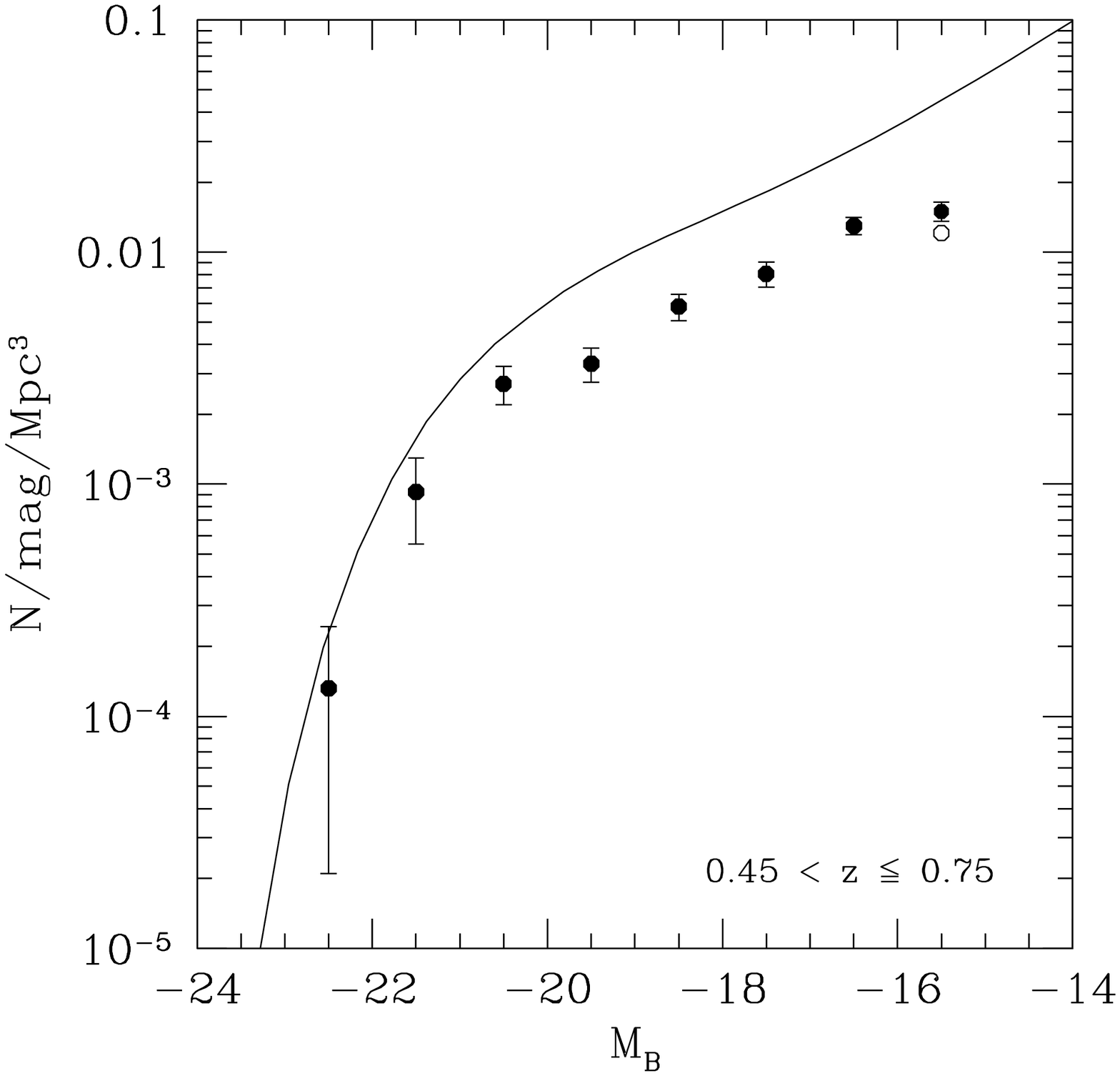}
\includegraphics[width=0.33\textwidth]{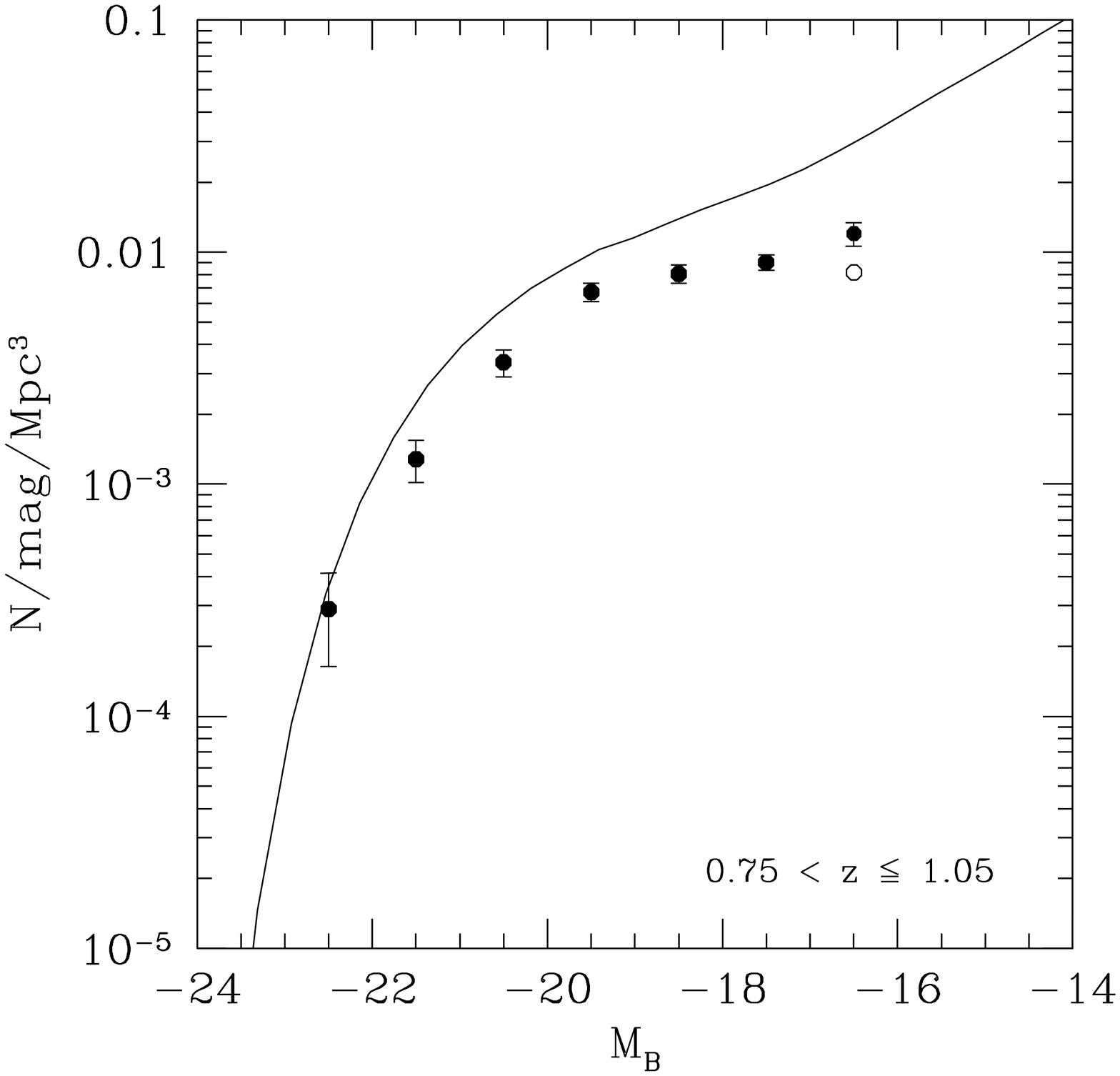}
\includegraphics[width=0.33\textwidth]{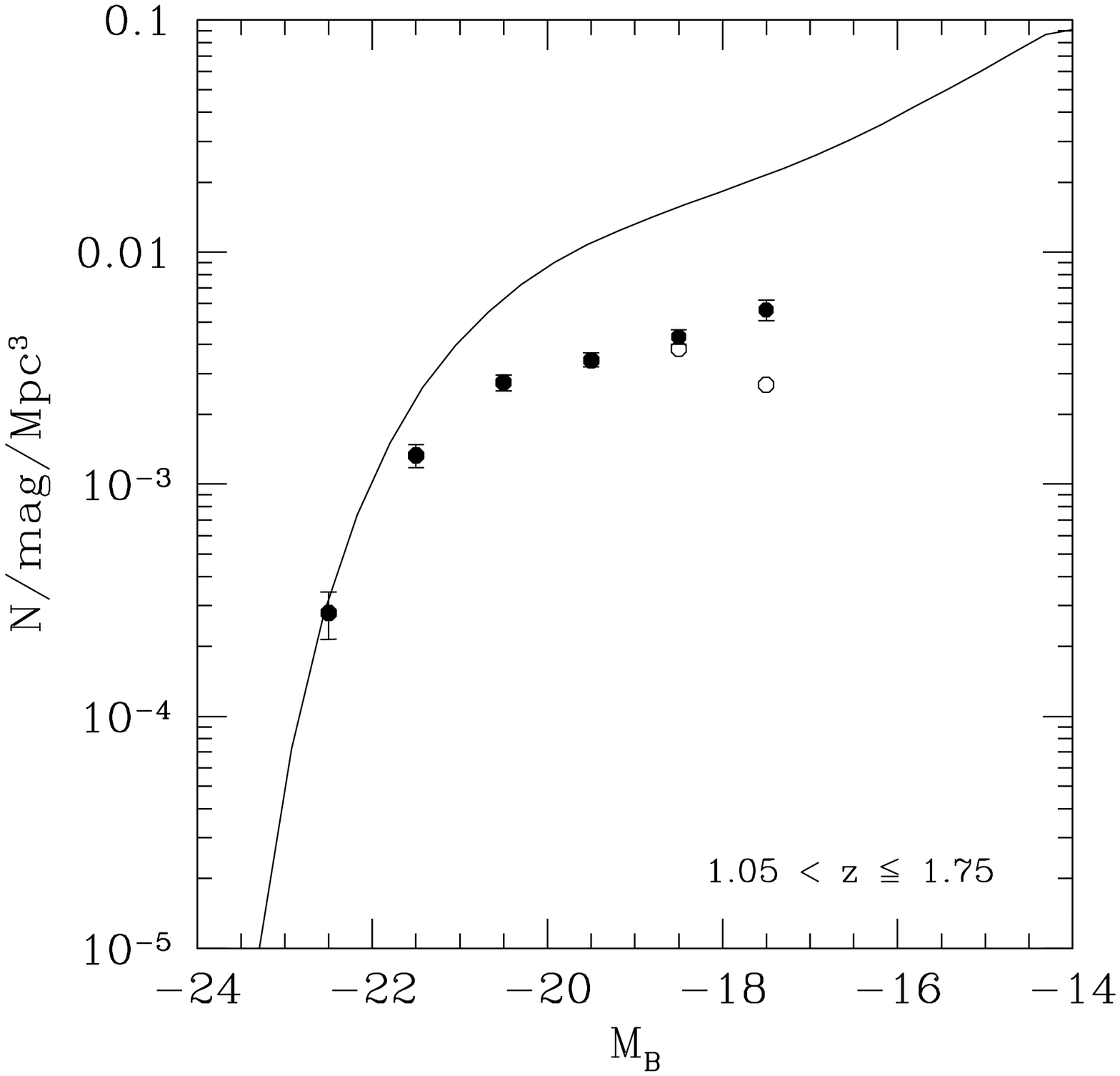}
\includegraphics[width=0.33\textwidth]{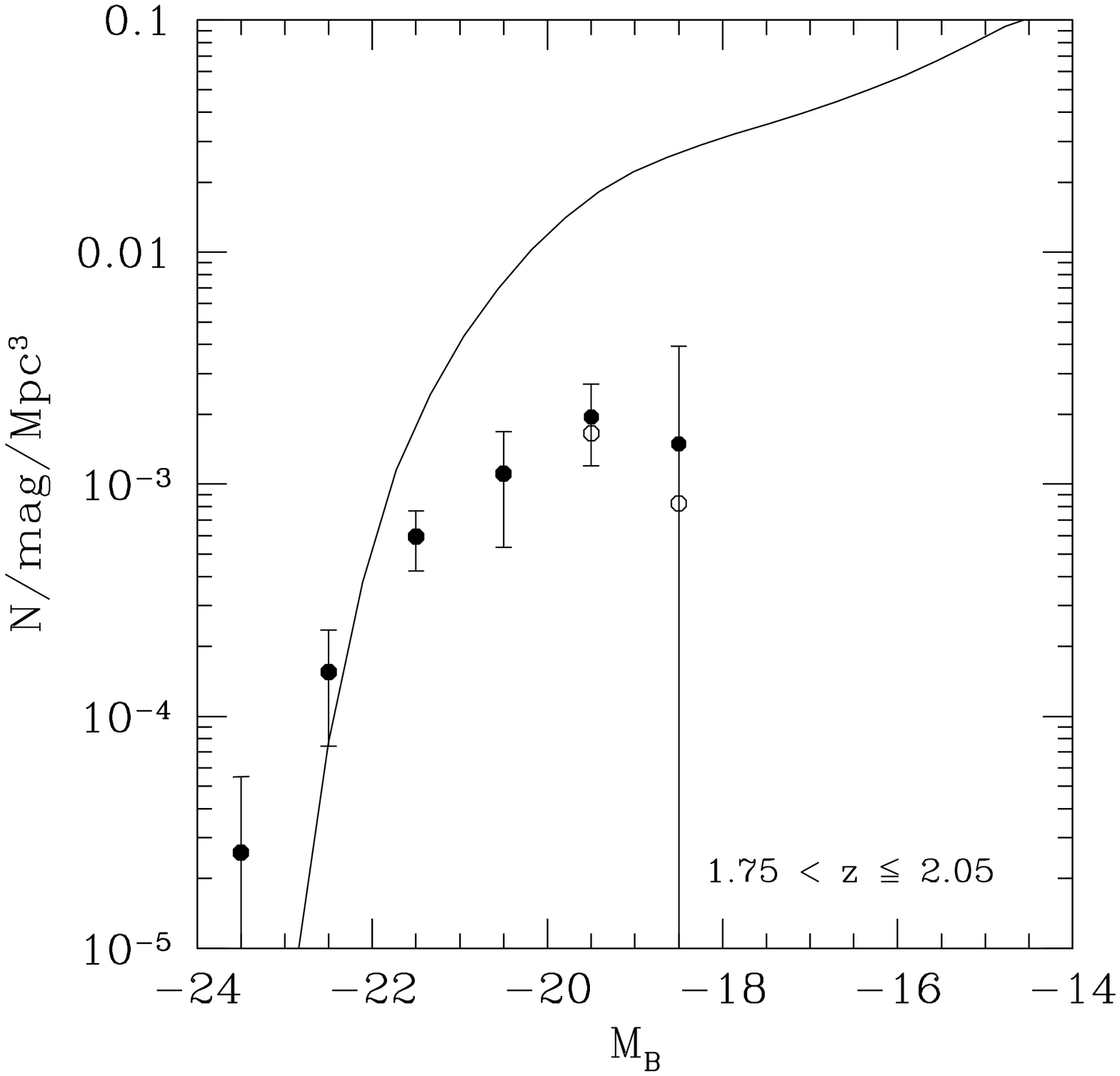}
\includegraphics[width=0.33\textwidth]{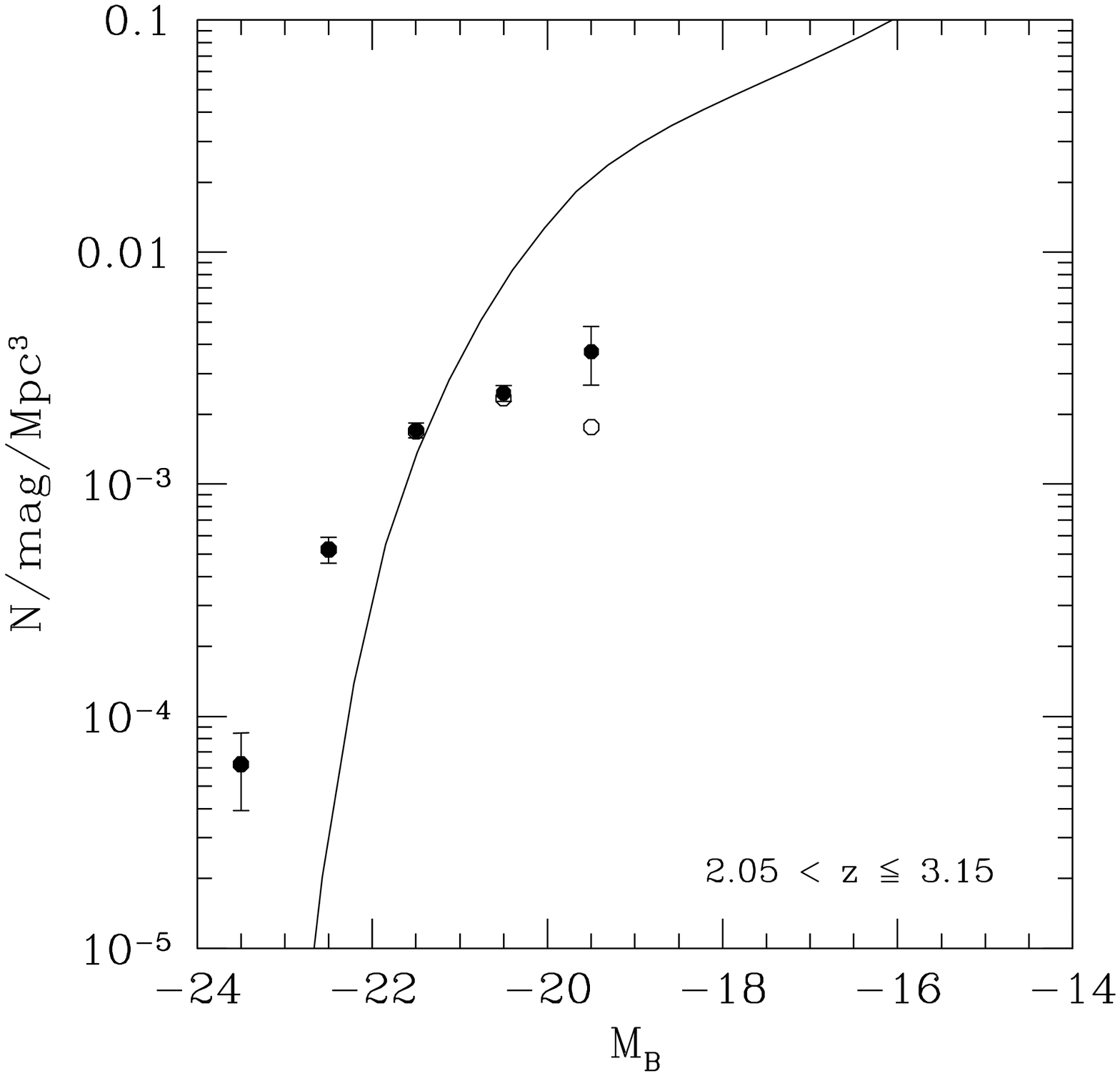}
\includegraphics[width=0.33\textwidth]{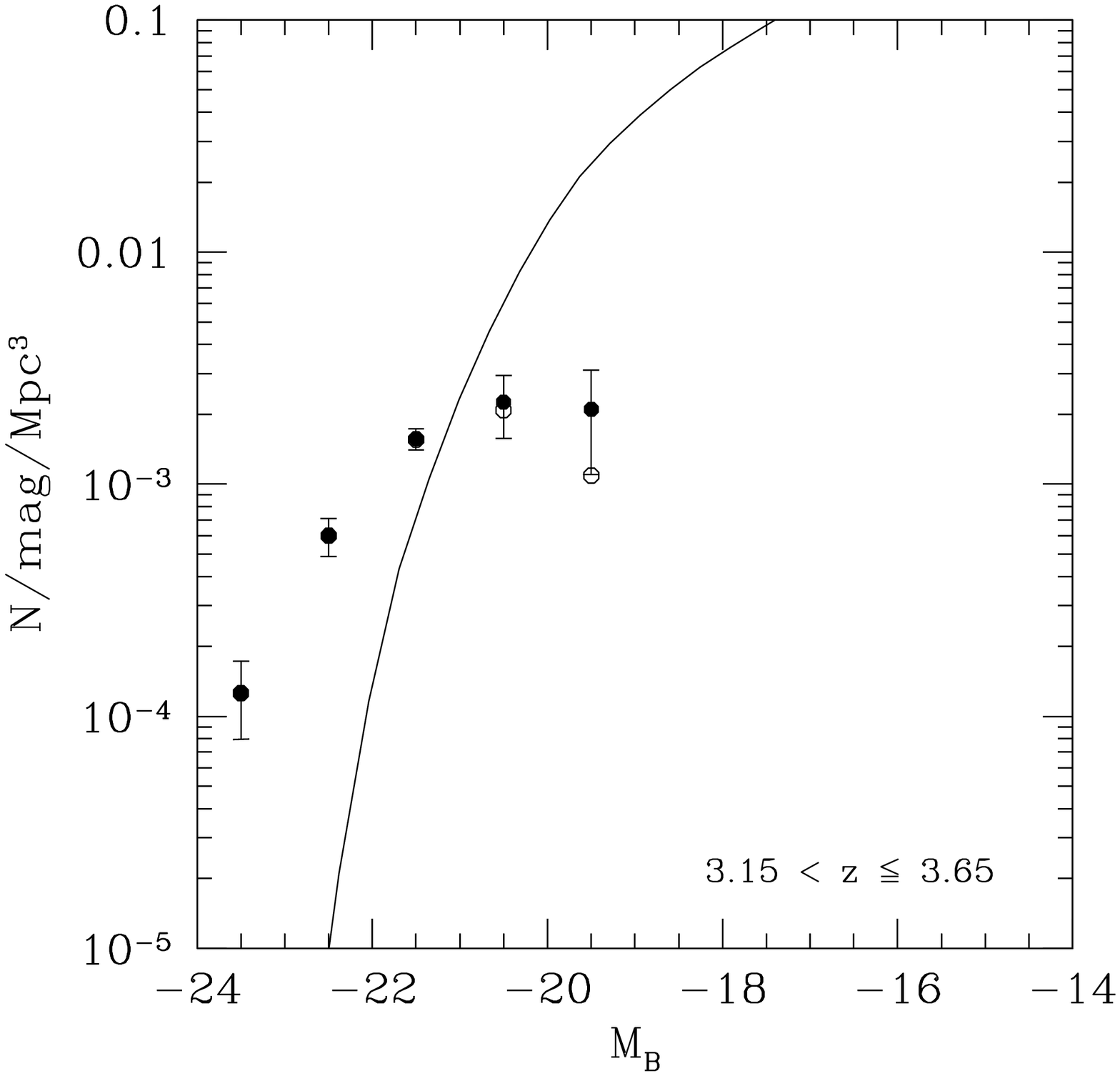}
\includegraphics[width=0.33\textwidth]{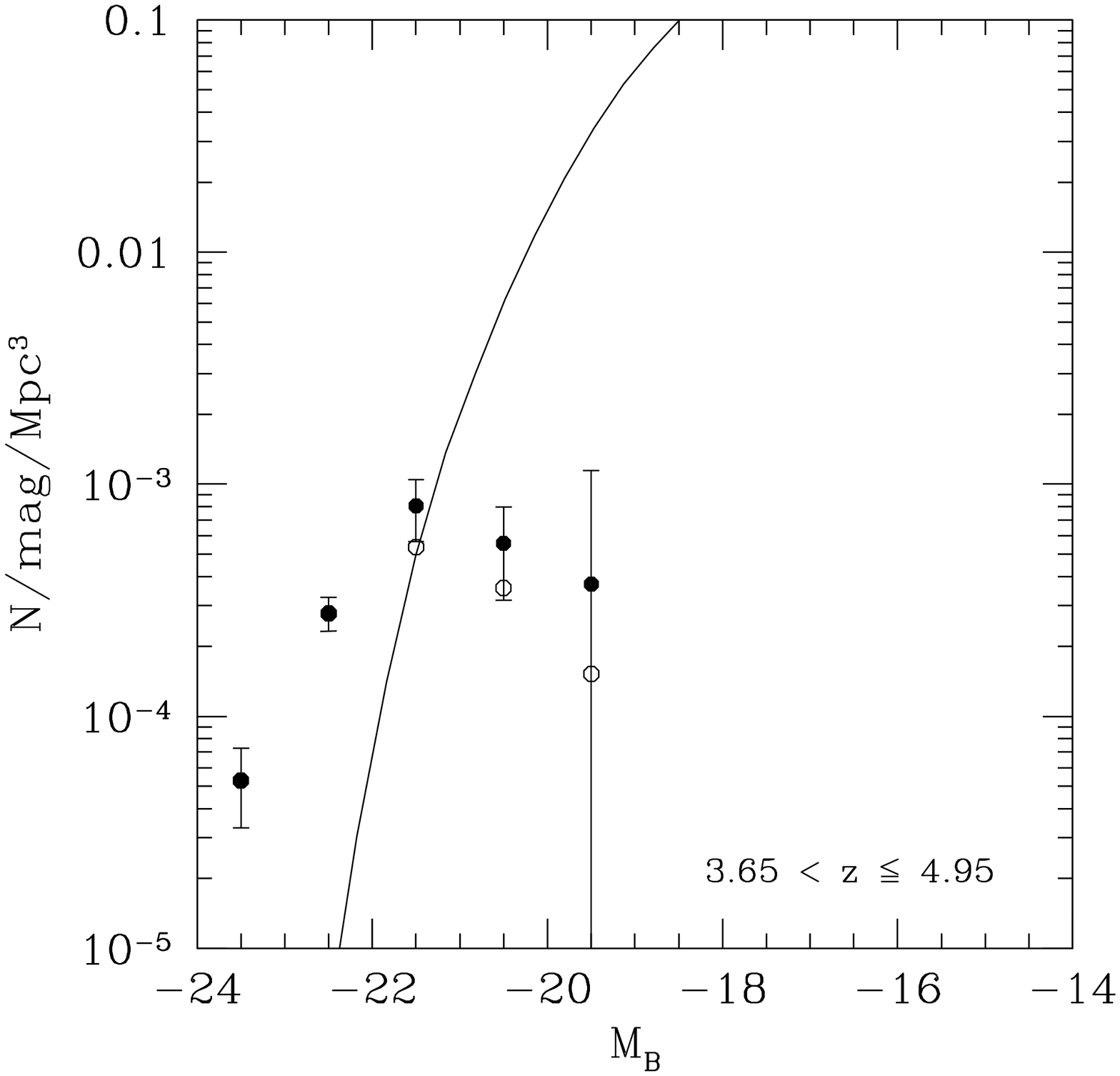}
\caption{\label{fig:lit_B_menci}
Comparison of the B-band luminosity function of the FDF with
predictions based on the CDM model of 
\textit{\citet{menci:1}} (private
communication; solid line):
\mbox{$\langle z \rangle\sim 0.3$},
\mbox{$\langle z \rangle\sim 0.6$},
\mbox{$\langle z \rangle\sim 0.9$},
\mbox{$\langle z \rangle\sim 1.4$},
\mbox{$\langle z \rangle\sim 1.9$},
\mbox{$\langle z \rangle\sim 2.6$},
\mbox{$\langle z \rangle\sim 3.4$}, and 
\mbox{$\langle z \rangle\sim 4.3$}
(from upper left to lower right panel). The filled (open) symbols show
the luminosity function corrected (uncorrected) for $V/V_{max}$.}
\end{figure*}

\noindent\textit{\citet{menci:1}:}\\
In Fig.~\ref{fig:lit_B_menci} we compare the B-band luminosity
functions of the FDF with  the semi-analytic model by
\citet{menci:1}  for the following redshifts:
\mbox{$\langle z \rangle\sim 0.3$},
\mbox{$\langle z \rangle\sim 0.6$},
\mbox{$\langle z \rangle\sim 0.9$},
\mbox{$\langle z \rangle\sim 1.4$},
\mbox{$\langle z \rangle\sim 1.9$},
\mbox{$\langle z \rangle\sim 2.6$},
\mbox{$\langle z \rangle\sim 3.4$}, and 
\mbox{$\langle z \rangle\sim 4.3$}.

 The agreement between the FDF data and the model in the lowest
  redshift bin \mbox{$\langle z \rangle\sim 0.3$} is very good, but
  this is again expected (see comment above). Moreover, if one
  focuses the comparison only on the higher luminosity bins considered
  by \citet{kauffmann:2}, similar acceptable agreement with the data is
  observed. However, at lower luminosities and higher redshifts, the
  galaxy density of the simulation is much higher than the observed
  one.
\\

\section{Summary and conclusions}
\label{sec:summary_conclusion}

We analyzed a sample of about 5600 I-band selected galaxies in the
FORS Deep Field down to a limiting magnitude of I $ = 26.8$ mag.

 A comparison with the very deep K-selected catalogue of
  \citet{labbe:1} shows, that more than 90~\% of their objects are
  brighter than our limiting I-band magnitude.  Therefore our
  scientific conclusions are not affected by this color bias.

Based on 9 filters we derived accurate photometric redshifts with
\mbox{$\Delta z / (z_{spec}+1) \approx 0.03 $} if compared with the
spectroscopic sample \citep{noll:1, boehm:1} of 362 objects. We
calculated and presented the luminosity functions in the UV (1500~\AA\ 
and 2800~\AA), u', B, and g' bands in the redshift range \mbox{$0.5 <
  z < 5.0$}.  The error budget of the luminosity functions includes
both the photometric redshift error as well as the Poissonian error.

The faint-end slope of the luminosity function does not show a large
redshift evolution and is compatible within $2\sigma$ with a constant
slope in most of the redshift bins and wavelengths considered here.
Furthermore, the slope in the 1500~\AA, 2800~\AA, and u' band is very
similar but differs from the slope in the g' and B band.  We derive a
best fitting slope of $\alpha=-1.07 \pm 0.04$ for the combined 1500
\AA , 2800~\AA\ and u' bands and $\alpha=-1.25 \pm 0.03$ for the
combined g' and B bands.  We find no evidence for a very steep slope
(\mbox{$\alpha \le -1.6$}) at $z \sim 3$ and 1700~\AA\ rest wavelength
as reported by other authors (e.g., \citealt{steidel:1},
\citealt{ouchi:3}). From our data we can exclude a slope of
\mbox{$\alpha \le -1.6$} at redshift \mbox{$\langle z \rangle\sim
  3.0$} and \mbox{$\langle z \rangle\sim 4.0$} at the $2\sigma$
level.

We investigate the evolution of M$^\ast$ and $\phi^\ast$ by means of a
redshift parameterization of the form \mbox{$M^\ast(z)= M^\ast_0 + a
  \ln(1+z)$} and $\phi^\ast(z)=\phi^\ast_0(1+z)^b$. We find a
substantial brightening of M$^\ast$ and a decrease of $\phi^\ast$ with
redshift in all analyzed wavelengths. If we follow the evolution of
the characteristic luminosity from \mbox{$\langle z \rangle\sim 0.5$}
to \mbox{$\langle z \rangle\sim 5$}, we find an increase of $\sim$~3.1
magnitudes in the UV, of $\sim$~2.6 magnitudes in the u' and of
$\sim$~1.6 magnitudes in the g' and B band.  Simultaneously the
characteristic density decreases by about 80 \% -- 90 \% in all
analyzed wavebands.

Moreover, we compare the luminosity function derived in the FDF with
previous observational datasets, mostly based on photometric results,
and discuss discrepancies. In general, we find good agreement at the
bright end, where their samples are complete.  Differences in the
faint-end slope in some cases can be attributed to the shallower
limiting magnitudes of most of the other surveys.  The only
observations which reach the same limiting magnitudes as the FDF
observations are those of \citet{poli:1, poli:3} and the K-selected
sample of \citet{kashikawa:1}.  
The FDF results for the faint-end
slope are in excellent agreement with those of \citet{kashikawa:1} 
but the slope of the Schechter function favored by \citet{poli:1,
  poli:3} is steeper than we would expect from the FDF.

The semi-analytical models predict luminosity functions which describe
(by construction) the data at low redshift quite well, but show
growing disagreement with increasing redshifts.

\begin{acknowledgements}
  We thank the referee, Dr.~A.~J.~Bunker, for his careful reading of the
  manuscript and several constructive comments which helped us to
  improve the presentation of the results.  Moreover, we thank
  Dr.~N.~Menci for providing an electronic version of his unpublished
  model calculation and for interesting remarks.  AG thanks
  Dr.~C.~Maraston, J.~Fliri and J.~Thomas for stimulating discussions
  as well as A.~Riffeser and C.~A.~G\"ossl for help dealing with their
  image reduction software. We acknowledge the support of the ESO
  Paranal staff during several observing runs. This work was supported
  by the \emph{Deut\-sche For\-schungs\-ge\-mein\-schaft, DFG}, SFB
  375 (Astro\-teil\-chen\-phy\-sik), SFB 439 (Galaxies in the young
  Universe) and Volkswagen Foundation (I/76\,520).
\end{acknowledgements}

\appendix

\section{Schechter parameters}
\label{sec:schechter_parameter}

\begin{figure*}[hhh!]
\includegraphics[width=0.33\textwidth]{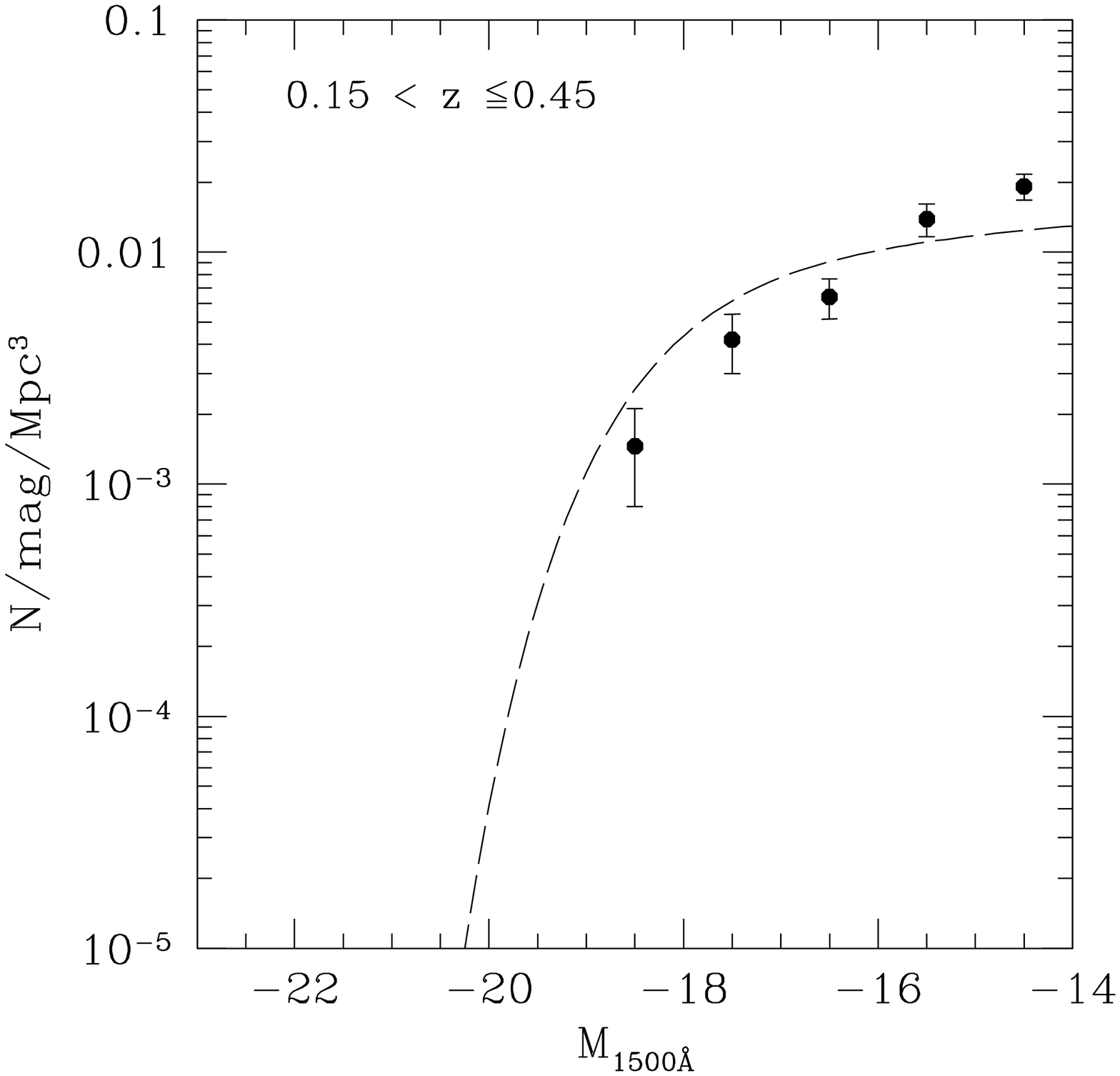}
\includegraphics[width=0.33\textwidth]{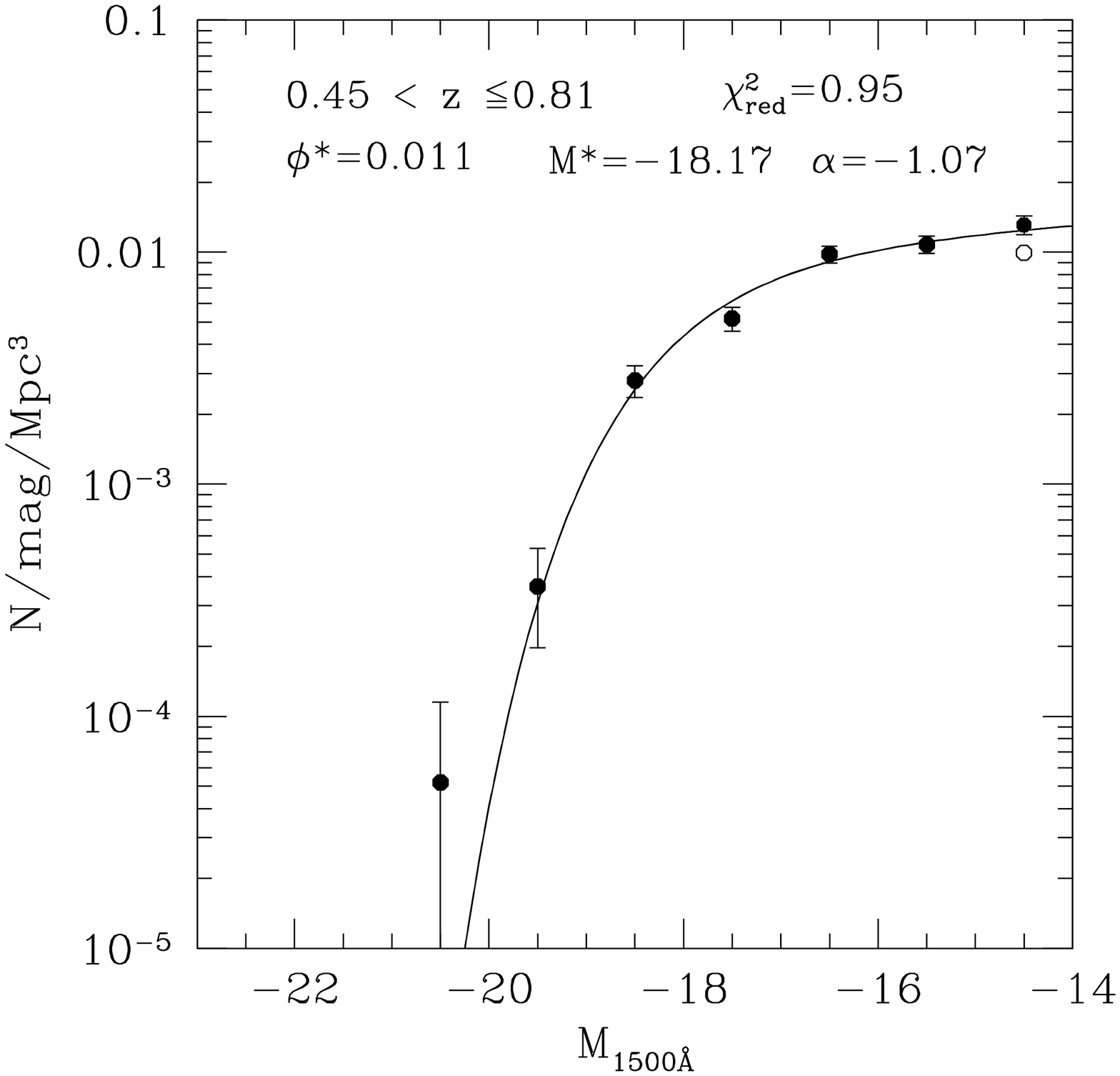}
\includegraphics[width=0.33\textwidth]{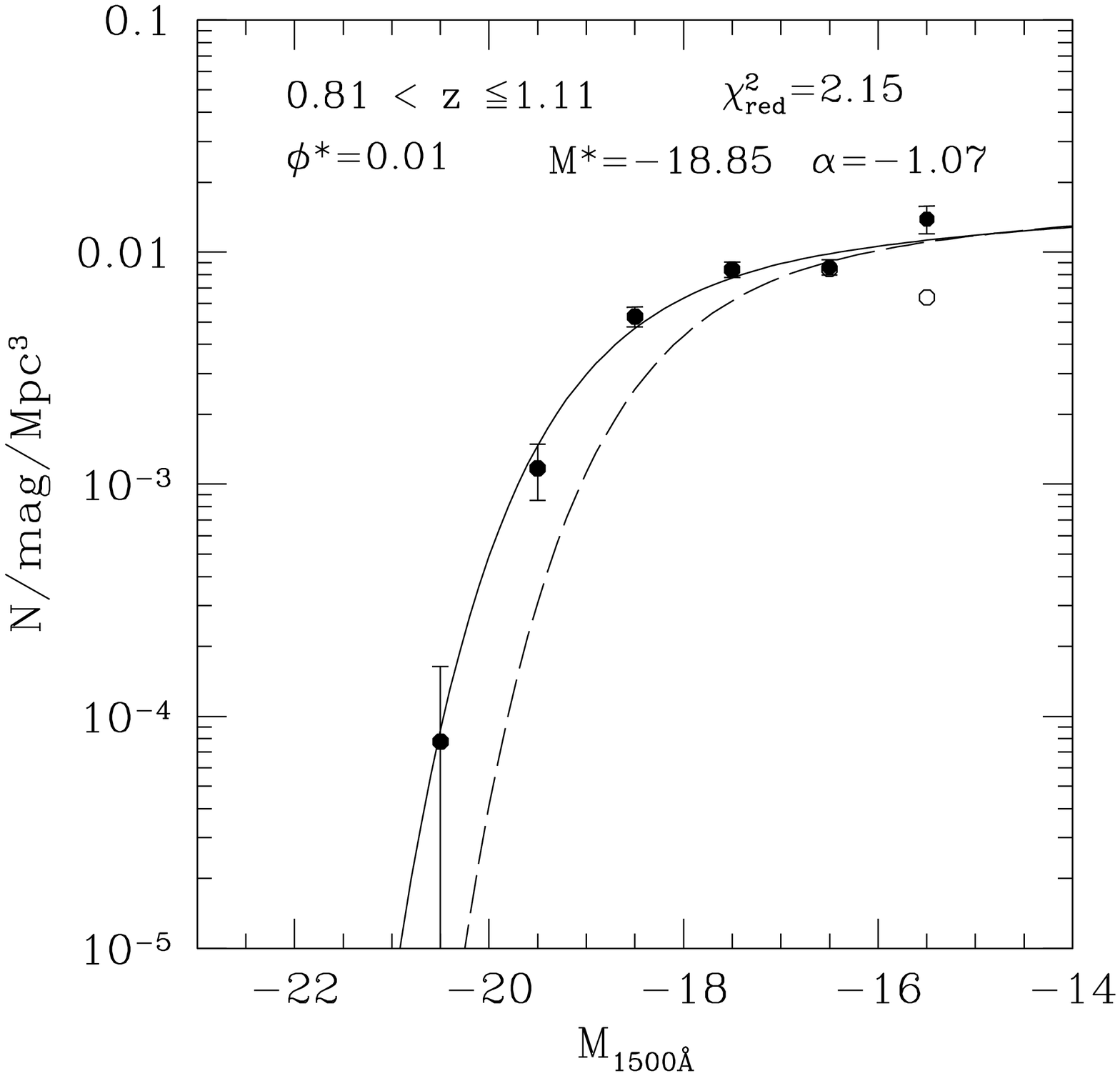}
\includegraphics[width=0.33\textwidth]{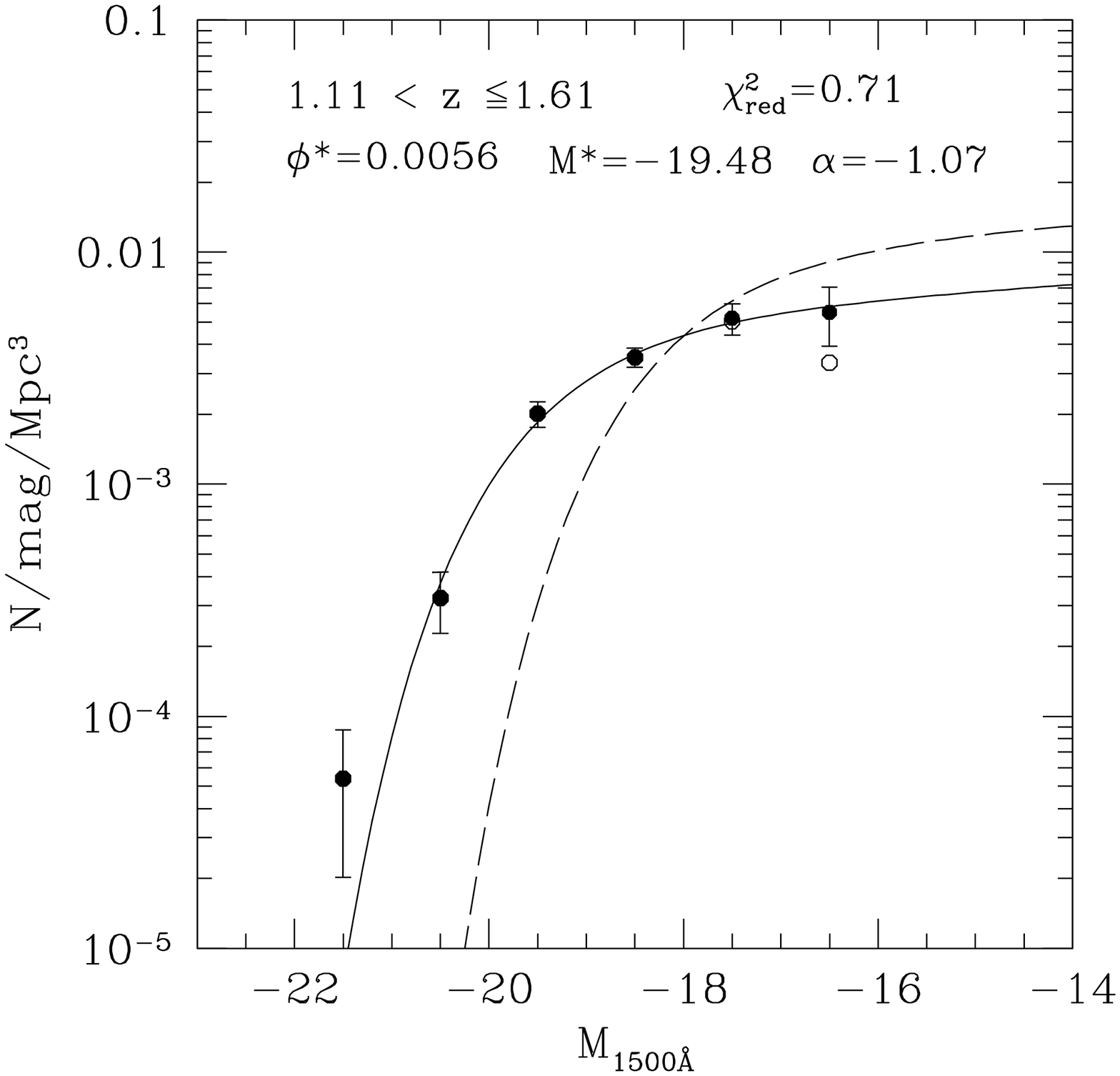}
\includegraphics[width=0.33\textwidth]{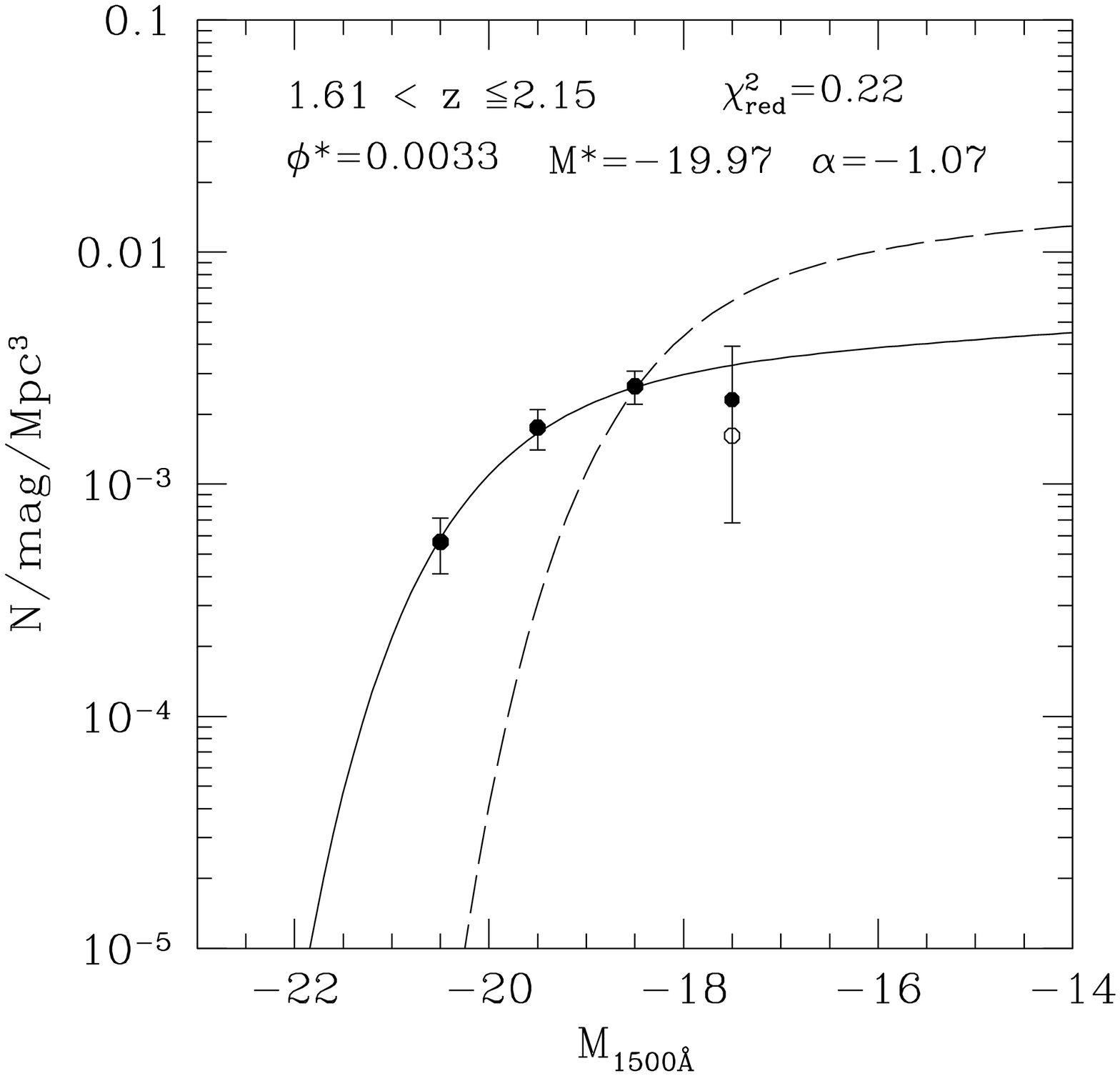}
\includegraphics[width=0.33\textwidth]{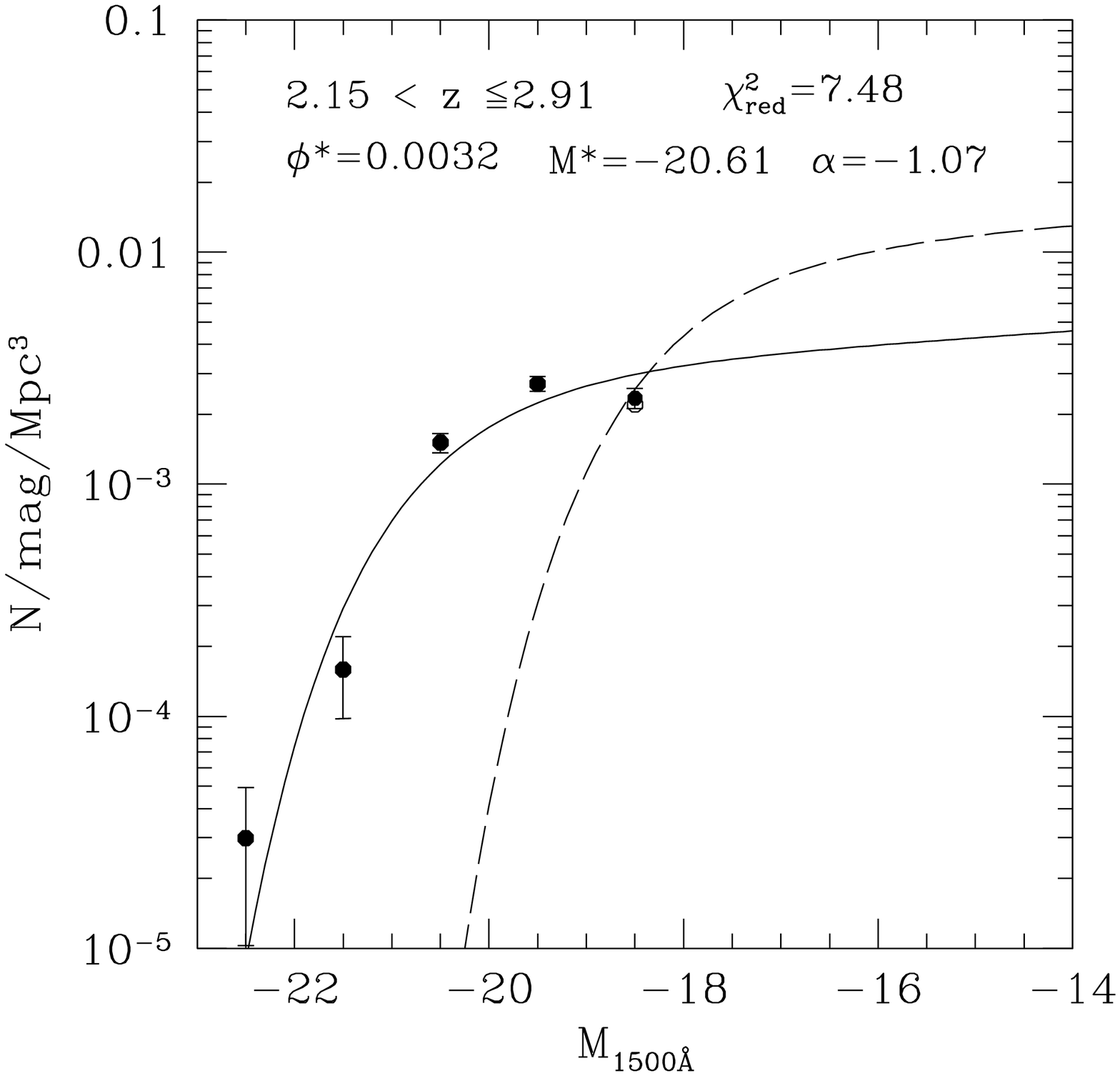}
\includegraphics[width=0.33\textwidth]{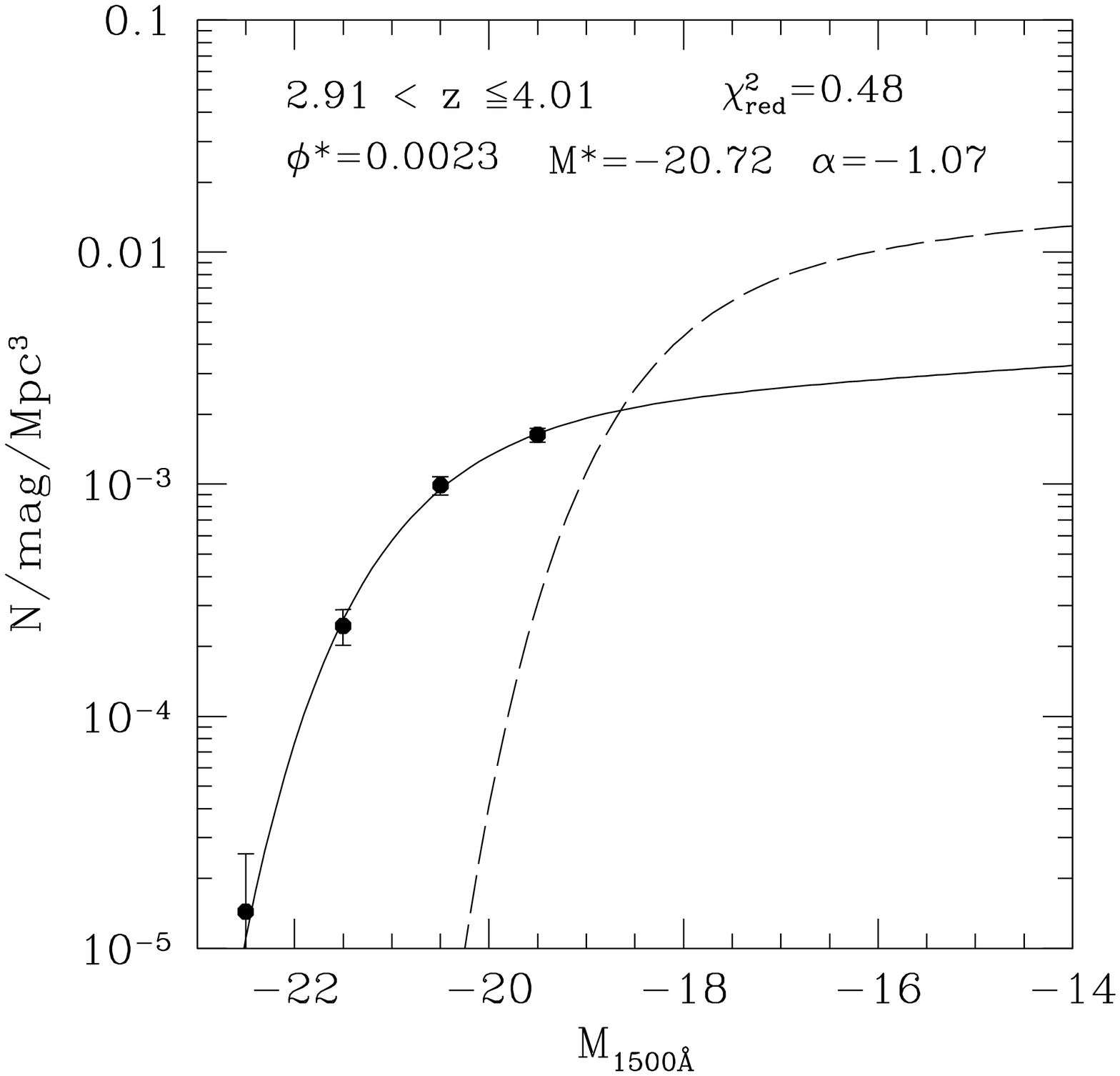}
\includegraphics[width=0.33\textwidth]{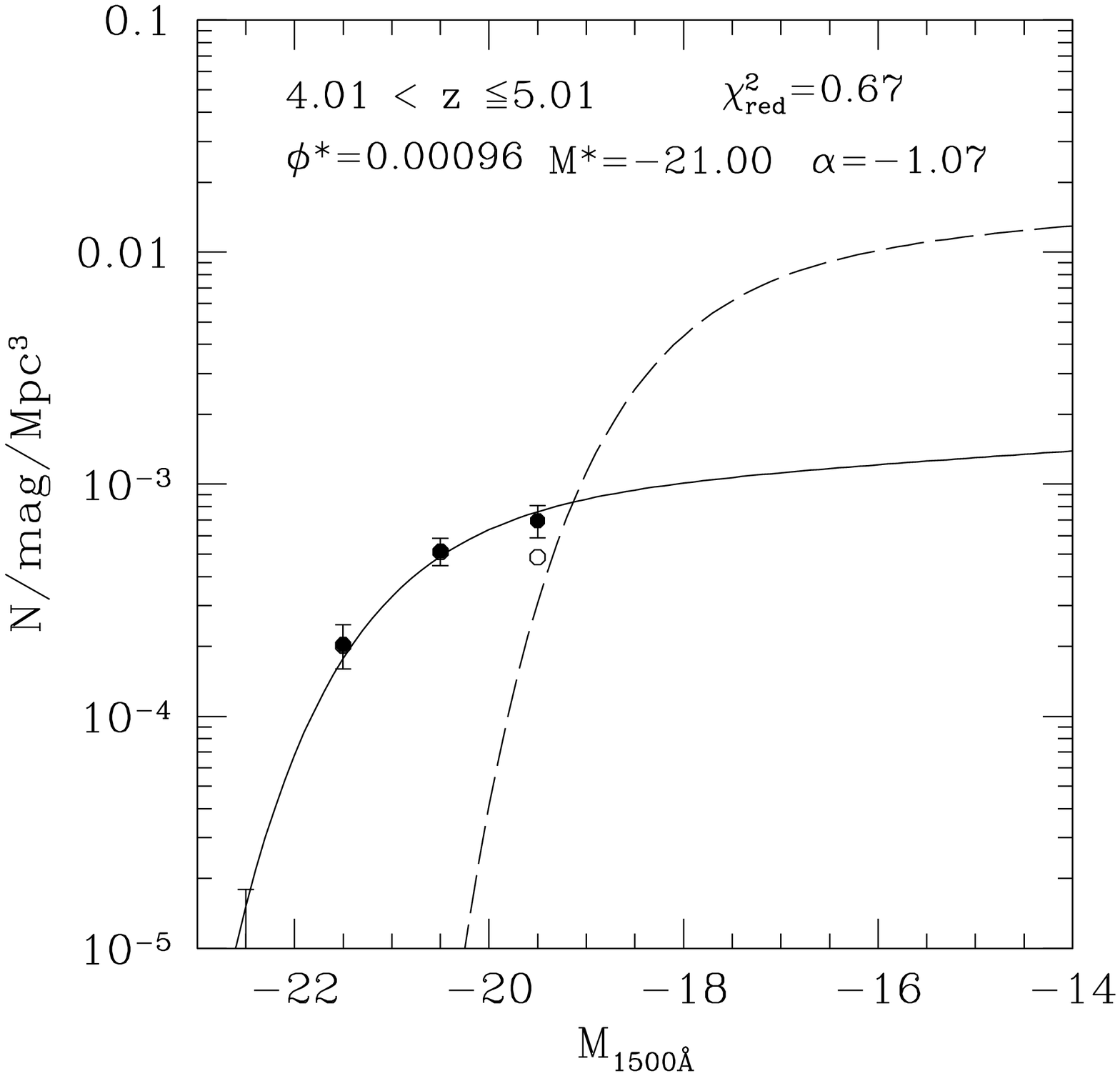}
\includegraphics[width=0.33\textwidth]{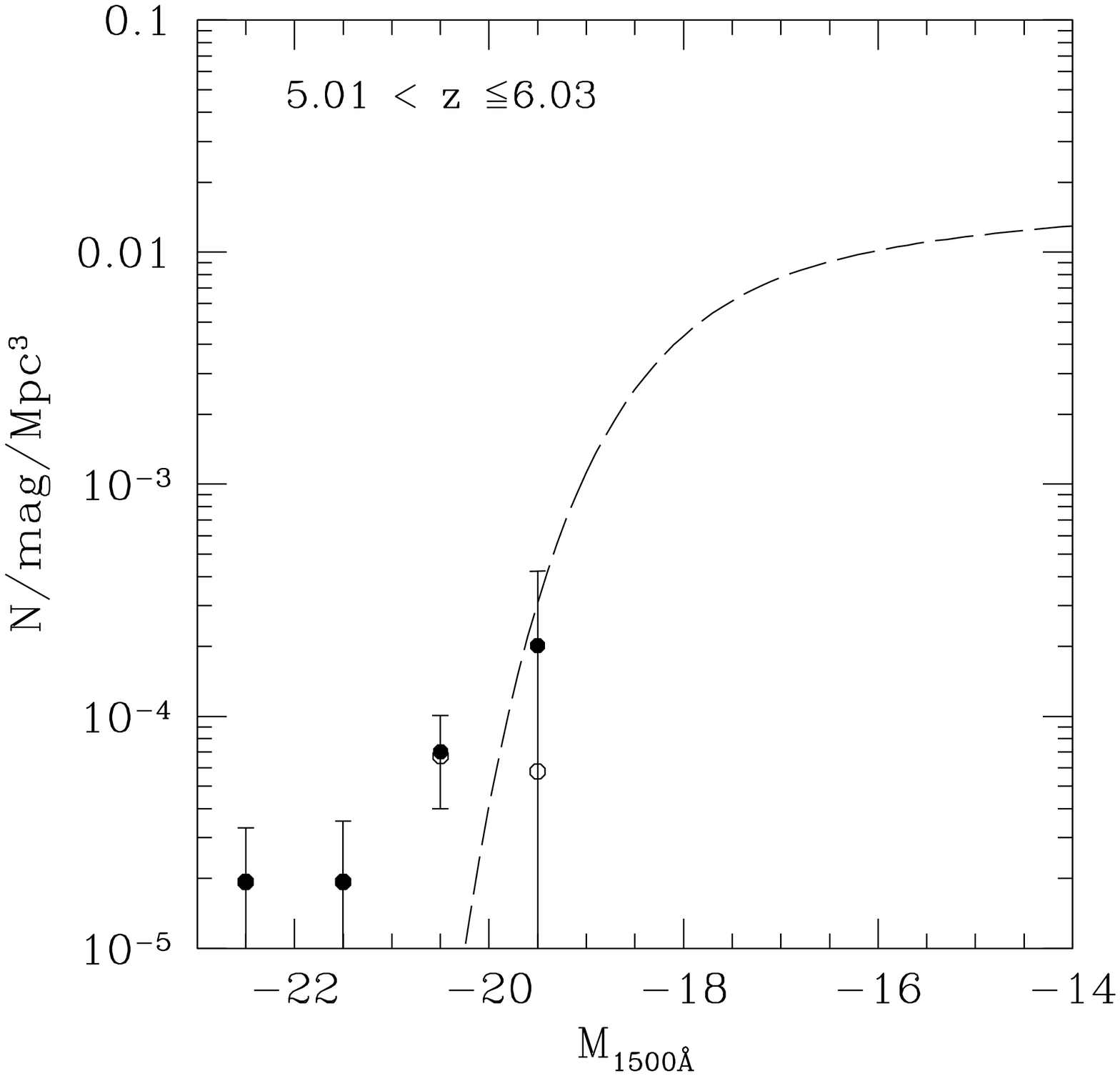}
\caption{\label{fig:lumfkt_fdf_1500}
  Luminosity functions at \textit{1500~\AA } from low redshift
  (\mbox{$\langle z\rangle=0.3$}, upper left panel) to high redshift
  (\mbox{$\langle z\rangle=5.5$}, lower right panel).  The filled
  (open) symbols show the luminosity function corrected (uncorrected)
  for $V/V_{max}$.  The fitted Schechter functions for a fixed slope
  $\alpha$ are shown as solid lines. Note that we only fit the
  luminosity functions from $\langle z\rangle=0.6$ to $\langle
  z\rangle=4.5$. The parameters of the Schechter functions can be
  found in Table~\ref{tab:schechter_fit_1500}. The Schechter fit for
  redshift $\langle z\rangle=0.6$ is indicated as dashed line in all
  panels.}
\end{figure*}
\begin{table*}[hhh!]
\caption[]{\label{tab:schechter_fit_1500}Schechter function fit at 1500~\AA}
\begin{center}
\begin{tabular}{c|c|c|c}
 redshift interval & M$^\ast$ (mag) & $\phi^\ast$ (Mpc$^{-3}$) & $\alpha$ (fixed)\\
\hline
0.45 -- 0.81 & $-$18.17 +0.11 $-$0.11 & 0.0110 +0.0007 $-$0.0006 &$-$1.07 \\  
0.81 -- 1.11 & $-$18.85 +0.10 $-$0.10 & 0.0103 +0.0006 $-$0.0006 &$-$1.07 \\  
1.11 -- 1.61 & $-$19.48 +0.11 $-$0.11 & 0.0056 +0.0006 $-$0.0005 &$-$1.07 \\  
1.61 -- 2.15 & $-$19.97 +0.22 $-$0.24 & 0.0033 +0.0006 $-$0.0006 &$-$1.07 \\  
2.15 -- 2.91 & $-$20.61 +0.09 $-$0.09 & 0.0032 +0.0002 $-$0.0002 &$-$1.07 \\  
2.91 -- 4.01 & $-$20.72 +0.09 $-$0.10 & 0.0023 +0.0002 $-$0.0002 &$-$1.07 \\  
4.01 -- 5.01 & $-$21.00 +0.15 $-$0.11 & 0.0010 +0.0001 $-$0.0001 &$-$1.07 \\  
\end{tabular}
\end{center}
\end{table*}

\begin{table*}[]
\caption[]{\label{tab:schechter_fit_2800}Schechter function fit at 2800~\AA}
\begin{center}
\begin{tabular}{c|c|c|c}
 redshift interval & M$^\ast$ (mag) & $\phi^\ast$ (Mpc$^{-3}$) & $\alpha$ (fixed)\\
\hline
0.45 -- 0.81 & $-$18.80 +0.15 $-$0.15 & 0.0104 +0.0007 $-$0.0007 &$-$1.07 \\  
0.81 -- 1.11 & $-$19.52 +0.09 $-$0.10 & 0.0089 +0.0005 $-$0.0005 &$-$1.07 \\  
1.11 -- 1.61 & $-$20.03 +0.09 $-$0.09 & 0.0053 +0.0004 $-$0.0004 &$-$1.07 \\  
1.61 -- 2.15 & $-$20.43 +0.18 $-$0.17 & 0.0029 +0.0005 $-$0.0004 &$-$1.07 \\  
2.15 -- 2.91 & $-$21.16 +0.09 $-$0.08 & 0.0030 +0.0002 $-$0.0002 &$-$1.07 \\  
2.91 -- 4.01 & $-$21.19 +0.10 $-$0.08 & 0.0021 +0.0002 $-$0.0001 &$-$1.07 \\  
4.01 -- 5.01 & $-$21.55 +0.17 $-$0.21 & 0.0009 +0.0001 $-$0.0001 &$-$1.07 \\  
\end{tabular}
\end{center}
\end{table*}

\begin{figure*}
\includegraphics[width=0.33\textwidth]{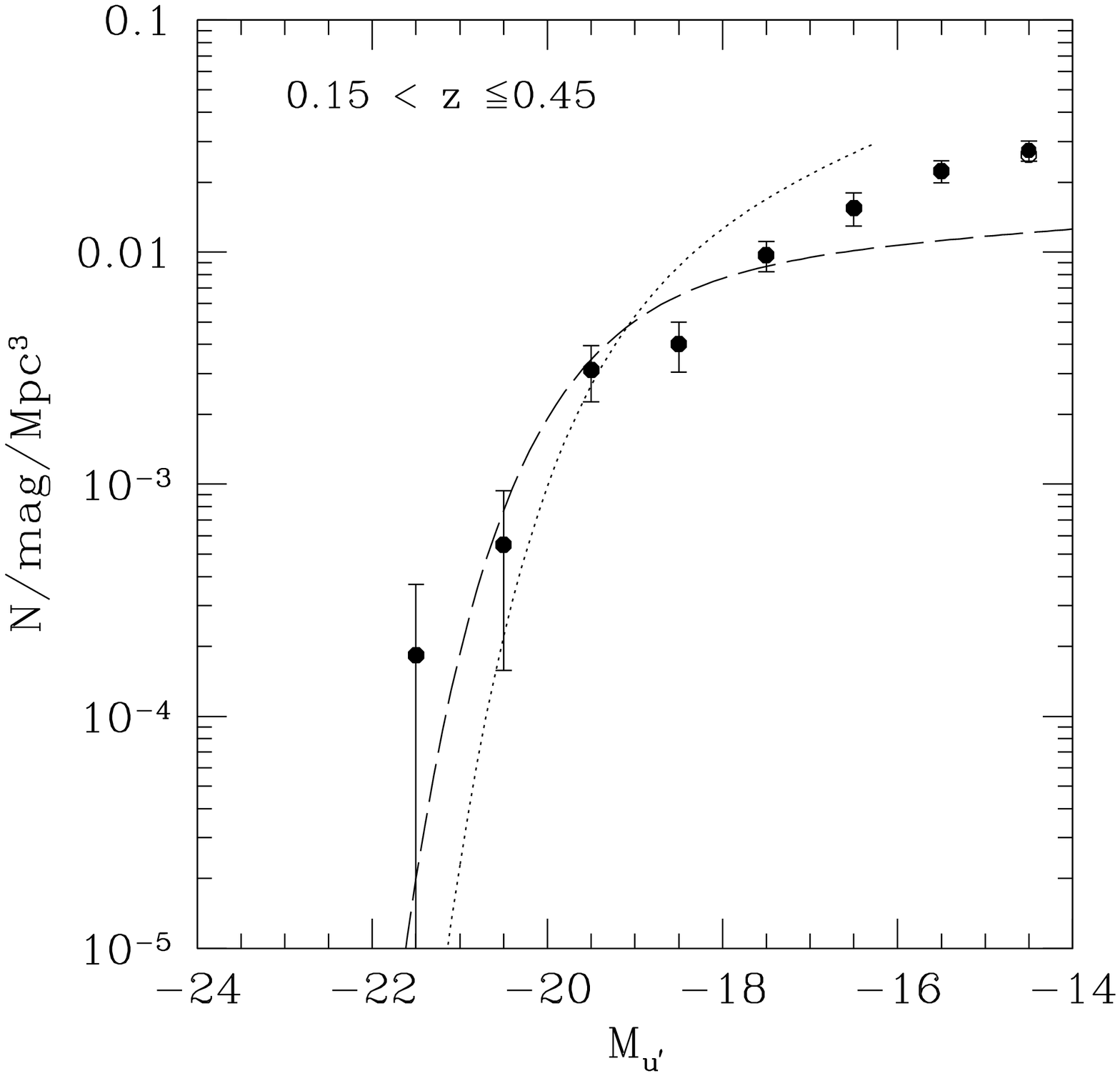}
\includegraphics[width=0.33\textwidth]{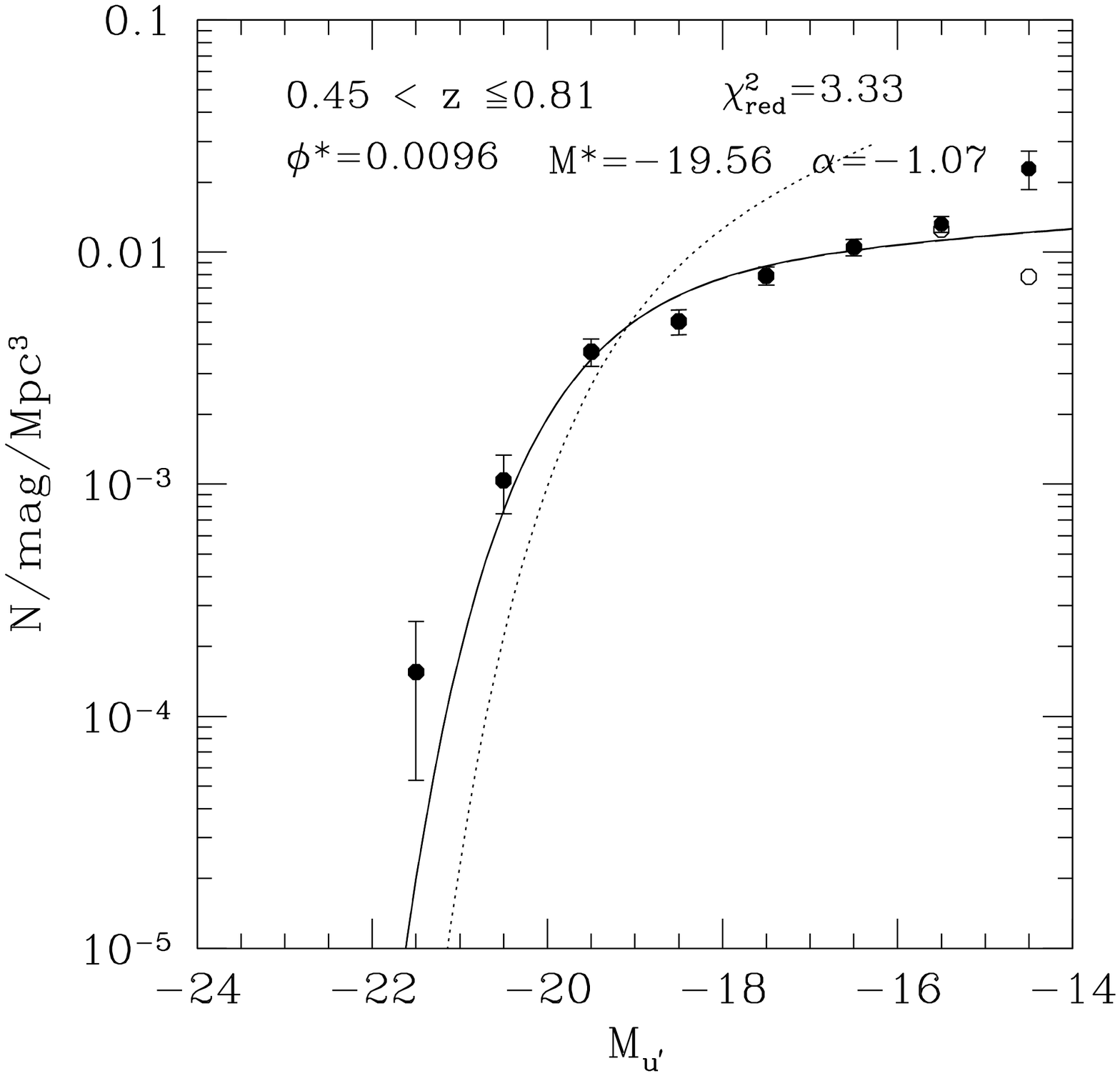}
\includegraphics[width=0.33\textwidth]{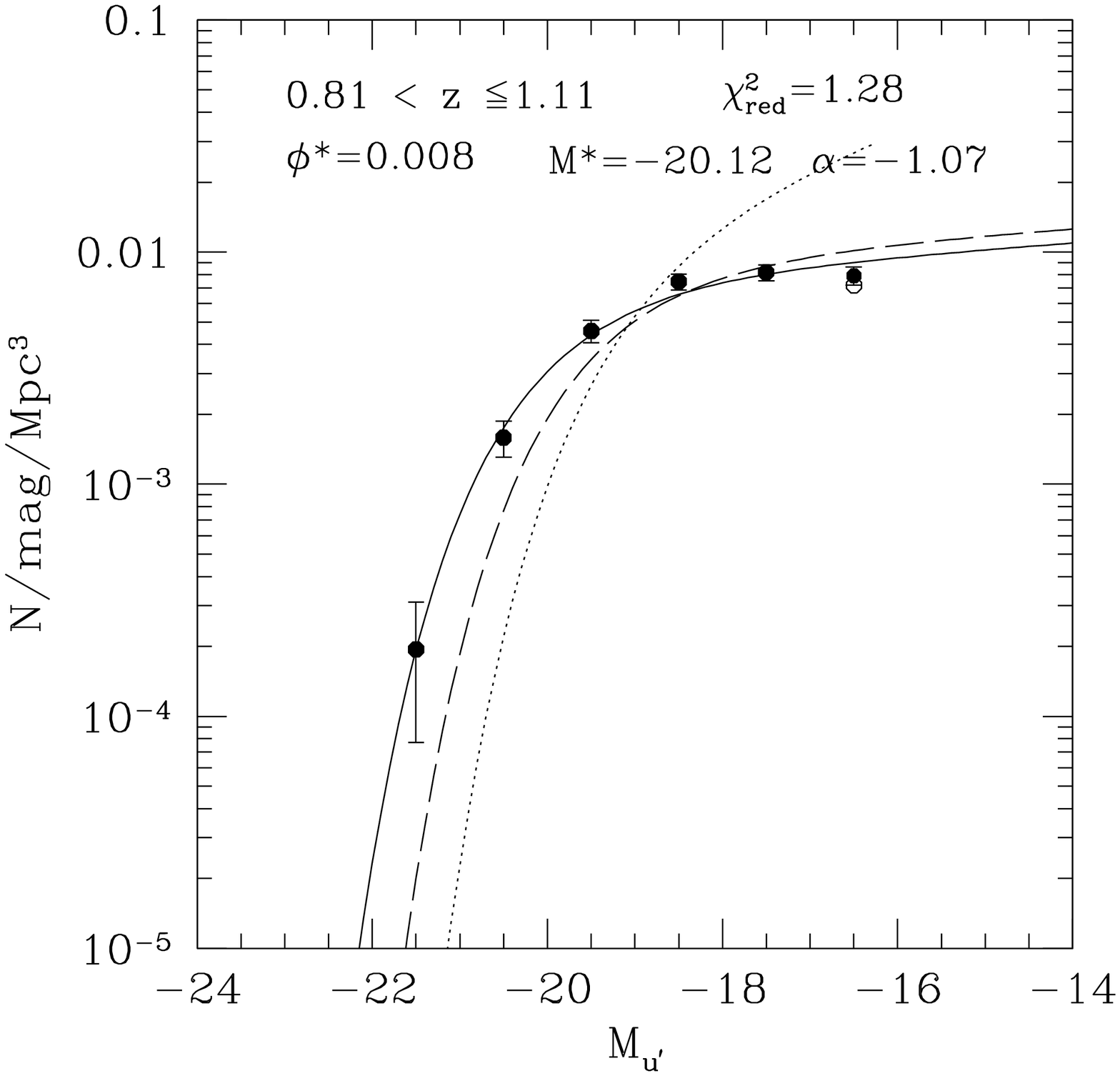}
\includegraphics[width=0.33\textwidth]{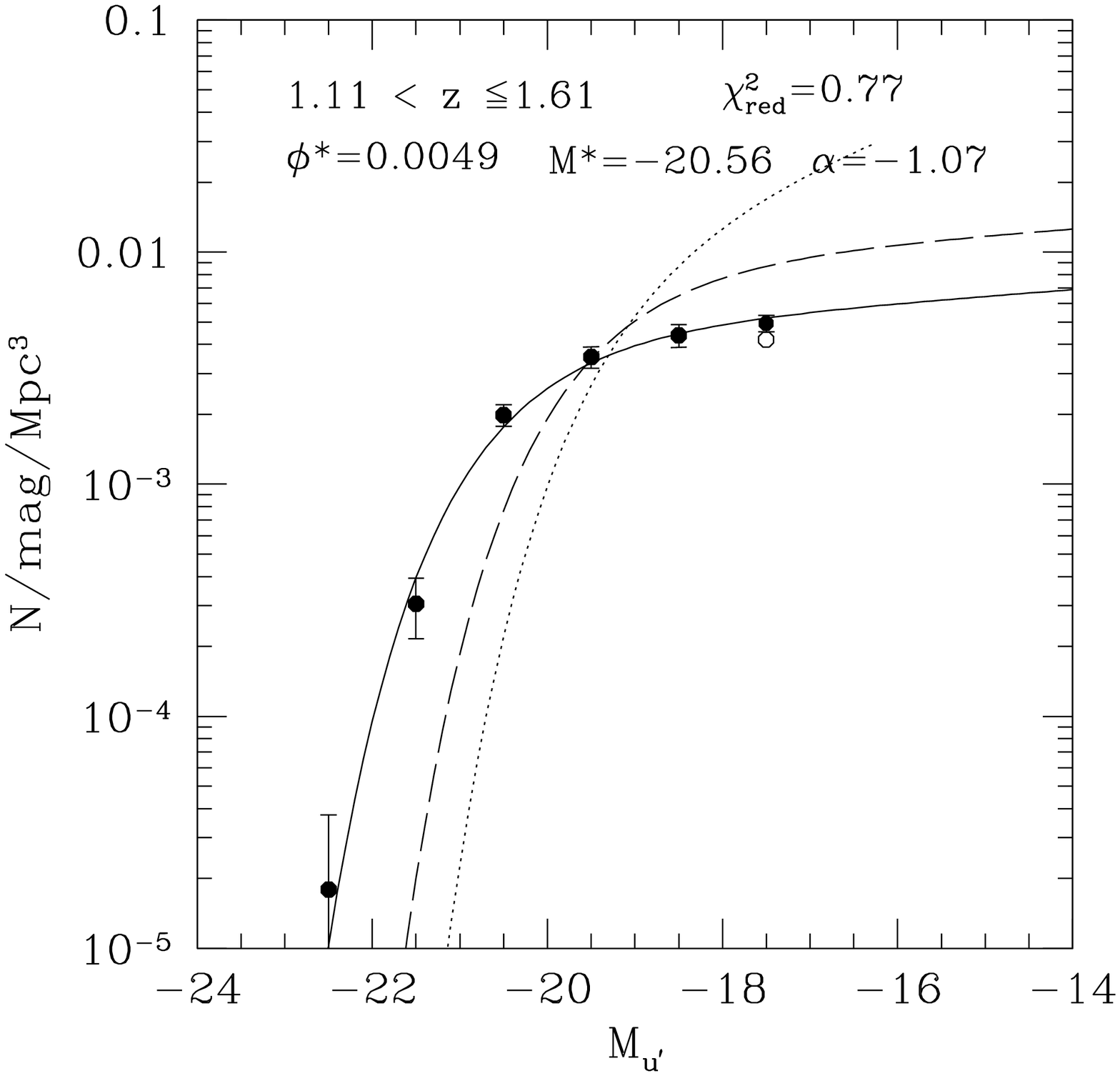}
\includegraphics[width=0.33\textwidth]{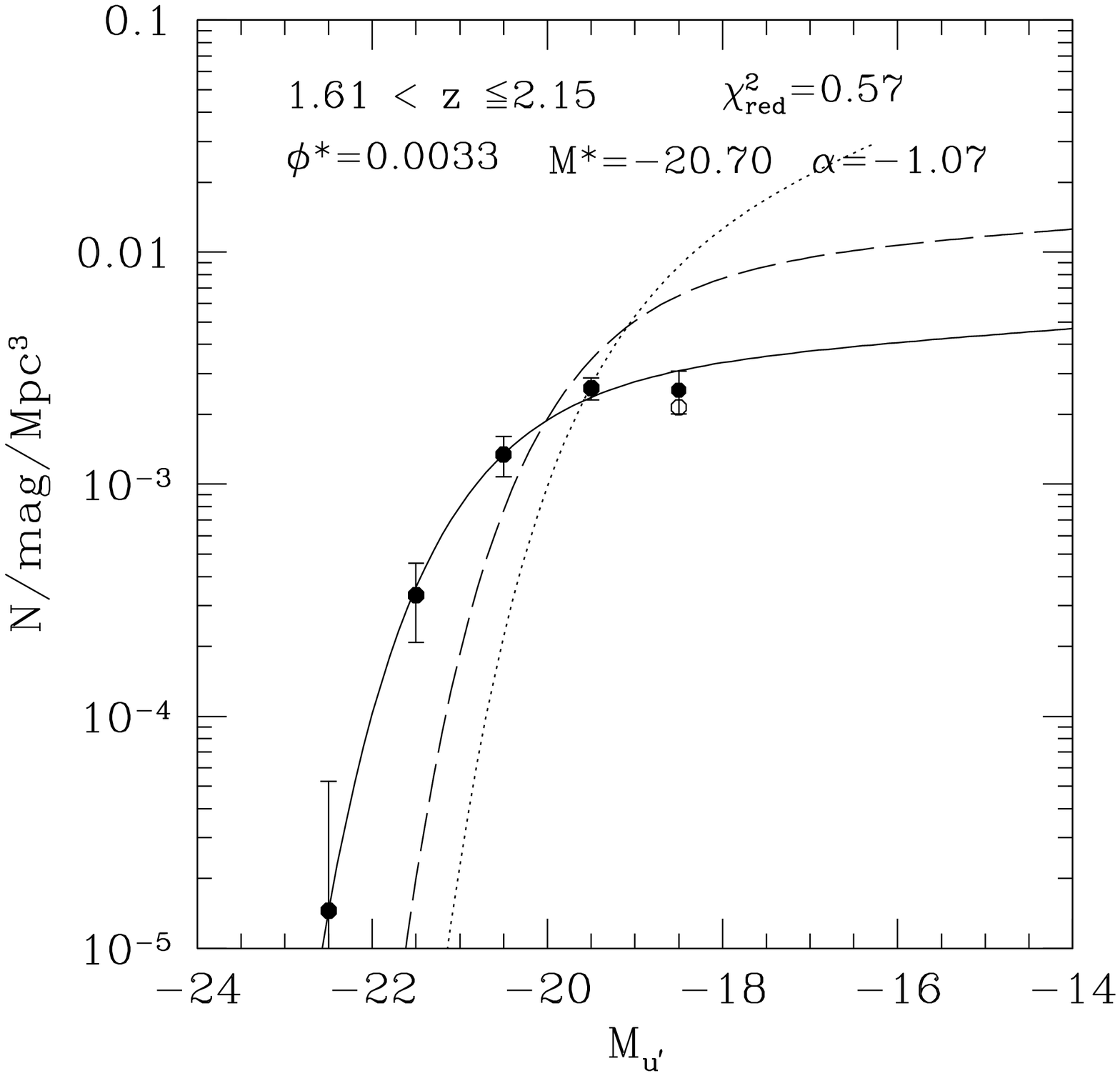}
\includegraphics[width=0.33\textwidth]{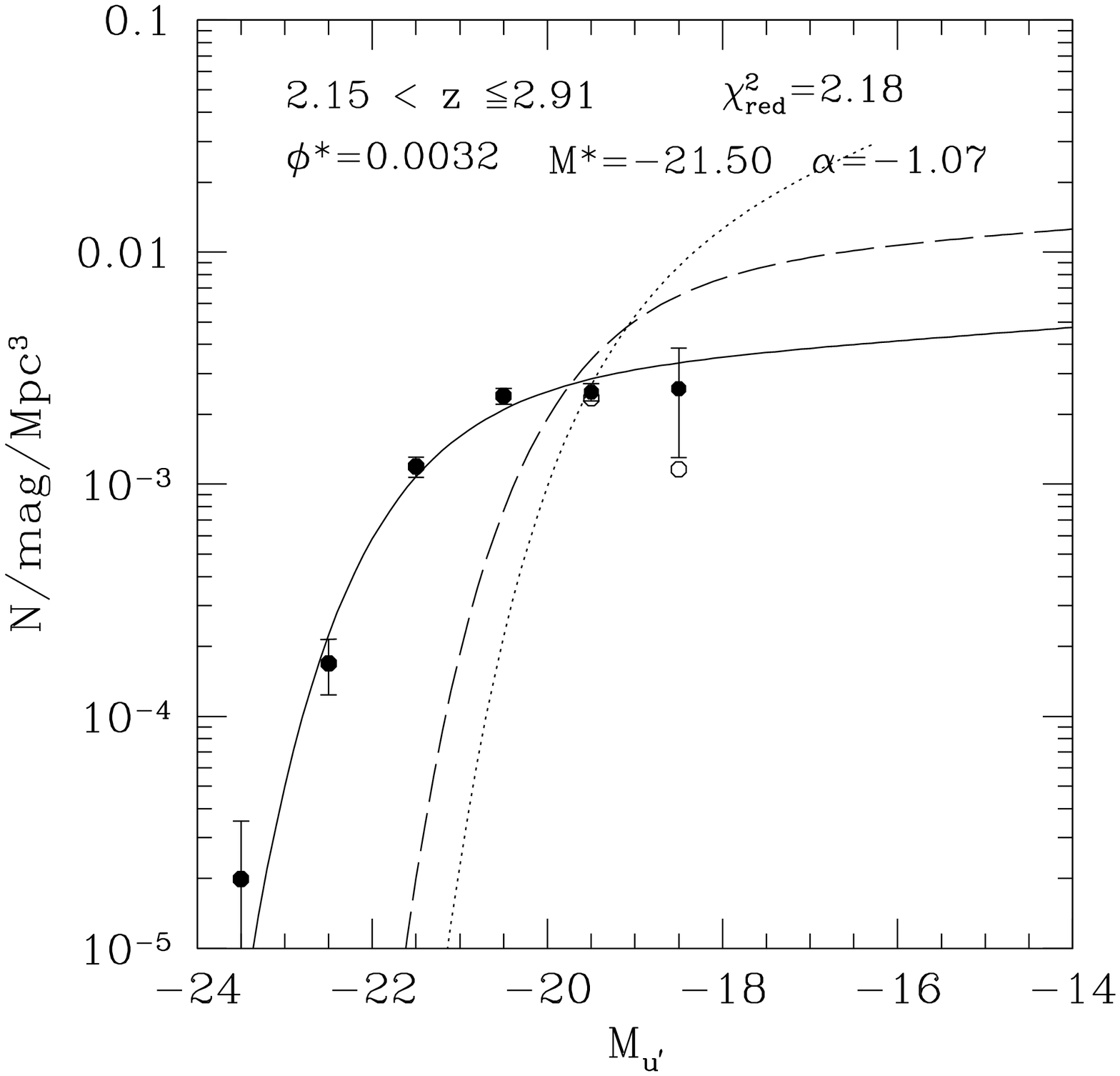}
\includegraphics[width=0.33\textwidth]{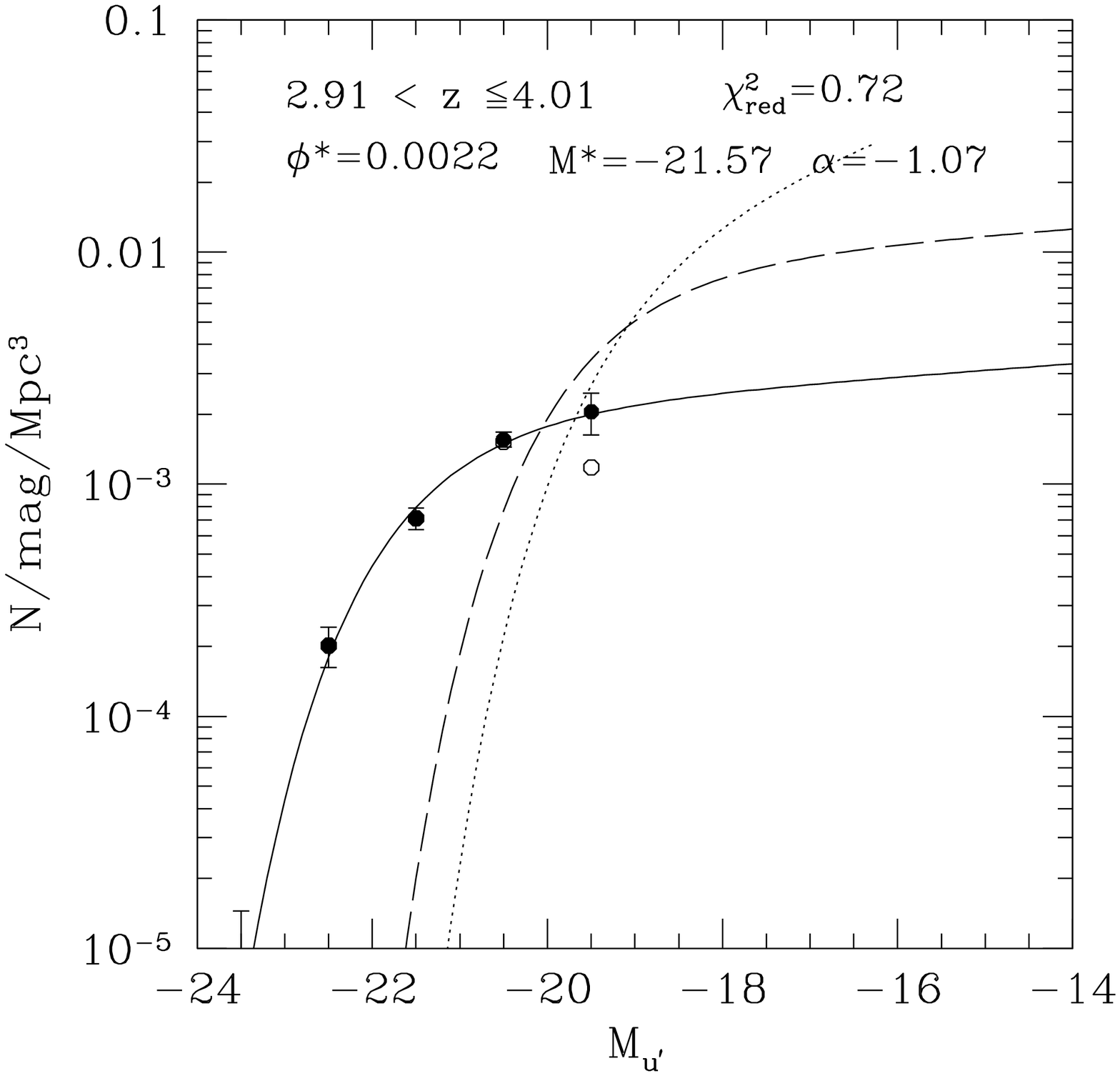}
\includegraphics[width=0.33\textwidth]{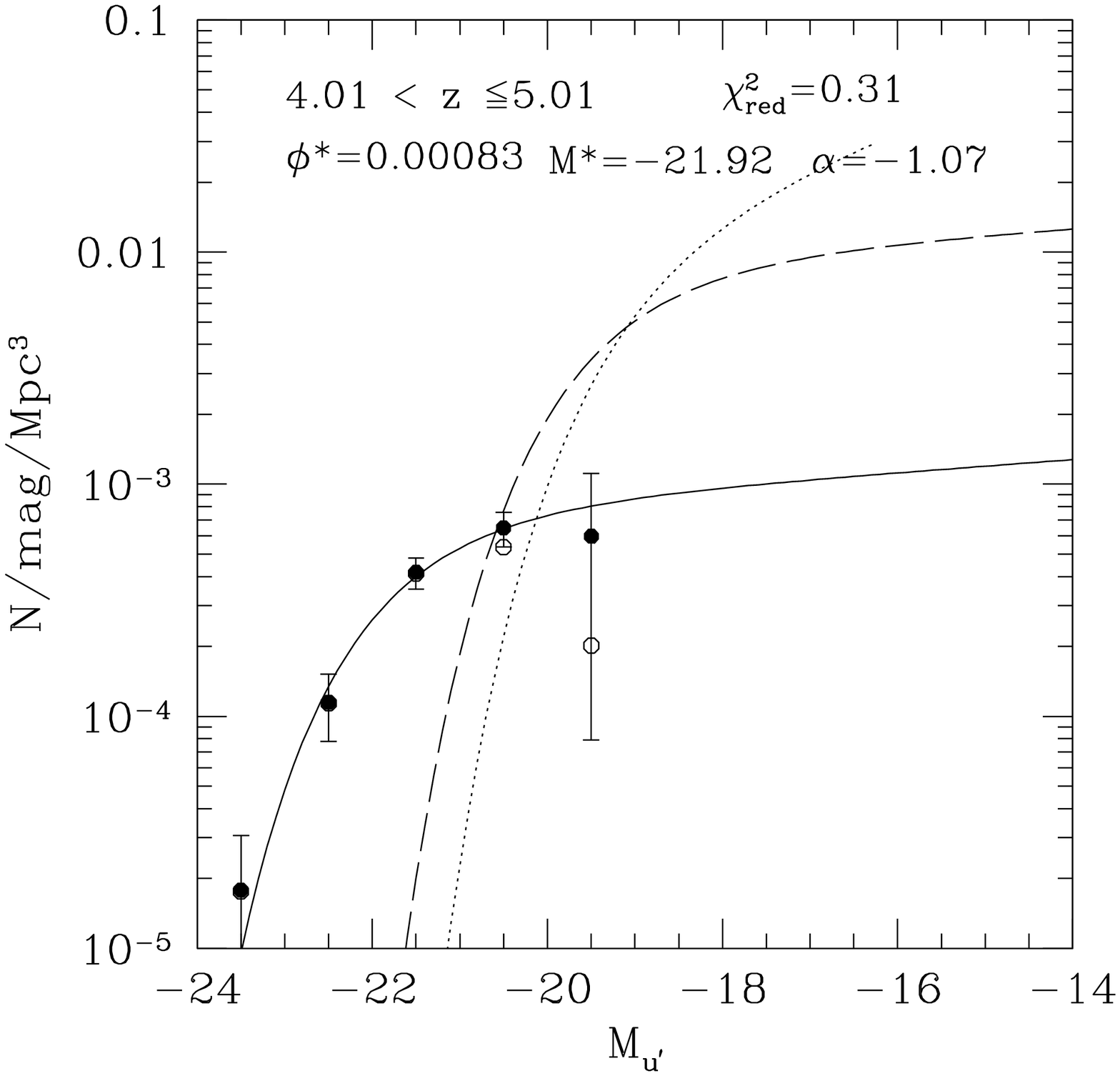}
\includegraphics[width=0.33\textwidth]{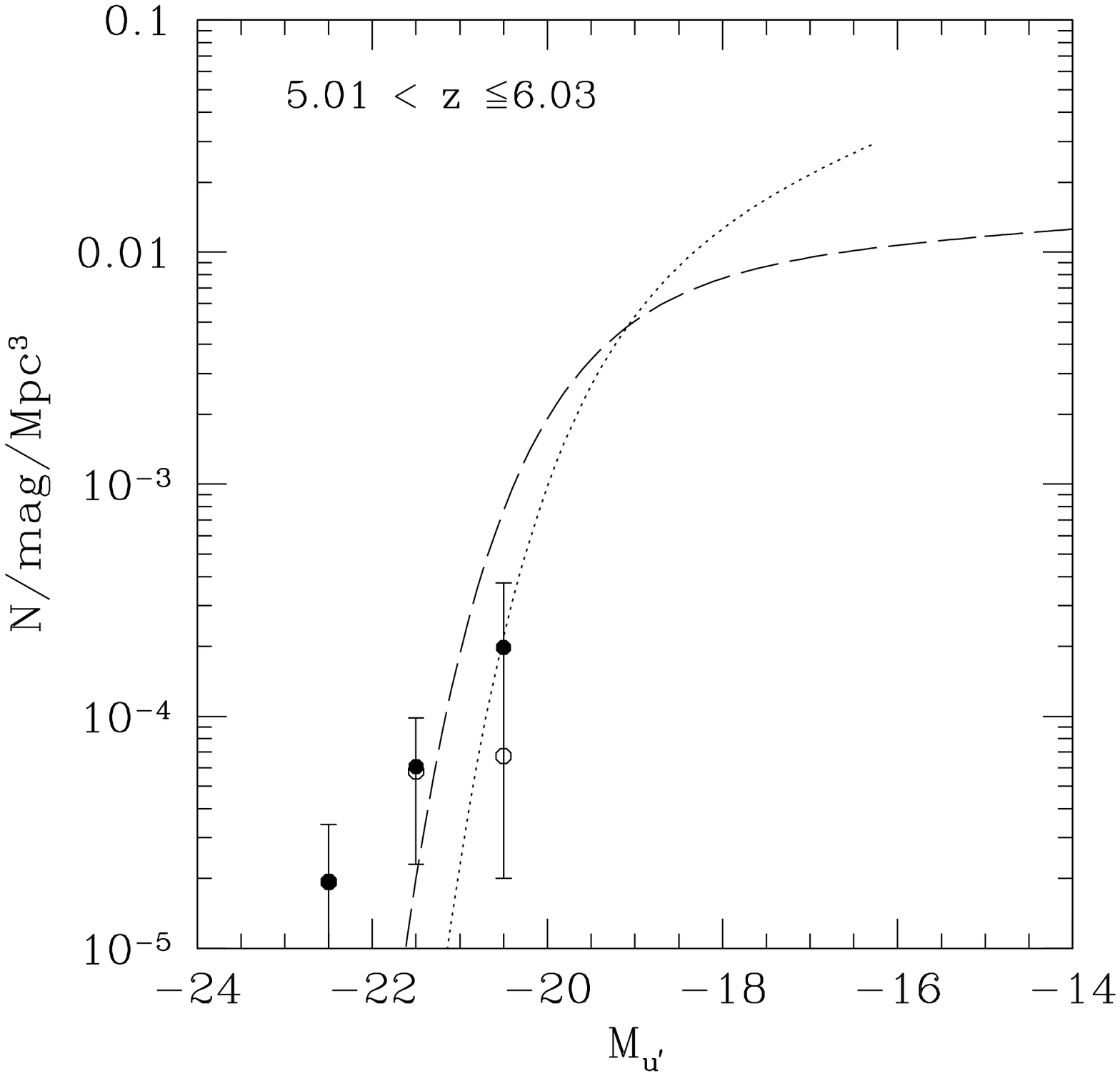}
\caption{\label{fig:lumfkt_fdf_u}
  Luminosity functions in the \textit{u'-band} from low redshift
  (\mbox{$\langle z\rangle=0.3$}, upper left panel) to high redshift
  (\mbox{$\langle z\rangle=5.5$}, lower right panel).  The filled
  (open) symbols show the luminosity function corrected (uncorrected)
  for $V/V_{max}$.  The fitted Schechter functions for a fixed slope
  $\alpha$ are shown as solid lines. Note that we only fit the
  luminosity functions from $\langle z\rangle=0.6$ to $\langle
  z\rangle=4.5$. The parameters of the Schechter functions can be
  found in Table~\ref{tab:schechter_fit_u}. The dotted line represents
  the local u'-band luminosity function derived from the SDSS
  \citep{blanton:1}. The Schechter fit for redshift $\langle
  z\rangle=0.6$ is indicated as dashed line in all panels.}
\end{figure*}
\begin{table*}[bh!]
\caption[]{\label{tab:schechter_fit_u}Schechter function fit in the u'-band}
\begin{center}
\begin{tabular}{c|c|c|c}
 redshift interval & M$^\ast$ (mag) & $\phi^\ast$ (Mpc$^{-3}$) & $\alpha$ (fixed)\\
\hline
0.45 -- 0.81 & $-$19.56 +0.16 $-$0.15 & 0.0096 +0.0006 $-$0.0006 &$-$1.07 \\  
0.81 -- 1.11 & $-$20.12 +0.10 $-$0.10 & 0.0080 +0.0004 $-$0.0004 &$-$1.07 \\  
1.11 -- 1.61 & $-$20.56 +0.08 $-$0.09 & 0.0049 +0.0003 $-$0.0003 &$-$1.07 \\  
1.61 -- 2.15 & $-$20.70 +0.18 $-$0.16 & 0.0033 +0.0004 $-$0.0004 &$-$1.07 \\  
2.15 -- 2.91 & $-$21.50 +0.08 $-$0.08 & 0.0032 +0.0002 $-$0.0002 &$-$1.07 \\  
2.91 -- 4.01 & $-$21.57 +0.11 $-$0.10 & 0.0022 +0.0002 $-$0.0002 &$-$1.07 \\  
4.01 -- 5.01 & $-$21.92 +0.24 $-$0.20 & 0.0008 +0.0002 $-$0.0001 &$-$1.07 \\  
\end{tabular}
\end{center}
\end{table*}

\begin{table*}[hhh!]
\caption[]{\label{tab:schechter_fit_g}Schechter function fit in the g'-band}
\begin{center}
\begin{tabular}{c|c|c|c}
 redshift interval & M$^\ast$ (mag) & $\phi^\ast$ (Mpc$^{-3}$) & $\alpha$ (fixed)\\
\hline
0.45 -- 0.81 & $-$21.47 +0.20 $-$0.20 & 0.0042 +0.0003 $-$0.0003 &$-$1.25 \\  
0.81 -- 1.11 & $-$21.72 +0.15 $-$0.15 & 0.0039 +0.0003 $-$0.0003 &$-$1.25 \\  
1.11 -- 1.61 & $-$22.01 +0.14 $-$0.14 & 0.0026 +0.0002 $-$0.0002 &$-$1.25 \\  
1.61 -- 2.15 & $-$21.82 +0.20 $-$0.20 & 0.0020 +0.0004 $-$0.0003 &$-$1.25 \\  
2.15 -- 2.91 & $-$22.62 +0.13 $-$0.10 & 0.0020 +0.0002 $-$0.0002 &$-$1.25 \\  
2.91 -- 4.01 & $-$22.51 +0.13 $-$0.14 & 0.0016 +0.0002 $-$0.0002 &$-$1.25 \\  
4.01 -- 5.01 & $-$23.12 +0.22 $-$0.23 & 0.0005 +0.0001 $-$0.0001 &$-$1.25 \\  
\end{tabular}
\end{center}
\end{table*}

\begin{figure*}
\includegraphics[width=0.33\textwidth]{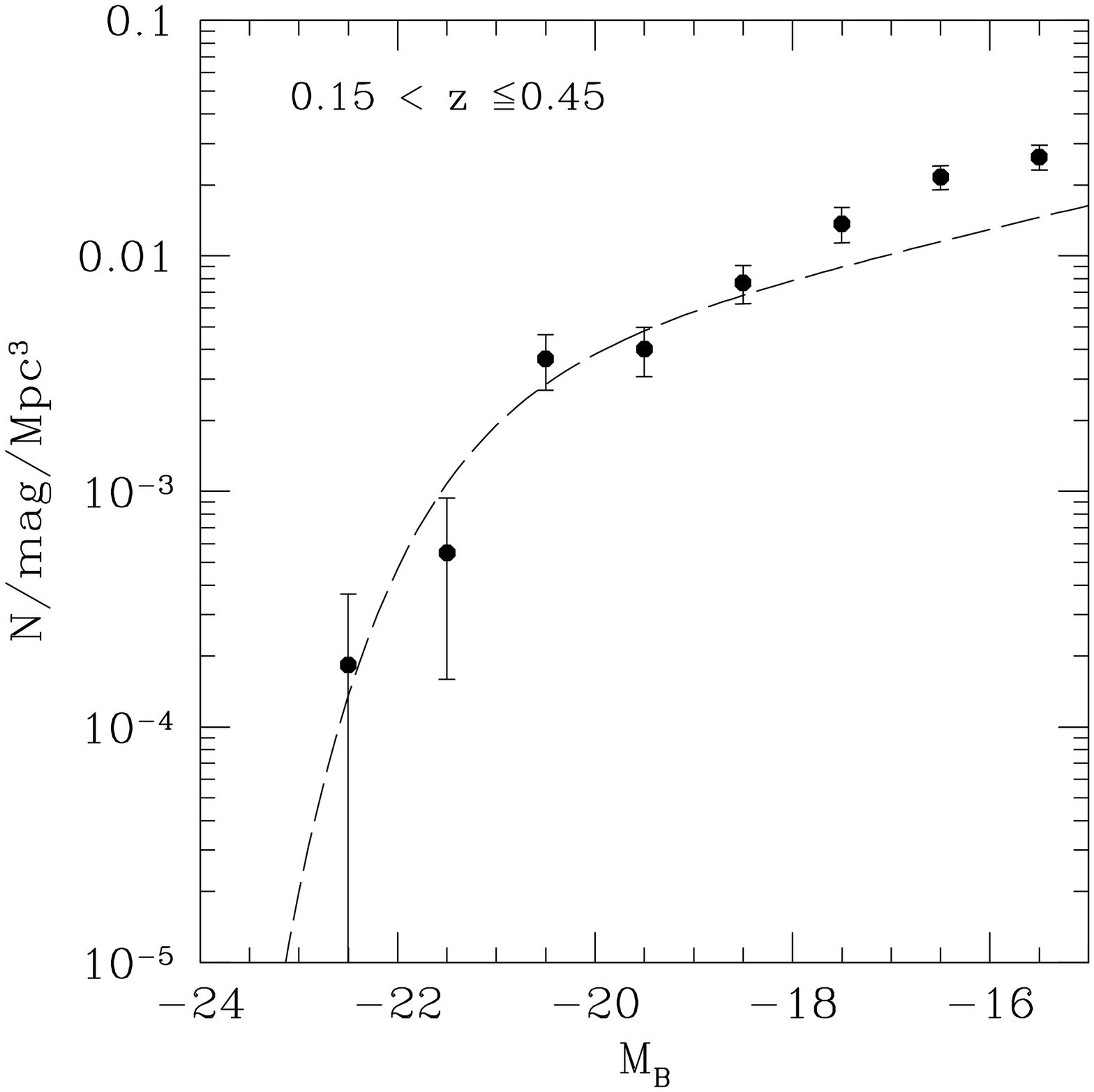}
\includegraphics[width=0.33\textwidth]{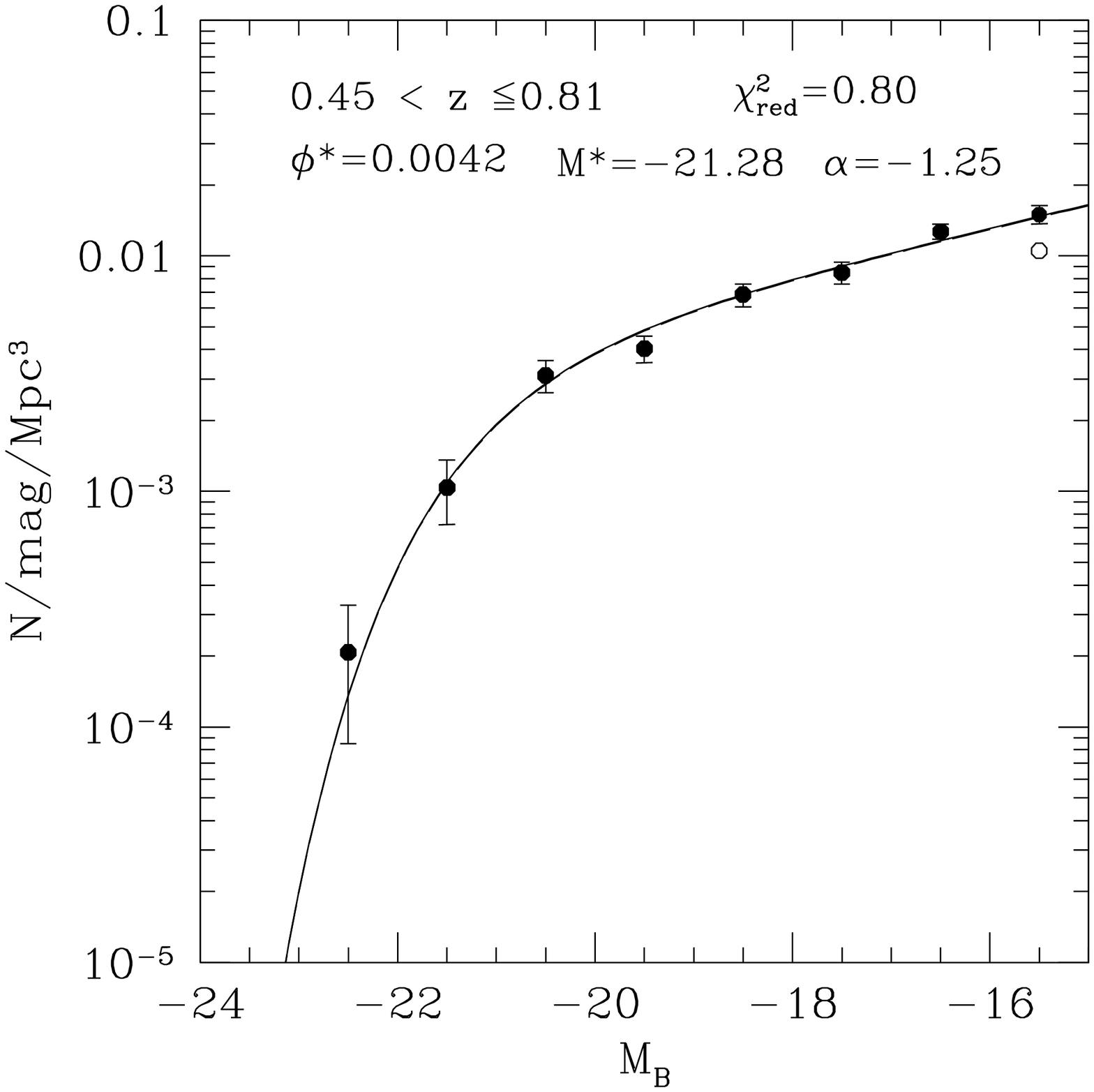}
\includegraphics[width=0.33\textwidth]{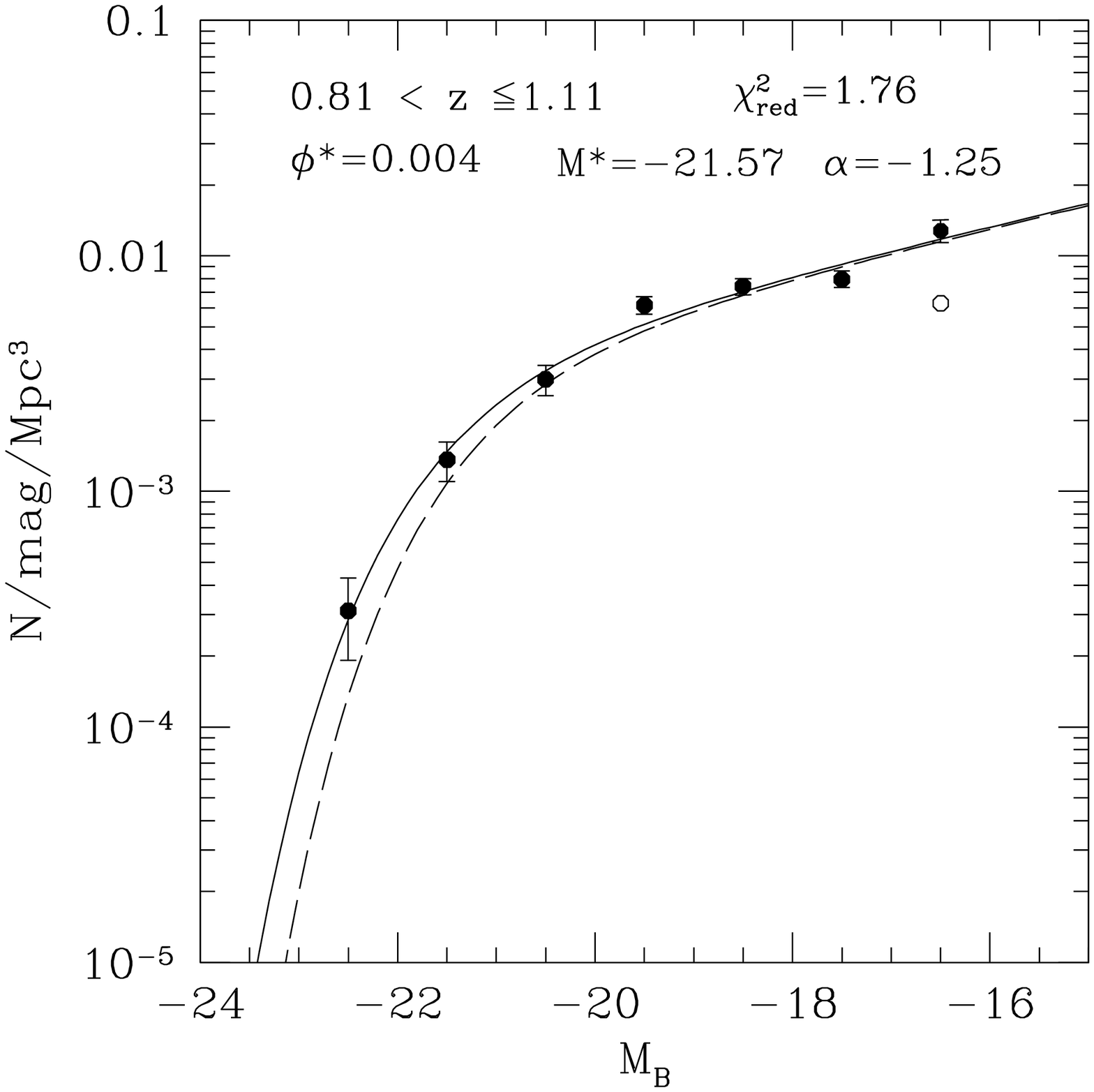}
\includegraphics[width=0.33\textwidth]{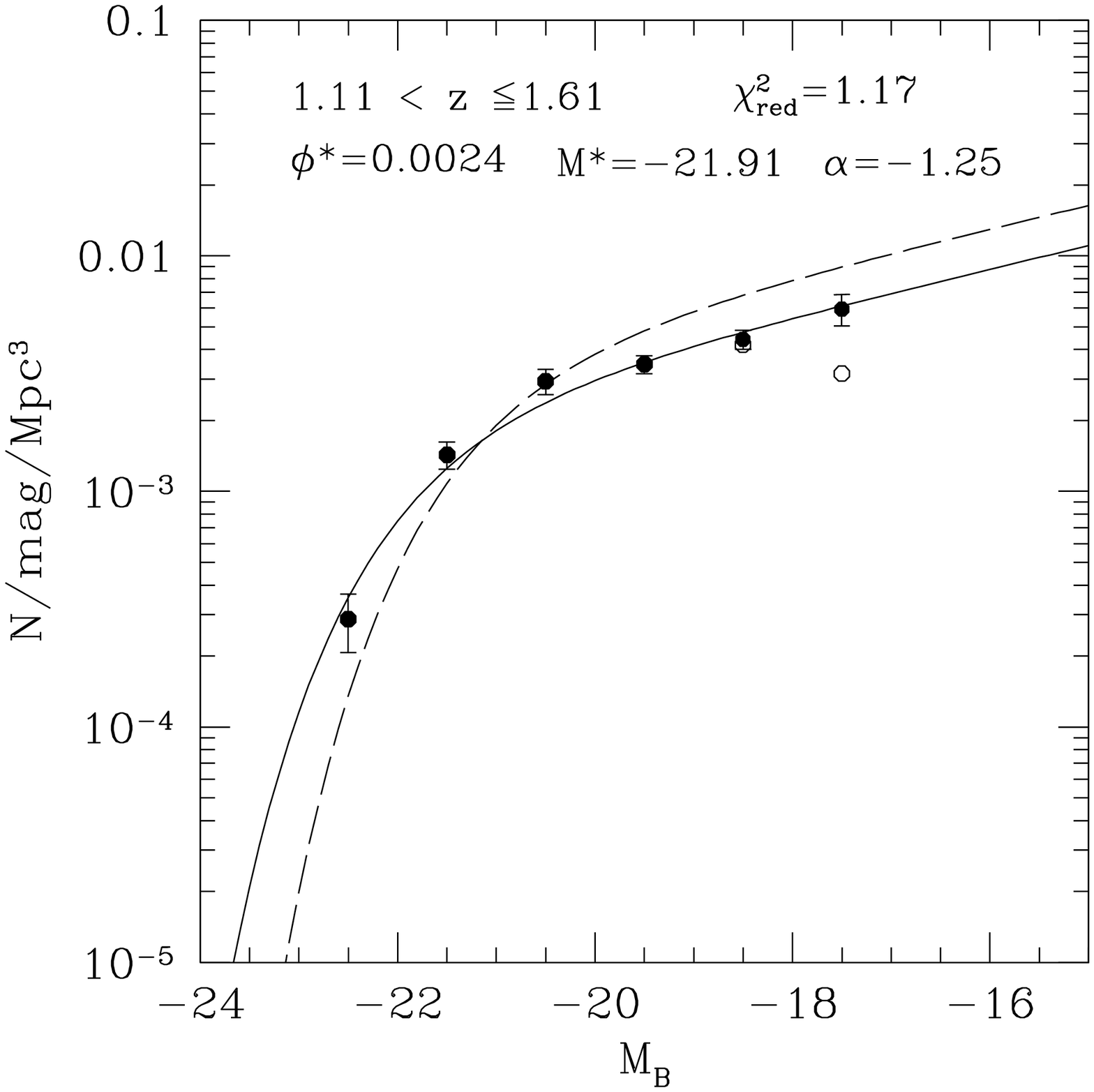}
\includegraphics[width=0.33\textwidth]{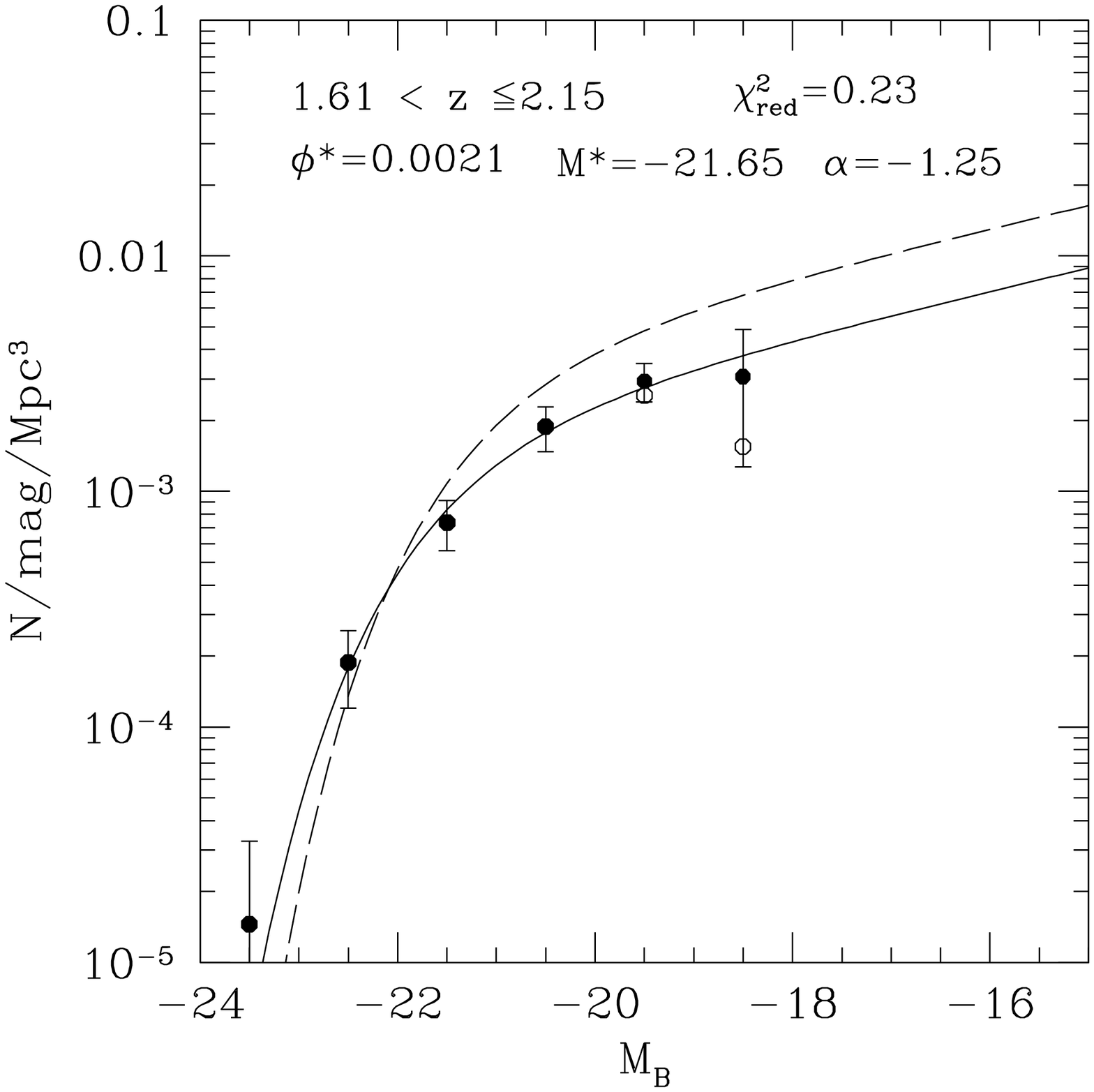}
\includegraphics[width=0.33\textwidth]{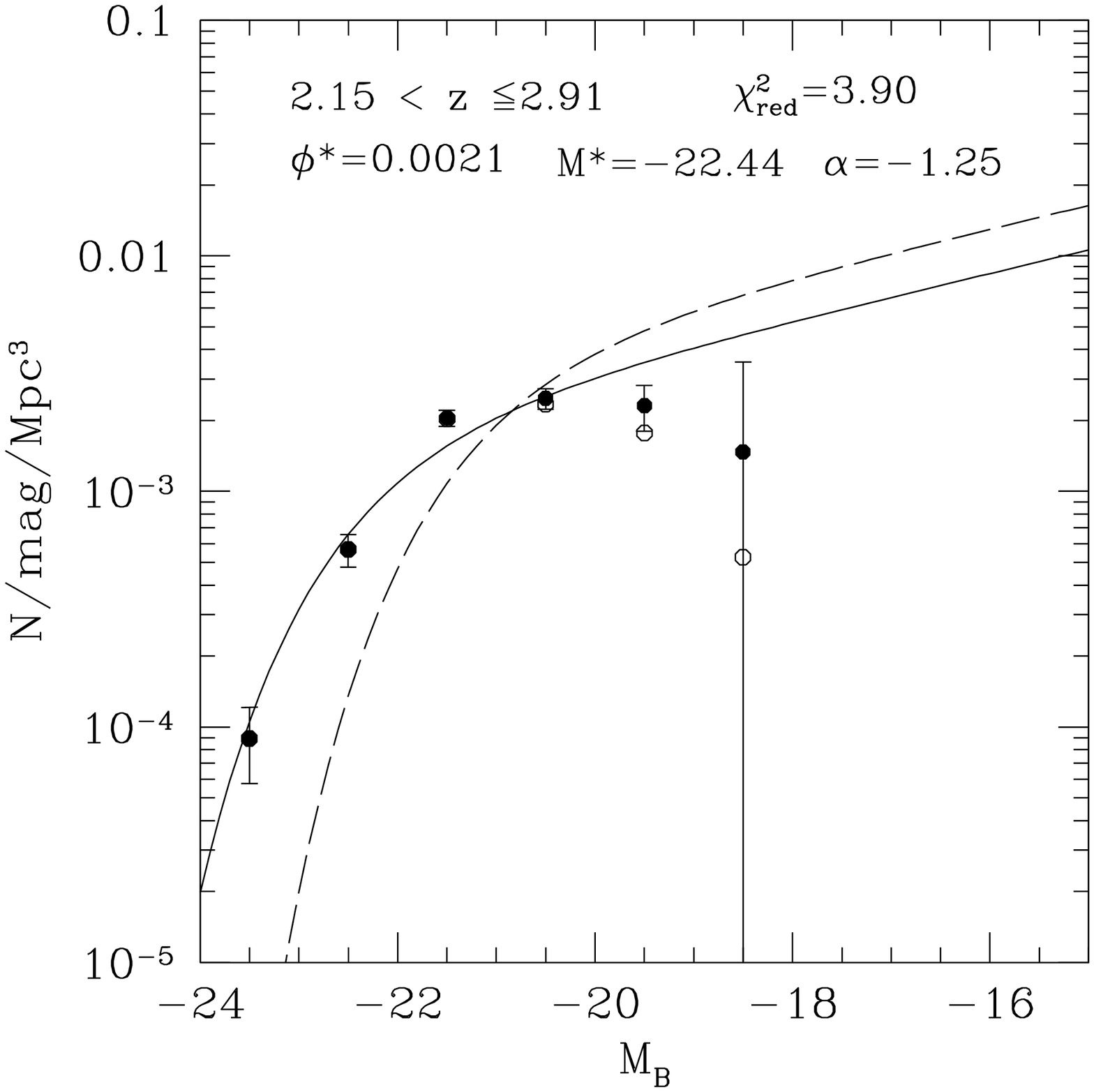}
\includegraphics[width=0.33\textwidth]{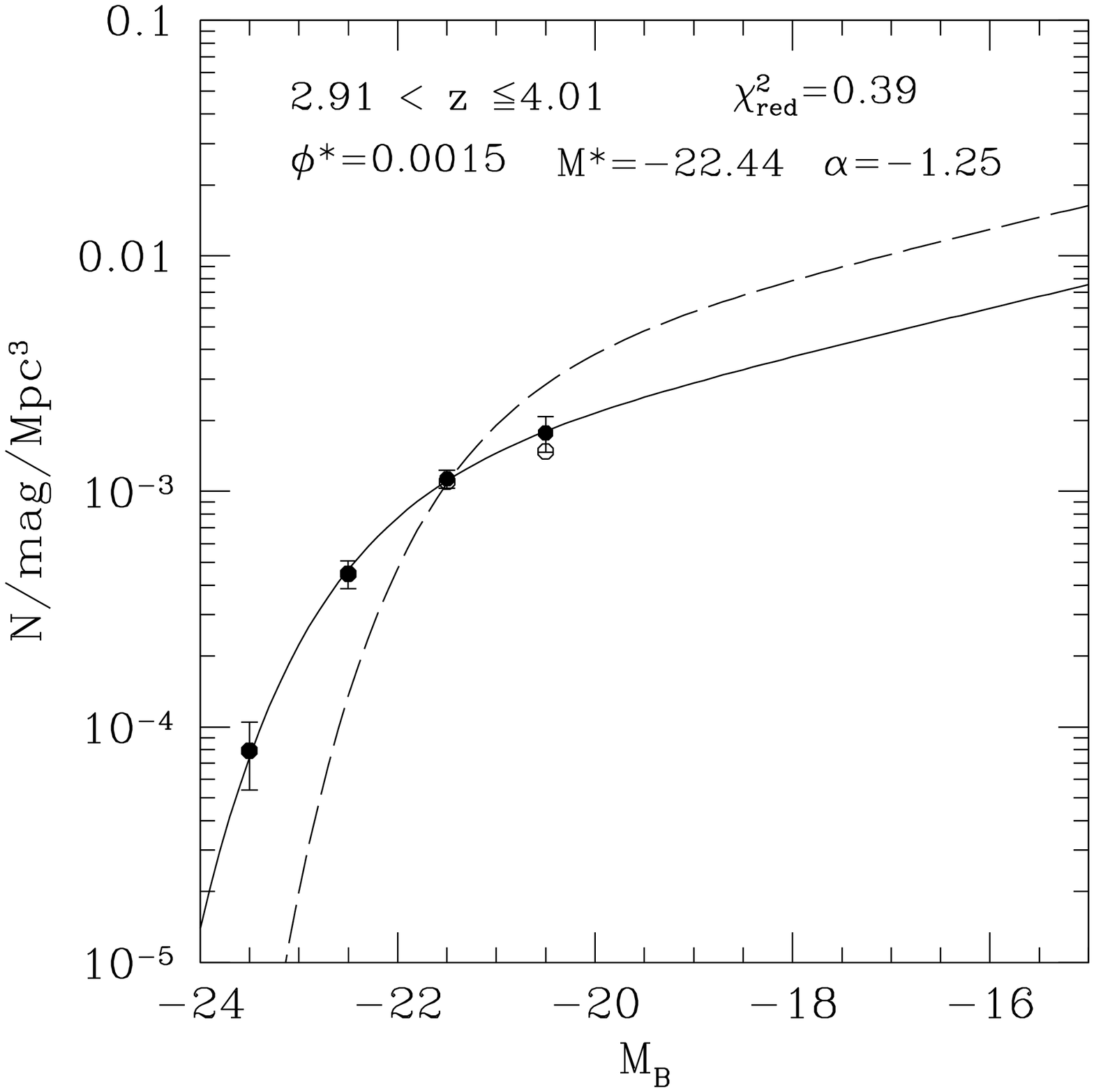}
\includegraphics[width=0.33\textwidth]{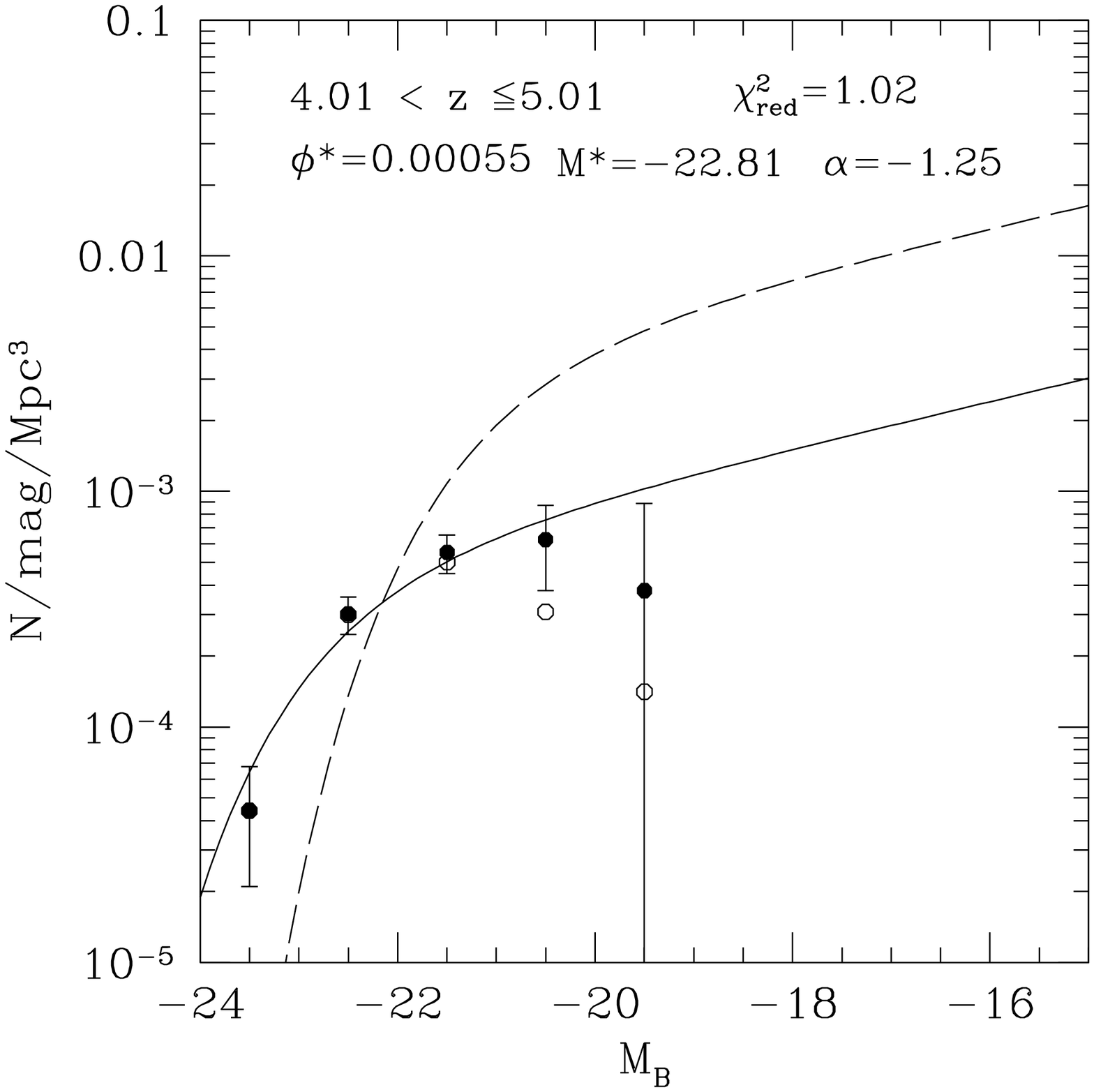}
\includegraphics[width=0.33\textwidth]{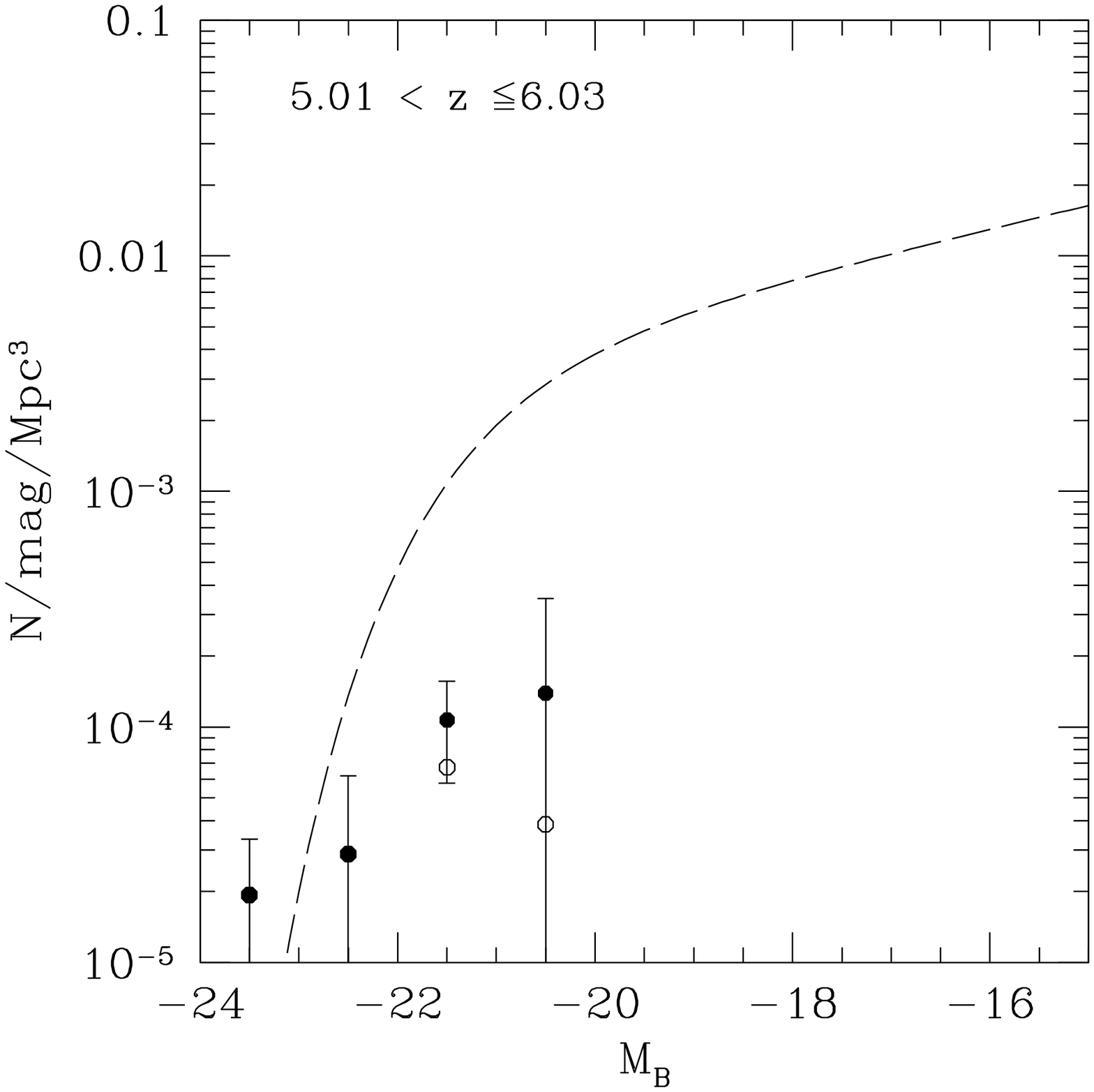}
\caption{\label{fig:lumfkt_fdf_b}
  Luminosity functions in the \textit{B-band} from low redshift
  (\mbox{$\langle z\rangle=0.3$}, upper left panel) to high redshift
  (\mbox{$\langle z\rangle=5.5$}, lower right panel). The filled
  (open) symbols show the luminosity function corrected (uncorrected)
  for $V/V_{max}$.  The fitted Schechter functions for a fixed slope
  $\alpha$ are shown as solid lines. Note that we only fit the
  luminosity functions from $\langle z\rangle=0.6$ to $\langle
  z\rangle=4.5$. The parameters of the Schechter functions can be
  found in Table~\ref{tab:schechter_fit_b}. The Schechter fit for
  redshift $\langle z\rangle=0.6$ is indicated as dashed line in all
  panels.}

\end{figure*}

\begin{table*}[]
\caption[]{\label{tab:schechter_fit_b}Schechter function fit in the B-band}
\begin{center}
\begin{tabular}{c|c|c|c}
 redshift interval & M$^\ast$ (mag) & $\phi^\ast$ (Mpc$^{-3}$) & $\alpha$ (fixed)\\
\hline
0.45 -- 0.81 & $-$21.28 +0.21 $-$0.18 & 0.0042 +0.0004 $-$0.0003 &$-$1.25 \\  
0.81 -- 1.11 & $-$21.57 +0.15 $-$0.13 & 0.0040 +0.0003 $-$0.0002 &$-$1.25 \\  
1.11 -- 1.61 & $-$21.91 +0.13 $-$0.13 & 0.0024 +0.0002 $-$0.0002 &$-$1.25 \\  
1.61 -- 2.15 & $-$21.65 +0.22 $-$0.22 & 0.0021 +0.0004 $-$0.0004 &$-$1.25 \\  
2.15 -- 2.91 & $-$22.44 +0.11 $-$0.09 & 0.0021 +0.0002 $-$0.0002 &$-$1.25 \\  
2.91 -- 4.01 & $-$22.44 +0.15 $-$0.14 & 0.0015 +0.0002 $-$0.0002 &$-$1.25 \\  
4.01 -- 5.01 & $-$22.81 +0.21 $-$0.25 & 0.0005 +0.0001 $-$0.0001 &$-$1.25 \\  
\end{tabular}
\end{center}
\end{table*}

\clearpage\newpage

\section{Comparison with literature}
\label{sec:comp_lit}

In this appendix we compare the luminosity functions derived in the
FDF with the results of further publications as introduced in
Sect.~\ref{sec:lit}. The filled (open) circles show the completeness
corrected (uncorrected) luminosity function as derived in the FDF
in the redshift bin listed in the lower right corner. The solid
line(s) represent the Schechter function given in the different papers
transformed to our cosmology.  To visualize the errors associated to
this Schechter function we perform a Monte-Carlo simulation using the
errors of the Schechter parameters reported in the specific paper (see
Sect.~\ref{sec:lit} for more details).  As the errors for all three
Schechter parameters ($\Delta$M$^\ast$, $\Delta\phi^\ast$, and
$\Delta\alpha$) are not always given in the paper, we denote in the
caption the errors used to perform the simulation.  The region wherein
\mbox{68.8 \%} of the realizations lie are shown as shaded region in
the plots and corresponds roughly to the 1$\sigma$ error due to the
Schechter errors reported in the figure captions. Moreover the cut-off
of the shaded region marks the limiting magnitude of the survey we
compare with.

\subsection{UV bands}
\label{sec:uv_restframe}

\begin{center}
\begin{figure}[]
\includegraphics[width=0.45\textwidth]{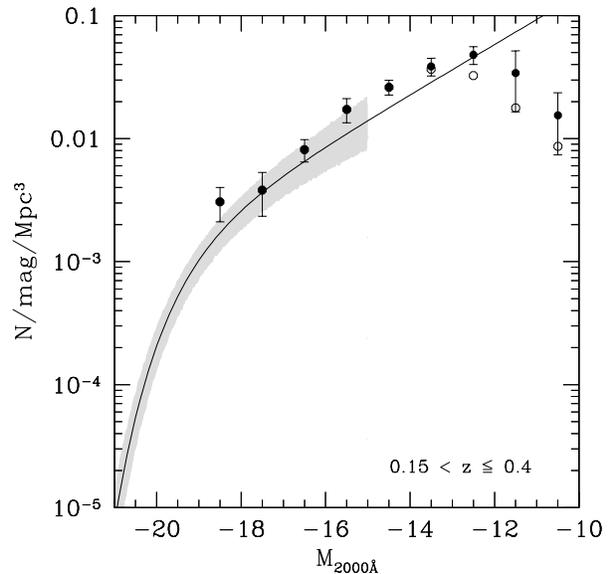}
\caption{\label{fig:lit_uv_sullivan} 
  Comparison of the luminosity function at 2000~\AA\ of the FDF with
  the Schechter function derived in \textit{\citet{sullivan:1}}
  ($z\simeq 0-0.4$).  The shaded region is based on $\Delta$M$^\ast$,
  $\Delta\phi^\ast$, and $\Delta\alpha$, where the cut-off at
  low luminosity indicates the limiting magnitude of the sample.}
\end{figure}
\end{center}

\textit{\citet{sullivan:1}:}\\
Although the volume of the FDF at low redshift is rather small, and
therefore is not well suited to properly sample the bright end of the
Schechter function, we compare for completeness in
Fig.~\ref{fig:lit_uv_sullivan} our luminosity function also with the
luminosity function derived in \citet{sullivan:1}.
Their sample contains 433 UV-selected sources within an area of $2.2
deg^2 $.  273 of these objects are galaxies and cover the redshift
range $z\simeq 0-0.4$.  The solid line in
Fig.~\ref{fig:lit_uv_sullivan} represents the luminosity function at
2000~\AA\ from \citet{sullivan:1} whereas the filled circles shows our
$V/V_{max}$ corrected luminosity function derived at $0.15 <z\le 0.4$.
Despite the small volume, the I-selected catalogue and the
extrapolated 2000~\AA\ luminosity function (see above) there is a
general agreement with only small systematic offsets (probably also
due to a known cluster at $z\sim 0.33$  \citep{noll:1}).  This
is an additional confirmation of the validity of our technique to
derive the luminosity function as
described in Sect.~\ref{sec:lumfkt:method}.\\

\begin{figure*}
\includegraphics[width=0.45\textwidth]{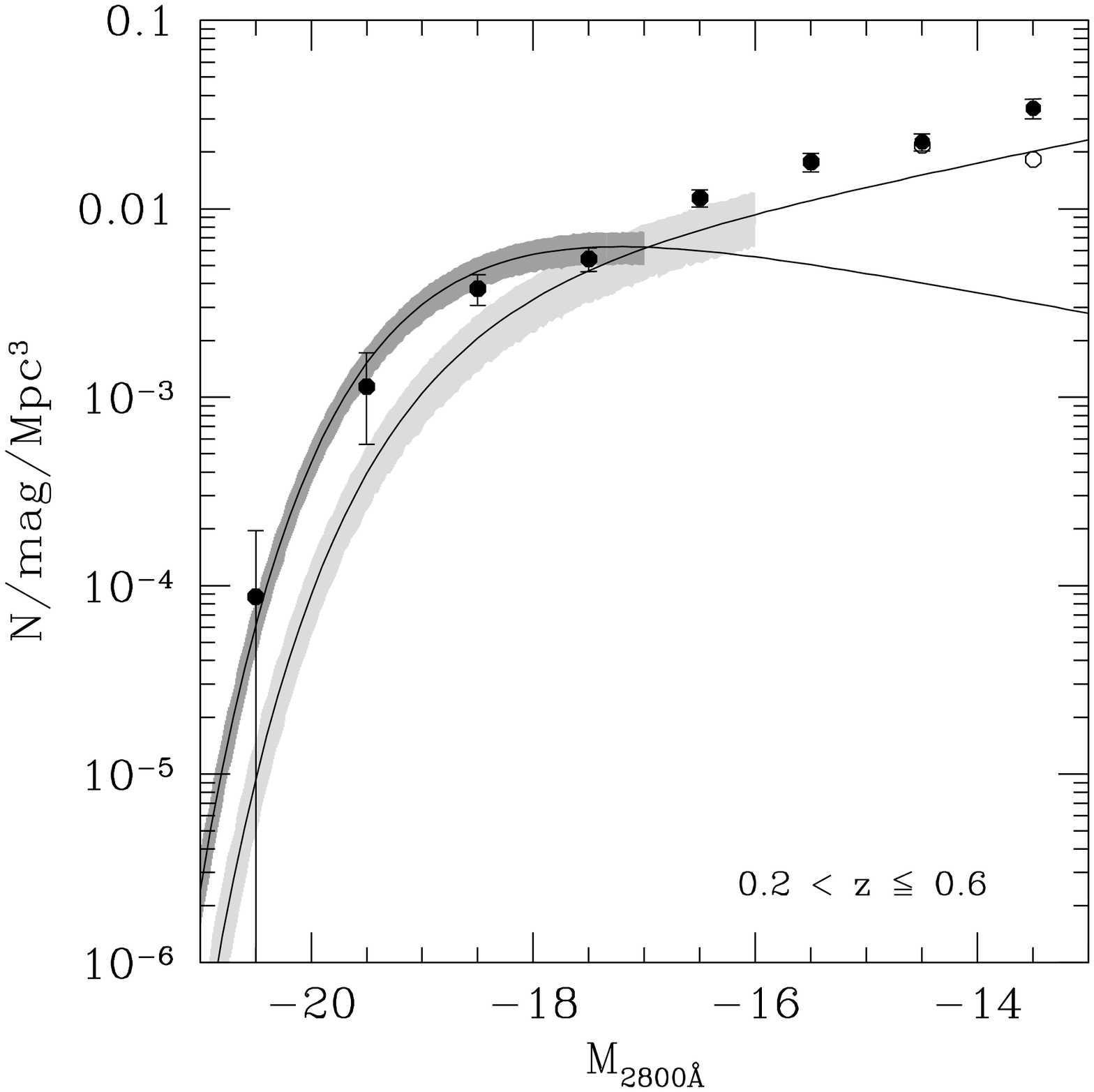}
\hfill
\includegraphics[width=0.45\textwidth]{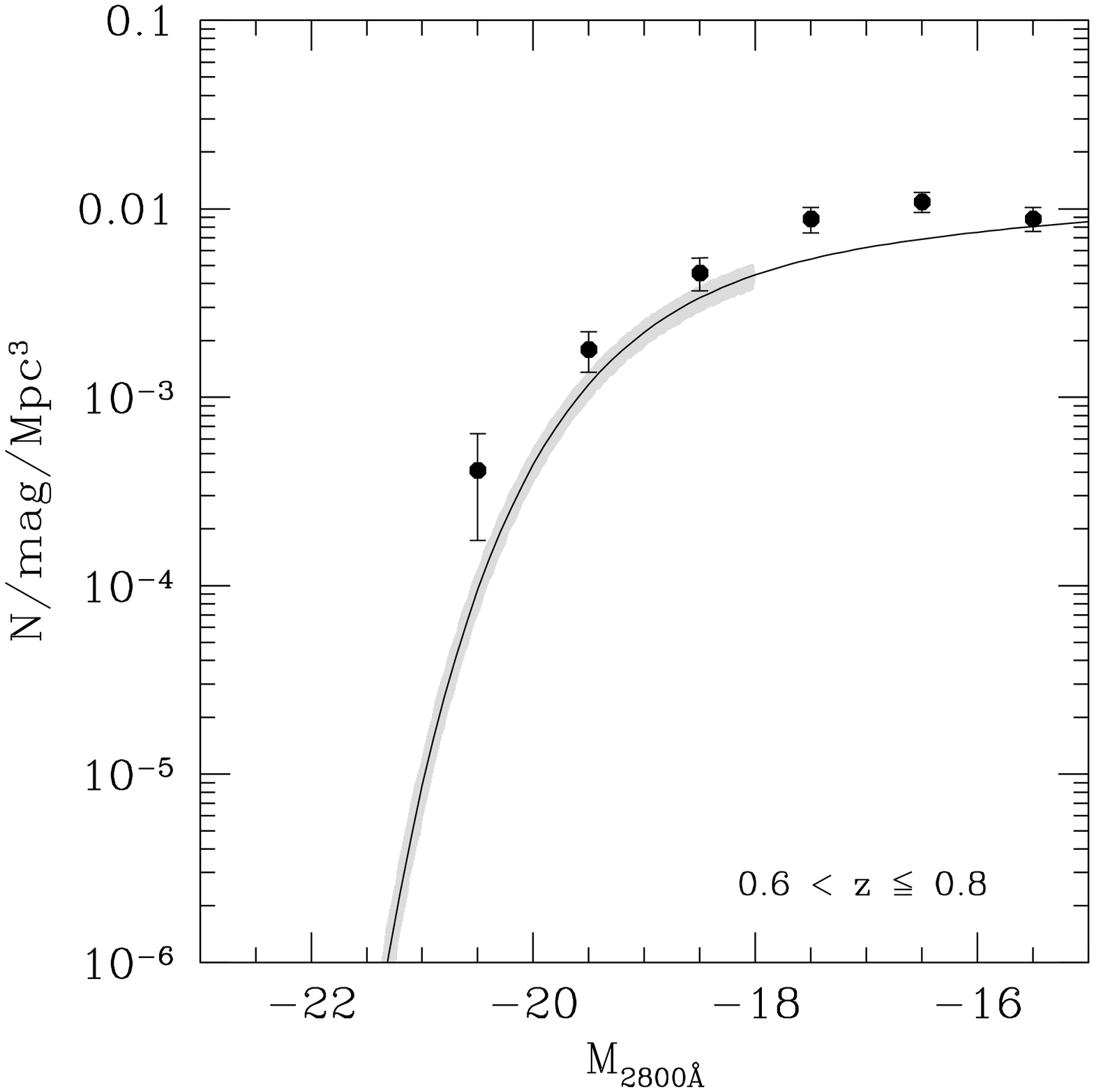}
\includegraphics[width=0.45\textwidth]{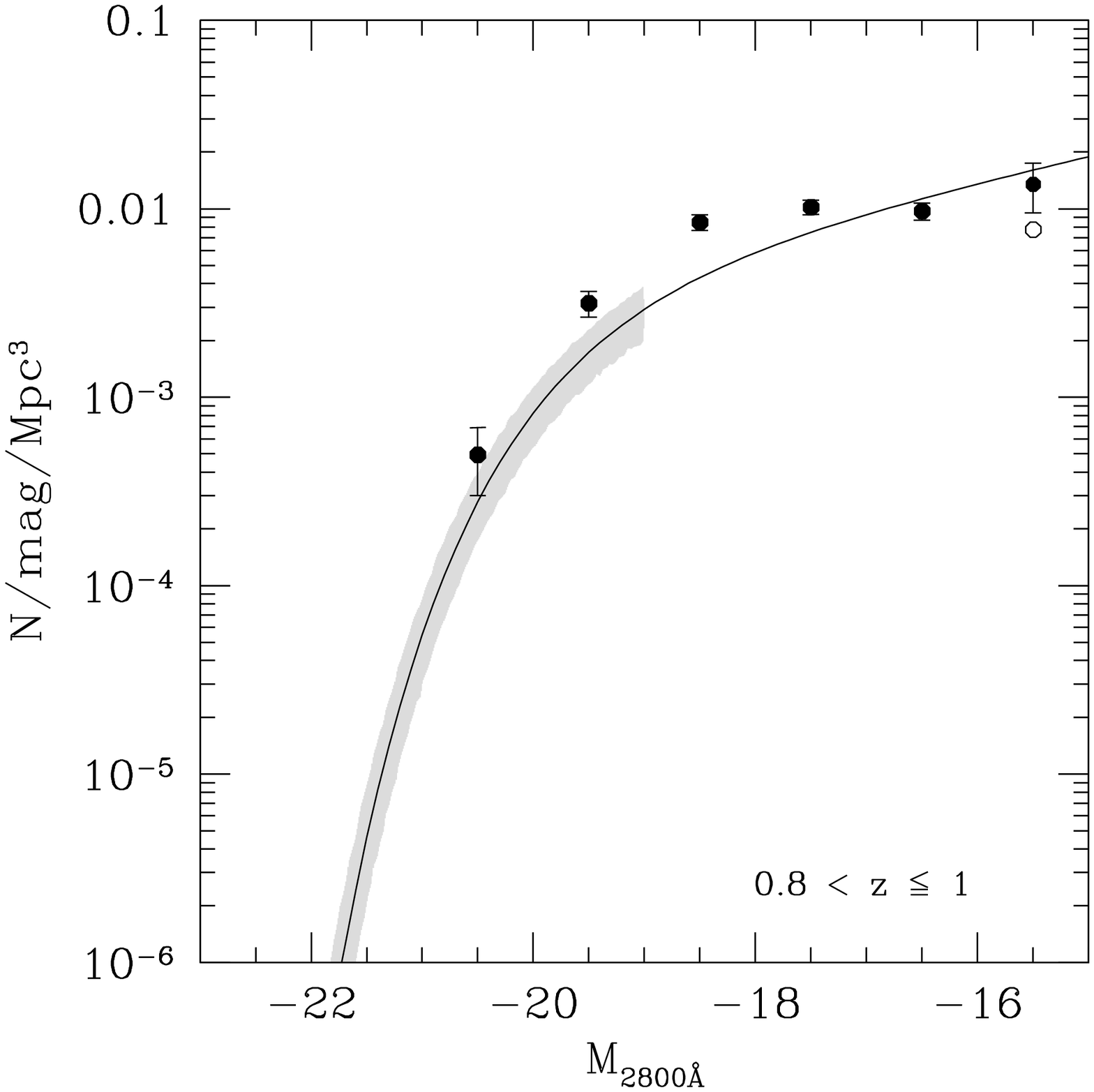}
\hfill
\includegraphics[width=0.45\textwidth]{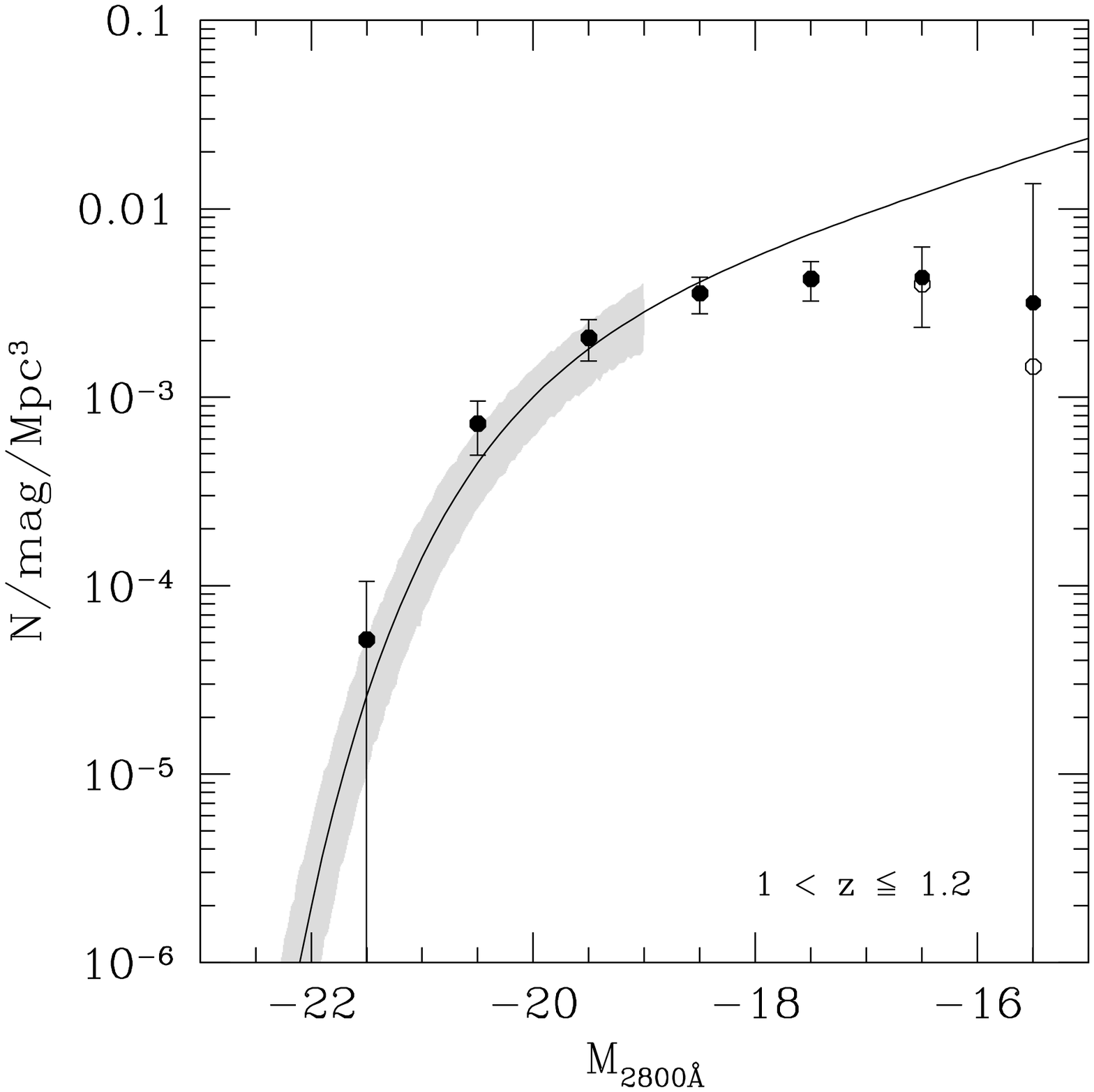}
\caption{\label{fig:lit_uv_combo}
Comparison of the luminosity function at 2800~\AA\ of the FDF 
with the Schechter function derived in
\textit{\citet{combo17:1}}:
\mbox{$0.2 <z\le 0.4$} (upper left panel, light gray), 
\mbox{$0.4 <z\le 0.6$} (upper left panel, dark grey), 
\mbox{$0.6 <z\le 0.8$} (upper right panel), 
\mbox{$0.8 <z\le 1.0$} (lower left panel), 
\mbox{$1.0 <z\le 1.2$} (lower right panel). 
The shaded regions of all plots are based on 
$\Delta$M$^\ast$, $\Delta\phi^\ast$, and $\Delta\alpha$, where the
  cut-off at low luminosity indicates the limiting magnitude
  of the sample.
}
\end{figure*}

\noindent\textit{\citet{combo17:1}:}\\
In Fig.~\ref{fig:lit_uv_combo} we compare the luminosity function at
\mbox{2800~\AA}\ of the FDF with the R-band selected luminosity
function derived in the COMBO-17 survey \citep{combo17:1} for
different redshift bins: 0.2 -- 0.6, 0.6 -- 0.8, 0.8 -- 1.0, 1.0 --
1.2.  Because of the limited sample size of the FDF at low redshift we
could not use the same local redshift binning as \citet{combo17:1}.
We compare therefore in Fig.~\ref{fig:lit_uv_combo} (upper left panel)
the COMBO17 Schechter function at \mbox{$\langle z \rangle\sim 0.3$}
(light gray) and \mbox{$\langle z \rangle\sim 0.5$} (dark gray) with
the FDF luminosity function derived at $0.2 <z\le 0.6$.  There is an
overall good agreement between the FDF data and the COMBO-17 survey at
all redshifts under investigations if we compare only the magnitude
range in common to both surveys (shaded region). Nevertheless the
number density of the FDF seems to be slightly higher which most
probably can be attributed to cosmic variance.  The \citet{combo17:1}
team derived the faint-end slope from relatively shallow data which
have only a limited sensitivity for the faint-end slope. Thus, the
disagreement between the much deeper FDF data and the \citet{combo17:1}
results at $z \sim 0.5$ and for $z > 1$ does not come as
a surprise.\\

\begin{figure*}
\includegraphics[width=0.45\textwidth]{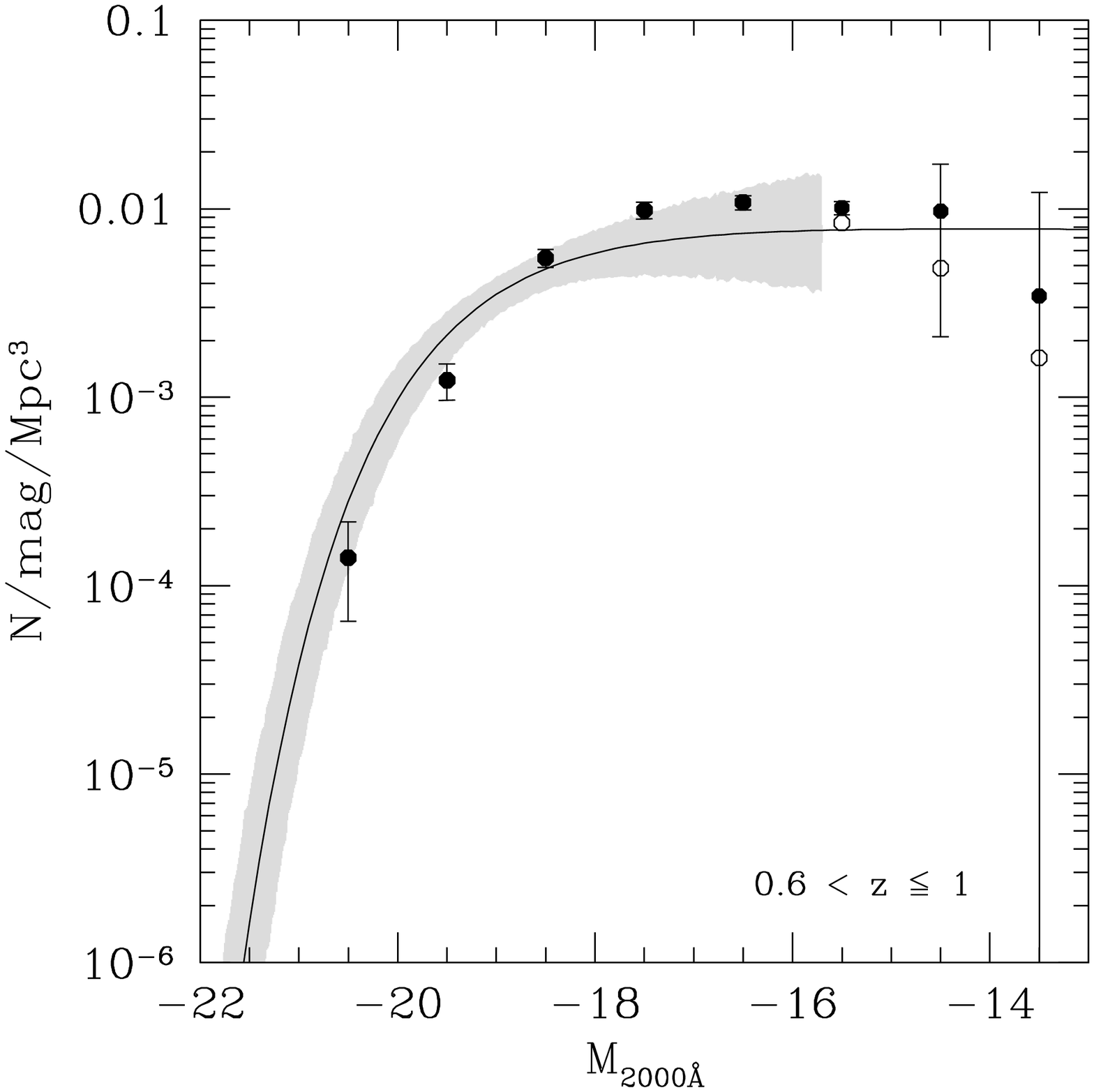}
\hfill
\includegraphics[width=0.45\textwidth]{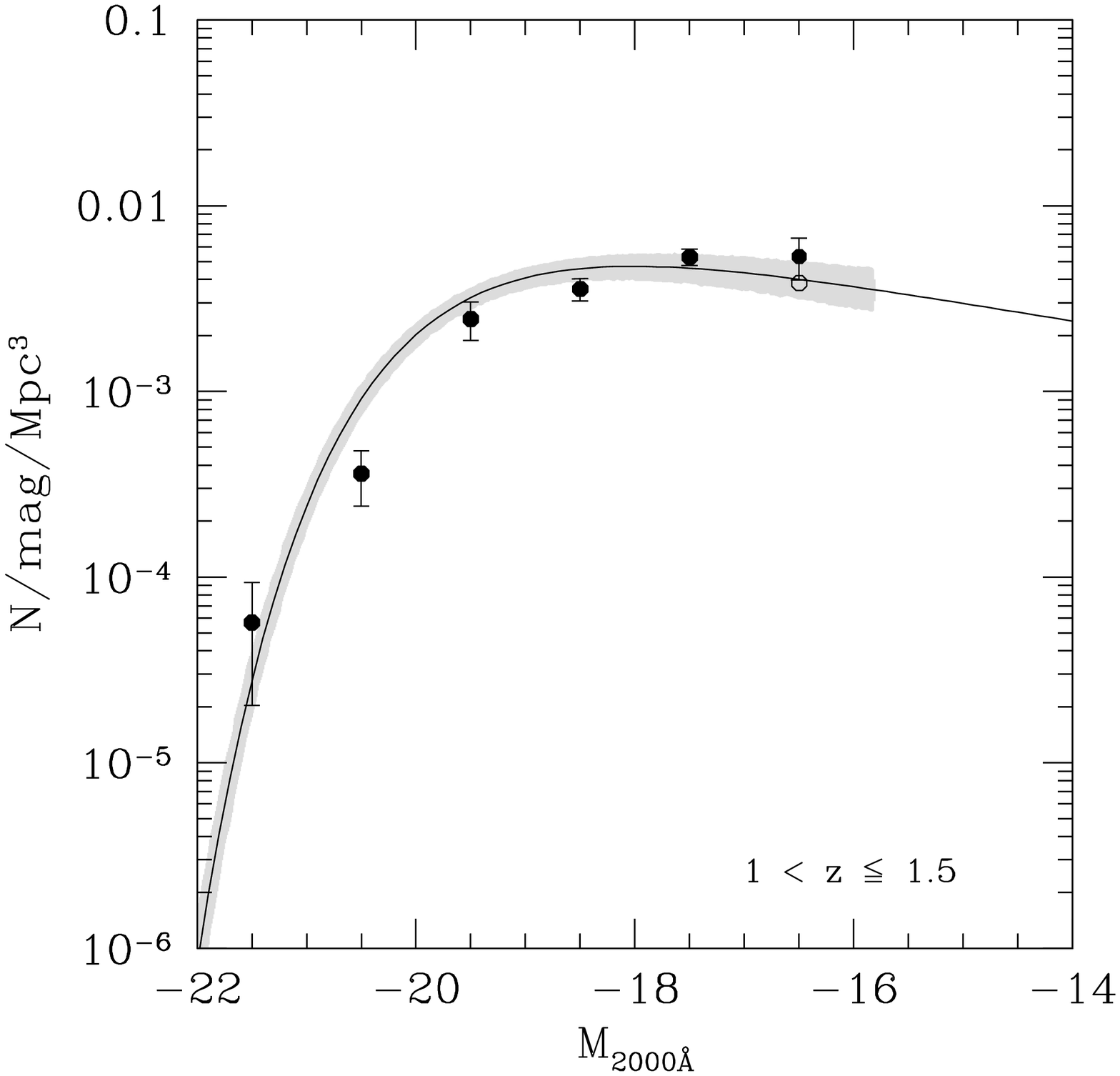}
\includegraphics[width=0.45\textwidth]{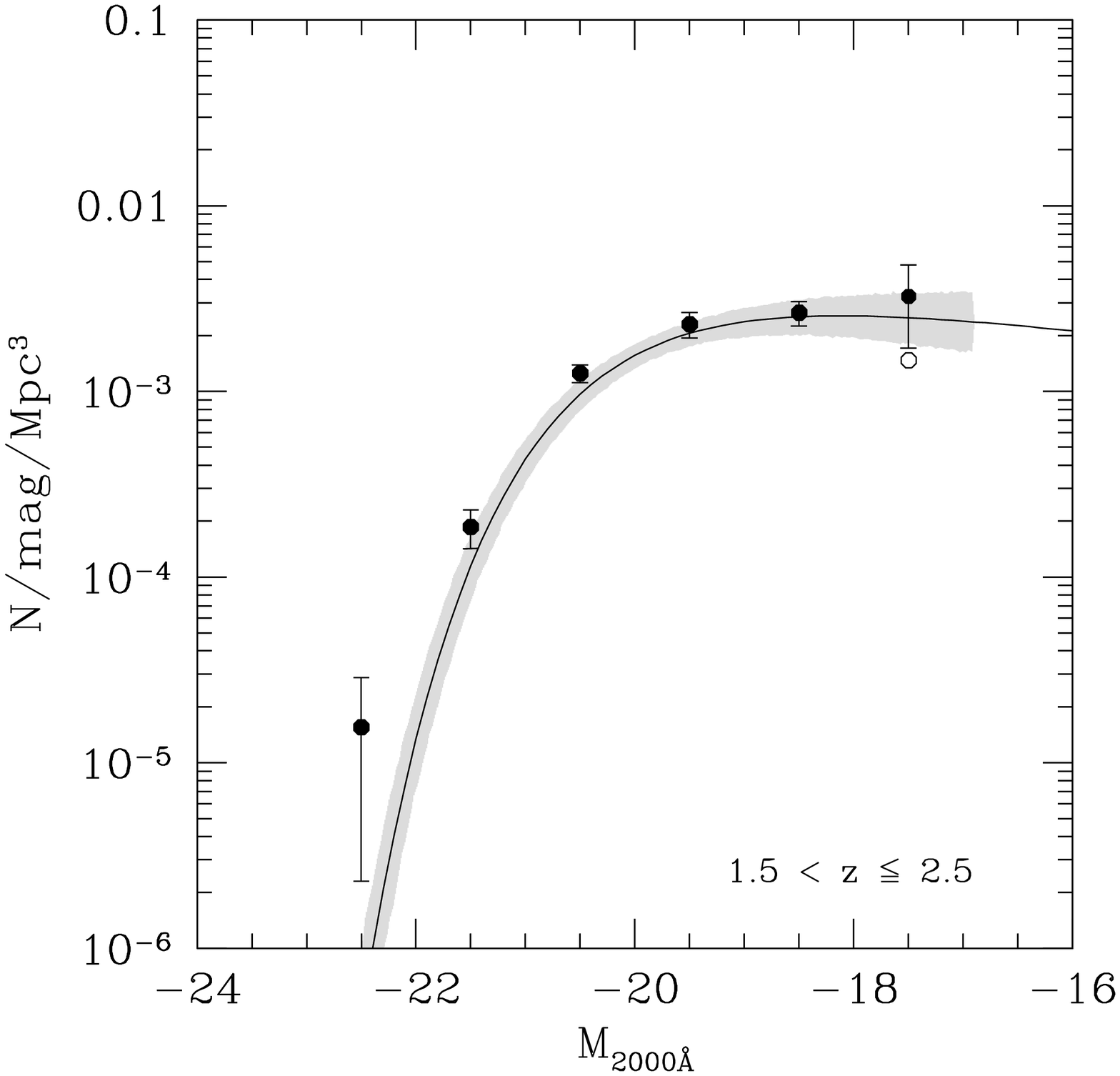}
\hfill
\includegraphics[width=0.45\textwidth]{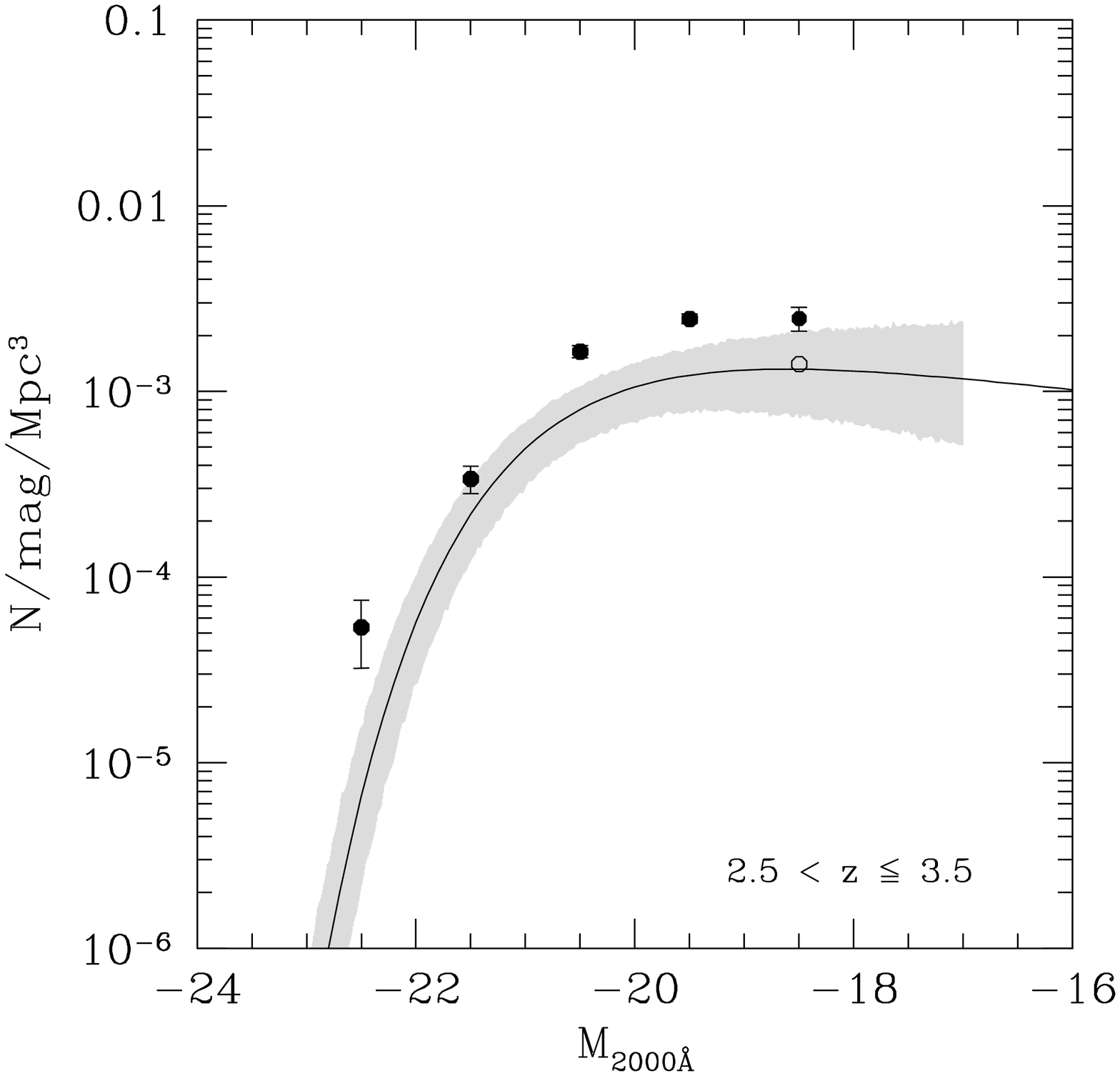}
\caption{\label{fig:lit_uv_kashikawa}
Comparison of the luminosity function at 2000~\AA\ of the FDF with the
Schechter function derived in 
\textit{\citet{kashikawa:1}}: 
\mbox{$0.6 <z\le 1.0$} (upper left panel), 
\mbox{$1.0 <z\le 1.5$} (upper right panel), 
\mbox{$1.5 <z\le 2.5$} (lower left panel), 
\mbox{$2.5 <z\le 3.5$} (lower right panel).
The shaded regions of all plots are based on 
$\Delta$M$^\ast$, $\Delta\phi^\ast$, and $\Delta\alpha$, where the
  cut-off at low luminosity indicates the limiting magnitude
  of the sample.
}
\end{figure*}

\noindent\textit{\citet{kashikawa:1}:}\\
In Fig.~\ref{fig:lit_uv_kashikawa} we compare our luminosity function
with the K-band selected 2000~\AA\ luminosity function of
\citet{kashikawa:1} derived in the Subaru Deep Survey. They used
photometric redshift to 
determine the distance for 439 field galaxies.
There is a good overall agreement of the luminosity functions 
in the redshift bins  
$0.6 <z\le 1.0$, $1.0 <z\le 1.5$, $1.5 <z\le 2.5$. Only in the
highest redshift bin ($2.5 <z\le 3.5$) the number density derived in 
\citet{kashikawa:1} is lower by a factor of about 2 when compared with
the FDF.\\

\begin{figure*}
\centering
\includegraphics[width=0.65\textwidth]{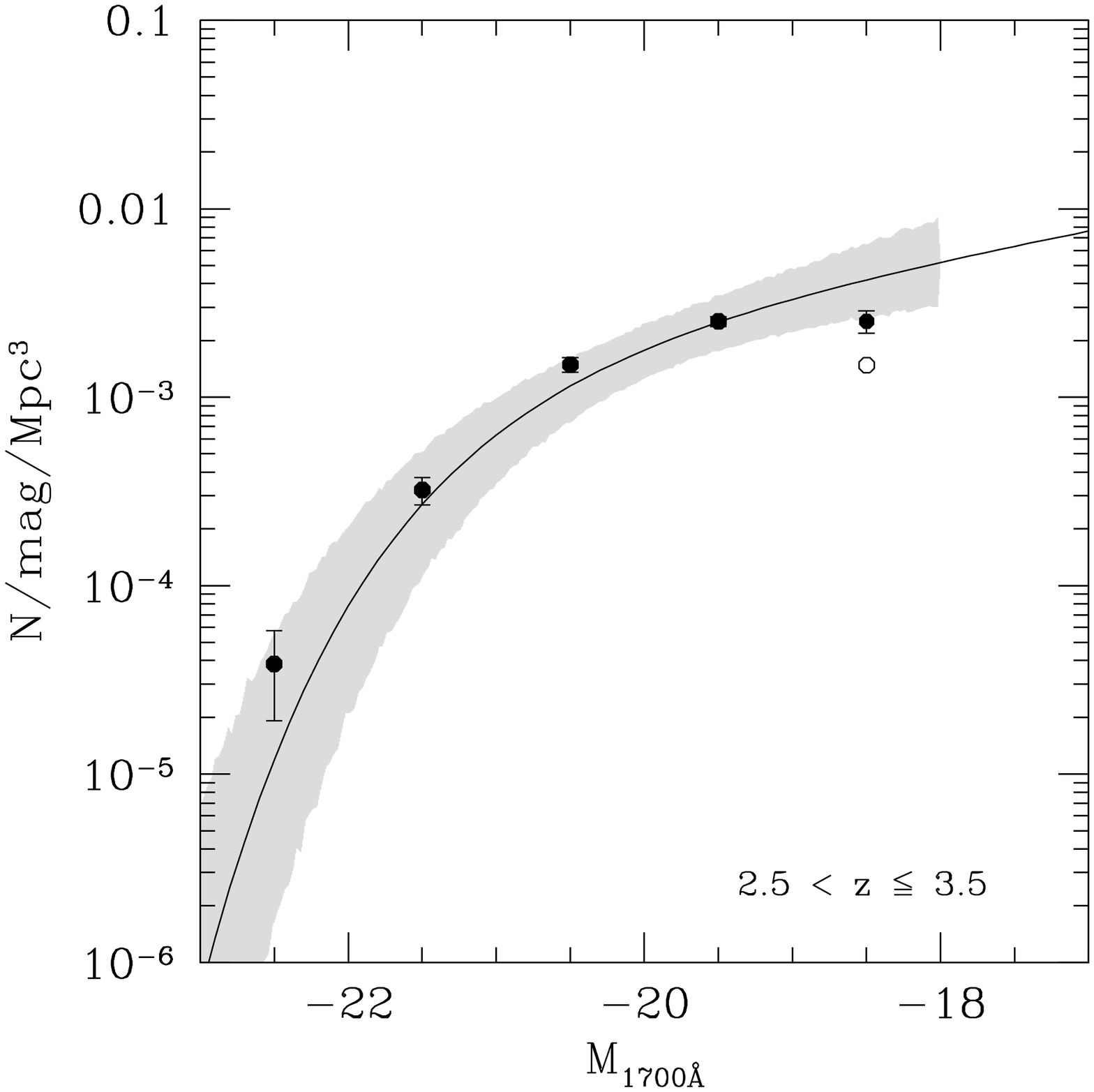}
\caption{\label{fig:lit_uv_poli_2001}
Comparison of the luminosity function at 1700~\AA\ of the FDF 
with the Schechter function derived in
\textit{\cite{poli:1}} (\mbox{$2.50 <z\le 3.50$ }).
The shaded region is based only on
$\Delta$M$^\ast$, and $\Delta\alpha$, where the
  cut-off at low luminosity indicates the limiting magnitude
  of the sample. 
}
\end{figure*}

\noindent\textit{\citet{poli:1}:}\\
\citet{poli:1} combined three pencil beam surveys as the HDFN, the
HDFS and the New Technology Telescope Deep Field
\citep{arnouts_ntt} reducing the influence of cosmic variance and
derived the 1700~\AA\ luminosity function at $2.5 <z\le 3.5$.  In
Fig.~\ref{fig:lit_uv_poli_2001} we compare the result with the
luminosity function in the FDF.  There is very good agreement
although the slope of the Schechter function ($\alpha=-1.37$) is
slightly steeper than we would expect
from the FDF.\\

\begin{figure*}
\includegraphics[width=0.45\textwidth]{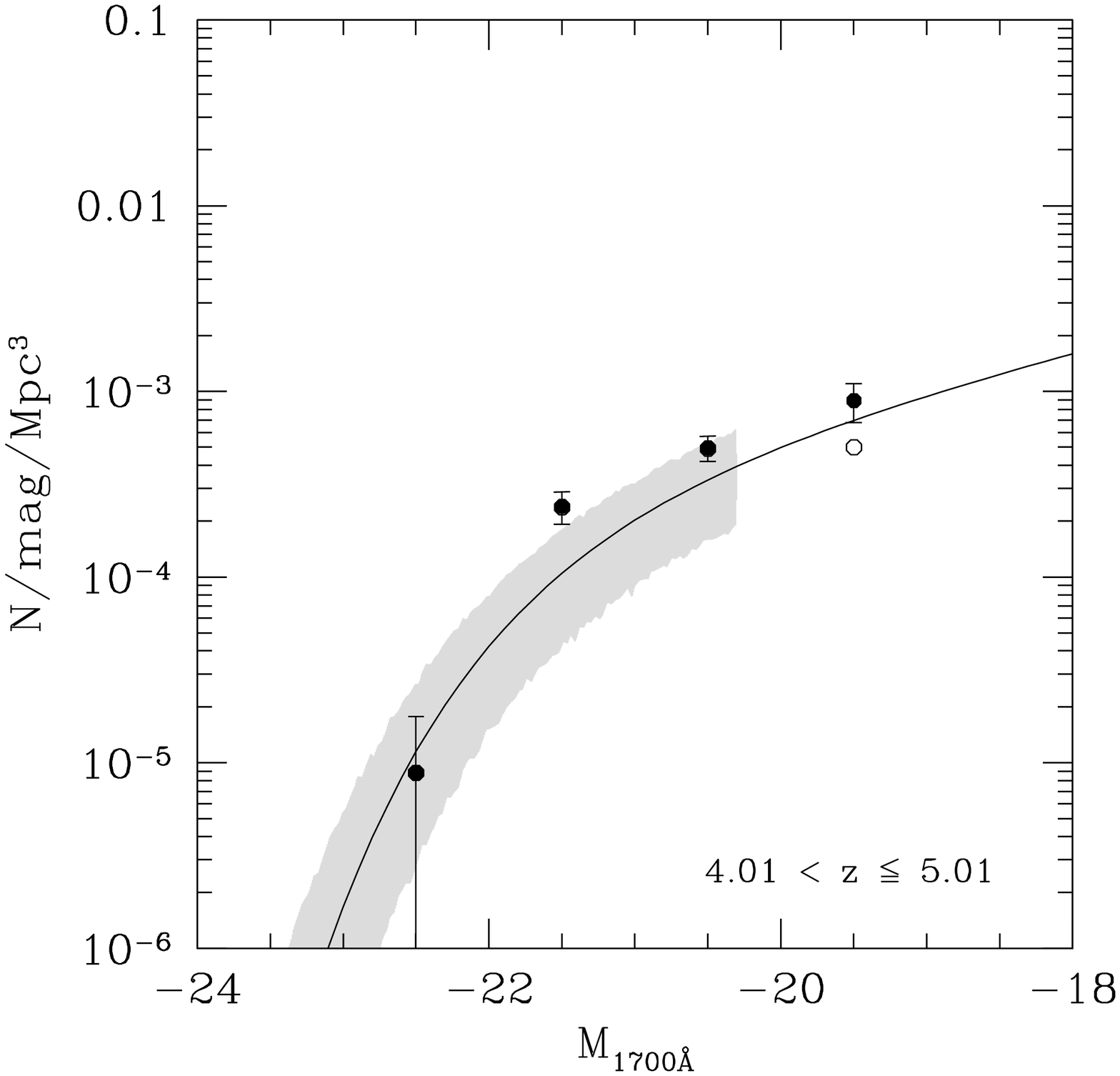}
\hfill
\includegraphics[width=0.45\textwidth]{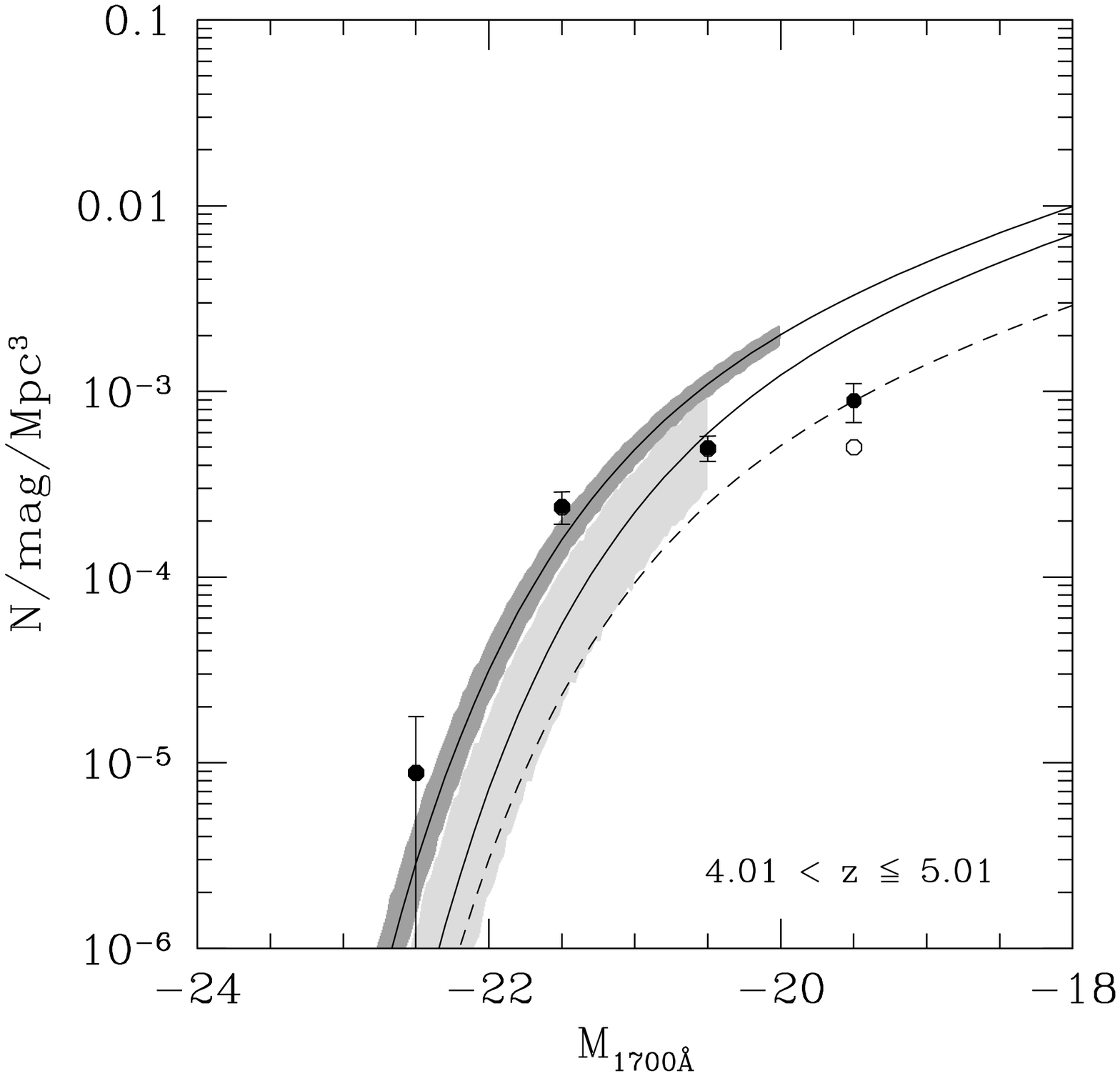}
\caption{\label{fig:lit_uv_iwata_ouchie_high_z}
Left panel:
Comparison of the luminosity function at 1700~\AA\ of the FDF
with the Schechter function derived in
\textit{\cite{iwata:1}} ($z\sim 5$). The shaded region is based only
on $\Delta$M$^\ast$, and $\Delta\phi^\ast$.
Right panel: Comparison of the luminosity function at 1700~\AA\ 
of the FDF with the Schechter functions derived in 
\textit{\cite{ouchi:3}}: 
$z=4.0 \pm 0.5$ (dark shaded), 
$z=4.7 \pm 0.5$ (light shaded), and 
$z=4.9 \pm 0.3$ (not shaded; dashed line).
Both shaded regions are based only on
$\Delta$M$^\ast$, and $\Delta\phi^\ast$, where the
  cut-off at low luminosity indicates the limiting magnitude
  of the sample. 
}
\end{figure*}

\noindent\textit{\citet{iwata:1}:}\\
\citet{iwata:1} analyzed about 300 galaxies in a 575 square-arcmin
field detected in the I and z band at redshift $z\sim 5$, selected by
means of the Lyman-break technique. They derived the luminosity
function at 1700~\AA\ statistically.  We analize Table~3 of
\citet{iwata:1} with the same method as described in
Sect.~\ref{sec:lumfkt:method} to get approximate errors for M$^\ast$
and $\phi^\ast$ for a fixed slope of $\alpha=-1.5$ (as given in the
paper). From these $\Delta$M$^\ast$ and $\Delta\phi^\ast$ we calculate
the shaded region of Fig.~\ref{fig:lit_uv_iwata_ouchie_high_z} (left
panel).  Fig.~\ref{fig:lit_uv_iwata_ouchie_high_z} (left panel)
compares the luminosity function of \citet{iwata:1} with the
luminosity function of the FDF derived at \mbox{$4.01 <z\le 5.01$}.
Although the number density of \citet{iwata:1} at $z\sim 5$ seems to
be slightly lower than the number density derived in the FDF at
\mbox{$\langle z \rangle\sim 4.5$} the overall agreement is rather
good.  On the other hand, part of this decrease in density may also be
due to intrinsic evolution between redshift \mbox{$\langle z
  \rangle\sim 4.5$} and \mbox{$\langle z \rangle\sim 5.0$}. According
to our evolution model as derived in Sect.~\ref{sec:evol_parameter} we
would expect a
decrease of $\phi^\ast$ of about 15 \%.\\

\noindent\textit{\citet{ouchi:3}:}\\
\citet{ouchi:3} investigated photometric properties of about 2600
Lyman-break galaxies at $z= 3.5 - 5.2$. 
Based on this sample they derived the luminosity function at 1700~\AA \
for three redshift bins: 
\mbox{$z=4.0 \pm 0.5$}, 
\mbox{$z=4.7 \pm 0.5$},
\mbox{$z=4.9 \pm 0.3$}.        
In Fig.~\ref{fig:lit_uv_iwata_ouchie_high_z} (right panel) we compare
their Schechter function for a fixed slope of $\alpha=-1.6$ with the
luminosity function of the FDF derived at $4.01 <z\le 5.01$. 
The Schechter function for $z=4.0 \pm 0.5$ is shaded in dark gray,  
the $z=4.7 \pm 0.5$ Schechter Function is shaded light gray
and  the $z=4.9 \pm 0.3$ Schechter Function is represented by the
dashed line (no errors reported).  
It is difficult to compare the results of \citet{ouchi:3} with the
FDF. Our data favor a less steep slope of the luminosity function
than advocated by \citet{ouchi:3}.

\subsection{SDSS bands (u', g', $^{0.1}$u, $^{0.1}$g)}

In this section we want to compare the luminosity function in the FDF
with the one from the SDSS.

\begin{figure*}
\includegraphics[width=0.45\textwidth]{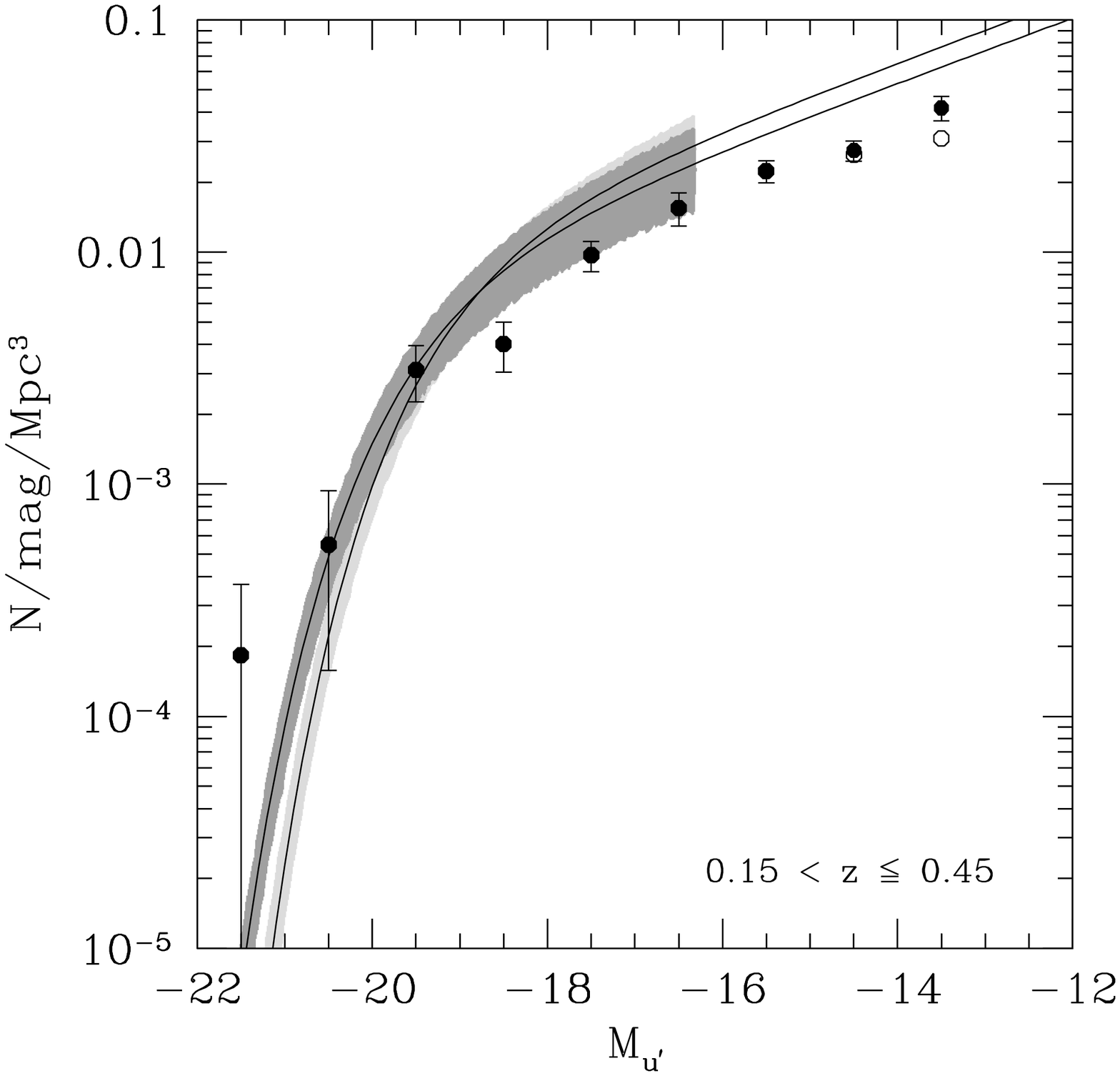}
\hfill
\includegraphics[width=0.45\textwidth]{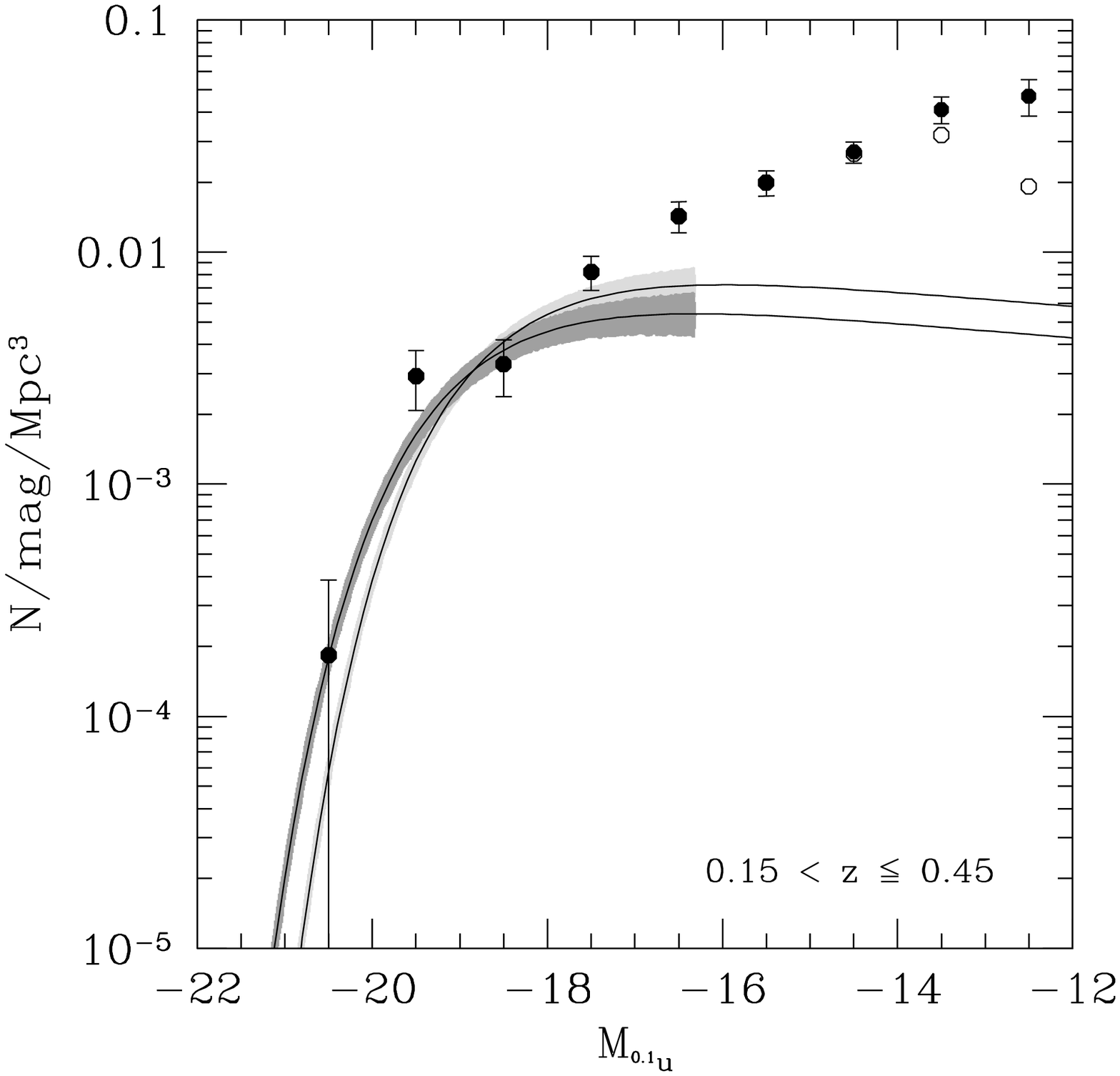}
\caption{\label{fig:lit_sdss_u}
Left panel: 
Comparison of the u'-band luminosity function  of the FDF 
with the Schechter function derived in
\textit{\cite{blanton:1}} at $z\sim 0.1$ (light shaded).
The dark shaded region shows the Schechter function of
\citet{blanton:1} evolved according to our
evolutionary model described in Sect.~\ref{sec:evol_parameter} to 
redshift $z\sim 0.3$.
The shaded regions are based on 
$\Delta$M$^\ast$, $\Delta\phi^\ast$, and $\Delta\alpha$.
Right panel: 
Comparison of the $^{0.1}$u-band luminosity function  of the FDF 
with the Schechter function derived in
\textit{\cite{blanton:2}} at $z\sim 0.1$ (light shaded).
The dark shaded region shows the Schechter function of
\citet{blanton:2} evolved according to our
evolutionary model described in Sect.~\ref{sec:evol_parameter} to 
redshift $z\sim 0.3$.
The shaded regions are based on 
$\Delta$M$^\ast$, $\Delta\phi^\ast$, and $\Delta\alpha$, where the
  cut-off at low luminosity indicates the limiting magnitude
  of the sample.
}
\end{figure*}

\begin{figure*}
\includegraphics[width=0.45\textwidth]{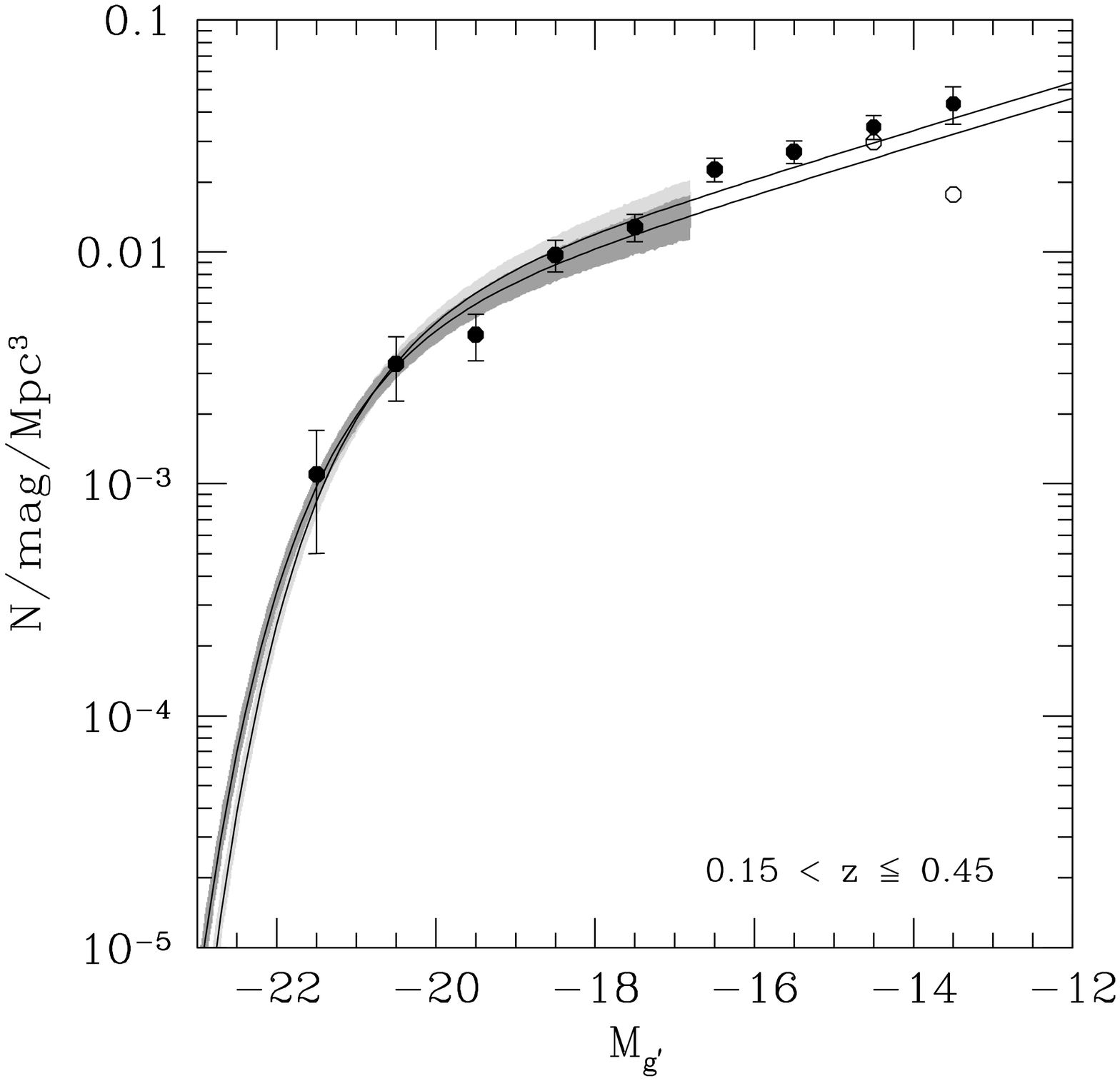}
\hfill
\includegraphics[width=0.45\textwidth]{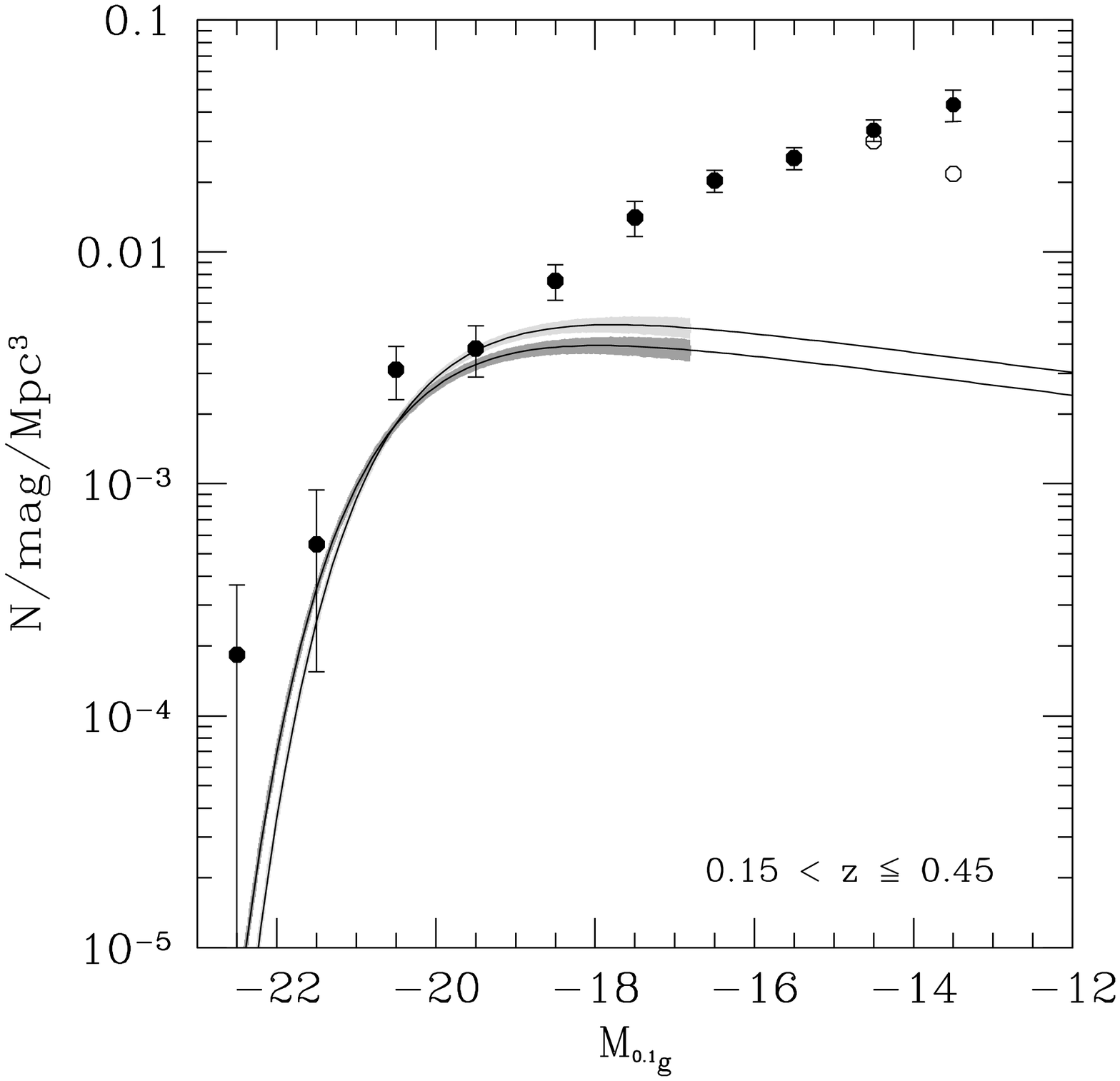}
\caption{\label{fig:lit_sdss_g}
  Left panel: Comparison of the g'-band luminosity function of the FDF
  with the Schechter function derived in \textit{\cite{blanton:1}} at
  $z\sim 0.1$ (light shaded).  The dark shaded region shows the
  Schechter function of \citet{blanton:1} evolved according to our
  evolutionary model described in Sect.~\ref{sec:evol_parameter} to
  redshift $z\sim 0.3$.  The shaded regions are based on
  $\Delta$M$^\ast$, $\Delta\phi^\ast$, and $\Delta\alpha$.  Right
  panel: Comparison of the $^{0.1}$g-band luminosity function of the
  FDF with the Schechter function derived in \textit{\cite{blanton:2}}
  at $z\sim 0.1$ (light shaded).  The dark shaded region shows the
  Schechter function of \citet{blanton:2} evolved according to our
  evolutionary model described in Sect.~\ref{sec:evol_parameter} to
  redshift $z\sim 0.3$.  The shaded regions are based on
  $\Delta$M$^\ast$, $\Delta\phi^\ast$, and $\Delta\alpha$, where the
  cut-off at low luminosity indicates the limiting magnitude
  of the sample.  }
\end{figure*}

In Fig.~ \ref{fig:lit_sdss_u} (left panel) and Fig.~\ref{fig:lit_sdss_g}
(left panel) we show the luminosity function derived in
\citet{blanton:1} for $z\sim 0.1$ in the u' and g' band,
respectively, as light shaded region. To make a more appropriate
comparison between our `local' results derived at $0.15 <z\le 0.45$,
we evolve the Schechter function of \citet{blanton:1} to 
\mbox{$\langle z \rangle\sim 0.3$} according to our
evolutionary model described in Sect.~\ref{sec:evol_parameter}. 
We use for the u'-band the parameter $a=-1.80$ and $b=-1.70$ whereas
for the g'-band we use $a=-1.08$ and $b=-1.29$. The evolved Schechter
function is shown as dark shaded region in in 
Fig.~\ref{fig:lit_sdss_u} (left panel) and Fig.~\ref{fig:lit_sdss_g} 
(left panel) for the u' and g' band, respectively.  
Despite the small volume of the FDF in the local redshift bin, the
agreement is very good in both bands and especially in the g'-band.  
We therefore conclude that there is no hint of a possible systematic
offset between the two datasets.\\ 

In Fig.~\ref{fig:lit_sdss_u} (right panel) and
Fig.~\ref{fig:lit_sdss_g} (right panel) we also show the luminosity
function derived in \citet{blanton:2} for the blue-shifted filter
$^{0.1}$u and $^{0.1}$g.  Again, the light shaded region represents
the \mbox{$\langle z \rangle\sim 0.1$} luminosity function whereas the
dark shaded region shows the luminosity function evolved to
\mbox{$\langle z \rangle\sim 0.3$}. We use the same evolution
parameter as derived for u' and g'. The approach used by
\citet{blanton:2} differs from those used in all other studies,
including ours and the previous SDSS \citep{blanton:1} results. It
is therefore beyond the scope of the paper to explain the
discrepancies.

\subsection{B-band}

\begin{figure*}
\includegraphics[width=0.45\textwidth]{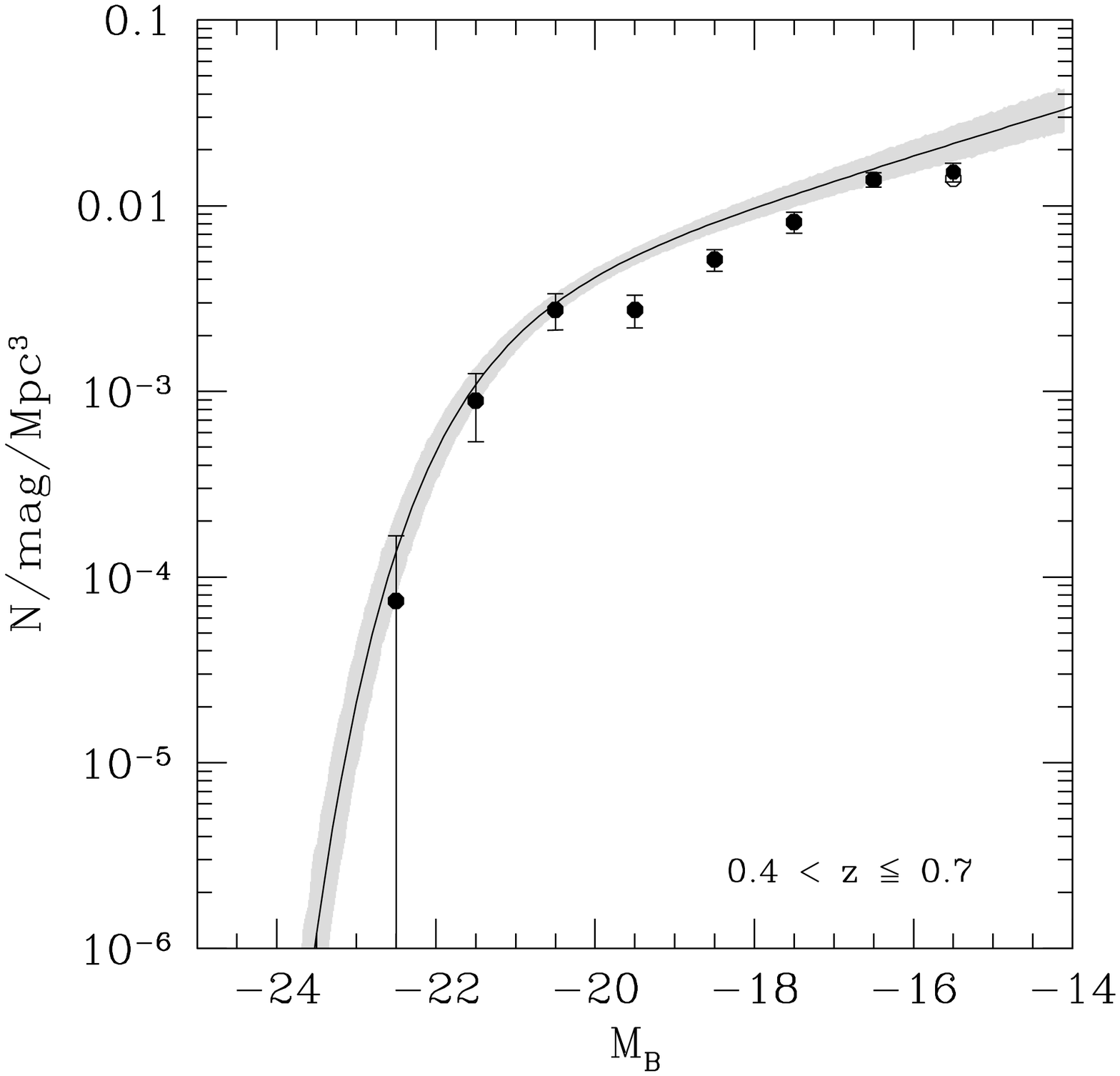}
\hfill
\includegraphics[width=0.45\textwidth]{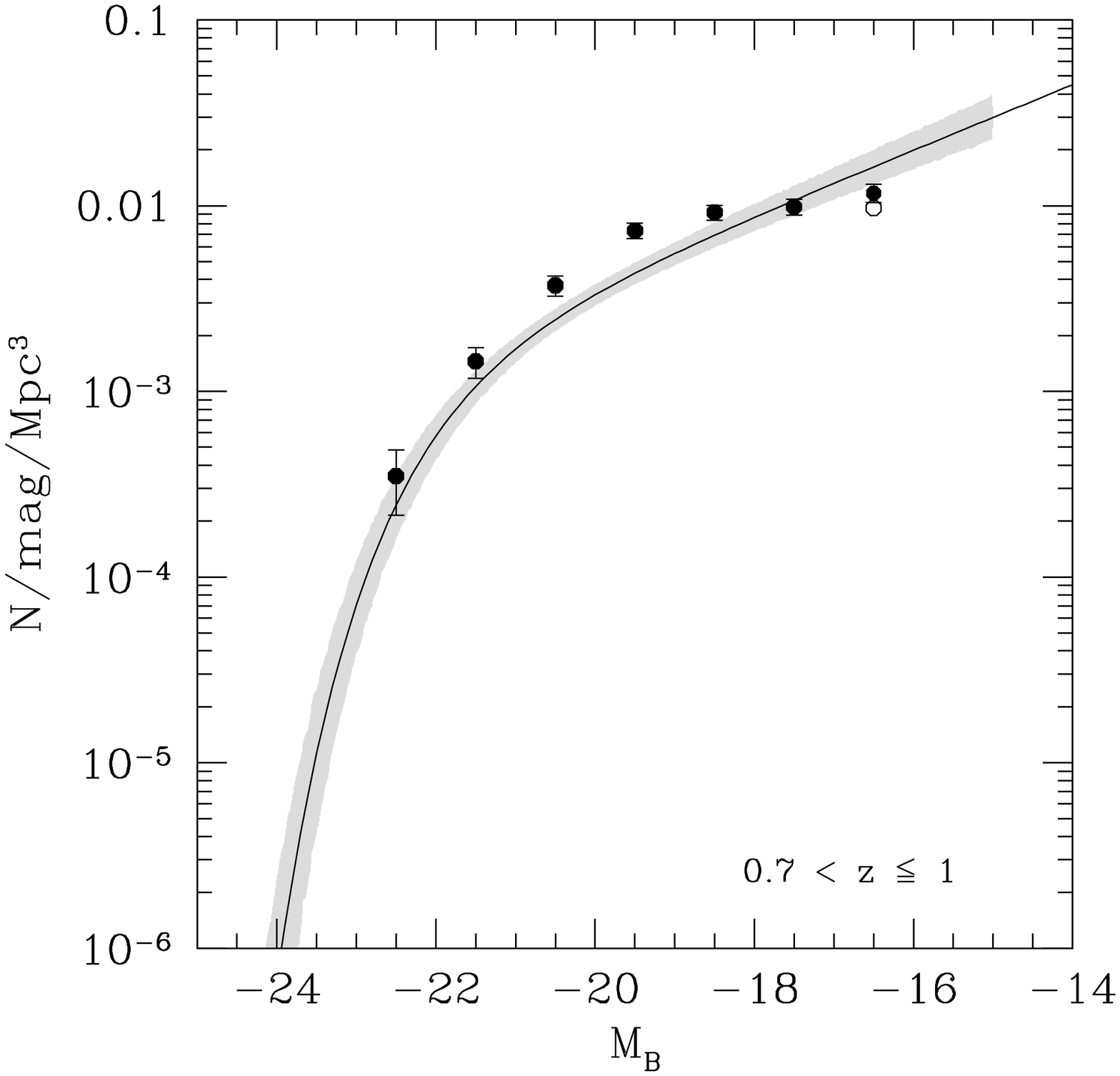}
\includegraphics[width=0.45\textwidth]{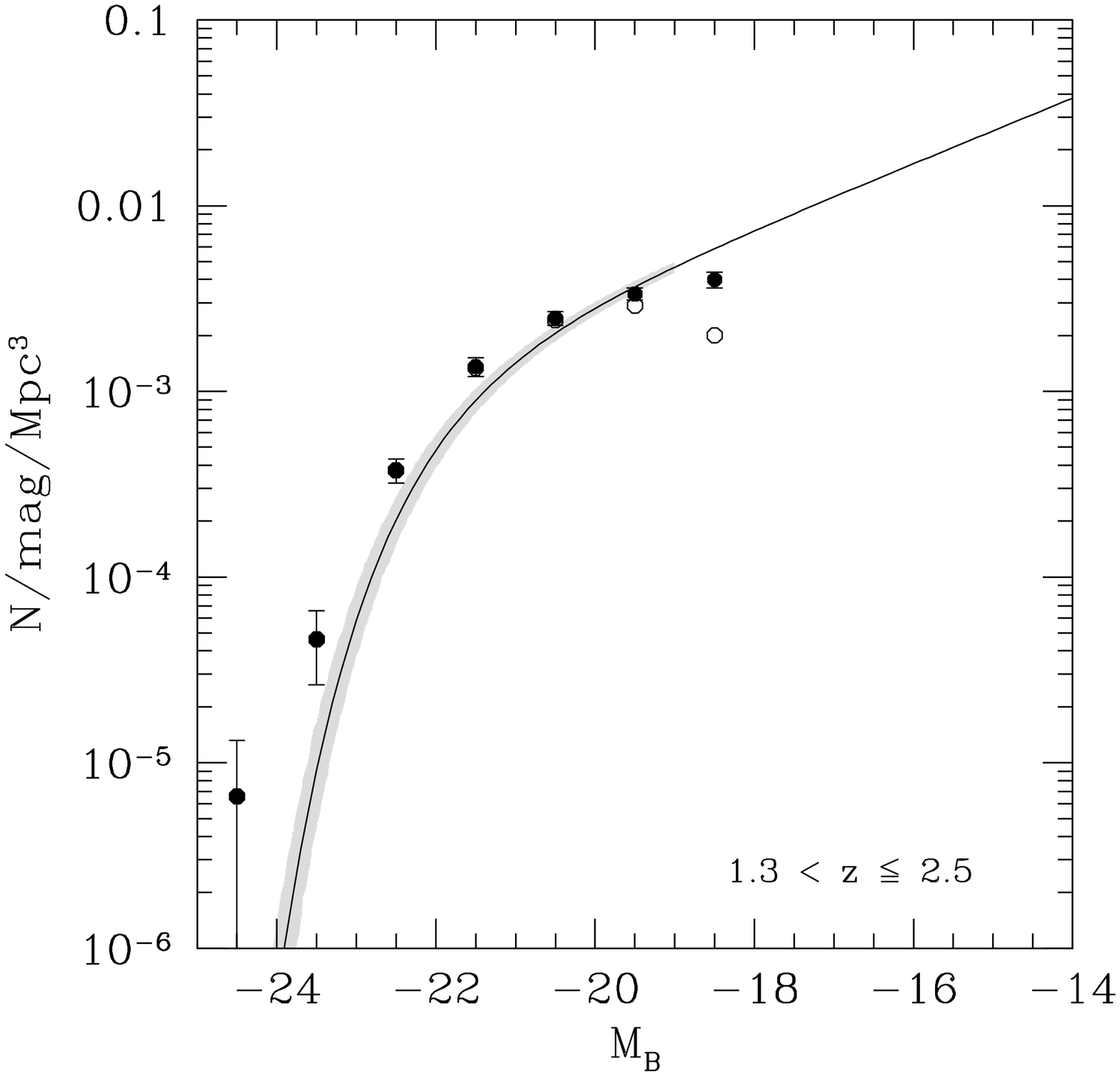}
\hfill
\includegraphics[width=0.45\textwidth]{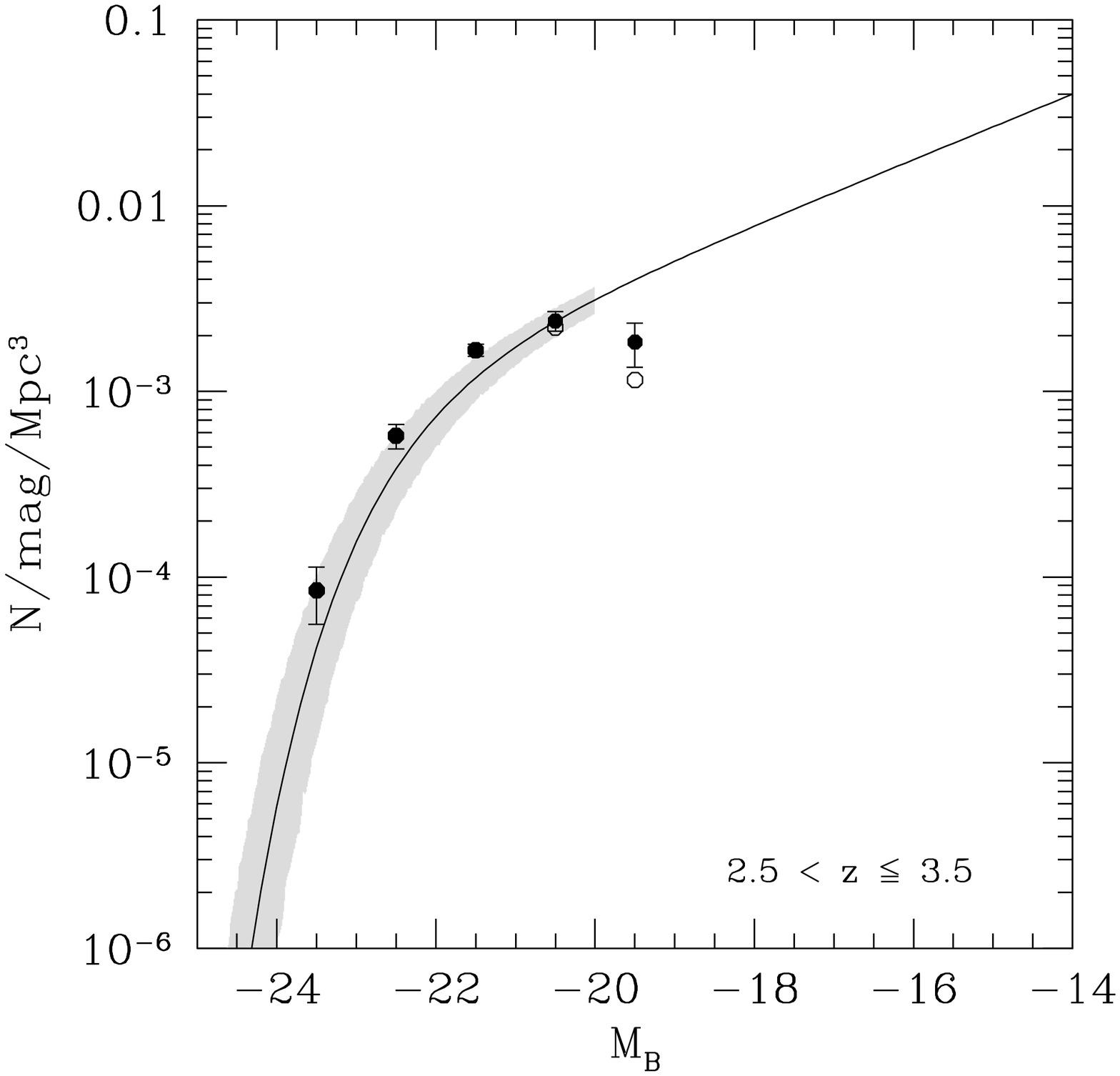}
\caption{\label{fig:lit_B_poli_2003}
Comparison of the B-band luminosity function of the FDF with the
Schechter function derived in
\textit{\citet{poli:3}}: 
\mbox{$0.4 <z\le 0.7$} (upper left panel), 
\mbox{$0.7 <z\le 1.0$} (upper right panel), 
\mbox{$1.3 <z\le 2.5$} (lower left panel), and  
\mbox{$2.5 <z\le 3.5$} (lower right panel). 
The shaded region is based only
on $\Delta$M$^\ast$, and $\Delta\alpha$  for 
\mbox{$0.4 <z\le 0.7$}, and 
\mbox{$0.7 <z\le 1.0$}, whereas for the 
\mbox{$1.3 <z\le 2.5$}, and
\mbox{$1.3 <z\le 2.5$}
the shaded region is based only on $\Delta$M$^\ast$, where the
  cut-off at low luminosity indicates the limiting magnitude
  of the sample. 
}
\end{figure*}

\noindent\textit{\citet{poli:3}:}\\
\citet{poli:3} analyzed 1541 I-selected and 138 K-selected galaxies to
construct the B-band luminosity function up to redshift  
\mbox{$\langle z \rangle\sim 3$}. 
A comparison between the luminosity function of \citet{poli:3} and the
FDF is shown in Fig.~\ref{fig:lit_B_poli_2003} for the redshift bins 
\mbox{$0.4 <z\le 0.7$} (upper left panel), 
\mbox{$0.7 <z\le 1.0$} (upper right panel), 
\mbox{$1.3 <z\le 2.5$} (lower left panel) and 
\mbox{$2.5 <z\le 3.5$} (lower right panel). 

In neither of the redshift bins  an error for $\phi^\ast$ is reported
in the paper and therfore could not be included in the simulation of
the shaded region. For the two lower redshift bins 
(\mbox{$0.4 <z\le 0.7$} and \mbox{$0.7 <z\le 1.0$}) the shaded region
is based on $\Delta$M$^\ast$ and $\Delta\phi^\ast$ whereas in the high
redshift bins (\mbox{$1.3 <z\le 2.5$} and \mbox{$2.5 <z\le 3.5$}) the
shown error of the Schechter function (shaded region) is based only on
$\Delta$M$^\ast$. If this is taken into account, the results of
\citet{poli:3} are in good agreement with the FDF, but again, the
slope of the Schechter function is too steep when compared with the
FDF luminosity function. On the other hand the brightening of M$^\ast$ 
with redshift is present in both samples.

\section{Confidence levels for the slope}
\label{sec:3_para_fit_slope}

\begin{figure*}[hb]
\includegraphics[width=0.5\textwidth]{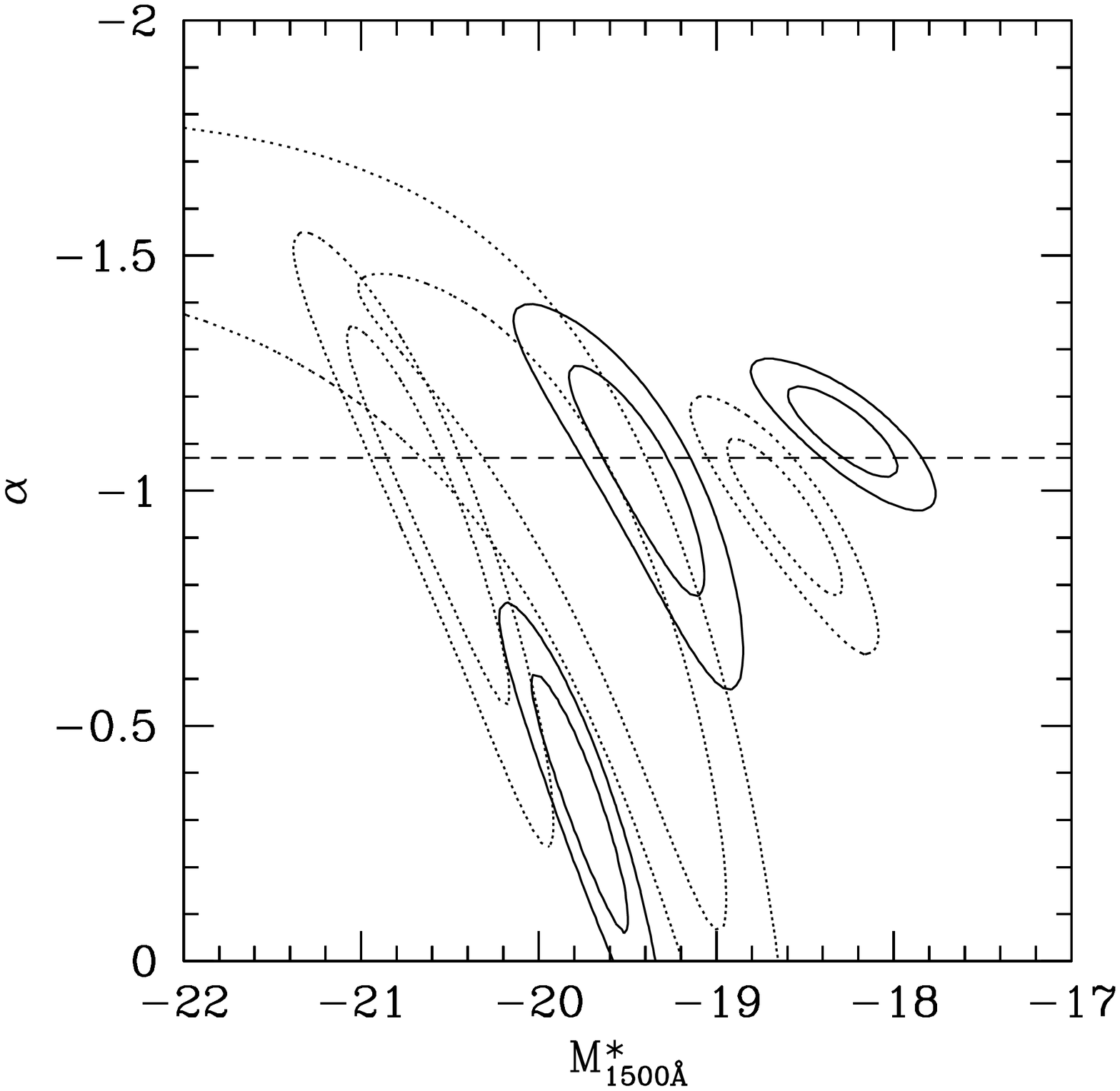}
\includegraphics[width=0.5\textwidth]{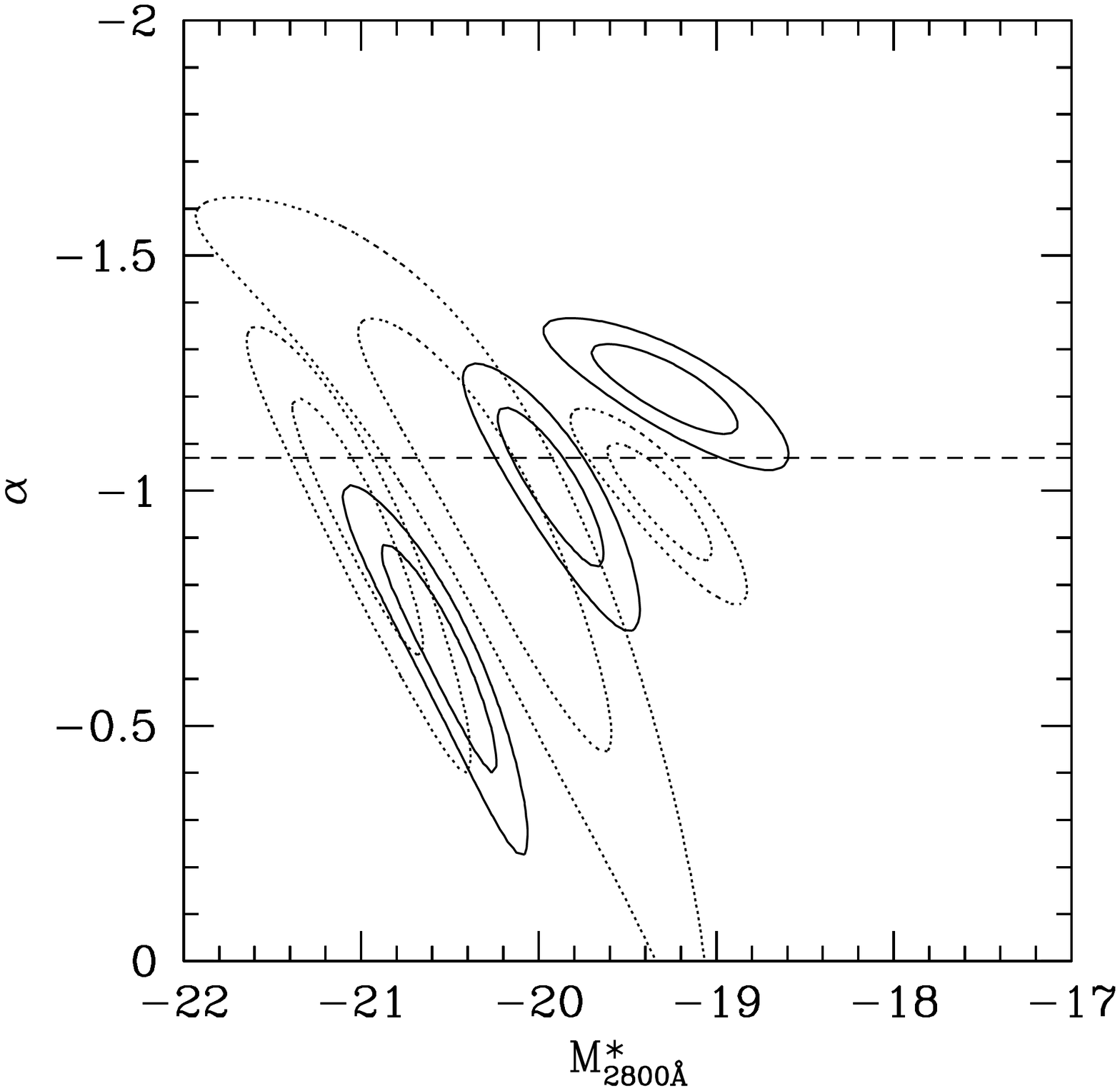}
\includegraphics[width=0.5\textwidth]{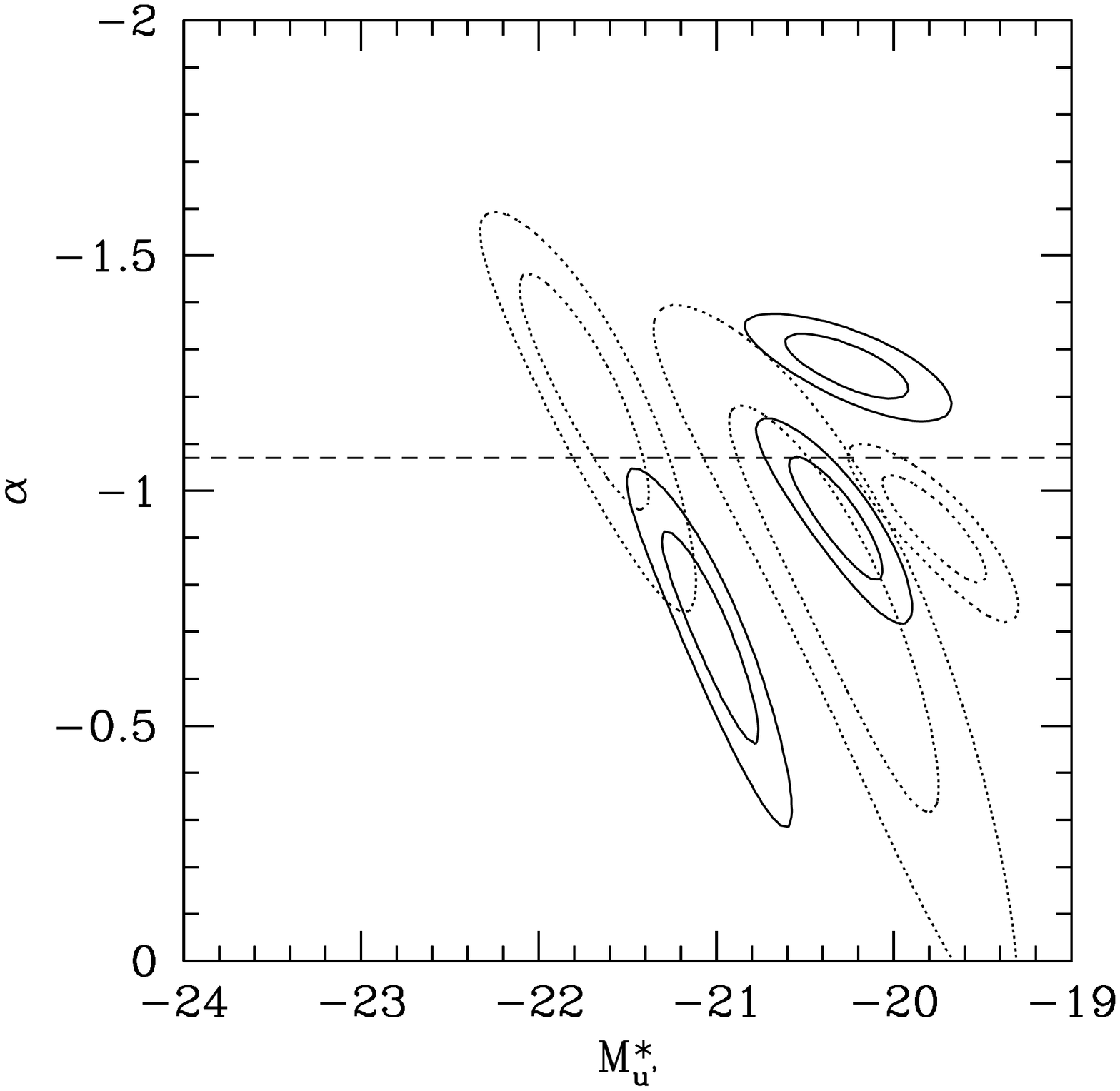}
\includegraphics[width=0.5\textwidth]{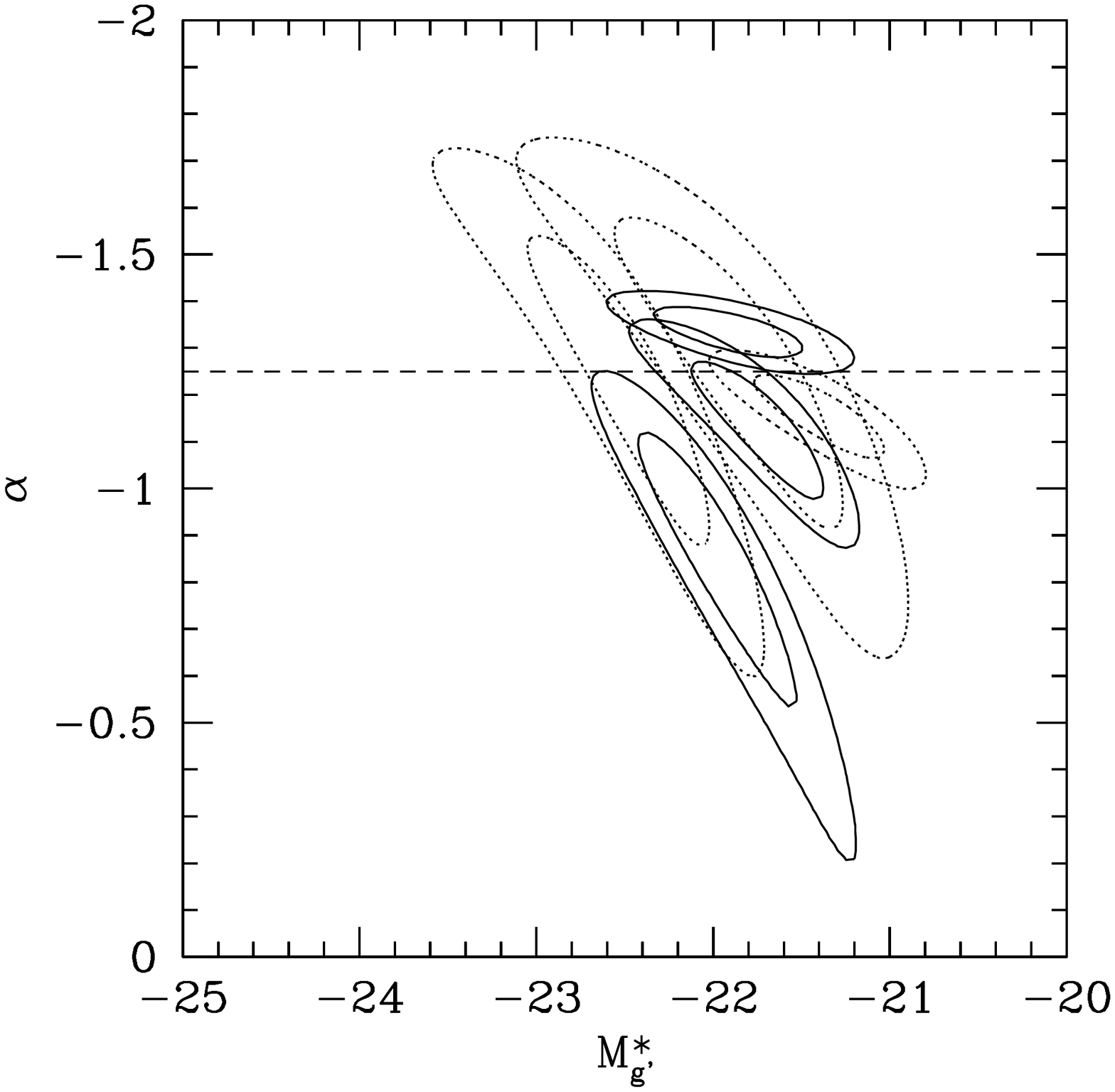}
\caption{\label{fig:contour_3param_alpha}
  \mbox{$1\sigma$} and \mbox{$2\sigma$} confidence levels in Schechter
  parameter space.  A Schechter function with three free parameters
  M$^\ast$, $\phi^\ast$, and $\alpha$ has been fitted to the
  luminosity function at 1500~\AA \ (\textit{upper left panel}),
  2800~\AA \ (\textit{upper right panel}), u' (\textit{lower left
    panel}) and g'-band (\textit{lower right panel}) and projected to
  the M$^\ast$ -- $\alpha$ plane. The various contours in each panel
  correspond to the different redshift bins, ranging from $\langle z
  \rangle = 0.6$ (low luminosity) to $\langle z \rangle = 3.5$ (high
  luminosity).  We alternate continuous and dotted lines for clarity.
  The dashed line marks the fixed slope ($\alpha(z) =
  \mathrm{const.}$) used to derive the luminosity functions in the
  different wavebands (see Table~\ref{tab:slope_fixed} lower part).  }
\end{figure*}

\end{document}